\documentclass[prd,nofootinbib,onecolumn,preprint,11pt]{revtex4-2}

\usepackage{graphicx}
\usepackage{bm}
\usepackage{float}
\usepackage{subfigure}
\usepackage{amsmath}
\usepackage{amsfonts}
\usepackage{color}
\usepackage{slashed}
\usepackage{fancyhdr}
\usepackage[colorlinks=true,linkcolor=blue,citecolor=blue,filecolor=blue,urlcolor=blue,unicode]{hyperref}
\usepackage{relsize} 
\usepackage{mathtools}

\newcommand{\includecode}[1]{
\begin{figure}[H]
	\includegraphics[width=\textwidth]{Figures/#1.png}
\end{figure}
}

\newcommand{\sizeincludecode}[2]{
	\begin{figure}[H]
		\includegraphics[width=#1 \textwidth]{Figures/#2.png}
	\end{figure}
}

\newcommand{\twopicincludecode}[2]{
	\begin{figure}[H]
		\includegraphics[width=\textwidth]{Figures/#1.png}
		\includegraphics[width=\textwidth]{Figures/#2.png}
	\end{figure}
}

\newcommand{\createlabel}[1]{
	\phantomsection
	\label{sent:#1}
}

\newcommand{\uselabel}[2]{
	\hyperref[sent:#1]{#2}
}

\newcommand{\boldtitle}[1]{{\bf #1} \qquad }

\newcommand{\dd}{\text{d}}

\newcommand{\TUD}[3]{#1^{#2}_{\phantom{#2}#3}}

\newcommand{\MMA}{{\sf Mathematica}}
\newcommand{\TC}{{\sf TensoriaCalc}}

\begin{document}

\title{TensoriaCalc: A User-Friendly Tensor Calculus Package for the Wolfram Language}

\author{Wei-Hao Chen$^{1}$, Yi-Zen Chu$^{1,2}$, Vaidehi Varma$^{3,4}$}
\affiliation{
	$\,^1$Department of Physics, National Central University, Chungli 32001, Taiwan \\
	$\,^2$Center for High Energy and High Field Physics (CHiP), National Central University, Chungli 32001, Taiwan \\
	$\,^3$Graduate Institute of Astronomy, National Central University, Chungli 32001, Taiwan \\
	$\,^4$Department of Physics, Maharaja Krishnakumarsinhji Bhavnagar University, Bhavnagar, Gujarat, India
}

\begin{abstract}
We describe \TC, a tensor calculus package written to be smoothly consistent with the Wolfram Language, so as to ensure ease of usage. It allows multiple metrics to be defined in a given session; and, once a metric is computed, associated standard differential geometry operations to be carried out -- covariant derivatives, Hodge duals, index raising and lowering, derivation of geodesic equations, etc. Other non-metric operations, such as the Lie and exterior derivatives, coordinate transformation on tensors, etc. are also part of its built-in functionality.
\end{abstract}

\maketitle

\newpage

\tableofcontents

\section{Introduction}
\label{Section_Introduction}

\MMA\ \cite{MMA} and its associated Wolfram language is widely used by physicists, likely at least in part due to their ease of use. To date, however, the Wolfram language does not offer a comprehensive range of built-in functionality needed for differential geometric calculations commonly encountered in General Relativity, classical and quantum field theory, and cosmology; though both general and somewhat more specialized packages do exist that attempt to fill this gap -- see, for instance, \cite{Shtabovenko:2023idz} \cite{Shoshany:2021iuc} \cite{xAct} \cite{Gravitas}. The goal of this article is to introduce \TC\ \cite{TC}, a package that performs concrete (as opposed to abstract) differential geometry calculations, whereby various operations such as covariant derivatives can be carried out once an explicit metric has been entered. In particular, to maximize as much as possible user-friendliness, we have undertaken serious effort to ensure that using \TC\ is very similar to using \MMA\ itself.

One of the key aspects of the Wolfram language is its symbolic computation ability. There is no programming overhead spent in terms of variable(s) declaration. Together with \MMA's multitude of in-built functions, this means mathematical calculations can often be accomplished with a single or few lines of code. For instance, one may perform an indefinite integral, say of $\sin x$ with respect to $x$ -- 
\includecode{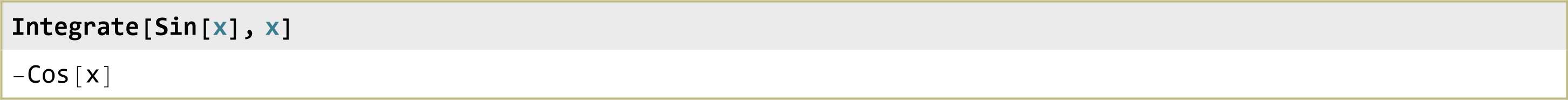}
-- without having to declare the integration variable $x$. (Strictly speaking, $x$ and the abstract variables used in this paper should be identified in the Wolfram Language as Symbol(s), in that MatchQ[$x$,$\_$Symbol$]=$ True or, equivalently, Head$[x]=$ Symbol.) Let us now illustrate how this straightforward style of symbolic computation works similarly for \TC, by defining the metric tensor -- the primary object in most differential geometry calculations.

\boldtitle{Defining a metric} \qquad In \TC, an arbitrary metric tensor may be defined by specifying its indices, explicit components, and the associated coordinate system. For example, the round $2-$sphere geometry with the usual spherical coordinates $\{0\le\theta\le\pi,\ 0\le\phi<2\pi\}$ is defined either by the sum of squares of infinitesimal displacements -- namely, in terms of $(\dd\theta)^2 = \text{DifferentialD}[\theta]^2$ and $(\dd\phi)^2 = \text{DifferentialD}[\phi]^2$ -- or by its matrix representation. (The $\dd$ can be entered via the keystrokes Esc dd Esc.)
\includecode{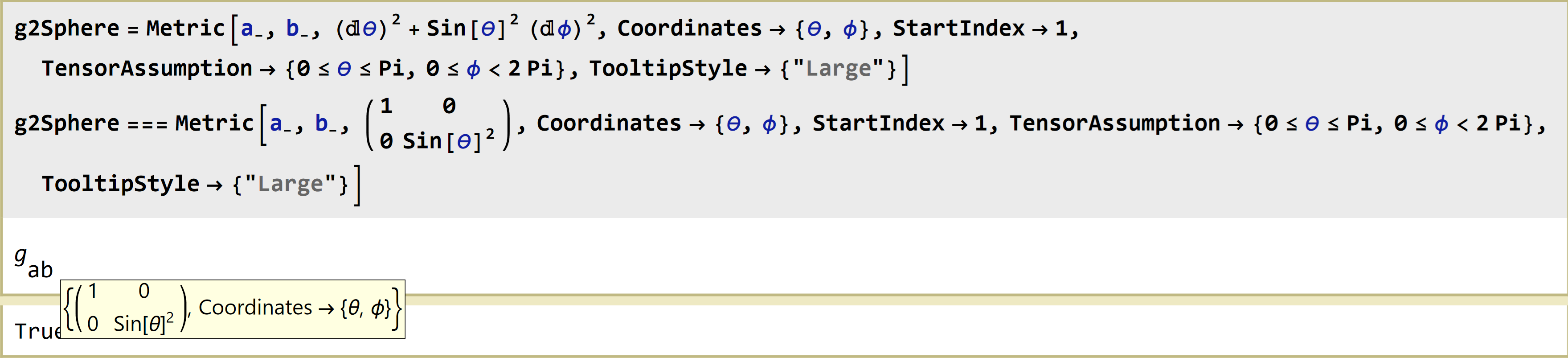}
The box displaying the metric components is triggered by hovering the cursor over the $g_{ab}$.

The same geometry may also be defined through the inverse metric, as a sum over the squares of the partial derivatives -- namely, superposition of $(\nabla \theta)^2 = \mathrm{Del}[\theta]^2 \; ( \equiv \partial_\theta \otimes \partial_\theta )
\ \text{and} \ 
(\nabla \phi)^2 = \mathrm{Del}[\phi]^2 \; ( \equiv \partial_\phi \otimes \partial_\phi )$ -- or by its matrix representation.
\includecode{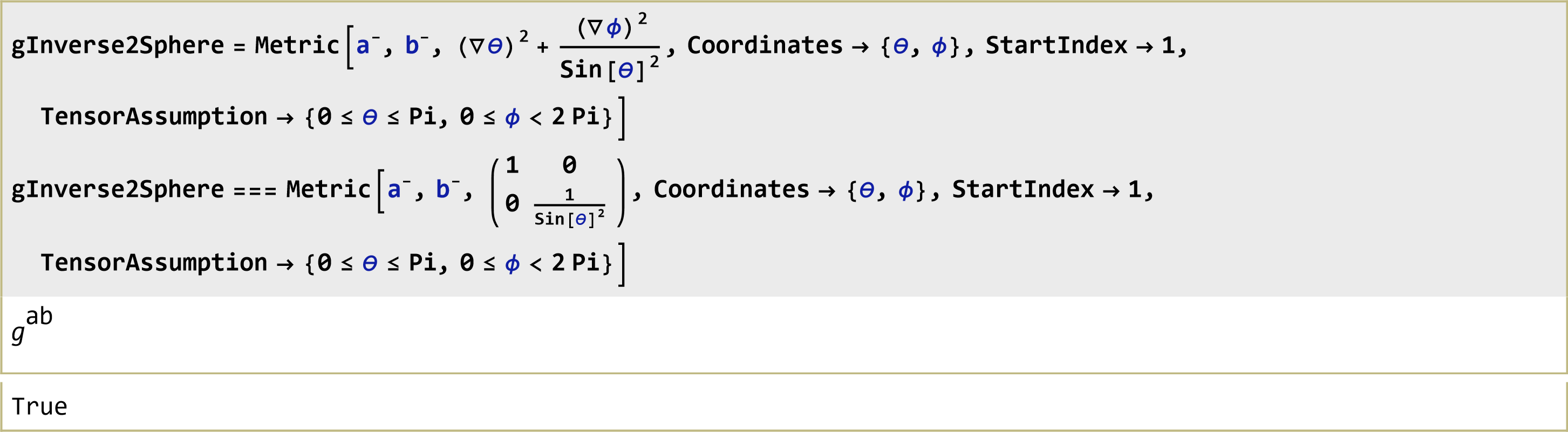}
Above, the $a$ and $b$ indices cannot belong to the given coordinate system but are otherwise arbitrary Symbol's. The SubMinus notation, $\text{SubMinus}[a] = a_-$ and $\text{SubMinus}[b] = b_-$, indicates they are lower indices; while the SuperMinus notation, $\text{SuperMinus}[a] = a^-$ and $\text{SuperMinus}[b] = b^-$, indicates they are upper indices. 

The StartIndex and TensorAssumption Rule's are optional arguments; with default values given by StartIndex $\to 0$ and TensorAssumption $\to \{\}$. Our notation here follows that of the Wolfram Language, where optional arguments are Rule's, with a predefined default value. In our online manual, whenever a default value for a function argument is given, that means the argument in question is an optional one. Note, however, even though Coordinates looks like an optional argument, it is actually mandatory. 

StartIndex refers to the integer associated with the first coordinate. For example, in the flat spacetime $\dd s^2 = \dd t^2 - \dd\vec{x}^2$, it is common practice to let the time $t$ be not only the leading coordinate, but also labeled as $t = x^0$; while $x^i = (x^1,\dots,x^D)$ are the spatial ones (where $D$ is the dimension of space). For the $2-$sphere metric at hand, StartIndex $\to 1$ means $(x^1,x^2) = (\theta,\phi)$. 

TensorAssumption allows the user to record properties of the parameters occurring within the metric components, which may then be invoked when, applying Simplify or FullSimplify on expressions. 

Other optional arguments associated with Metric are described in the online manual; which shall be released in the near future.

Once a metric (or the inverse metric) has been defined, its inverse (or the metric) can be readily obtained by feeding it back into the Metric function.
\includecode{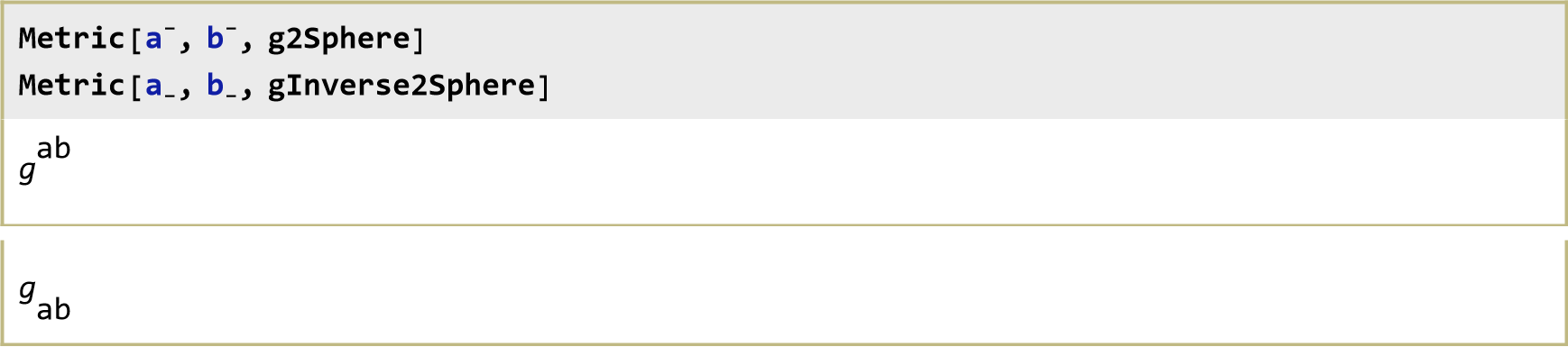}
Upon the definition of a geometry through Metric, \TC\ then outputs a Tensor object with arguments consisting of Rule's containing the metric components, indices, coordinates, Christoffel symbols, Ricci scalar, Ricci tensor and Riemann tensor, etc. -- see below. \TC\ has built in functions to extract these geometric information. This allows for different geometries to be defined and computed within the same \MMA\ session.
\newpage
Specifically, the g2Sphere reads as follows.
\includecode{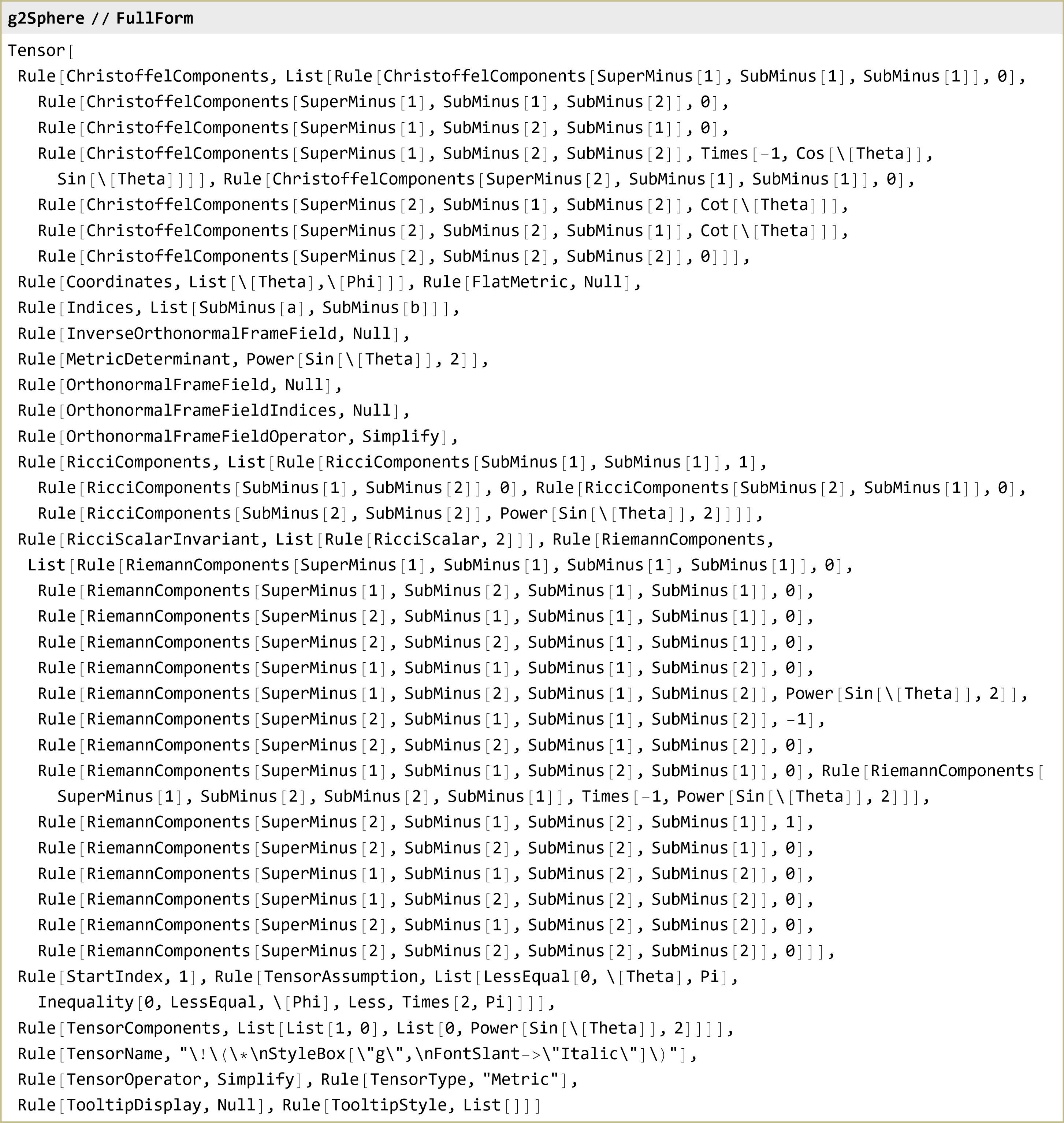} 
\newpage
This is analogous to, say, the output of any graphics related function in the Wolfram Language; where a Graphics object is generated, with its arguments containing the (text-based) relevant information that would allow \MMA\ to display it correctly. For example,
\includecode{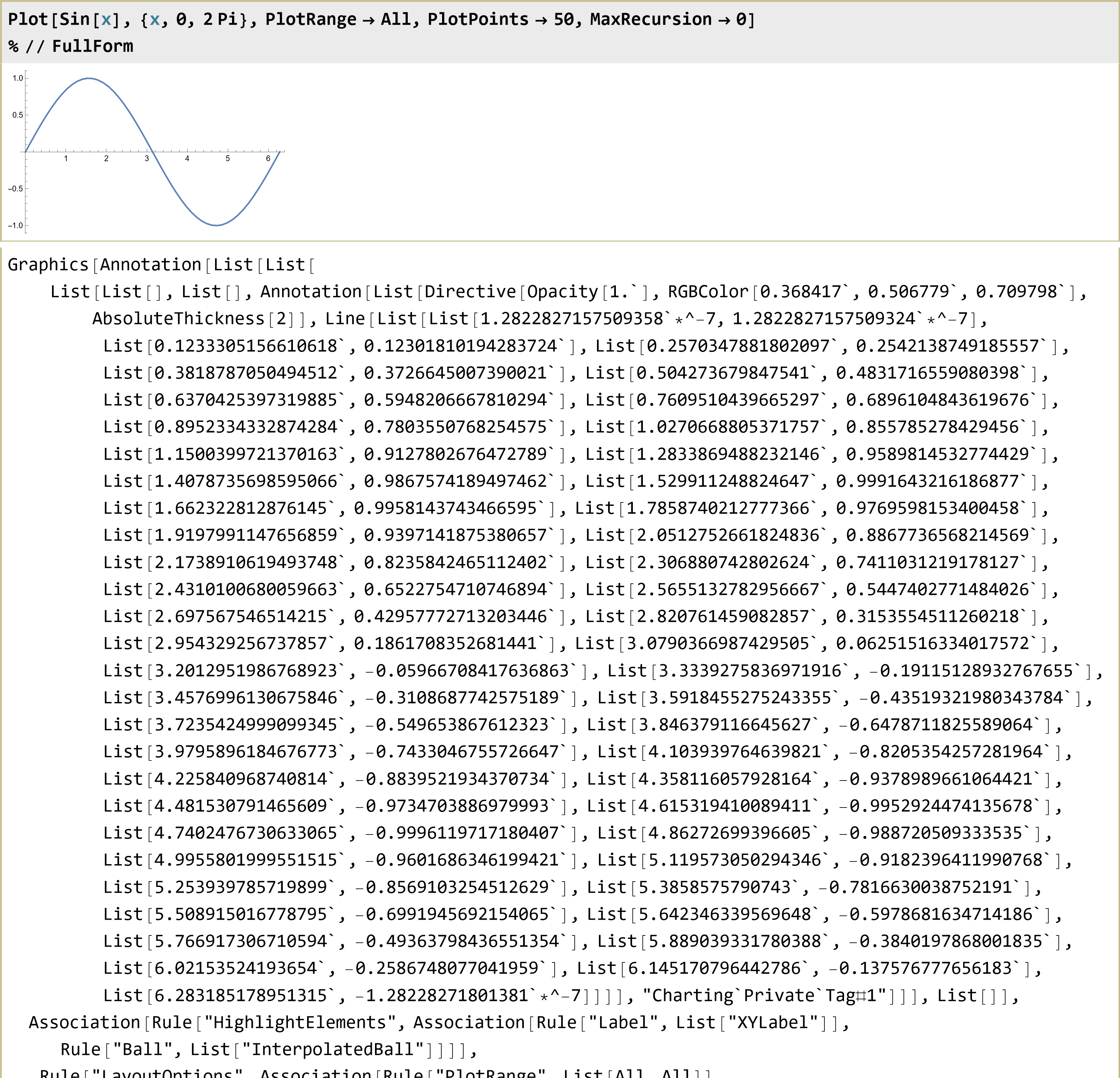} 
\newpage
To extract the information stored within the metric Tensor object g2Sphere, we simply apply to it the associated functions: Indices, Coordinates, Christoffel, Riemann, Ricci, RicciScalar and the (metric) Determinant.
\includecode{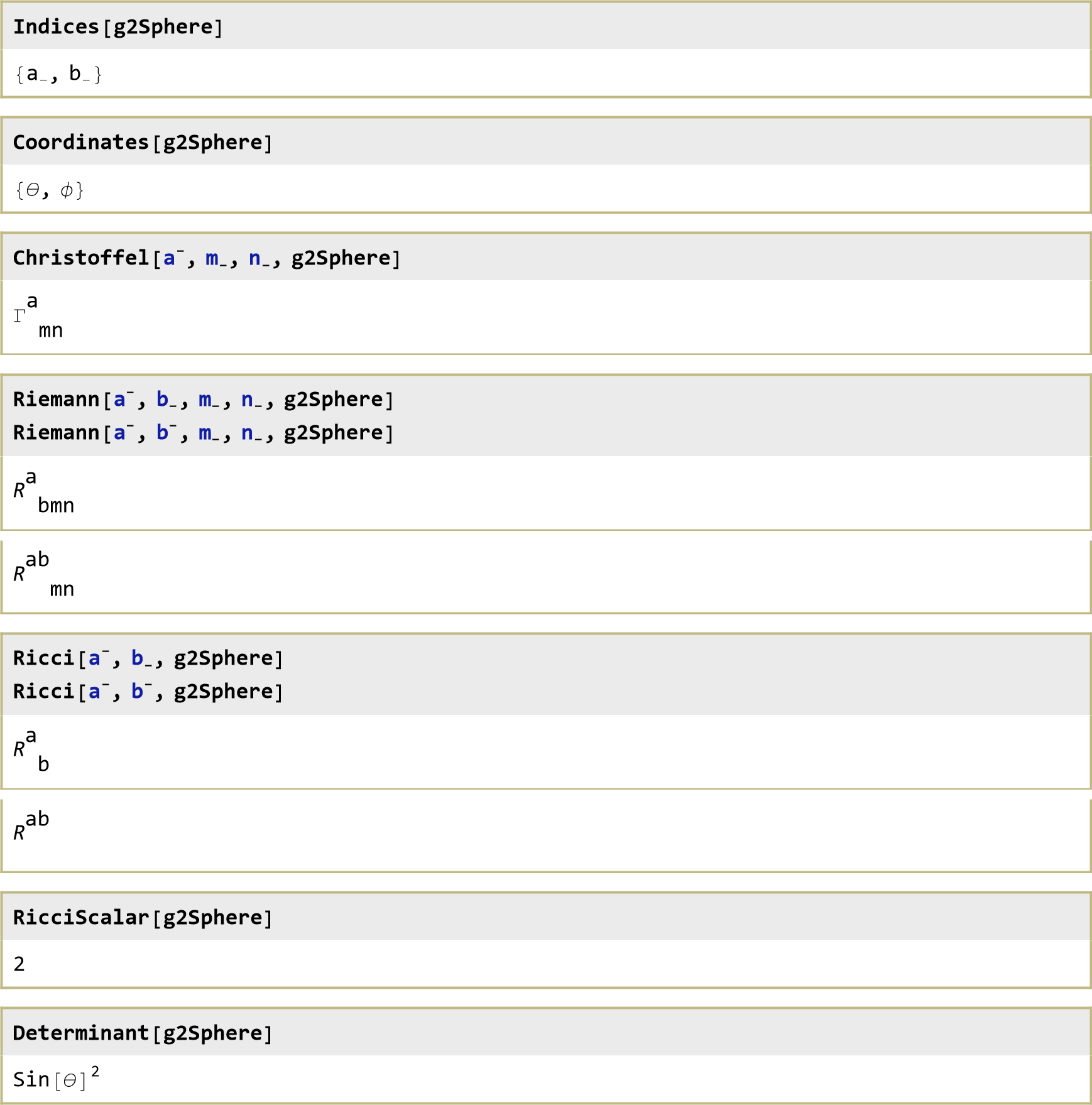}  
\newpage
TensorComponents applied to any Tensor object returns its components, usually a List. For instance, the components of the metric or inverse metric expressed as a square matrix (List of List's) can themselves be extracted with TensorComponents.
\includecode{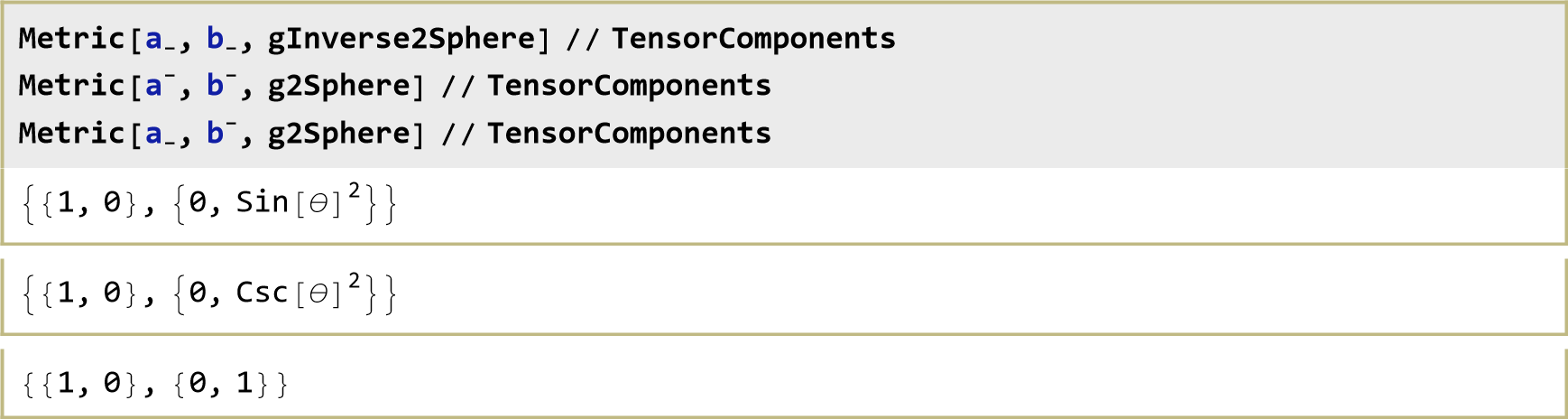}
The $2$-sphere is a maximally symmetric spacetime with three Killing vectors -- the maximum allowed for a two-dimensional ($2D$) space. One feature of such a spacetime is that its Riemann tensor can be constructed from a tensor product of the metric: $R_{abmn}=\frac{R}{2}g_{a[m}g_{n]b}$, where $[mn]$ denotes antisymmetrization; for e.g., $T_{[mn]} \equiv T_{mn} - T_{nm}$.
\includecode{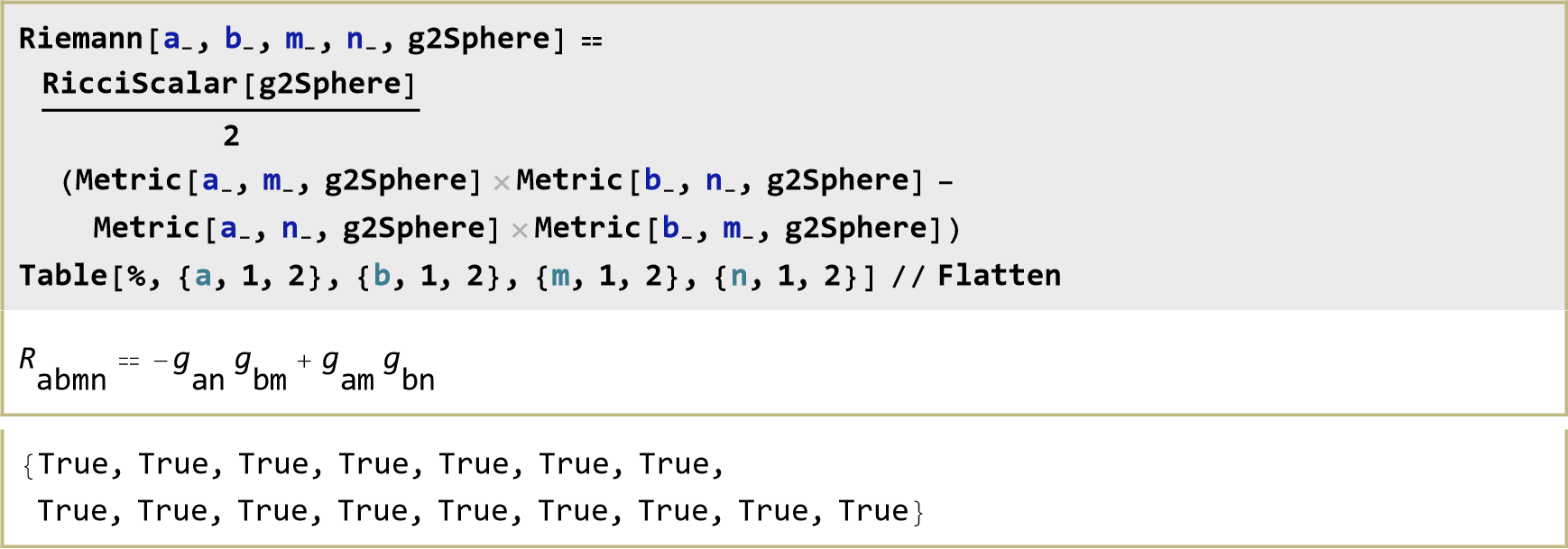}
\boldtitle{Generic Tensor objects} \qquad Let us now turn, briefly, to defining other (i.e., non-metric) tensors. Just as the above $2$-sphere metric was really a Tensor object, non-metric tensors may simply be directly entered as Tensor objects. A scalar $\varphi$, vector $V^{a}$, and rank-2 tensor $\TUD{T}{m}{n}$ are defined as follows.
\includecode{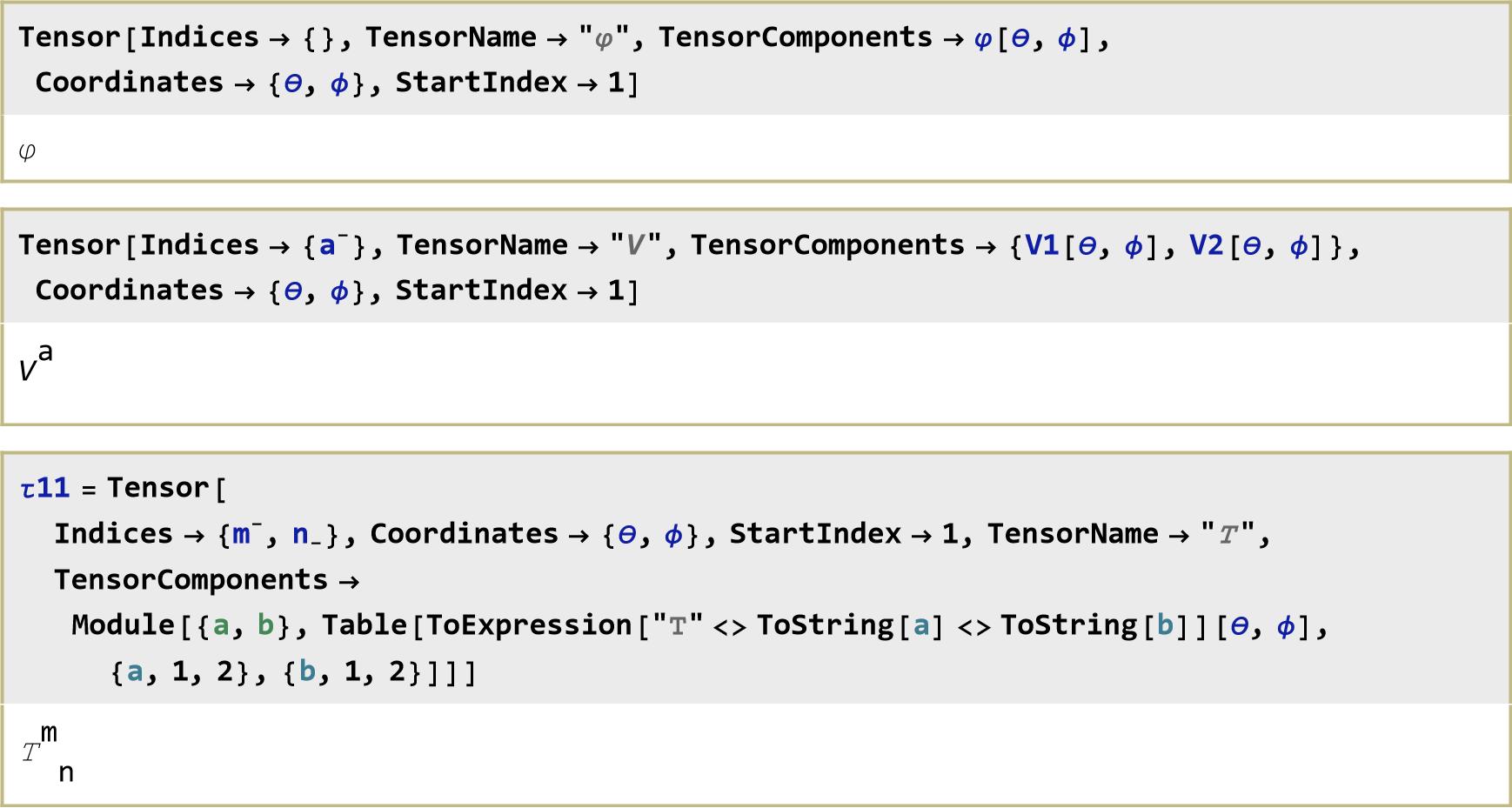} 
Note that the order of the arguments of Tensor is immaterial; we have set its Attributes to be Orderless.
\includecode{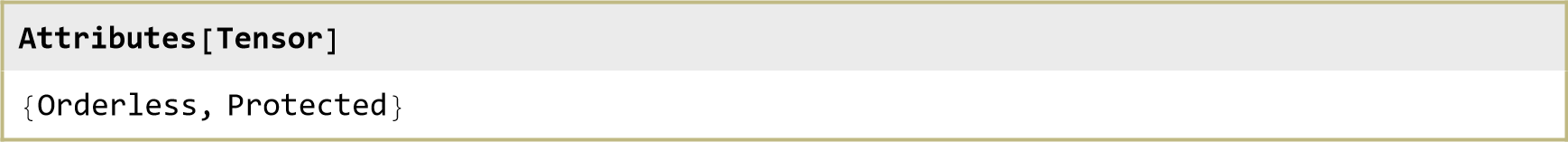}
The information stored in an arbitrary Tensor object can be extracted in a similar way as we did for the metric Tensor. Specifically, the left hand sides of all replacement Rule's occurring in a Tensor can be used an extraction function, in that applying it to the Tensor object returns the corresponding right hand side.
\includecode{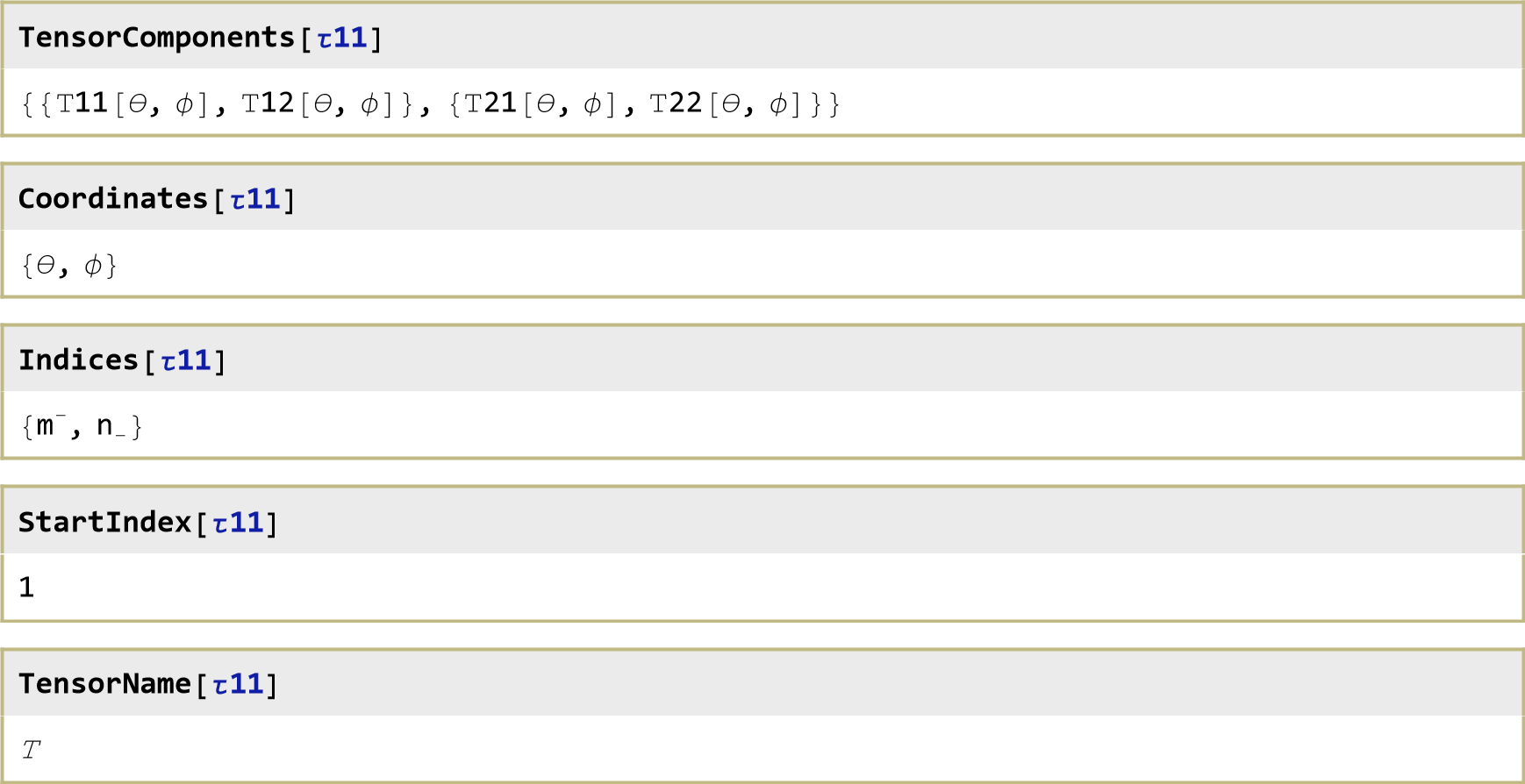}
Thus far, we have dealt with tensors with abstract indices. If we wish to extract specific components of tensors, however, we would need to evaluate their indices at the desired coordinates or their associated numerical values. For the $2$-sphere example at hand, with StartIndex $\to 1$, we may re-construct the metric component-by-component.
\includecode{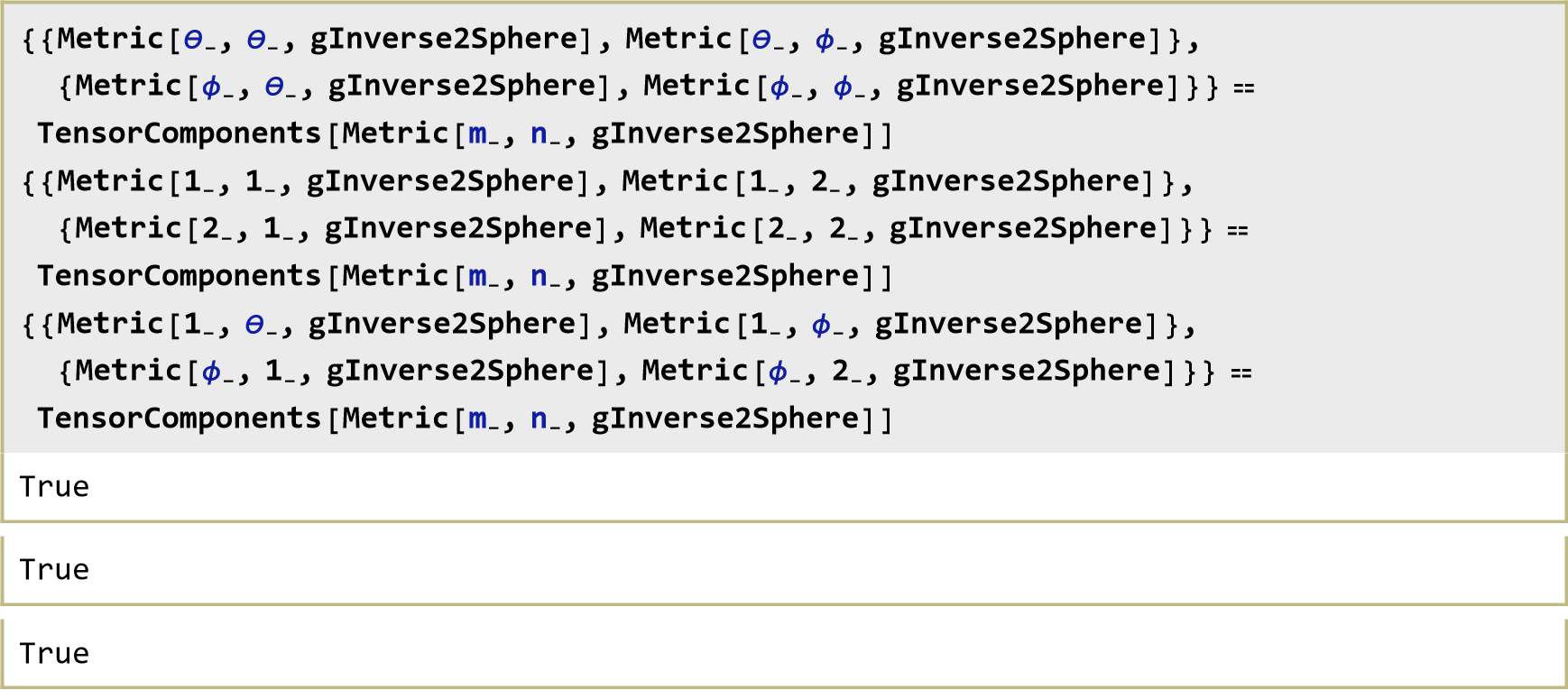}
Or, extract the $2 \times 2$ matrix given by the Christoffel symbol $\TUD{\Gamma}{\theta}{mn}=\TUD{\Gamma}{1}{mn}$
\includecode{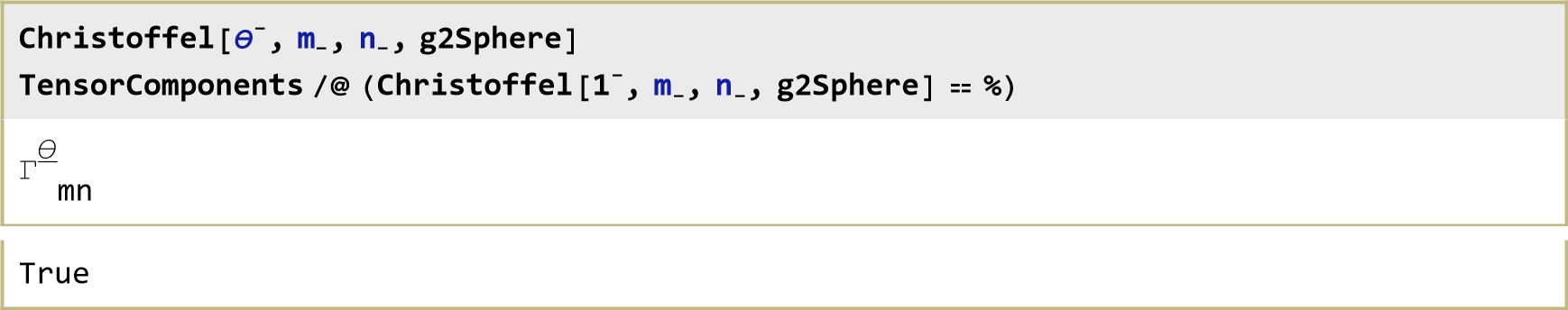} 
Or, the $2 \times 2$ matrix given by the Riemann tensor components $R_{\phi m \theta n} = R_{2 m 1 n}$
\includecode{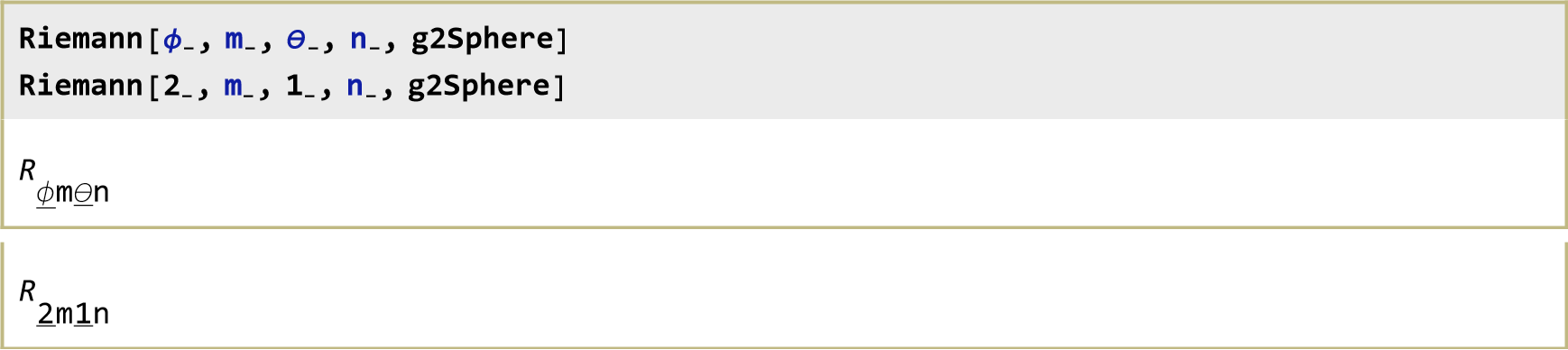}
The $\theta n \equiv 1 n$ components of the rank $(1,1)$ tensor above are $\{\TUD{T}{1}{1}, \TUD{T}{1}{2}\}$
\includecode{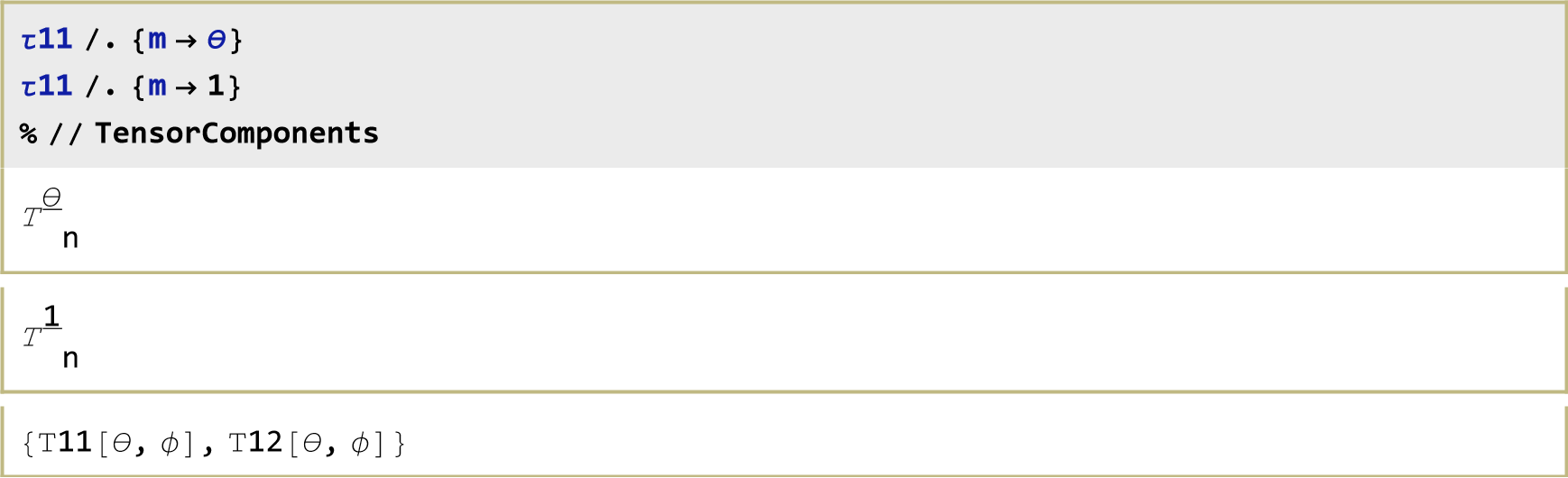}
Notice, upon evaluation at a specific coordinate, an index becomes underlined -- i.e., UnderBar is applied to it. This is to indicate its now inert nature.

\boldtitle{Einstein summation} \qquad We close this section by highlighting, \TC\  implements Einstein's summation convention within a single Tensor object -- namely, any pair of up-down repeated indices is automatically summed over.\footnote{In this paper (and the online User Guide), we shall adhere to the convention that Greek indices start at zero (the time coordinate) followed by the spatial ones over $\{1,\ldots,D-d-1\}$, where $D$ is the dimension of space and $d$ is that of spacetime. Latin/English alphabets run over only the spatial dimensions $\{1,\ldots,D\}$. Of course, when using \TC, the user is free to use whatever notation she desires.} The $2$-sphere metric obeys
\includecode{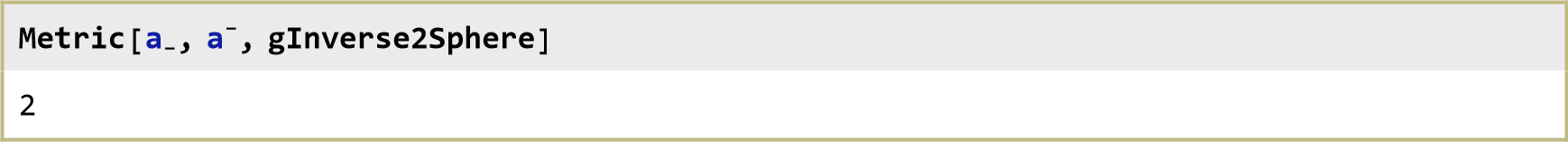}
The Ricci tensor is the result of contracting the first and third indices of Riemann:
\includecode{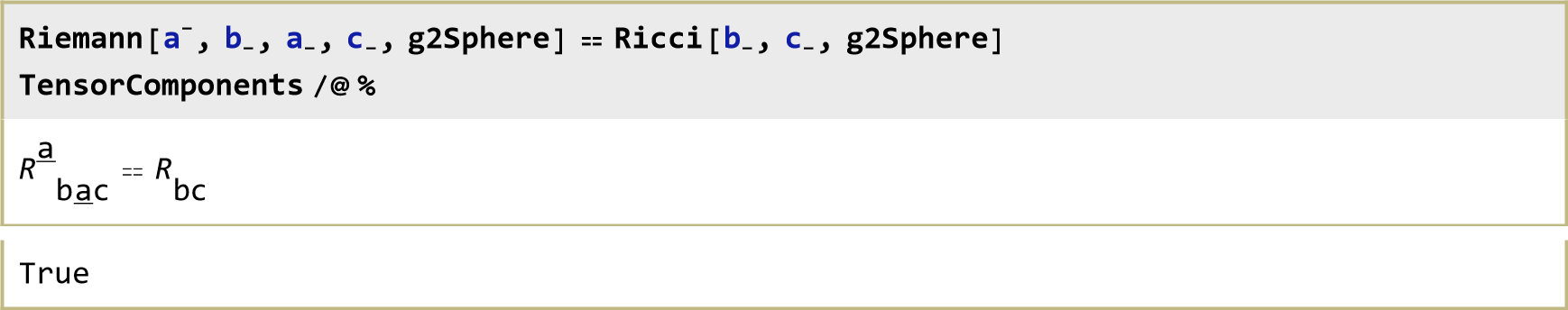}
With this basic introduction to \TC, we now move on to illustrate its functionality through examples of relevance to physics. Before doing so, however, let us first collect the definitions of the various geometry-related objects and operations carried out by \TC.

\section{Basic Definitions}


The orthonormal frame field $\varepsilon^{\hat{\alpha}}{}_{\mu}$ and its inverse $\varepsilon_{\hat{\alpha}}{}^{\mu}$ are defined through their relation to the metric itself:
\[
\eta_{\hat{\alpha}\hat{\beta}}\,\varepsilon^{\hat{\alpha}}{}_{\mu}\,
\varepsilon^{\hat{\beta}}{}_{\nu}
= g_{\mu\nu},
\qquad \text{and} \qquad
\eta^{\hat{\alpha}\hat{\beta}}\,\varepsilon_{\hat{\alpha}}{}^{\mu}\,
\varepsilon_{\hat{\beta}}{}^{\nu}
= g^{\mu\nu},
\]
where $\eta_{\hat{\alpha}\hat{\beta}}$ is the flat space(time) metric.

The Christoffel symbols are:
\[
\Gamma^{\alpha}{}_{\mu\nu}
=\frac{1}{2}\, g^{\alpha\lambda}
\left(
\partial_{\mu} g_{\nu\lambda}
+\partial_{\nu} g_{\mu\lambda}
-\partial_{\lambda} g_{\mu\nu}
\right).
\]

The covariant derivative of a vector $V$ is
\[
\nabla_{\alpha} V^{\gamma}
=\partial_{\alpha} V^{\gamma}
+\Gamma^{\gamma}{}_{\alpha\beta} V^{\beta},
\]
and that of a 1-form $A$ is
\[\nabla_{\alpha} A_{\beta}
=\partial_{\alpha} A_{\beta}
-\Gamma^{\gamma}{}_{\alpha\beta} A_{\gamma}.
\]
The Riemann curvature tensor is:
\[R^{\alpha}{}_{\beta\mu\nu}
=\partial_{\mu} \Gamma^{\alpha}{}_{\nu\beta}
-\partial_{\nu} \Gamma^{\alpha}{}_{\mu\beta}
+\Gamma^{\alpha}{}_{\sigma\mu}\Gamma^{\sigma}{}_{\nu\beta}
-\Gamma^{\alpha}{}_{\sigma\nu}\Gamma^{\sigma}{}_{\mu\beta}.
\]
The Ricci tensor is:
\[
R_{\beta\nu} \equiv R^{\alpha}{}_{\beta\alpha\nu}.
\]
The Ricci scalar is:
\[
R \equiv g^{\alpha\beta} R_{\alpha\beta}.
\]
The Hodge dual of a rank $N \le d$ tensor $T$ in $d$ dimensional
space(time) is
\[
(\star T)^{\mu_{1}\ldots\mu_{d-N}}
\equiv
\tilde{T}^{\mu_{1}\ldots\mu_{d-N}}
=\frac{1}{N!}\,
\tilde{\epsilon}^{\mu_{1}\ldots\mu_{d-N}\nu_{1}\ldots\nu_{N}}
\,T_{\nu_{1}\ldots\nu_{N}}.
\]
Here, the covariant Levi-Civita pseudo-tensor itself is defined as
\[
\tilde{\epsilon}_{\mu_{1}\ldots\mu_{d}}
\equiv|g|^{1/2}\,\epsilon_{\mu_{1}\ldots\mu_{d}},
\]
where $|g|^{1/2}$ is the square root of the absolute value of the metric determinant; and the $\epsilon_{\mu_{1}\ldots\mu_{d}}$ is the fully anti-symmetric Levi-Civita symbol, with
\[
\epsilon_{0\ 1\ \ldots\ d-1} \equiv 1.
\]
The Lie derivative of a vector $V$ along a vector $W$ is
\[
(\pounds_{W} V)^{\alpha}
=W^{\beta}\partial_{\beta} V^{\alpha}
-V^{\beta}\partial_{\beta} W^{\alpha},
\]
and that of a 1-form $A$ is
\[
(\pounds_{W} A)_{\alpha}
=W^{\beta}\partial_{\beta} A_{\alpha}
+A_{\beta}\partial_{\alpha} W^{\beta}.
\]
The exterior derivative on an $N$-form
$B_{\mu_{1}\ldots\mu_{N}}$ is
\[
(\mathrm{d}B)_{\nu\mu_{1}\ldots\mu_{N}}
=\frac{1}{N!}\,
\partial_{[\nu} B_{\mu_{1}\ldots\mu_{N}]},
\]
where the square bracket denotes anti-symmetrization.

We denote index symmetrization and anti-symmetrization with $\{\}$ and
$[\ ]$. For example,
\[
T_{\{\alpha\beta\gamma\}}
=T_{\alpha\beta\gamma}
+T_{\alpha\gamma\beta}
+T_{\beta\alpha\gamma}
+T_{\beta\gamma\alpha}
+T_{\gamma\beta\alpha}
+T_{\gamma\alpha\beta},
\]
and
\[
T_{[\alpha\beta\gamma]}
=T_{\alpha\beta\gamma}
-T_{\alpha\gamma\beta}
-T_{\beta\alpha\gamma}
+T_{\beta\gamma\alpha}
-T_{\gamma\beta\alpha}
+T_{\gamma\alpha\beta}.
\]

\section{3D Euclidean Space and the 2-Sphere}

In this section, we shall study the Euclidean group consisting of spatial translation and rotational symmetries of 3-dimensional (3D) flat space from a differential geometry point of view. 

\subsection{Isometries}

The most general active coordinate transformation that leaves the flat geometry
\includecode{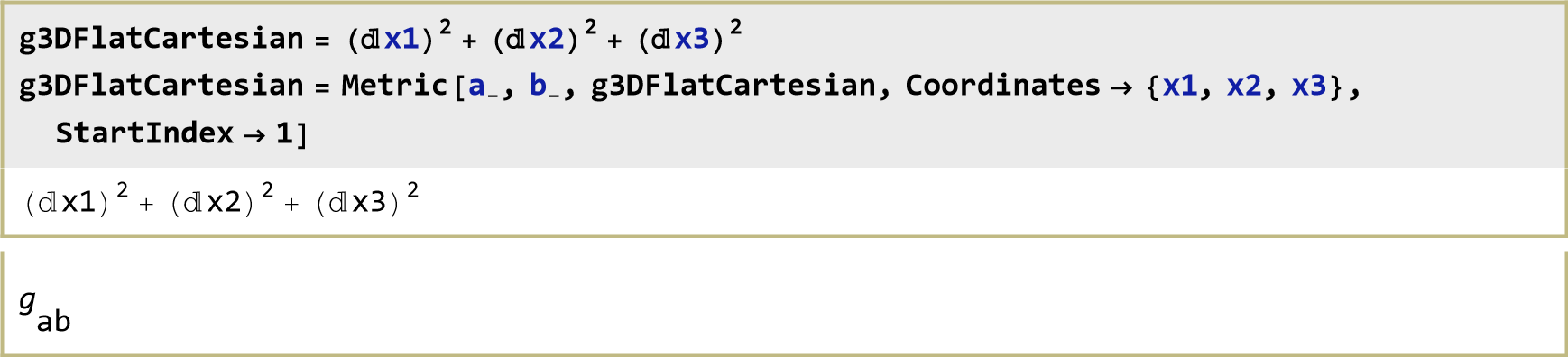}
invariant (namely, $\delta_{ij} \to \delta_{ij}$) is given by
\[
x^{i} \to R^{i}{}_{j} x^{j} + a^{i},
\]
for the space-independent orthogonal transformation $R^{i}{}_{j}$ obeying
\[
R^{a}{}_{i} R^{b}{}_{j} \,\delta_{ab} = \delta_{ij},
\]
and space-independent spatial translation vector $a^{i}$.

Spatial translations are generated by the three ``linear momentum'' operators
\[
P_{i} = P^{i} = -\mathrm{i}\,\partial_{x^{i}} .
\]
Spatial rotations
\[
R = \exp\!\left[-\mathrm{i}\,\vec{\theta}\cdot\vec{L}\right]
\approx 1 - \mathrm{i}\,\vec{\theta}\cdot\vec{L} + \cdots
\]
around the $k$-th axis are generated by the ``angular momentum'' operator
\[
L^{k} = -\mathrm{i}\,(\vec{x}\times\vec{\nabla})^{k}
= (\vec{x}\times\vec{P})^{k}.
\]
Let us first define both the linear and angular momentum operators in \TC\ using NonMetricTensor, whose mandatory arguments are the Indices,
TensorComponents, TensorName, and Coordinates.
\twopicincludecode{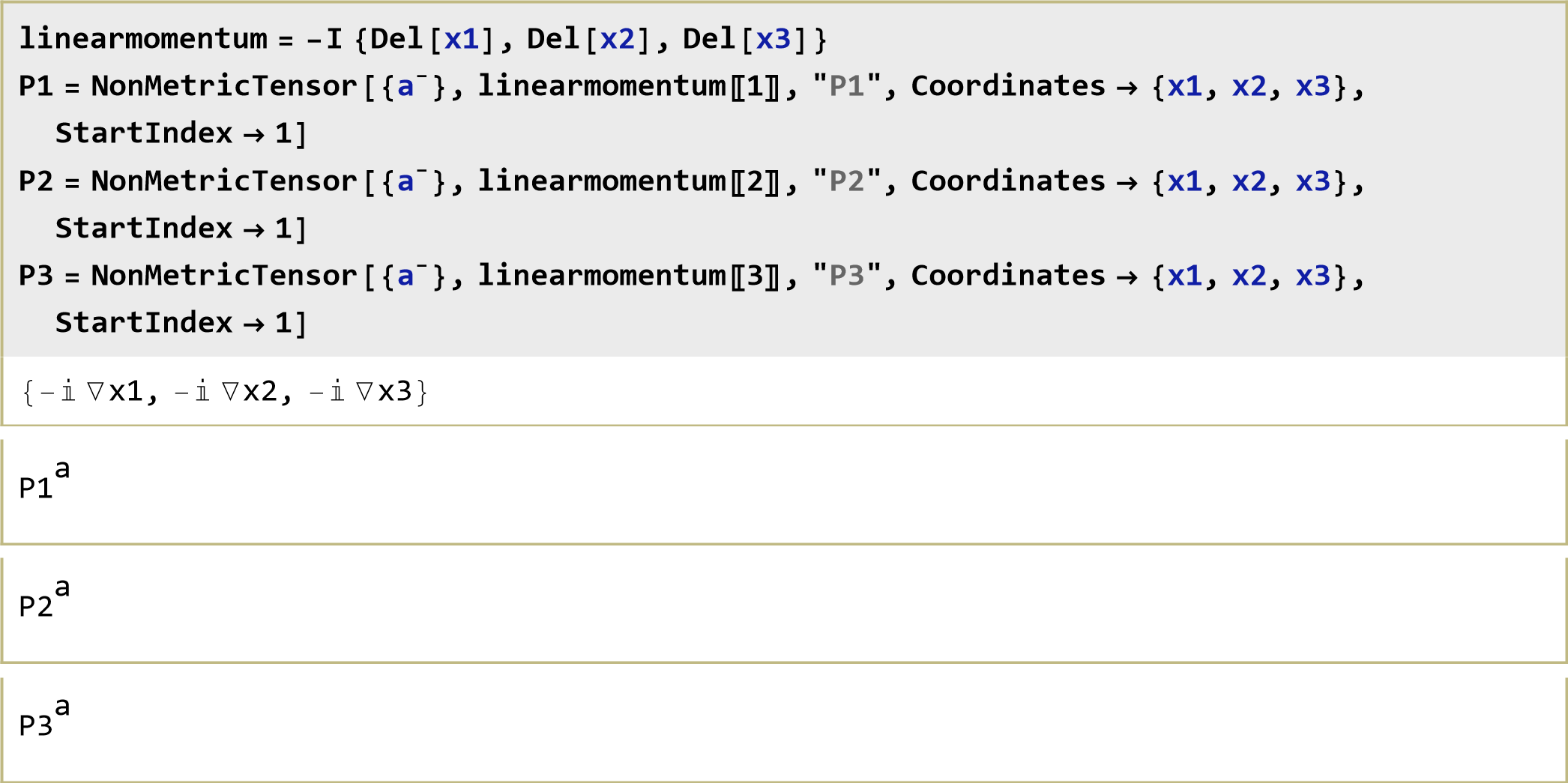}{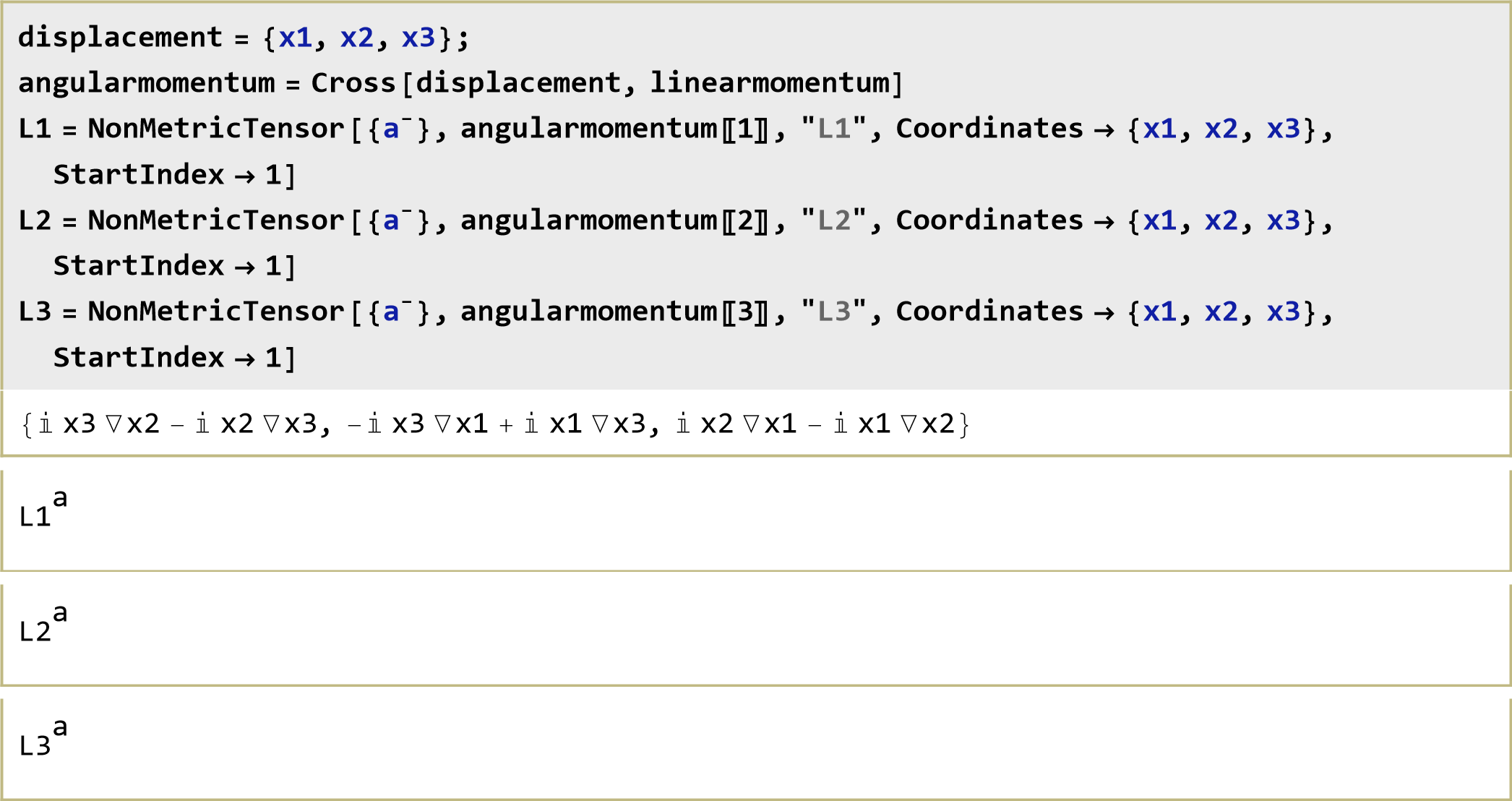}
The Lie derivative of a scalar $\varphi $ along some vector field \textit{V} is simply the latter contracted into the partial derivatives of the
former; namely, \(\pounds _V\varphi  = V^a\partial _a\varphi\). In \TC, the Lie derivative of a scalar or tensor \textit{B} along the vector
\textit{V} can be implemented as LieD[V,B]. The output Indices are simply the Indices of \textit{B}.

Under translations, $\varphi $[\(\overset{\to }{x}\)] $\rightarrow $ $\varphi $[\(\overset{\to }{x}+\overset{\to }{a}\)] = $\varphi $[\(\overset{\to}{x}\)] + \(i\overset{\to }{a}\cdot \overset{\to }{P}\) $\varphi $[\(\overset{\to }{x}\)] + ...
\includecode{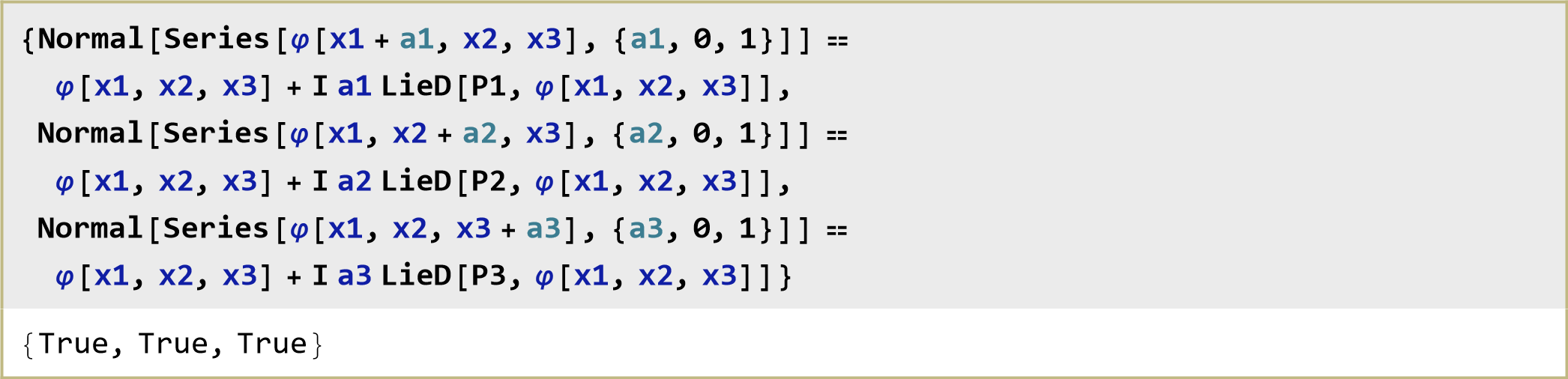}
Under rotations, $\varphi $[\(\overset{\to }{x}\)] $\rightarrow $ $\varphi $[R$\cdot $\(\overset{\to }{x}\)] = $\varphi $[\(\overset{\to }{x}\)]
+ \(i\overset{\to }{\theta }\cdot \overset{\to }{L}\) $\varphi $[\(\overset{\to }{x}\)] + ... To be specific, consider a \textit{counter-clockwise}
infinitesimal rotation around the 1-, 2-, and 3-axes.
\includecode{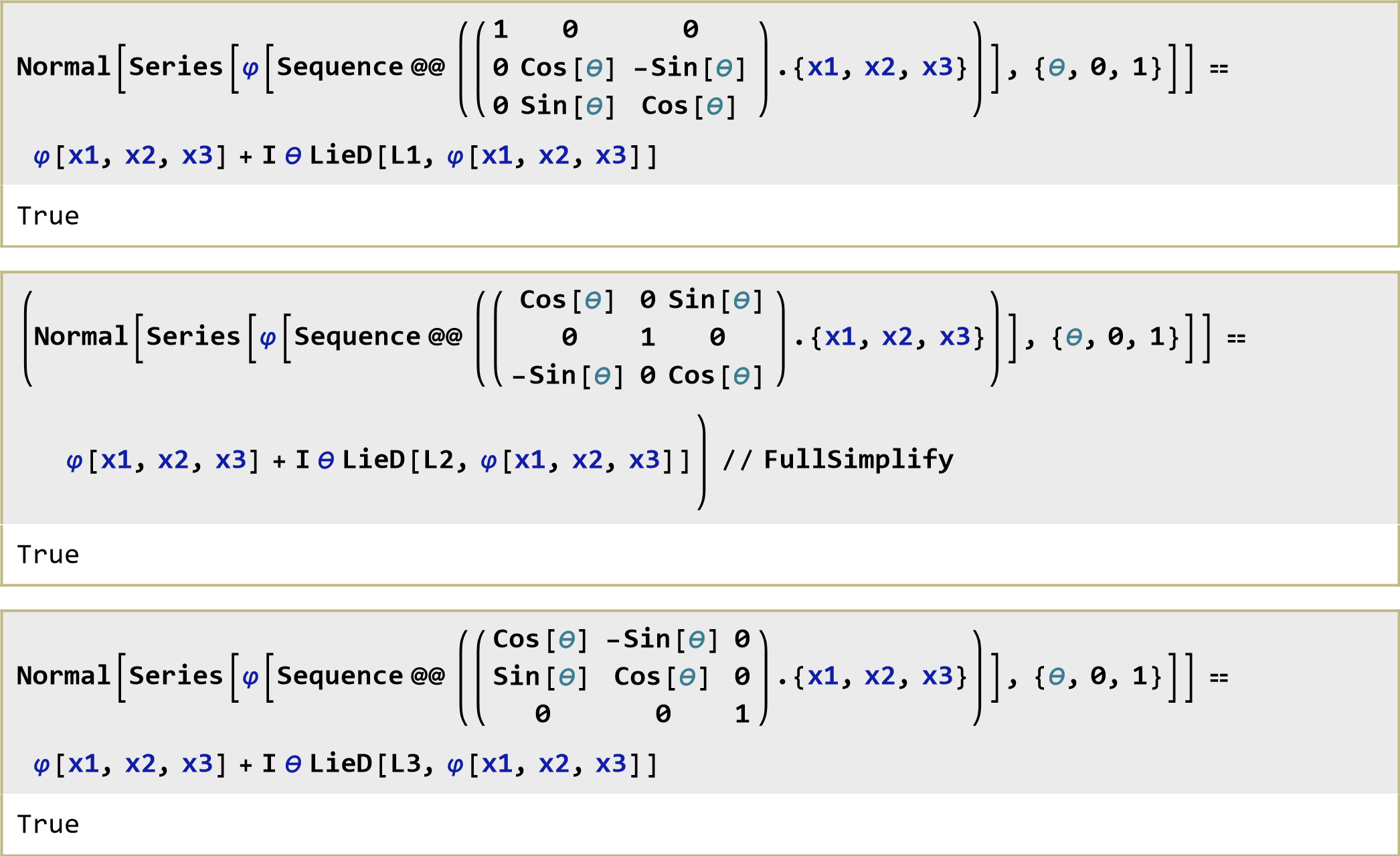}
Furthermore, the negative Laplacian is simply the {``}square{''} of the linear momentum operator.
\includecode{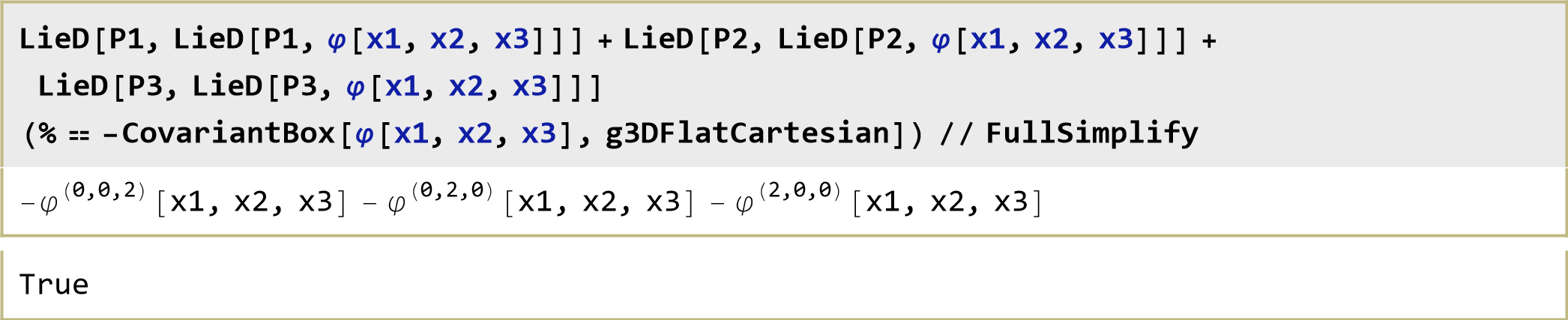}
CovariantBox[stuff, \textit{M}$\_$Tensor] returns a Tensor object whose components are \(\nabla _{\alpha }\nabla ^{\alpha }(\text{stuff})\), where the covariant derivative is with respect to the metric Tensor \textit{M}, while ``stuff" can be either a regular expression or a Tensor object.

That the infinitesimal coordinate transformations generated by these vector fields leave the flat 3D metric invariant means the Lie derivative of the metric along them is zero -- these vectors satisfy Killing{'}s equations.
\includecode{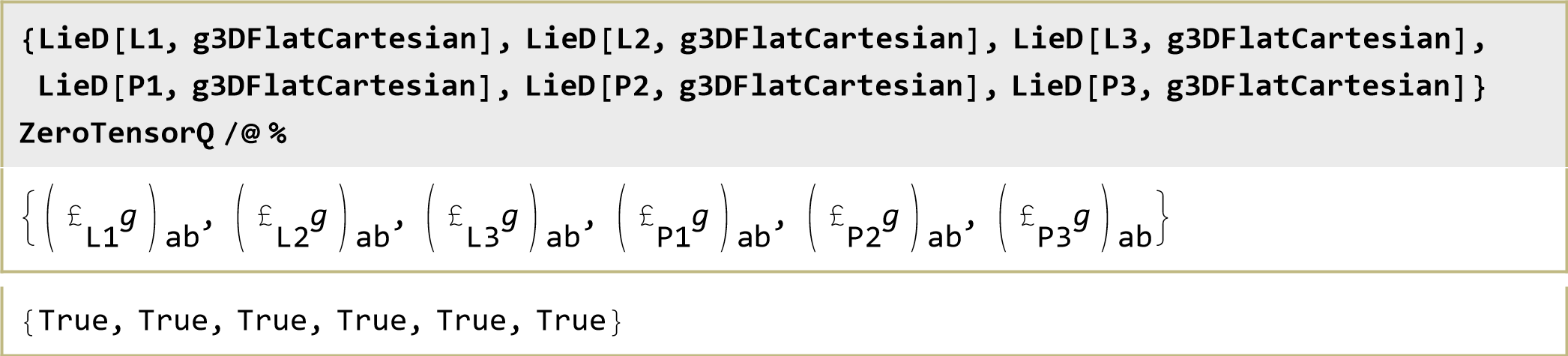}
The \(\text{so}_3\) Lie algebra of the rotation group is given by
\[
\left[L^a,L^b\right]=\text{i$\epsilon $}^{\text{abc}}L^c,
\]
where \(\epsilon ^{\text{abc}}\) is the Levi-Civita symbol with \(\epsilon ^{123}\equiv 1\). In quantum mechanics, these are the familiar commutation relations between the angular momentum operators.

Moreover, the Lie derivative of a vector B with respect to vector A is simply their Lie bracket; namely, \(\pounds _AB=[A,B].\) 

This means we may simply verify the \(\text{so}_3\) Lie algebra using Lie derivatives.
\includecode{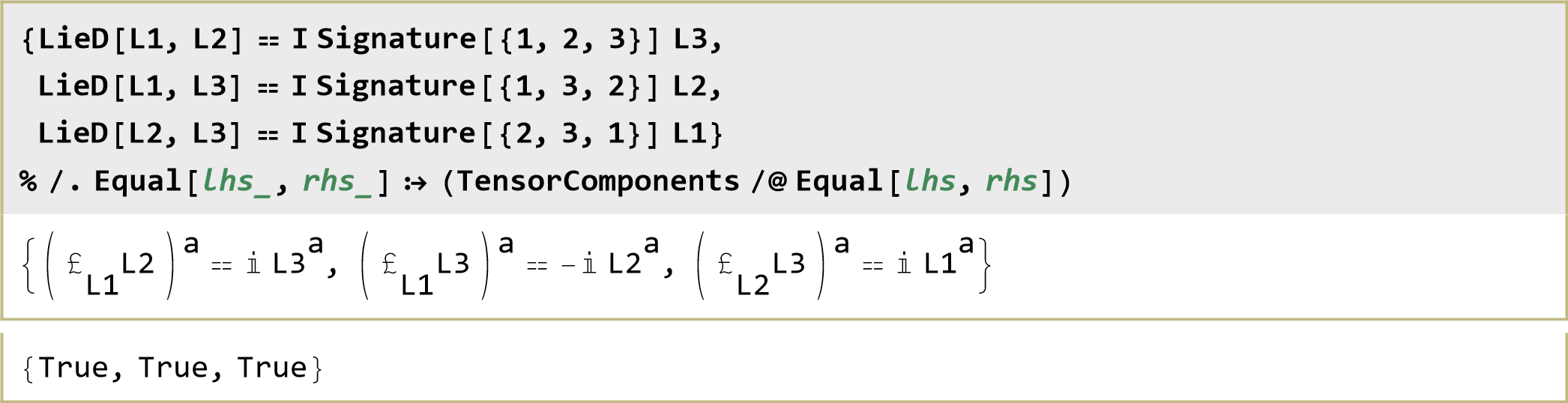}
That the momentum operator \(P_i\) transforms as a vector under rotations imply its commutator with the angular momentum operators must obey the
following relation:
\[
\left[ L^a,P^b \right] = i\epsilon ^{\text{abc}}P^c.
\]
\includecode{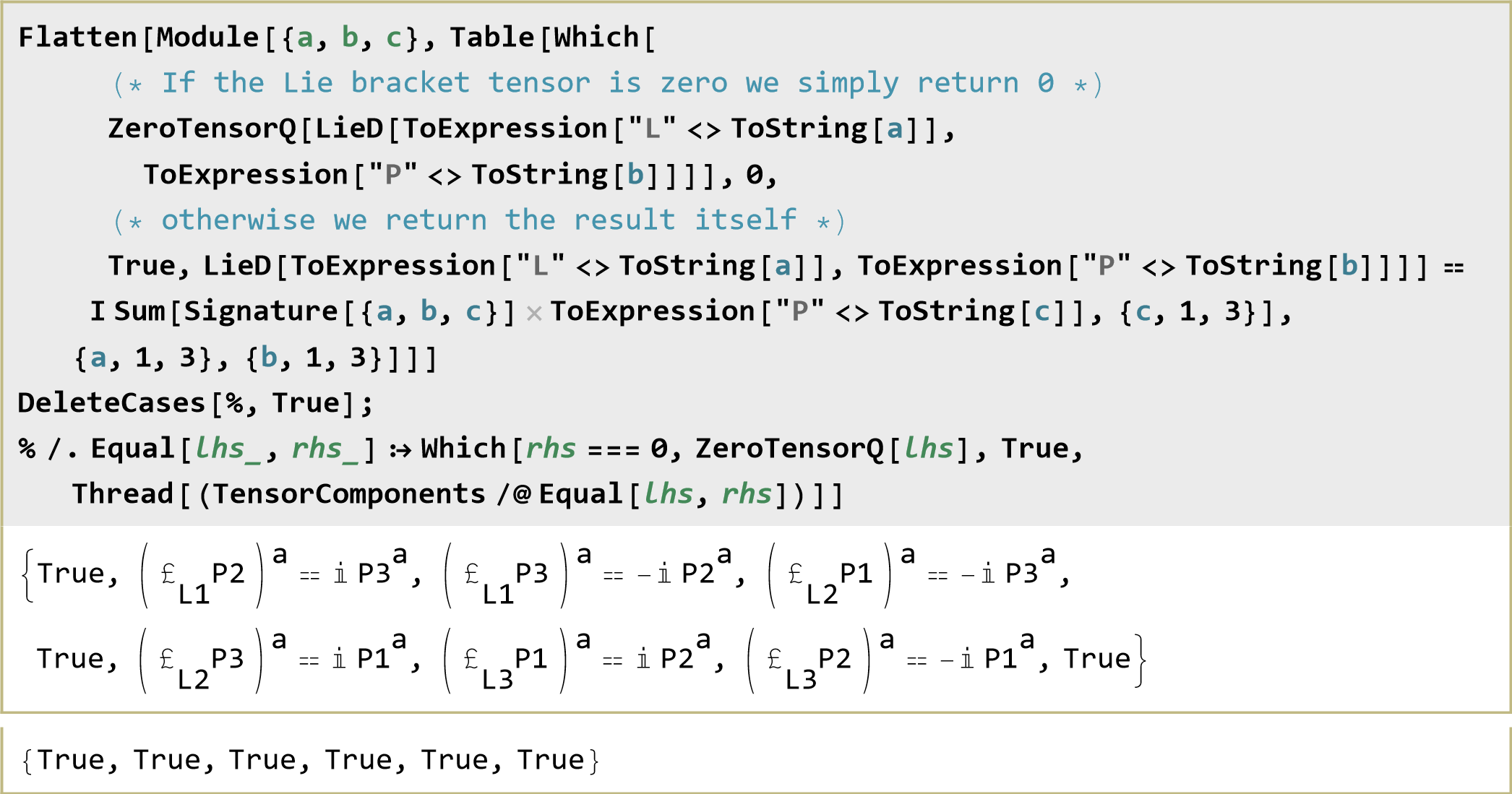}
The linear momentum operators commute amongst themselves.
\includecode{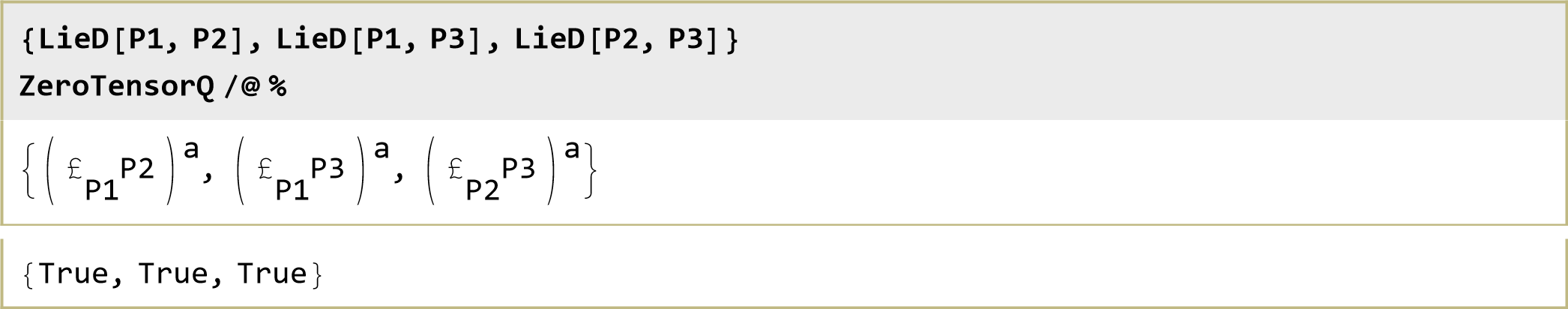}
Altogether, we may now gather the Lie algebra of the 3D Euclidean group consisting of rotations and translations:
\[
\left[ L^a,L^b\right]  = i\epsilon ^{\text{abc}}L^c, \quad
\left[ L^a,P^b\right]  = i\epsilon ^{\text{abc}}P^c, \quad
\left[ P^a,P^b\right]  = 0.
\]

\subsection{Round 2 Sphere and Spherical Harmonics} 

Let us focus now on the rotation subgroup of the 3D Euclidean group. Since the $\{$\(L^a\)$\}$ generate rotations, we expect that these {``}angular momentum{''} vector fields are strictly tangent to the round 2-sphere. To facilitate the ensuing analysis, therefore, let us introduce spherical coordinates $\{$\textit{r} $\geq $ 0, 0 $\leq $ $\theta $ $\leq $ $\pi $, 0 $\leq $ $\phi $ $<$ 2$\pi \}$.
\includecode{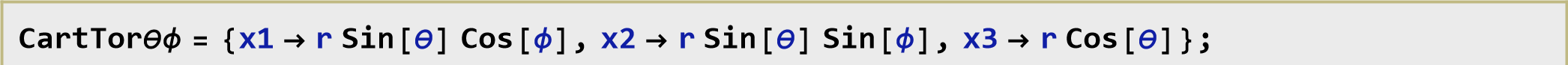}\TC{'}s CoordinateTransformation[$\{$x1 $\to $ f1[y1,y2,...], x2 $\to $ f2[y1,y2,...],...$\}$,$\{$y1,y2,...$\}$] returns a List of Rule{'}s containing the input coordinate transformations; together with the transformations of the infinitesimal displacements
\begin{align*}
	&\mathrm{d}x1
	\to
	\frac{\partial f1}{\partial y1}\,\mathrm{d}y1
	+
	\frac{\partial f1}{\partial y2}\,\mathrm{d}y2
	+ \cdots,
	\\
	&\mathrm{d}x2
	\to
	\frac{\partial f2}{\partial y1}\,\mathrm{d}y1
	+
	\frac{\partial f2}{\partial y2}\,\mathrm{d}y2
	+ \cdots,
	\\
	&\text{etc}.
\end{align*}
as well as the transformations of the basis partial derivatives
\begin{align*}
	&\nabla x1
	\to
	\frac{\partial y1}{\partial x1}\,\nabla y1
	+
	\frac{\partial y2}{\partial x1}\,\nabla y2
	+ \cdots,
	\\
	&\nabla x2
	\to
	\frac{\partial y1}{\partial x2}\,\nabla y1
	+
	\frac{\partial y2}{\partial x2}\,\nabla y2
	+ \cdots,
	\\
	&\text{etc}.
\end{align*}
where the $\partial $y/$\partial $x{'}s are the components of the inverse of the Jacobian matrix formed from the $\partial $f/$\partial $y{'}s. (If the second List $\{$y1,y2,...$\}$ in CoordinateTransformation is not provided, \TC\ will assume the transformation is an active one, from
$\{$x1,x2,...$\}$ to itself.) It is important to enter a single Symbol on the left hand side of each Rule; and while the order of the x1 $\to $ ..., x2 $\to $ ,... Rule{'}s in the first argument is immaterial, the order of the output/new coordinates in $\{$y1, y2, ...$\}$ is significant.

As a specific example, the Cartesian to spherical transformation reads as follows.
\includecode{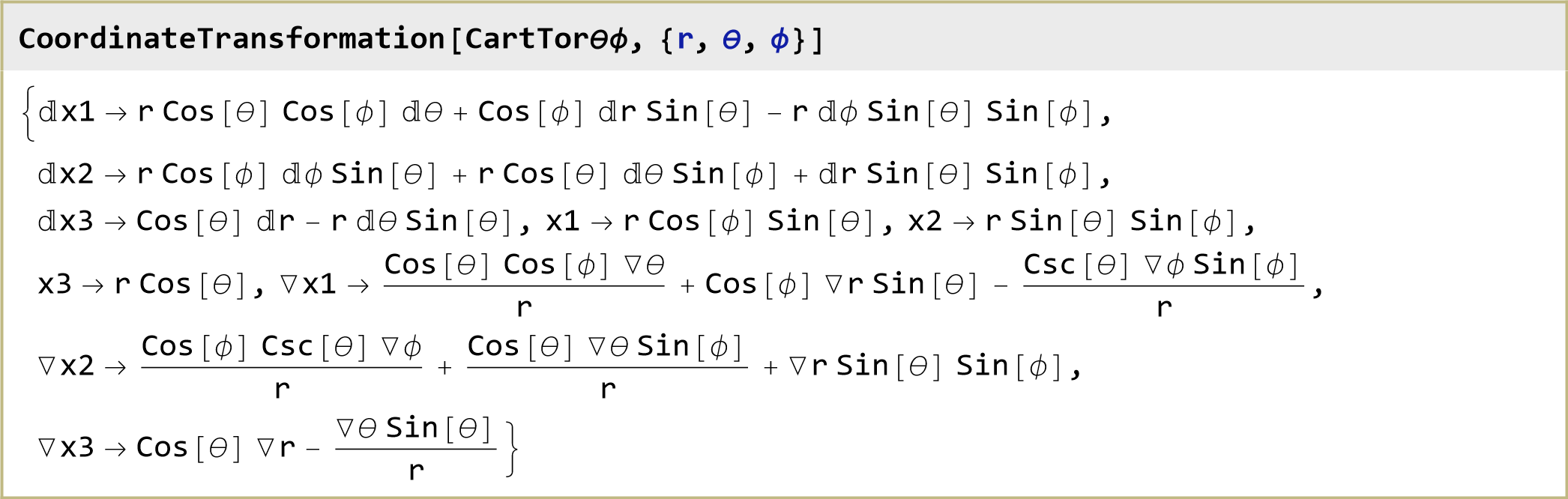}
We may readily use CoordinateTransformation to convert the Cartesian expression for flat 3D space into its spherical coordinates counterpart. First, we apply ToExpressionForm to g3DFlatCartesian to convert it into its quadratic form, as a sum over the squares of infinitesimal displacements. (More generally, ToExpressionForm[$\tau \_$Tensor] converts the Tensor object $\tau $ into the corresponding sum over its basis 1-forms and/or partial derivatives.) Then, we apply the Rule{'}s of coordinate transformations.
\includecode{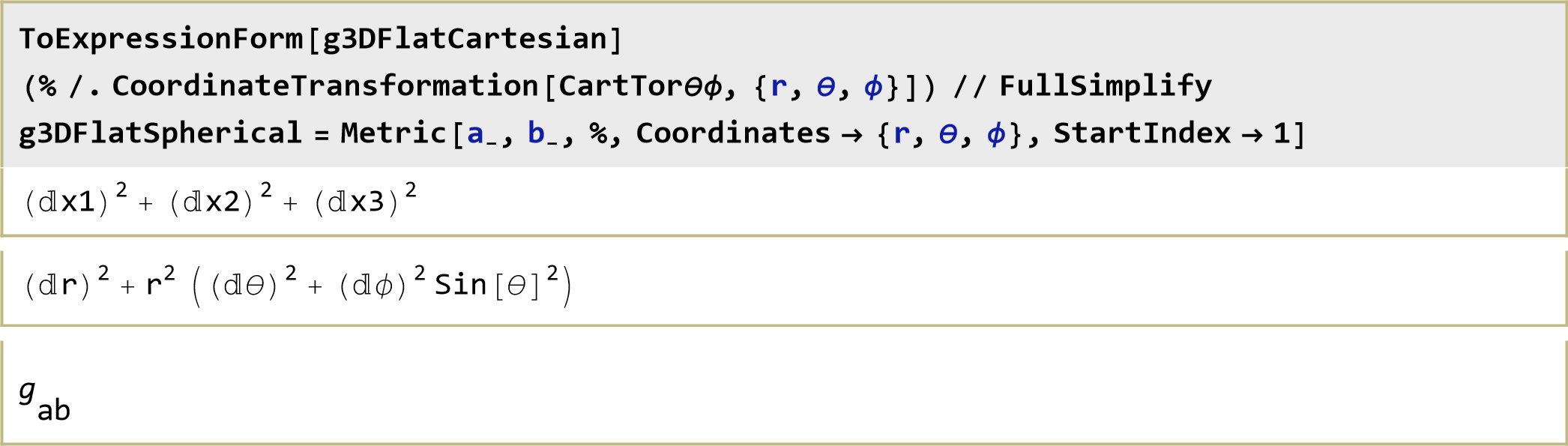}
Somewhat more directly, we may use CoordinateTransformation[inputtensor, $\{$x1 $\to $ f1[y1,y2,...], x2 $\to $ f2[y1,y2,...],...$\}$,$\{$y1,y2,...$\}$] to transform the Tensor object inputtensor from the coordinate system $\{$x1,x2,...$\}$ into $\{$y1,y2,...$\}$. For example, to transform the Cartesian 3D flat space into spherical coordinates, we may do the following.
\includecode{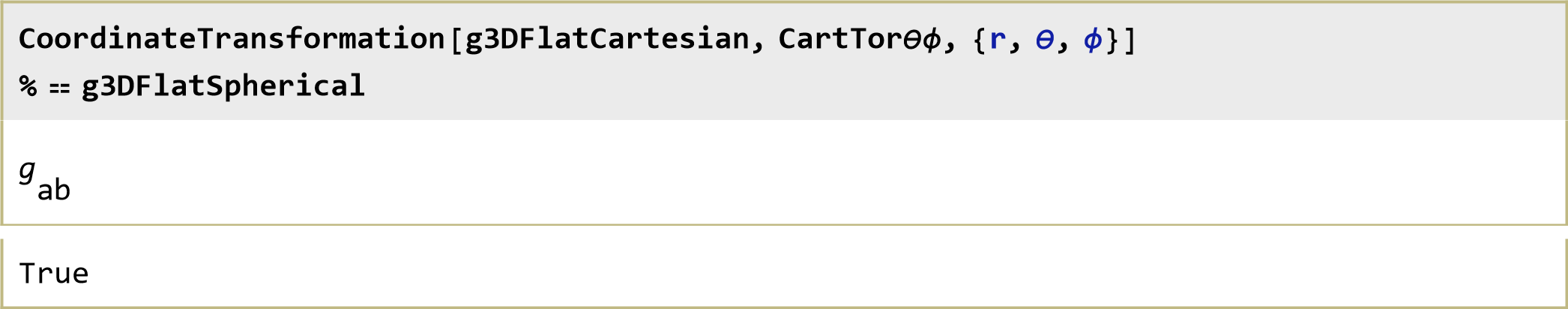}
Let us now coordinate transform the angular momentum operators -- really generators of 3D rotations -- from Cartesian to the spherical coordinate basis
\phantomsection
\label{sent:LIn3C}
\includecode{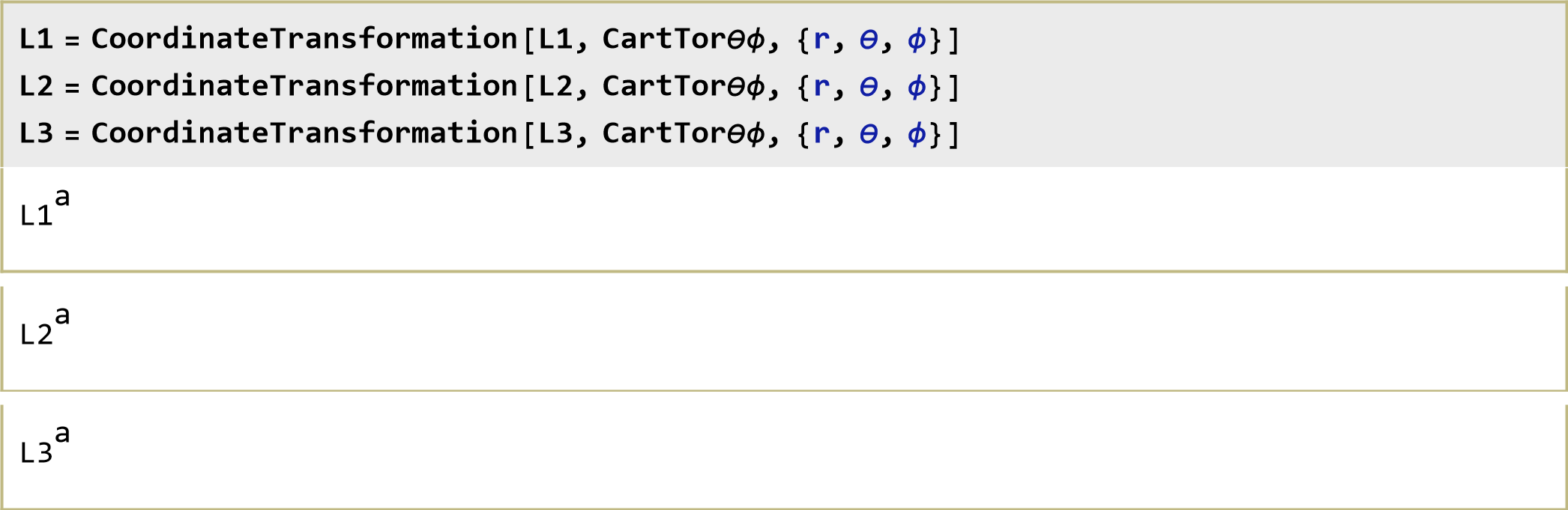}
We may confirm that the first ($r-$)components of these vectors are zero; i.e., they are tangent to the surface of the 2-sphere.
\includecode{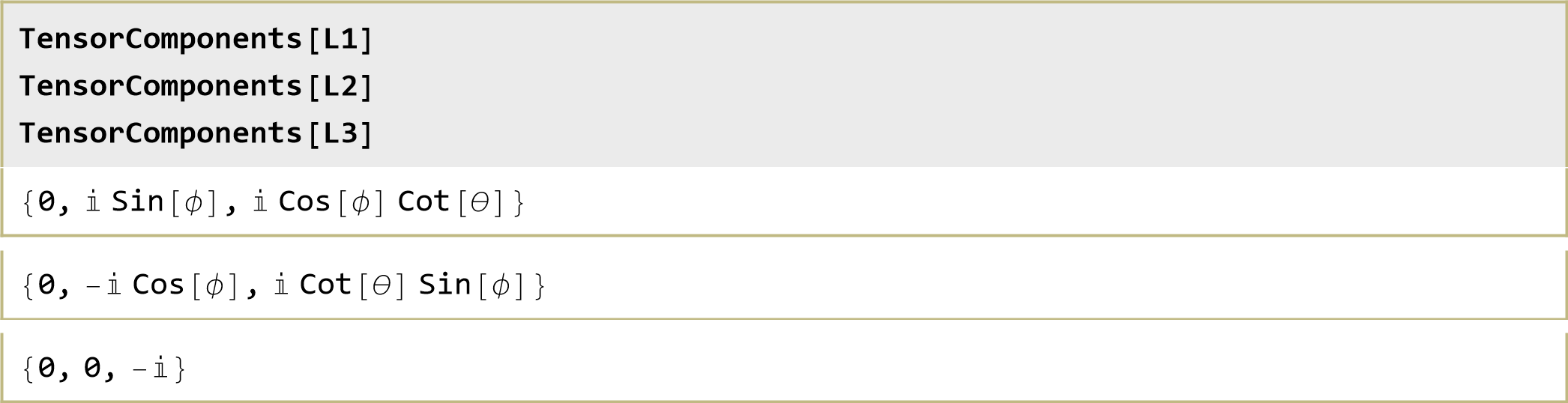}
The {``}square{''} of the angular momentum operators \(\overset{\to }{L}\cdot \overset{\to }{L}\) ought to be a rotationally invariant object. Given that it is built out of angular derivatives only, it is reasonable to guess it may be related to the angular portion (as opposed to the radial derivatives) of the 3D Laplacian. An explicit calculation would reveal that 
\[
-\overset{ }{\overset{\to }{\nabla }_{3D}^2}\varphi = r^{-2}\overset{\to }{L}\cdot \overset{\to }{L}\varphi  - \partial
_r\left(r^2\partial _r\varphi \right)/r^2.
\]
\includecode{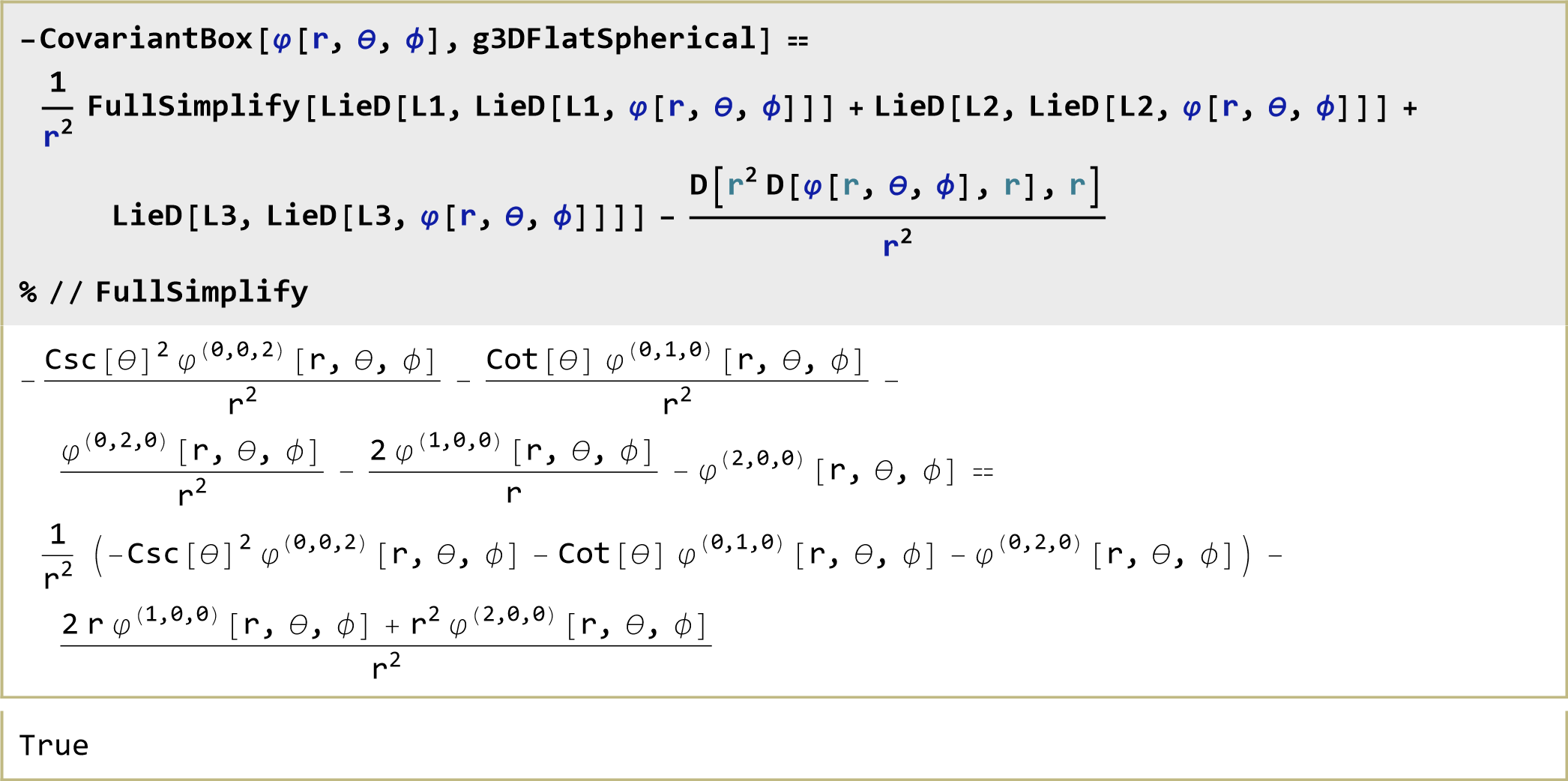}
In fact, we shall verify that \(\overset{\to }{L}\cdot \overset{\to }{L}\) is nothing but the negative Laplacian -\(\nabla _i\)\(\nabla ^i\) on the round unit 2-sphere. To this end, let us first compute the latter{'}s metric by simply taking the 3D flat metric and setting \textit{r} = 1 (and, hence, $\dd r = 0$).
\includecode{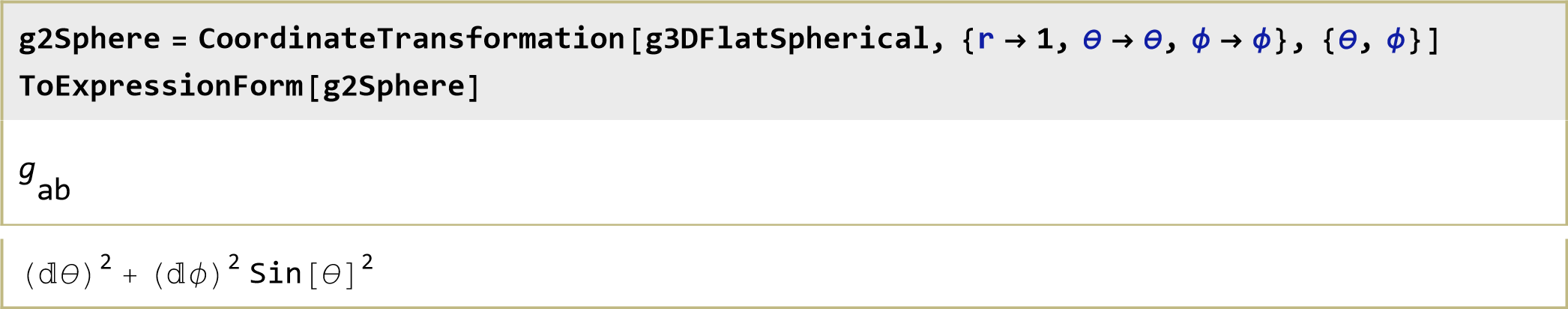}
As we see above, when employing CoordinateTransformation[RulesI$\_$List, RulesII$\_$List], it is possible to use fewer variables on the {``}right hand side{''} RulesII than on the {``}left hand side{''} RulesI defining the coordinate transformation(s).

A direct comparison between \(\overset{\to }{L}\cdot \overset{\to }{L}\)$\varphi $ and -\(\nabla _i\)\(\nabla ^i\)$\varphi $ on the 2-sphere verifies their equivalence.
\includecode{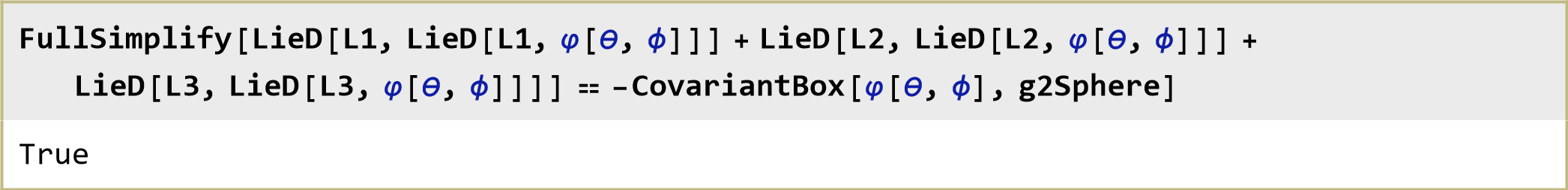}
Each \(L^a\) and the negative Laplacian \(\overset{\to }{L}\cdot \overset{\to }{L}\) = -\(\nabla _i\)\(\nabla ^i\) are Hermitian operators. We know that the $\{$\(L^a\)$\}$ do not mutually commute; but rotational symmetry of the Laplacian means [\(L^a\),-\(\nabla _i\)\(\nabla ^i\)] = 0, as we may readily verify. 
\includecode{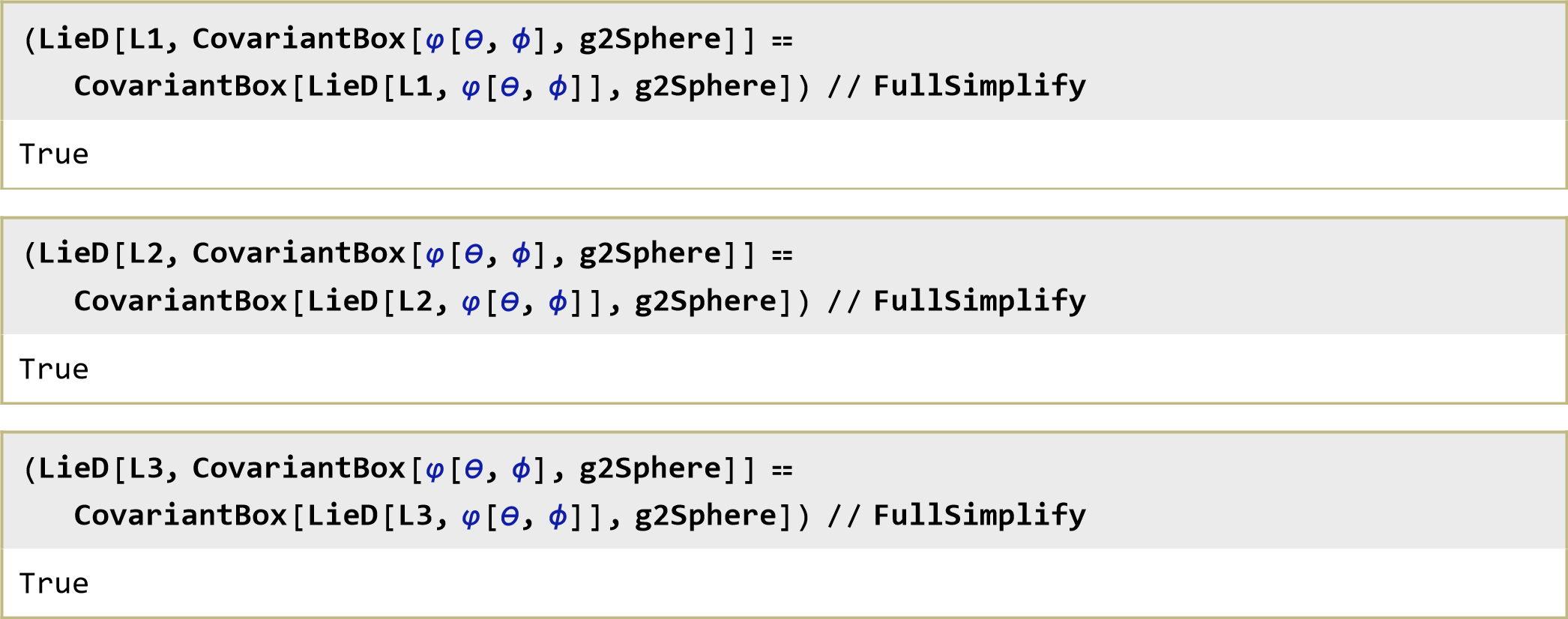}
We may therefore seek the simultaneous eigenfunctions of \(L^3\) and \(\overset{\to }{L}\cdot \overset{\to }{L}\). To find the eigenfunctions of
the latter, we first turn to the homogeneous solution $\varphi $ of Laplace{'}s equation in 3D flat space. 
\[
-\overset{ }{\overset{\to }{\nabla }_{3D}^2}\varphi = 0
\]
If we perform a Taylor expansion of $\varphi $ in Cartesian coordinates,
\[
\varphi [\overset{\to }{x}] = \Sigma _{\ell }(1/\ell !)x^{i_1}\text{...}x^{i_{\ell }}\partial _{i_1}\text{...}\partial _{i_{\ell}}\varphi \left[\overset{\to }{0}\right],
\]
we may recognize the terms containing precisely $\ell $ powers of Cartesian coordinates to be proportional to \(r^{\ell }\) when re-expressed in
spherical coordinates. That is, the $\ell$-th term of the Taylor-expanded Laplace's equation then reads as follows.
\includecode{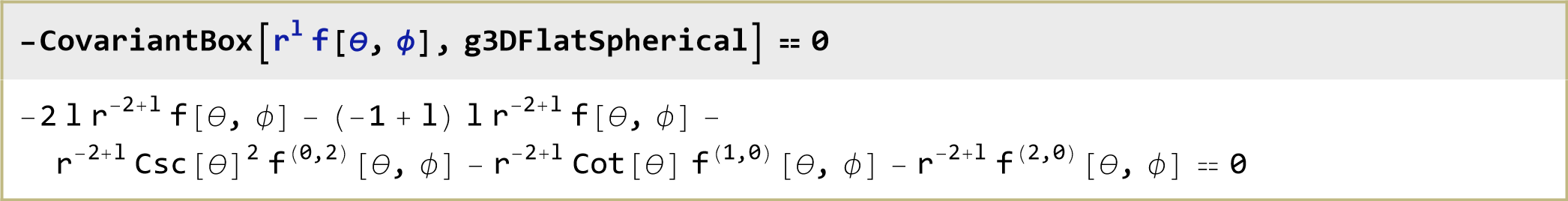}
Moreover, we may now postulate that $f$[$\theta $,$\phi $] must be expressible as a superposition over the eigenfunctions of \(\overset{\to }{L}\cdot\overset{\to }{L}\). Let such an eigenfunction be $Y$ and \(\overset{\to }{L}\cdot \overset{\to }{L}\)$Y$ = $-\lambda $ \(\overset{\to }{L}\cdot \overset{\to}{L}\)Y. Imposing this on Laplace{'}s equation, we uncover the well known fact that $\lambda = \ell (\ell+1)$.
\includecode{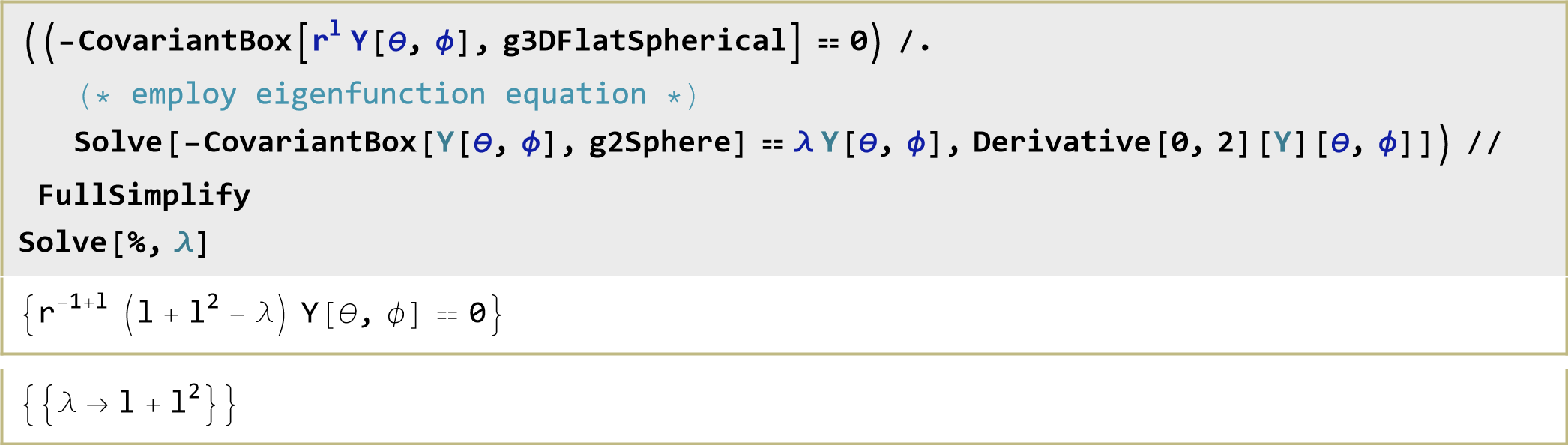}
As for the eigenfunction of \(L^3\), assuming its eigenvalue to be \textit{m}, we need to solve the following.
\includecode{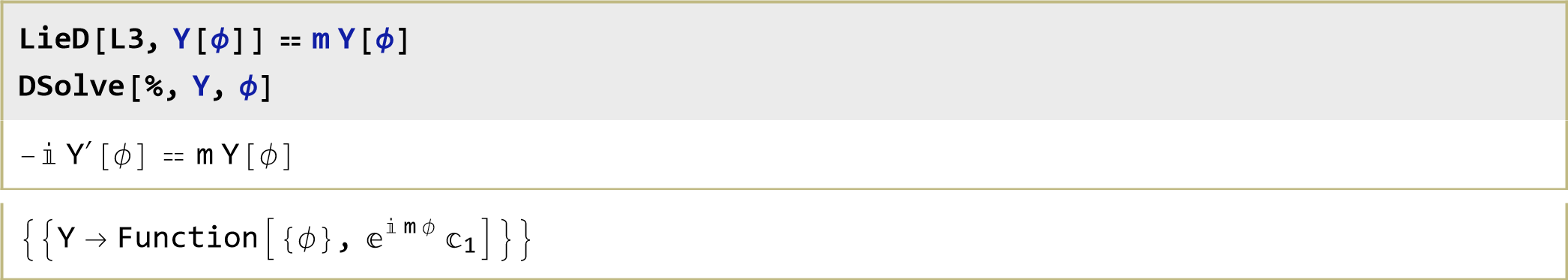}
Since $\phi $ and $\phi $+2$\pi $ must refer to the same location for a fixed $\theta $, this implies \textit{m} must be integer. Additionally, one may argue that no more than $\ell $ powers of exp[$\pm $i $\phi $] may appear in the combination \(r^{\ell }f[\theta ,\phi ]\) since it arose from $\ell $ powers of the Cartesian coordinates. This implies $|m| \leq \ell$.

Incidentally, if we define the raising ($+$) and lowering ($-$) operators \(L_{\pm }\equiv L^1\pm  i L^2\), then \(L^3\) \(L_{\pm }\) = [\(L^3\), \(L_{\pm}\)] + m \(L_{\pm }\) when acting on an eigenstate. To verify \(L_{\pm }\) are in fact raising and lowering operators acting on the \textit{m} azimuthal eigenvalue -- namely, \(L^3\) \(L_{\pm }\) = \((m\pm 1)\) \(L_{\pm }\) whenever this product of operators are acting on an eigenstate -- we want to check that [\(L^3\), \(L_{\pm }\)] = $\pm $ \(L_{\pm }\).
\includecode{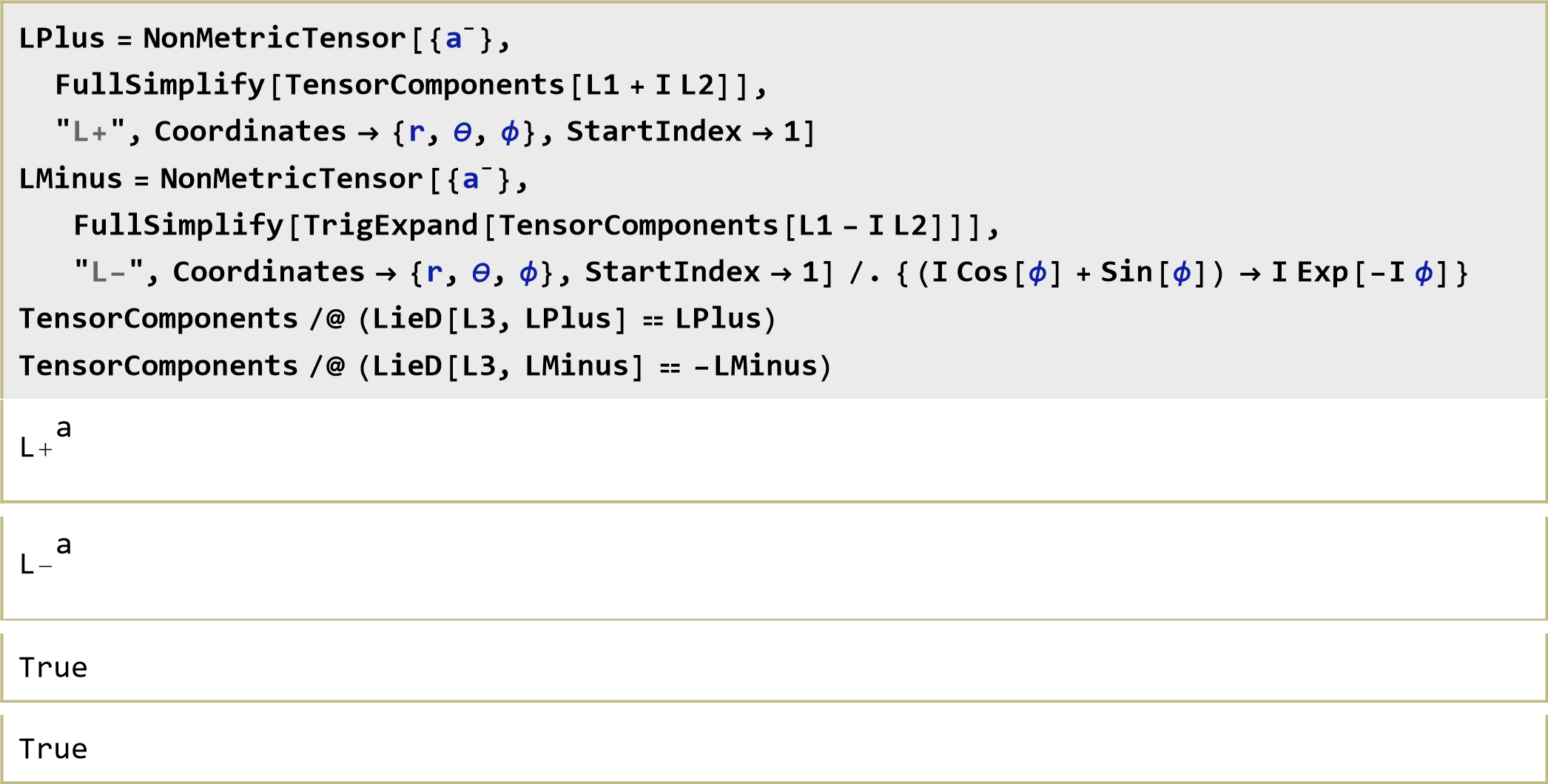}
At this point we have justified a separation-of-variables ansatz for the eigenfunction equation; and therefore $Y =$ Exp[I $m \phi $] P[cos $\theta$]. The $\theta $-dependence, in turn, leads to the associated Legendre equation in terms of $\chi $ $\equiv $ cos $\theta $. 
\includecode{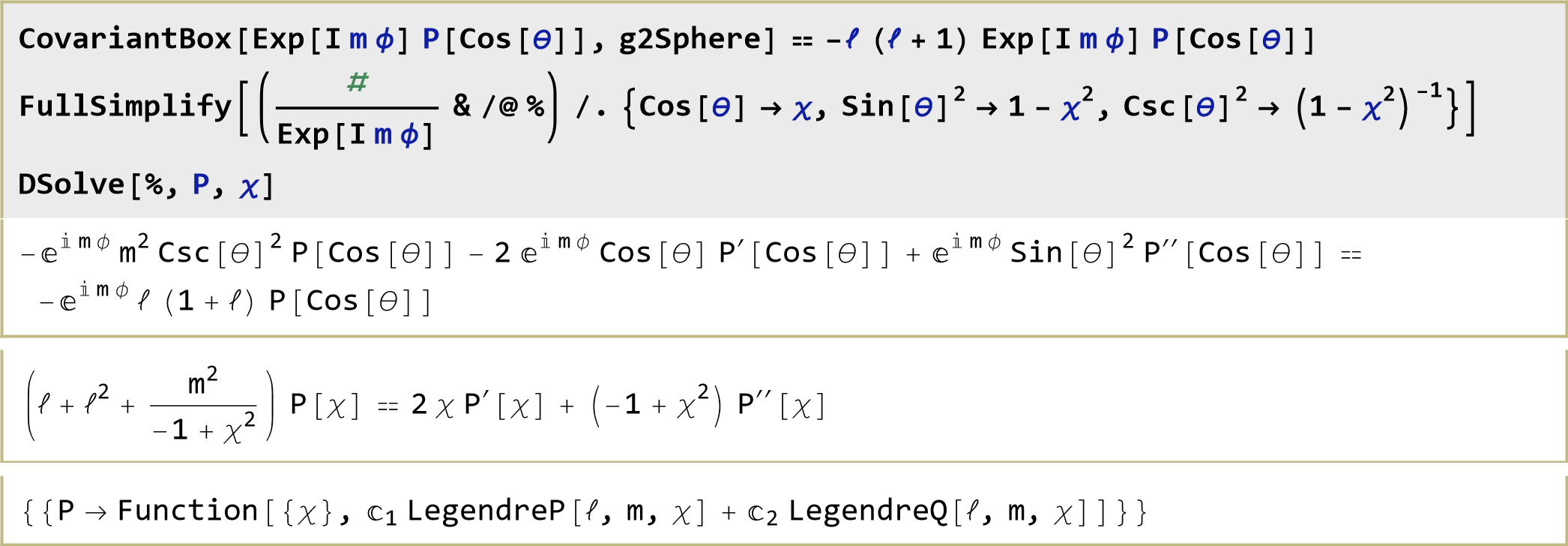}
The two linearly independent solutions to the associated Legendre equation are \(P_{\ell }^m\) and \(Q_{\ell }^m\), though the \(Q_{\ell }^m\)[$\cos \theta$] is singular at the poles $\theta \in \{ 0,\pi \}$ and must therefore be rejected.
\includecode{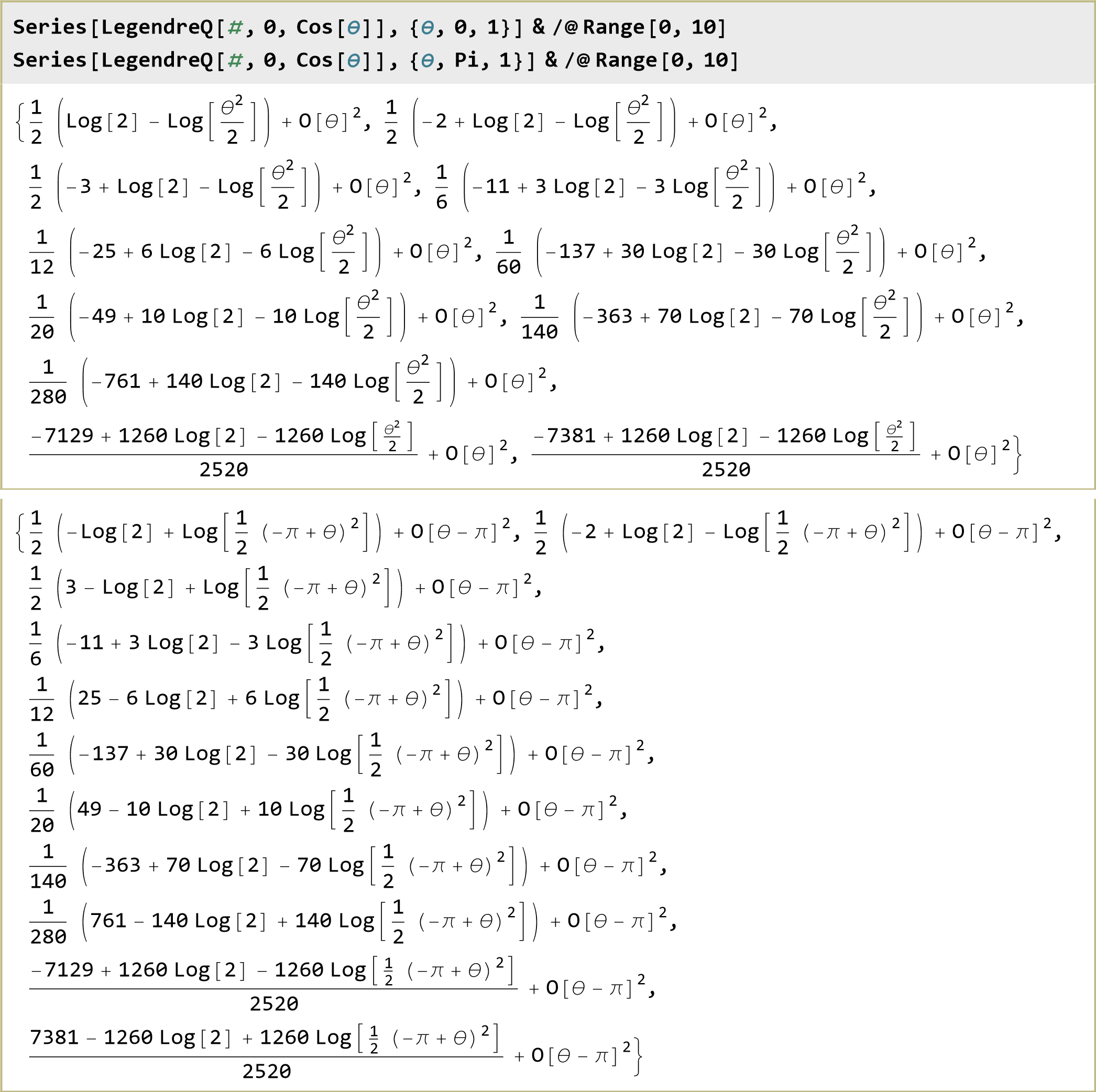}
These eigenfunctions of \(\overset{\to }{L}\cdot \overset{\to }{L}\) are commonly dubbed spherical harmonics \(Y_{\ell }^m\). We have thus argued that \(Y_{\ell }^m\) $\propto $ \(P_{\ell }^m\)[$\cos \theta$] exp[i $m \phi$], where $\ell $ $\in $ $\{$0,1,2,3,...$\}$ and ( $|$m$|$ $\leq $ $\ell$ )$\&\&$(m $\in $ Integers). \MMA{'}s inbuilt version is named SphericalHarmonicY. 

\boldtitle{Basis Vector Fields on 2-Sphere} \qquad Using these $\{$\(Y_{\ell }^m\)$\}$ we may construct basis vector fields on the round 2-sphere by simply taking their gradients and the Hodge duals of these gradients; namely, $\{$\(\nabla ^aY_{\ell }^m\), \(\tilde{\epsilon }^{\text{ab}}\)\(\nabla _bY_{\ell }^m\)$\}$. We first turn these $\{$\(Y_{\ell}^m\)$\}$ into Tensor objects.
\includecode{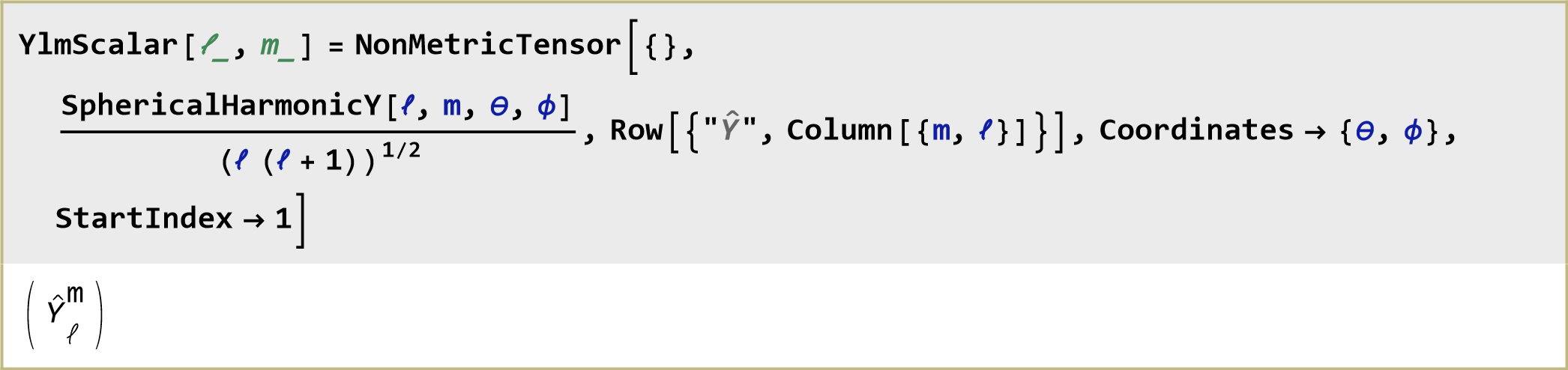}
For example, the $\ell=0$ spherical harmonics are the following.
\includecode{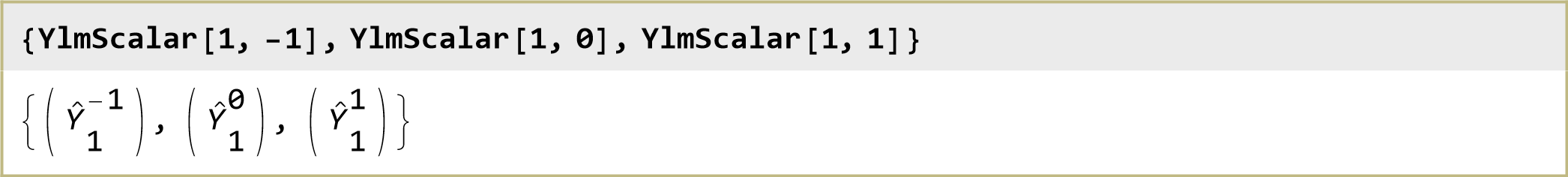}
We may take the gradient using the covariant derivative operator CovariantD, where\\
CovariantD[\(a_-,\text{$\tau \_$Tensor},\text{M$\_$Tensor}\)]
returns a Tensor object corresponding to \(\nabla _a\)$\tau $; namely, covariant derivative on $\tau $ with respect to the geometry M.
\includecode{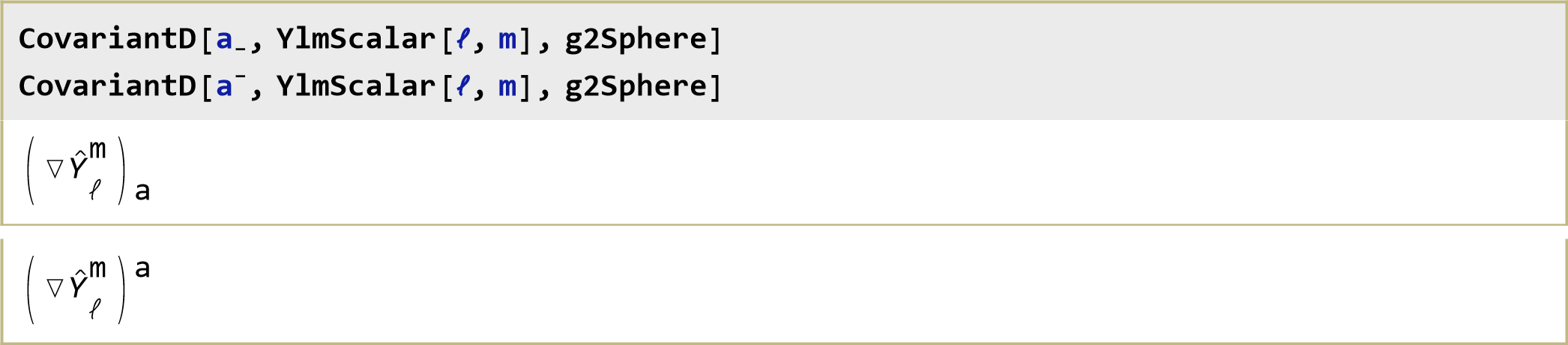}
Additionally, in D dimensional space with metric \textit{M}, the Hodge dual of a rank-N tensor $\tau $ may be computed as CovariantHodgeDual[\(\text{a1}^-,\text{...},\text{ak}^-,\text{$\tau \_$Tensor},\text{M$\_$Tensor}\)], where the output Indices are $\{$ \(\text{a1}_-\),..., \(\text{ak}_-\) $\}$, and $k = D - N$.
\includecode{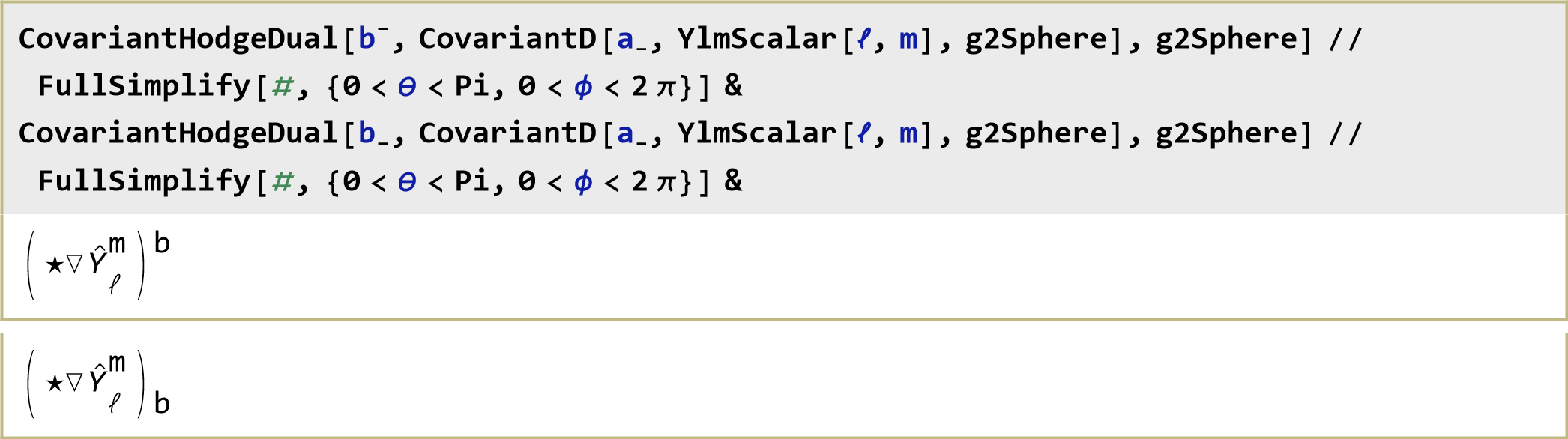}
These basis vector fields $\{$\(\nabla ^aY_{\ell }^m\), \(\tilde{\epsilon }^{\text{ab}}\)\(\nabla _bY_{\ell }^m\)$\}$ are in fact the eigenfunction
of \(\nabla _c\)\(\nabla ^c\) with eigenvalue $1 - \ell(\ell+1)$ for fixed $\ell$ and $m$. Below, we compute \(\nabla _c\)\(\nabla ^c\) \(\nabla ^aY_{\ell }^m\) $=$ -$\lambda $ \(\nabla ^aY_{\ell }^m\) and \(\nabla _c\)\(\nabla ^c\) \(\tilde{\epsilon }^{\text{ab}}\nabla _bY_{\ell}^m\) $=$ $-\lambda $ \(\tilde{\epsilon }^{\text{ab}}\nabla _bY_{\ell }^m\); then employ the scalar eigensystem relation \(\nabla _c\)\(\nabla ^c\) \(Y_{\ell }^m\) = $-\ell $($\ell $+1) \(Y_{\ell }^m\) and its $\theta $- and $\phi $-derivative to solve for $\lambda $.
\includecode{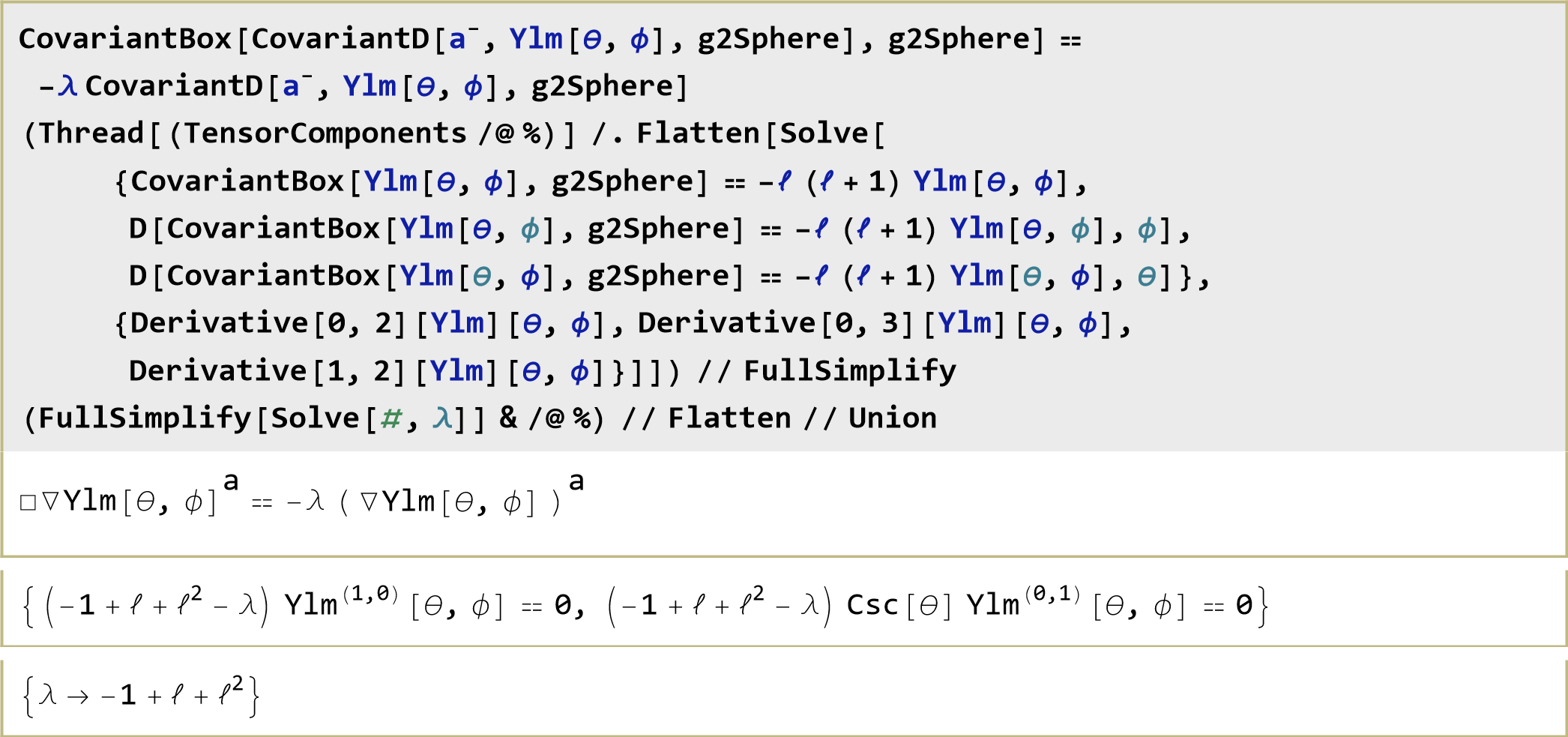}
\includecode{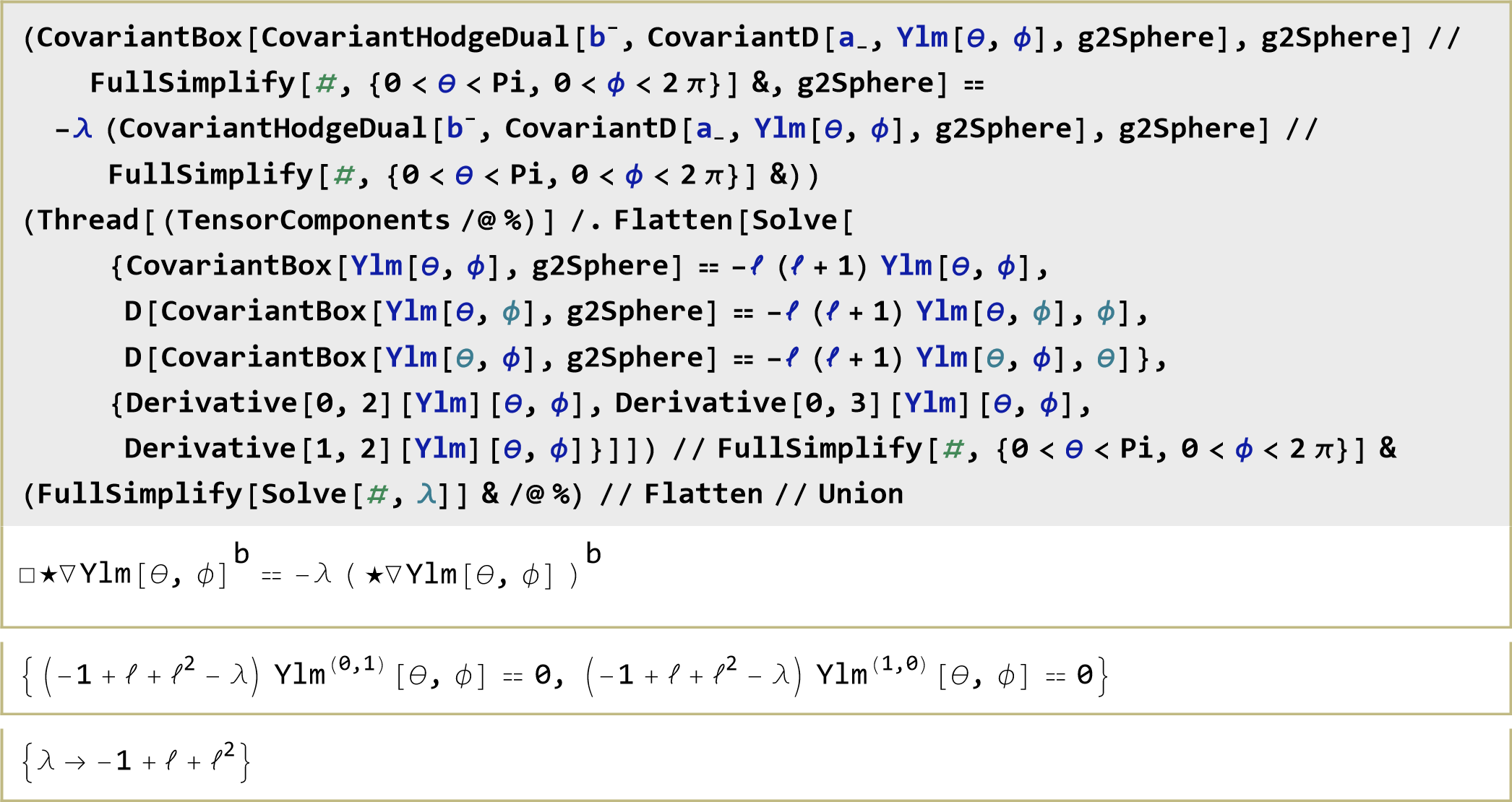}
Given a Tensor object $\tau $, we may move its Indices to some user-specified ones idx (a List of Indices) with MoveIndices[$\tau \_$Tensor,idx$\_$List,M$\_$Tensor], which returns $\tau $ but with its indices moved using the metric contained in M.

Hence, if we wish to create a rank-1 Tensor object with arbitrary position for its index, we further apply MoveIndices to the above definitions.
\includecode{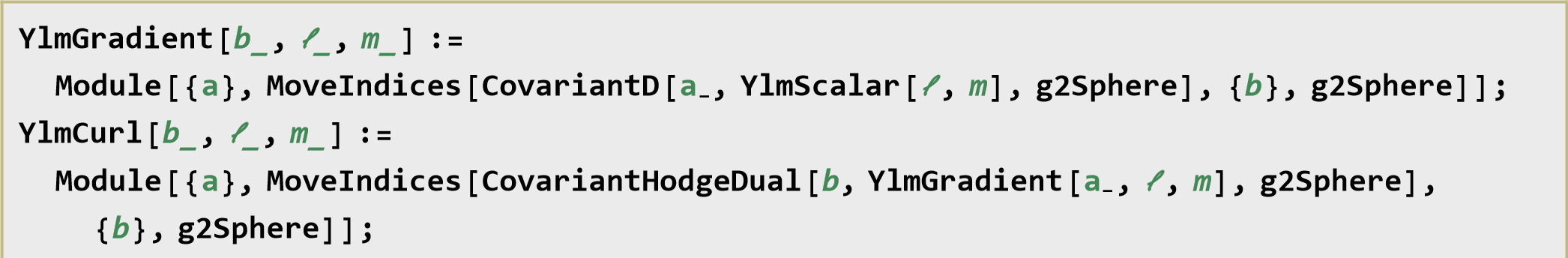}
\includecode{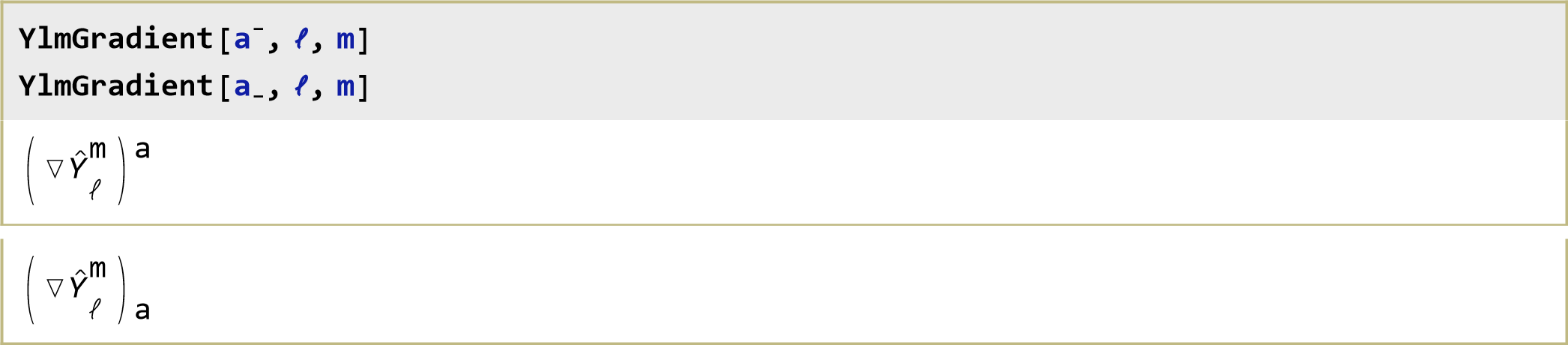}
\includecode{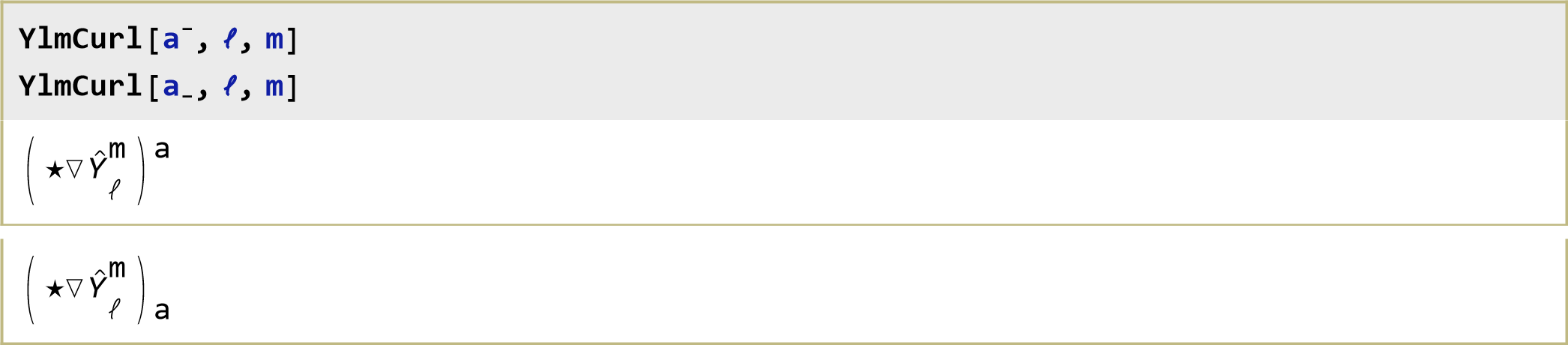}
As its name suggests, ContractTensors contract Tensor objects together whenever they share repeated Indices. Whenever the result is a rank-0 Tensor, the output would be a regular expression.
\includecode{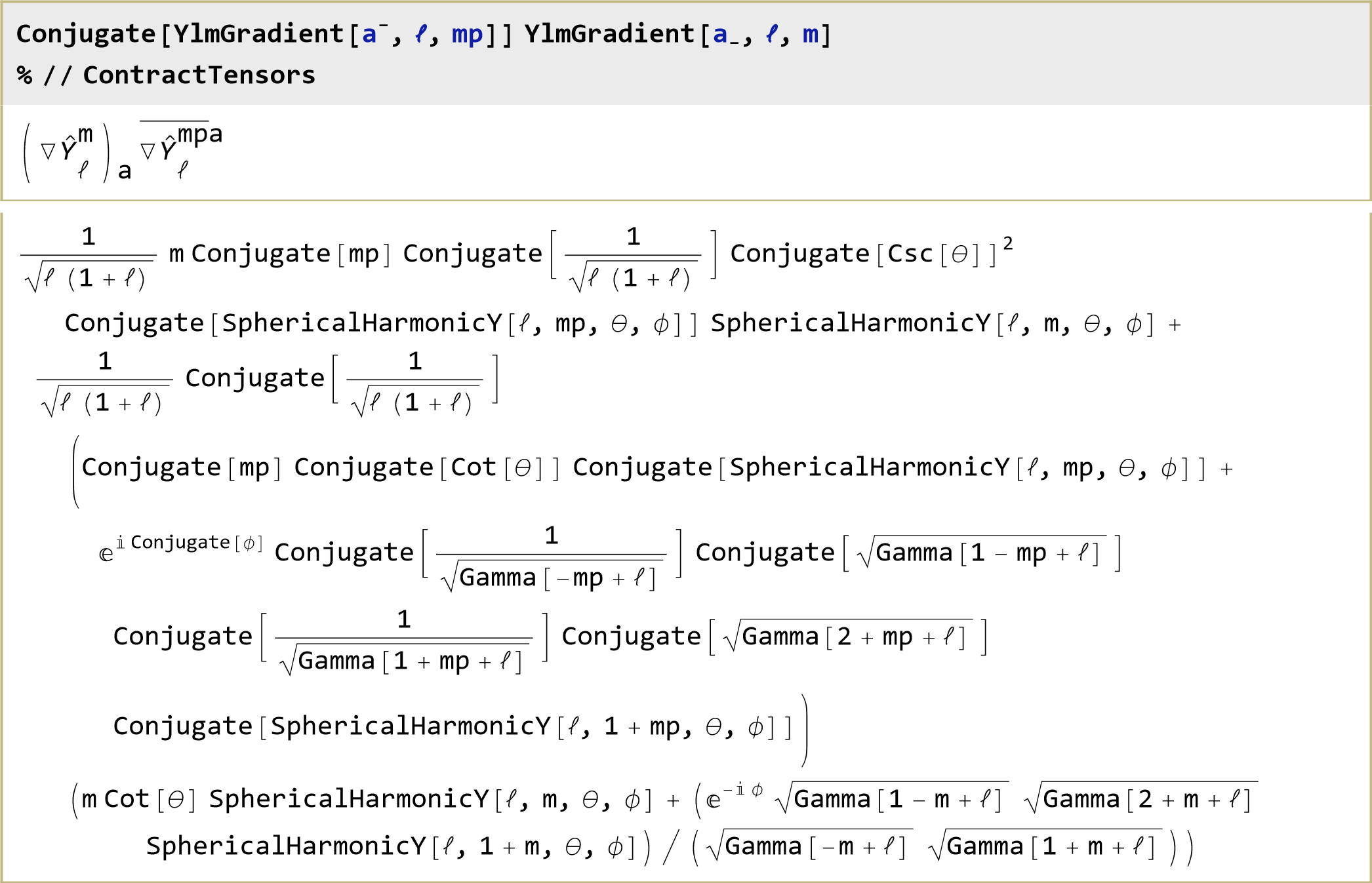}
The gradients involving distinct $\{\ell $,m$\}$ labels are mutually orthonormal. Let us verify 
\[
\int _{S^2}\overline{\nabla _aY_{\ell }^{m'}}\nabla ^aY_{\ell }^m\frac{d{}^2\Omega }{\ell (\ell +1)} = \delta
^{m'}{}_m
\]
by running the inner product over $\ell \in \{1,...,5\}$ and $|m|$, $|m'| \leq \ell$.
\newpage
\includecode{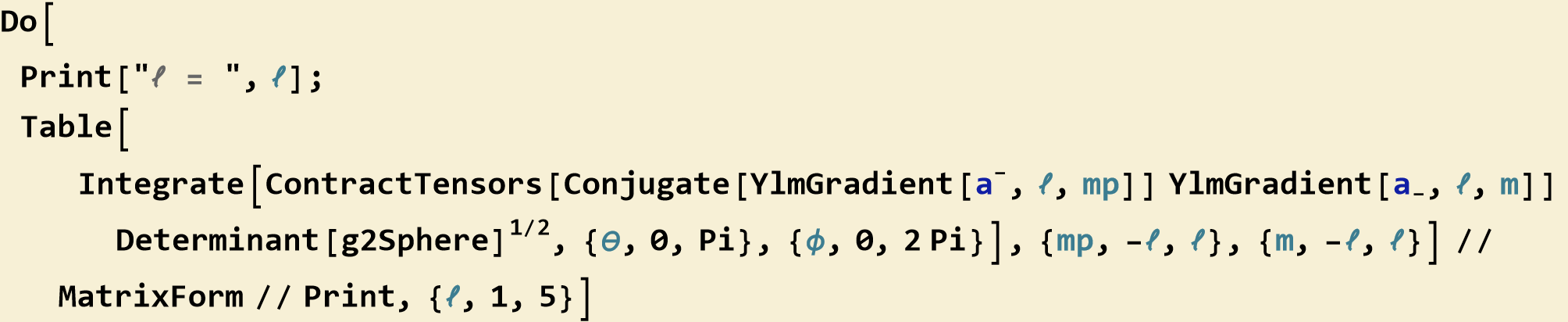}
\begin{figure}[H]
	\includegraphics[width=0.25\textwidth]{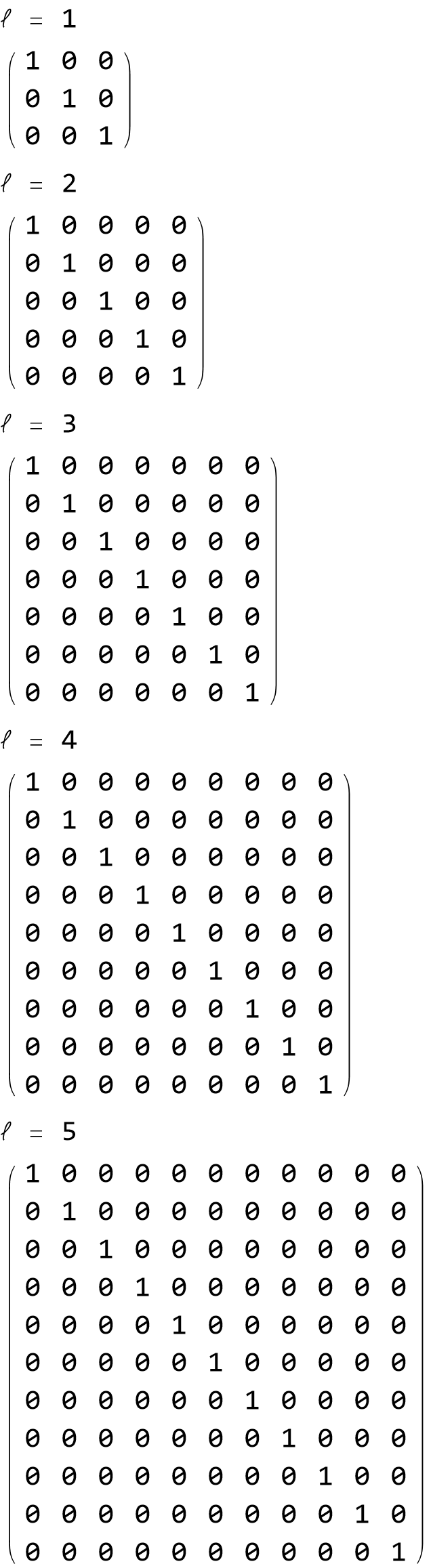}
\end{figure}
The Hodge duals of the gradients involving distinct $\{\ell $,m$\}$ labels are, too, mutually orthonormal:
\[
\int _{S^2}\overline{\tilde{\epsilon }^{\text{ab}}\nabla _bY_{\ell }^{m'}} \tilde{\epsilon }_{\text{ac}}\nabla ^cY_{\ell}^m\frac{d{}^2\Omega }{\ell (\ell +1)} = \delta ^{m'}{}_m
\]
\newpage
\includecode{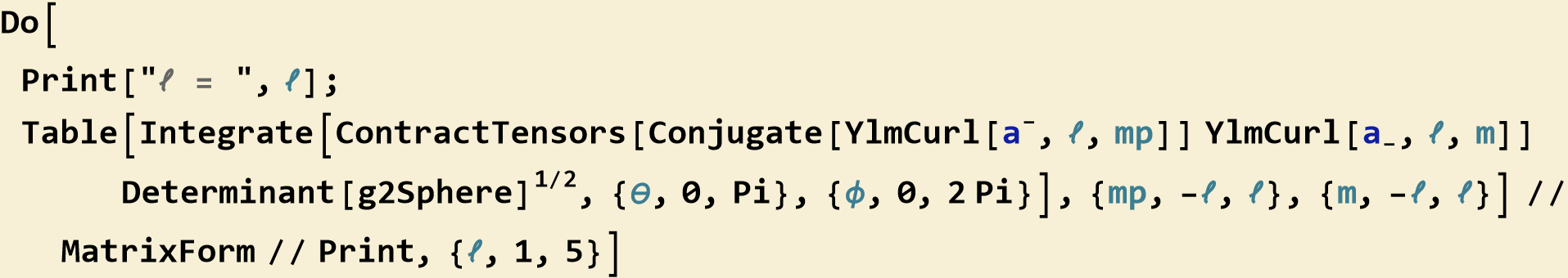}
\begin{figure}[H]
	\includegraphics[width=0.25\textwidth]{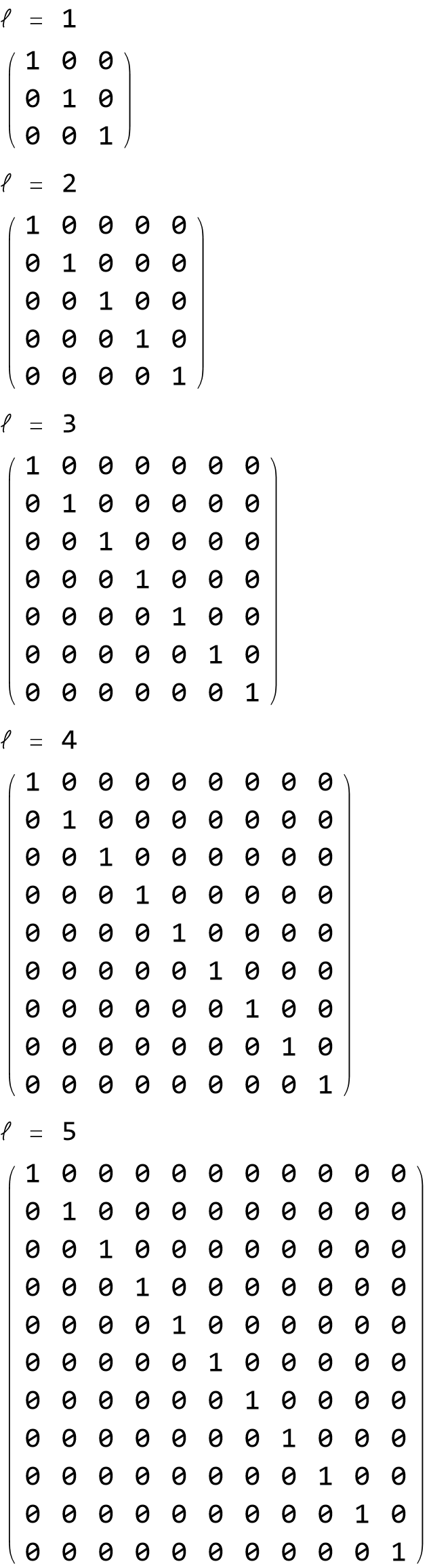}
\end{figure}
The gradients are always orthogonal to their Hodge duals:
\[
\int _{S^2}\overline{\tilde{\epsilon }^{\text{ab}}\nabla _bY_{\ell }^{m'}}\nabla _aY_{\ell }^m\frac{d{}^2\Omega}{\ell (\ell +1)} = 0
\]
\includecode{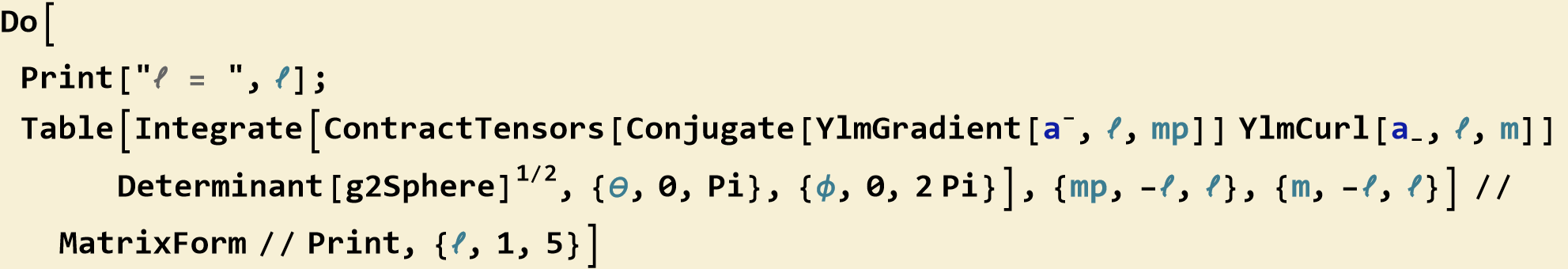}
\begin{figure}[H]
	\includegraphics[width=0.25\textwidth]{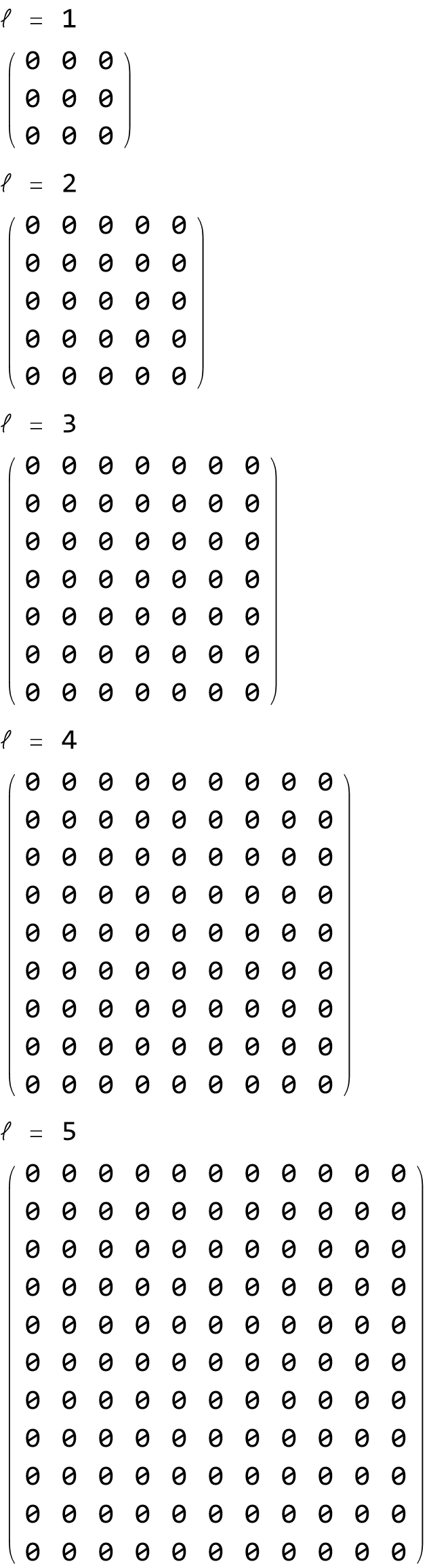}
\end{figure}

\section{3D Vector Calculus}

One way to learn differential geometry itself is to see how 3D multi-variable vector calculus is completely contained with the differential geometry of flat space written in arbitrary coordinate systems. We summarize here the main operations:

Gradient of a scalar $\varphi $, yielding a vector, is \(\nabla ^a\)$\varphi $.

Laplacian of a scalar $\varphi $, yielding a scalar, is \(\nabla _a\)\(\nabla ^a\)$\varphi $.

Divergence of a vector \(A^c\), yielding a scalar, is \(\nabla _a\)\(A^a\).

Curl of a vector \(A^c\), yielding a vector, is \(\tilde{\epsilon }^{\text{abc}}\) \(\nabla _b\)\(A_c\) = \(\frac{1}{2}\)\(\tilde{\epsilon }^{\text{abc}}\) \((dA)_{\text{bc}}\) = \(\left(\, ^{\star }(dA)\right)^a\), where \(\tilde{\epsilon }^{\text{abc}}\) is the covariant Levi-Civita pseudo-tensor. Here, we have used the exterior derivative of a 1-form \((dA)_{\text{ab}}=\partial _aA_b-\partial _bA_a\equiv \partial _{[a}A_{b]}\).

\subsection{Cartesian Coordinates}

First we begin with Cartesian coordinates $\{ x^1,x^2,x^3 \in \mathbb{R} \}$, where the coordinate basis coincides with the orthonormal basis since the metric is already a sum of squares consisting solely of infinitesimal displacements.
\includecode{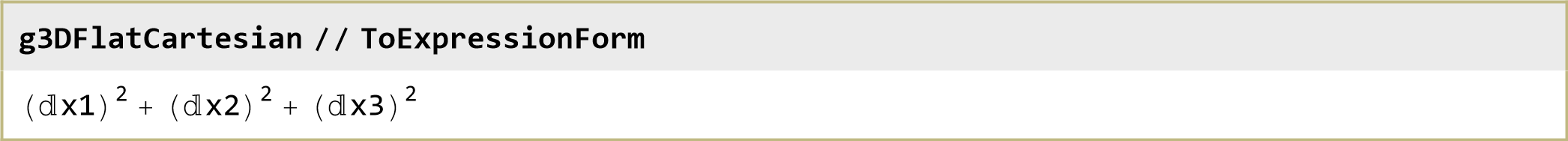}
\boldtitle{Gradient on scalar}
\includecode{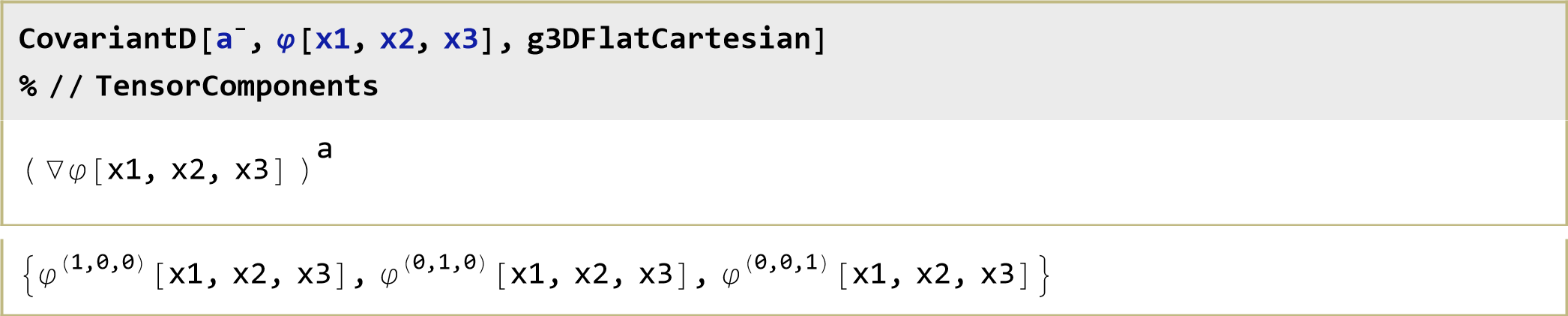}
\boldtitle{Laplacian on scalar}
\includecode{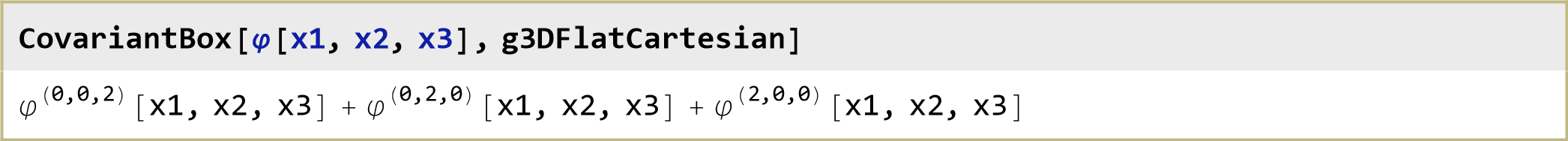}
\boldtitle{Divergence of a vector}
First we define the vector A.
\includecode{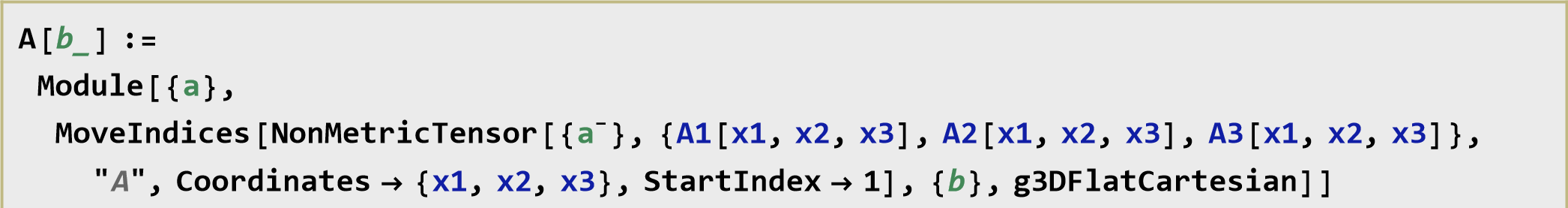}
Then we take its divergence.
\includecode{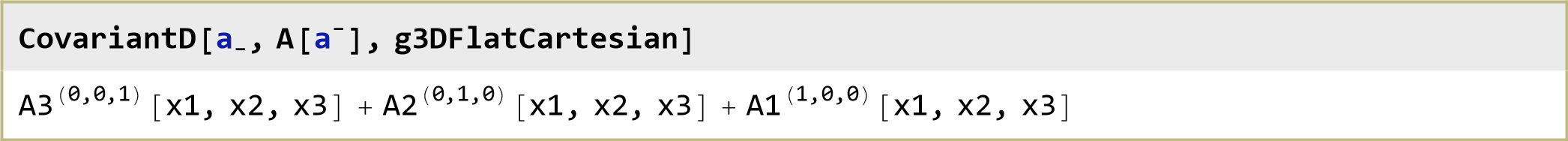}
\boldtitle{Curl of a vector} \qquad In \TC, ExteriorD[a,$\tau \_$Tensor] (short form \(d_a\tau\)) produces the exterior derivative of the Tensor object $\tau $; the \textit{a} is an index pre-pended to those of $\tau $. If the first argument is omitted \TC\ would generate a random unique index. For instance, we may take the exterior derivative of the 1-form \textit{A} in the following ways.
\includecode{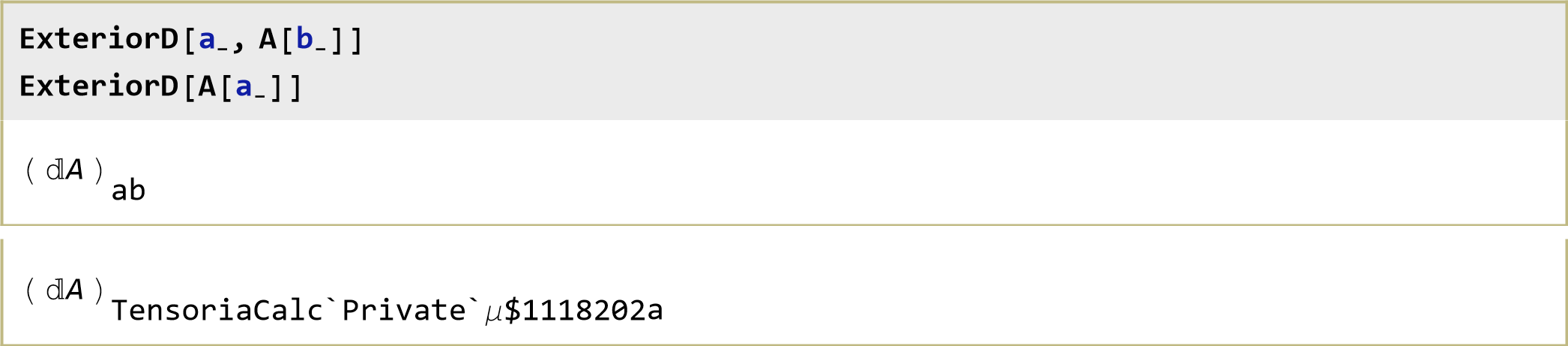}
\includecode{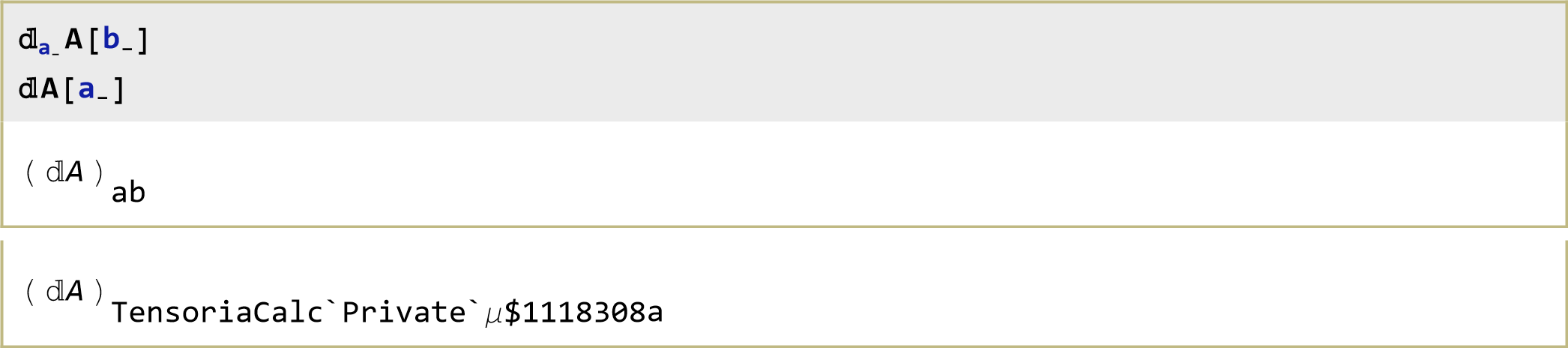}
Let us proceed to take the vector calculus curl of the vector \textit{A} by computing the Hodge dual of its exterior derivative. 
\includecode{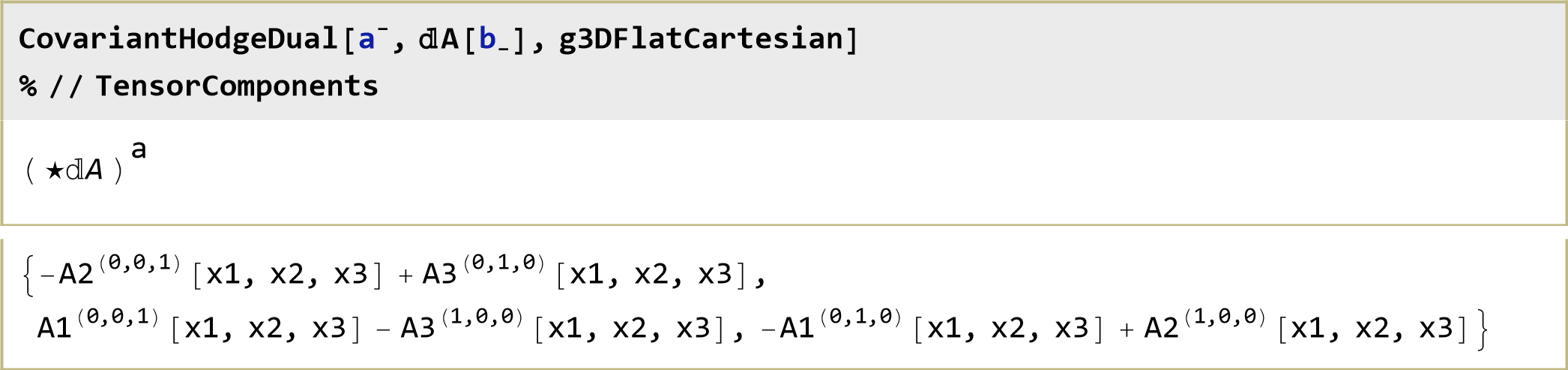}
\boldtitle{Eigensystem of Laplacian} \qquad The eigenfunction of the Hermitian Laplacian in infinite flat space is the plane wave \(E^{I \overset{\rightharpoonup}{k}.\overset{\rightharpoonup}{x}}\), with the corresponding eigenvalue -\(k^2\equiv -\overset{\rightharpoonup}{k}\cdot \overset{\rightharpoonup}{k}\). It is also the eigenfunction of the generators of spatial translations $\{$-I \(\partial _{x^i}\)$\}$. The set of plane waves with different wave vectors \(\overset{\rightharpoonup}{k}\) form the set of basis functions underlying Fourier analysis.
\includecode{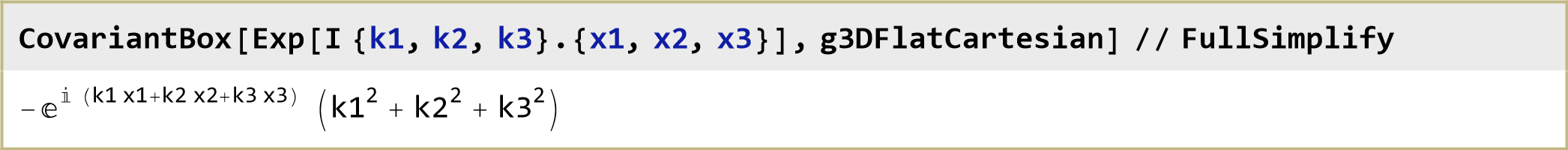}

\subsection{Polar Coordinates}

Next we turn to polar coordinates $\{ \rho \geq 0, 0 \leq \phi < 2\pi, z \in \mathbb{R} \}$.
\includecode{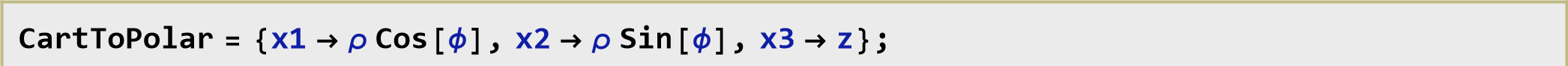}
Use CoordinateTransformation to convert the 3D flat metric to the polar coordinate system.
\includecode{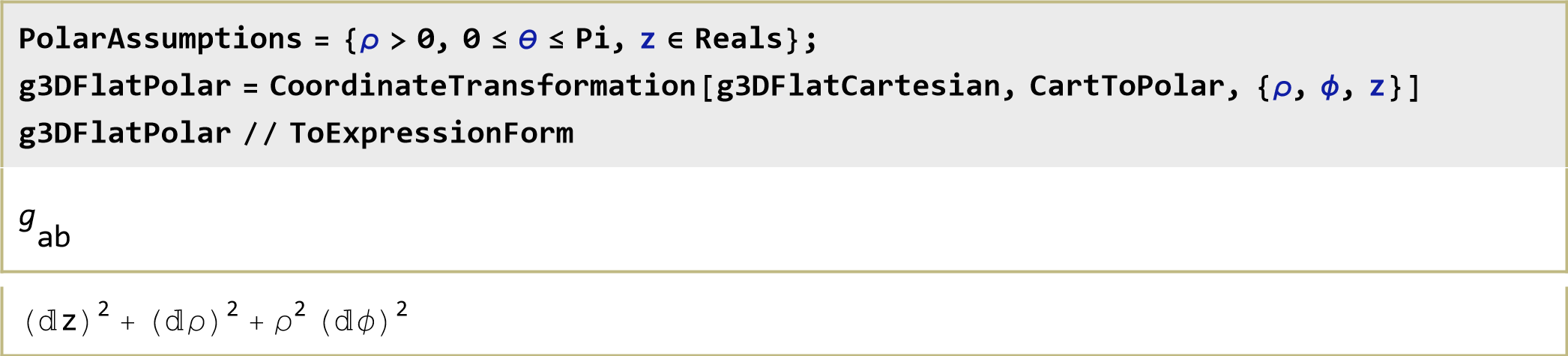}
\boldtitle{Orthonormal Frame Fields} \qquad To compute the orthonormal frame fields and append them to the metric Tensor object g3DFlatPolar, we apply OrthonormalFrameField to g3DFlatPolar, with the FlatMetric $\{$1,1,1$\}$ and Indices $\{$\(\hat{a}\),b$\}$ -- \textit{a} refers to the orthonormal index and \textit{b} the coordinate one on the orthonormal frame fields; for e.g., \(\varepsilon ^{\hat{a}}{}_b\) or \(\varepsilon _{\hat{a}}{}^b\). A given component of the input FlatMetric tells \TC\ what sign (+ or -) is to be multiplied to the corresponding diagonal component of the metric before taking the square root to form the orthonormal frame field matrix; in our case, the non-zero components are \(\varepsilon ^{\hat{\rho }}{}_{\rho }=\left(+g_{\rho \rho }\right){}^{1/2}\), { }\(\varepsilon ^{\hat{\phi }}{}_{\phi }=\left(+g_{\phi \phi }\right){}^{1/2}\), { }\(\varepsilon ^{\hat{z}}{}_z=\left(+g_{\text{zz}}\right){}^{1/2}\). \TC\ currently does not support automatic computation of orthonormal frame fields for non-diagonal metrics.
\includecode{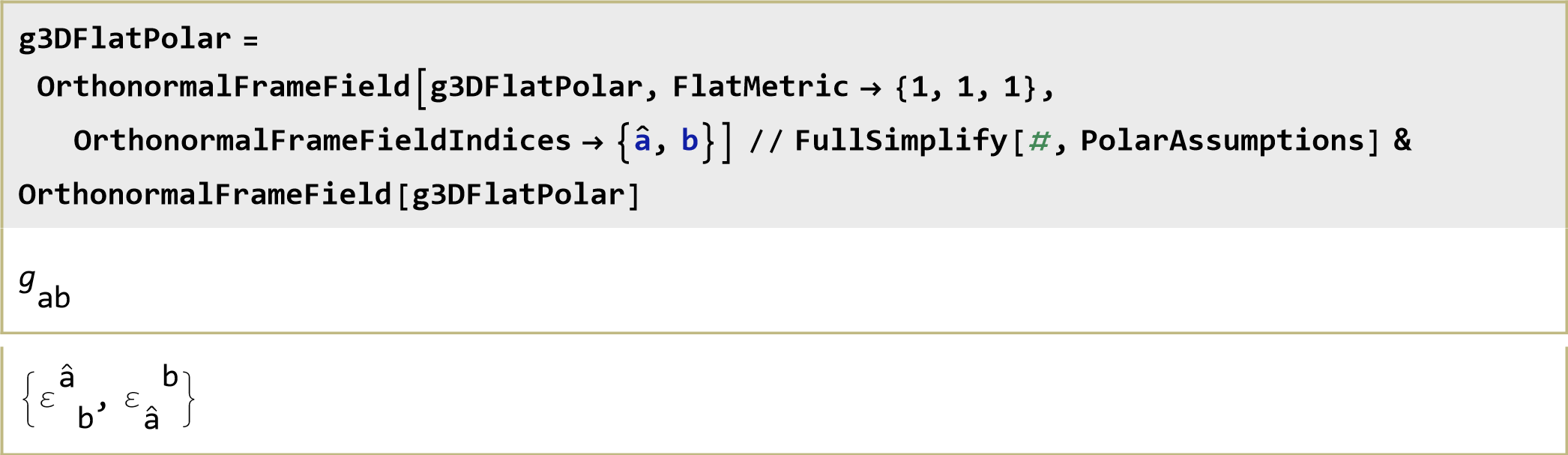}
\boldtitle{Gradient on scalar} \qquad We may now use MoveIndices to express the CovariantD on a scalar in either a coordinate or orthonormal basis.
\includecode{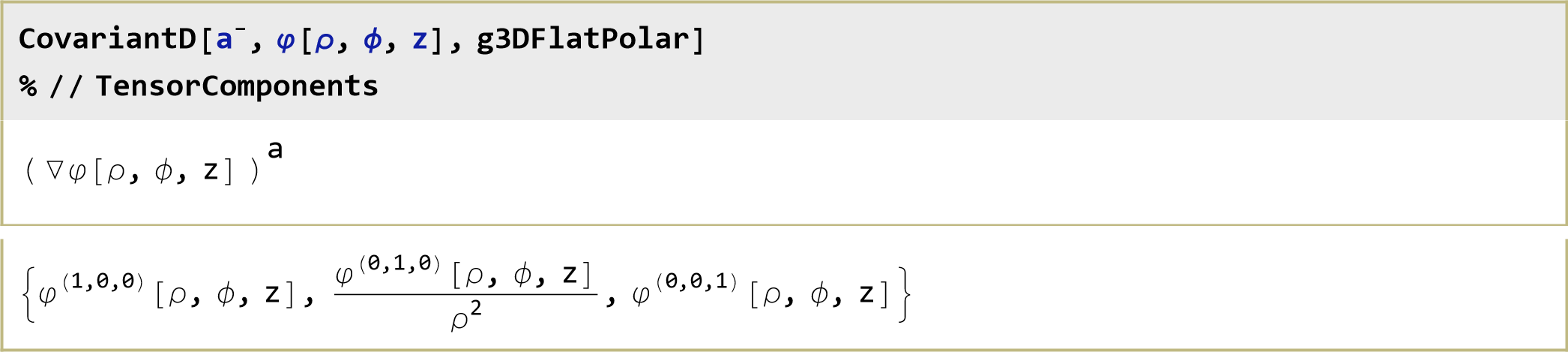}
\includecode{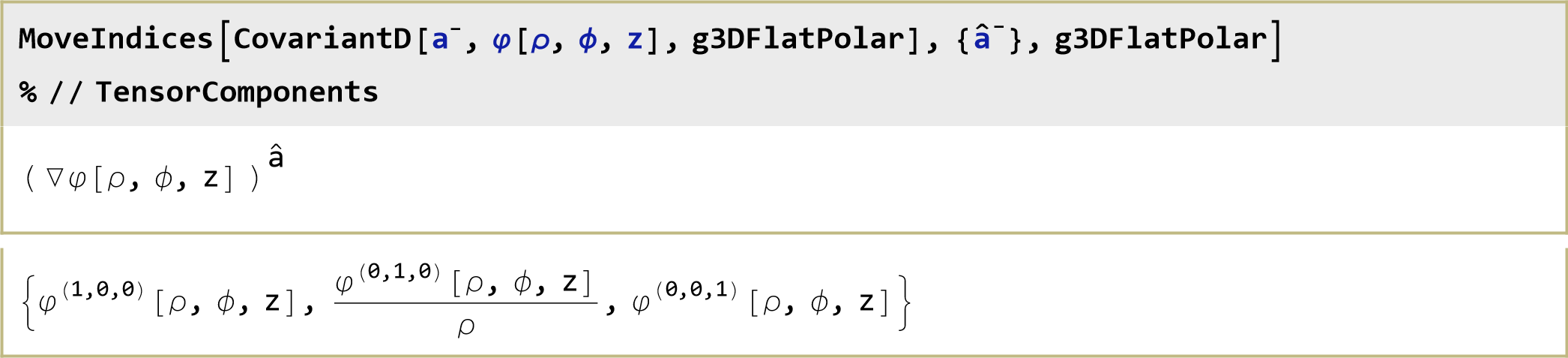}
\boldtitle{Laplacian on scalar}
\includecode{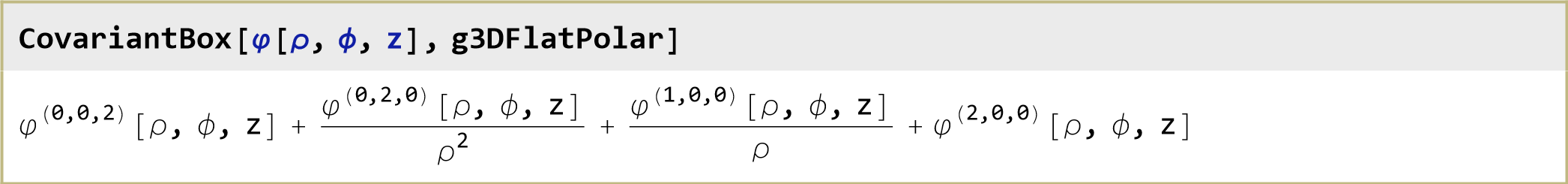}
\boldtitle{Divergence of a vector} \qquad To derive the vector calculus formulas provided in J.D.Jackson{'}s electromagnetism textbook \cite{Jackson:1998nia}, we first re-express the vector \textit{A} in terms of its orthonormal components, but still remaining within the coordinate basis. We append the names of orthonormal frame components with a {``}h{''}; for e.g., $\{$A$\rho $h, A$\phi $h, Azh$\}$ corresponds to the components of \(A^{\hat{a}}\).

Let us first define the vector A in a coordinate basis.
\includecode{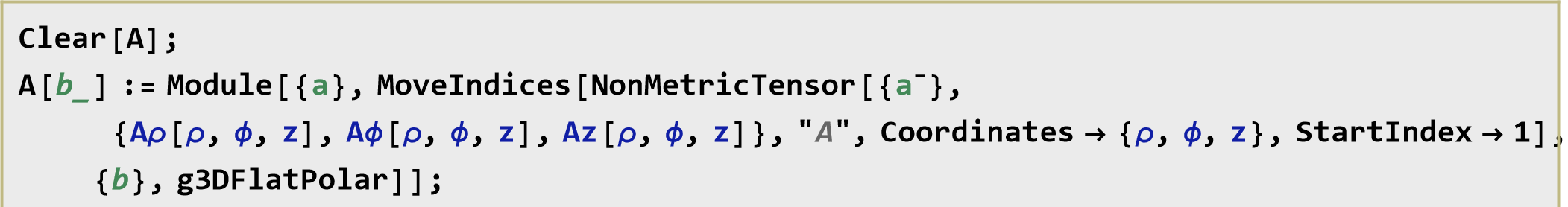}
We proceed to use MoveIndices to convert it to an orthonormal basis, then derive from it the relations between the coordinate and orthonormal components.
\includecode{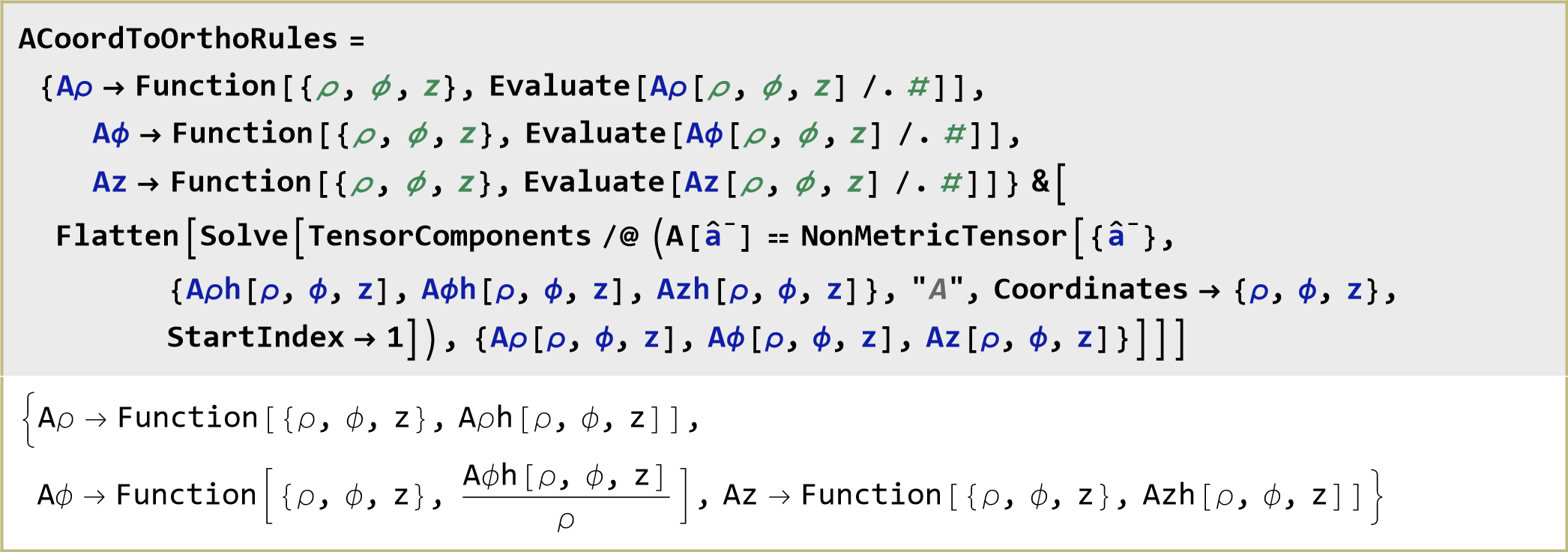}
Let{'}s check that the orthonormal components are recovered.
\includecode{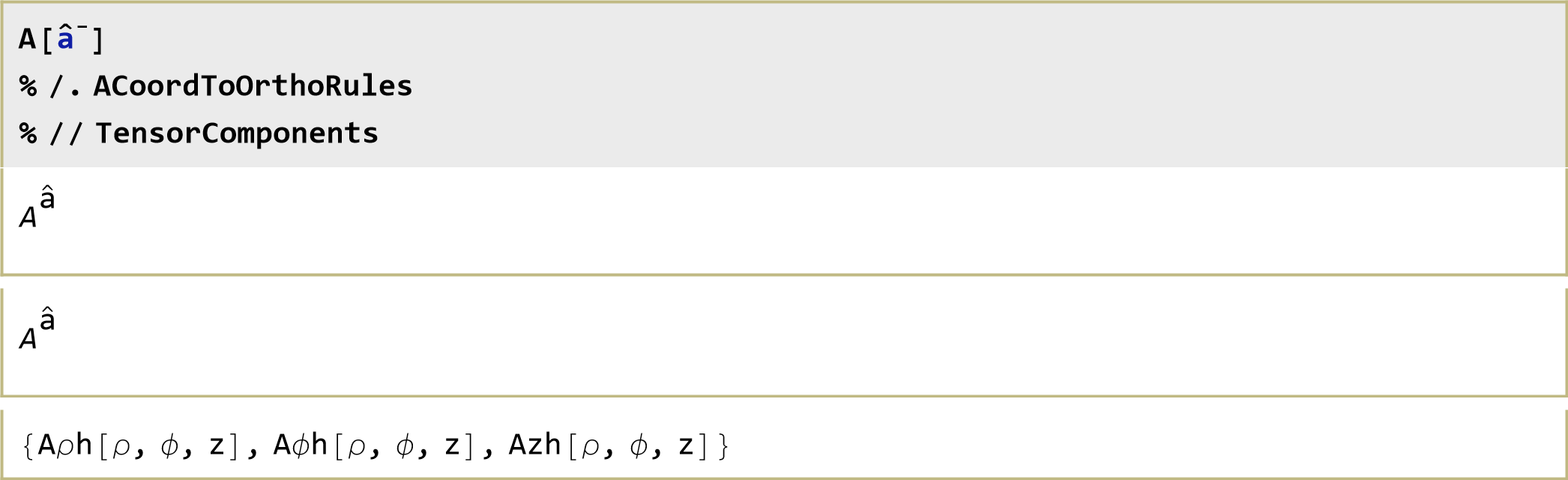}
Now, we take the divergence of \textit{A}.
\includecode{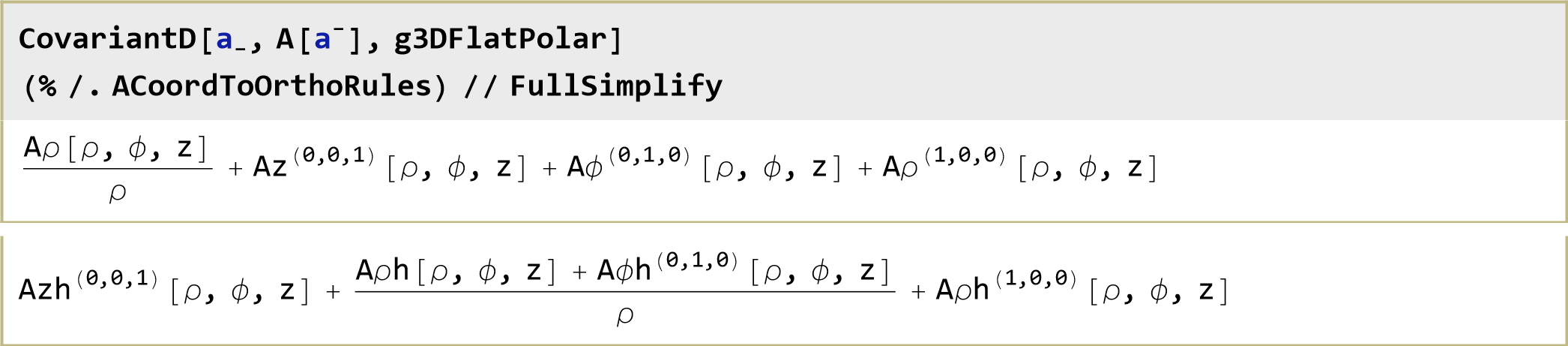}
\boldtitle{Curl of a vector}
Compute the curl of \textit{A}, first in the coordinate basis
\includecode{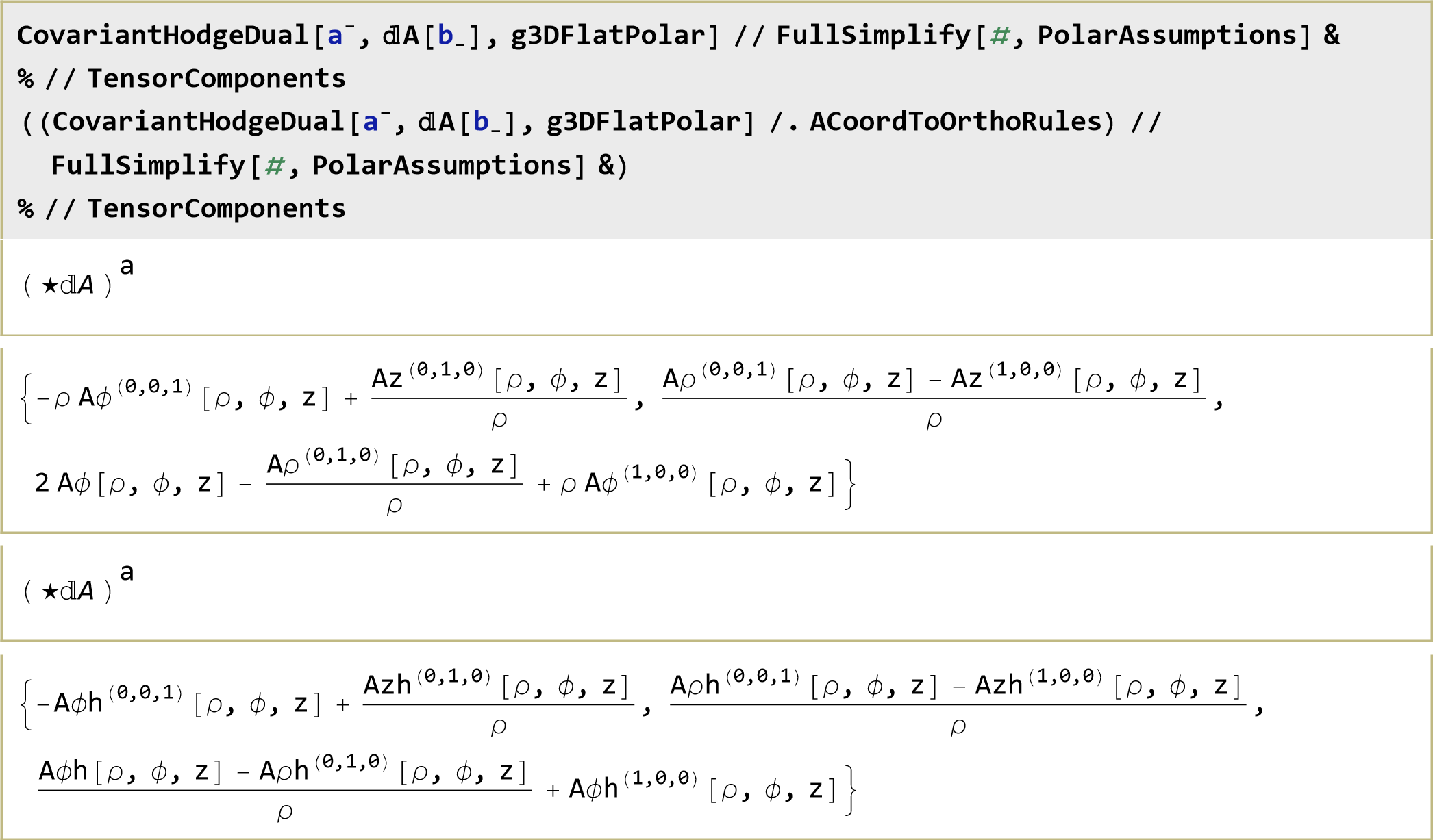}
and next in the orthonormal basis
\includecode{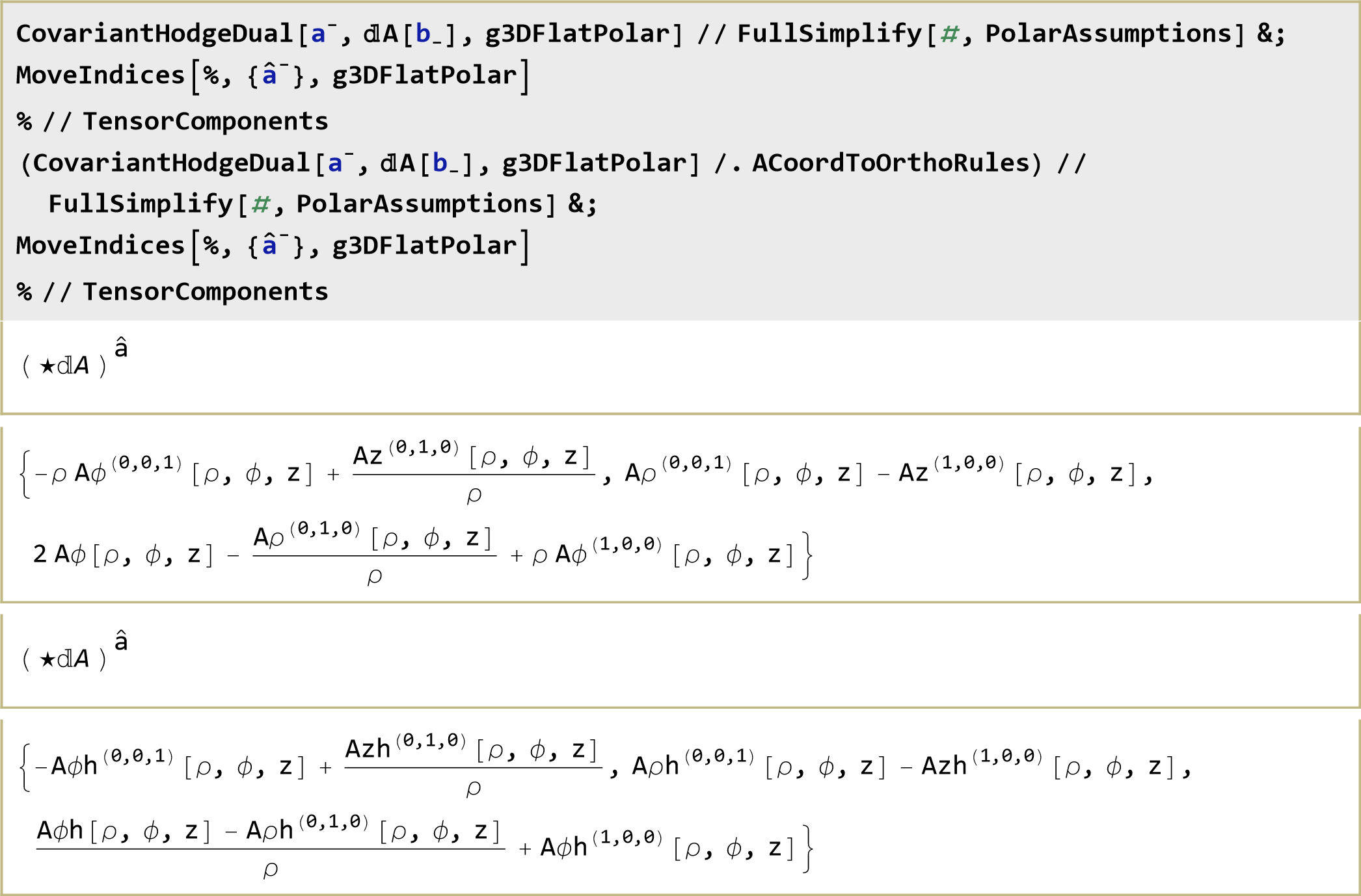}
\boldtitle{Eigensystem of Laplacian} \qquad The generator of rotations on the (1,2)-plane -I \(\partial _{\phi }\) (with eigenfunctions $\{$\(E^{I m \phi }\)$\}$, m $\in $ Integers), the generator of spatial translations along the 3-axis -I \(\partial _z\) (with eigenfunctions $\{$\(E^{I q z}\)$\}$, \textit{q} $\in $ Reals), and the 3D Laplacian (with eigenvalues $\{$-\(k^2\)$\}$) are mutually commuting Hermitian operators. To find their simultaneous eigenfunctions we postulate a separation-of-variables ansatz: R[$\rho $] \(E^{I m \phi }\) \(E^{I q z}\). This yields the Bessel differential equation for the radial mode function R[$\rho $].
\includecode{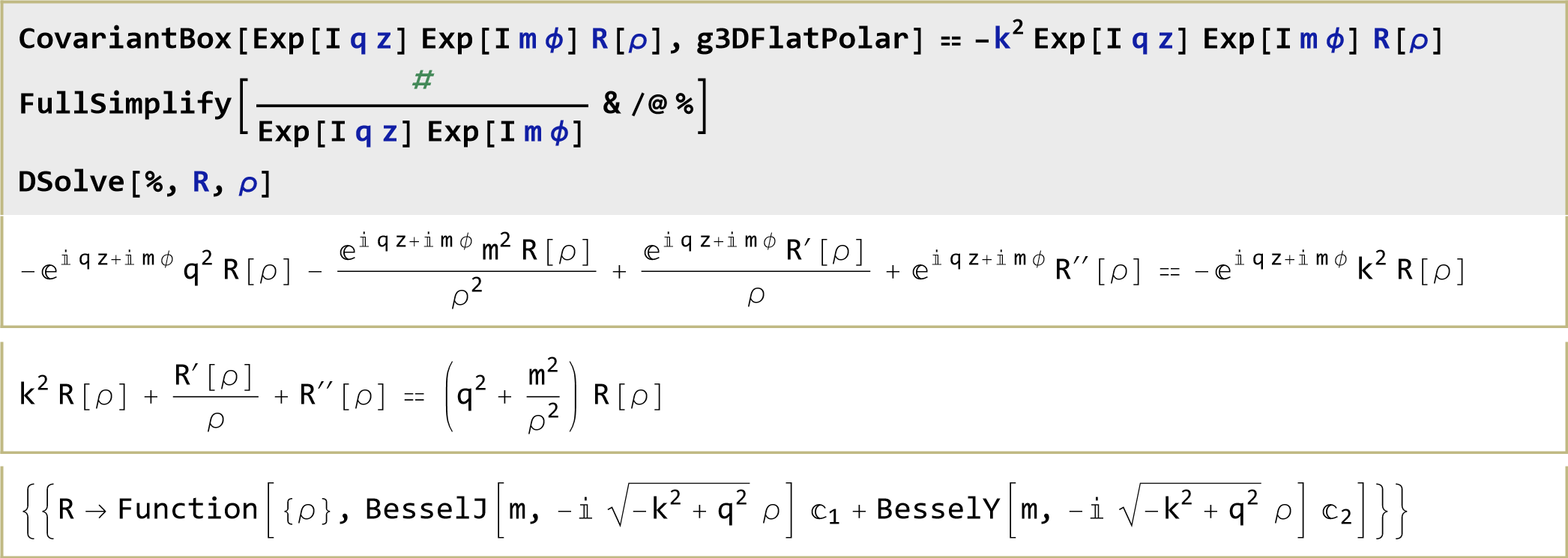}
The simultaneous eigenfunction of the generator of rotations, the generator of spatial translations along the 3-axis, and the 3D Laplacian is therefore proportional to BesselJ[m, \(\left(k^2-q^2\right)^{1/2}\)$\rho $] \(E^{I m \phi }\) \(E^{I q z}\) because the SphericalBesselY blows up as \textit{r} $\to $ 0.
\includecode{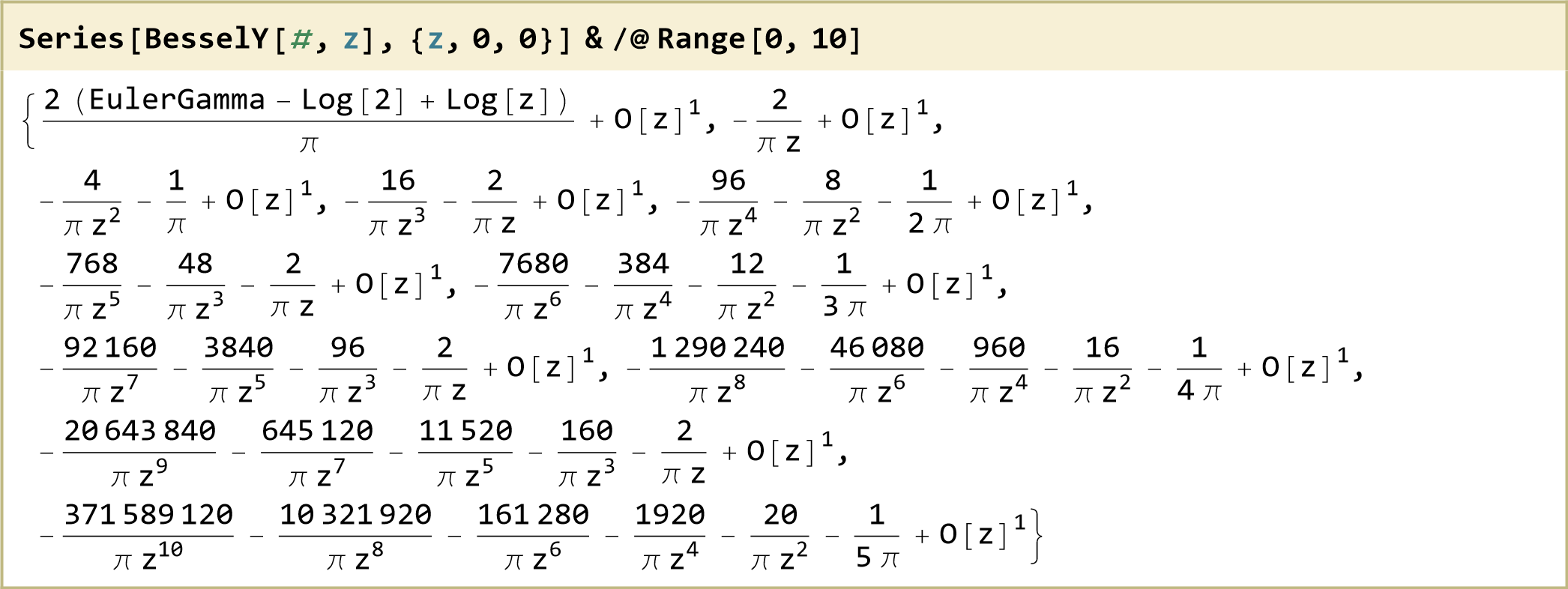}

\subsection{Spherical Coordinates}

Next we turn to spherical coordinates $\{r \geq 0, 0 \leq \theta \leq \pi, 0 \leq \phi < 2\pi \}$.
\includecode{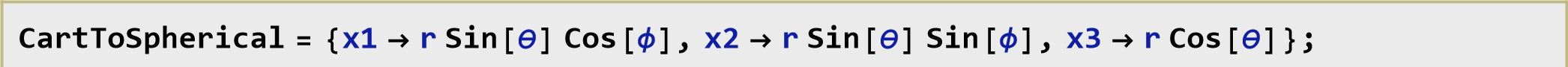}
Use CoordinateTransformation to convert the 3D flat metric to the spherical coordinate system.
\includecode{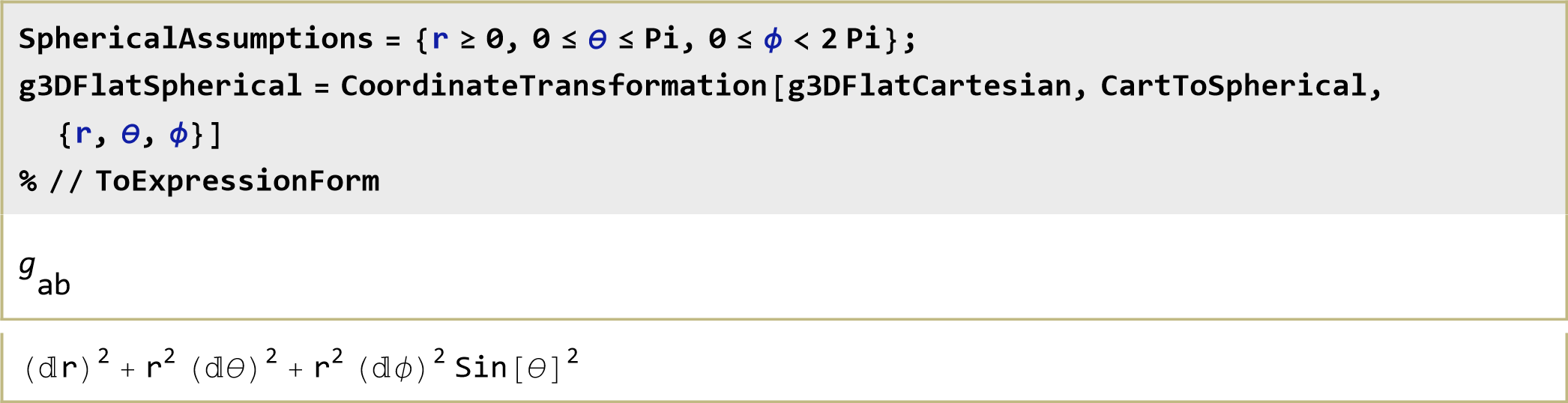}
Use OrthonormalFrameField to compute the orthonormal frame fields.
\includecode{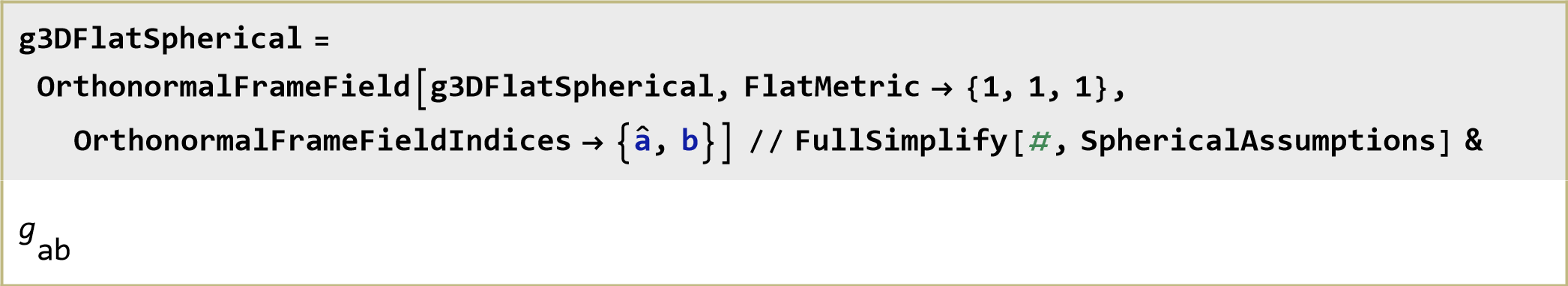}
\boldtitle{Gradient on scalar} \qquad We may now use MoveIndices to express the CovariantD on a scalar in either a coordinate or orthonormal basis.
\includecode{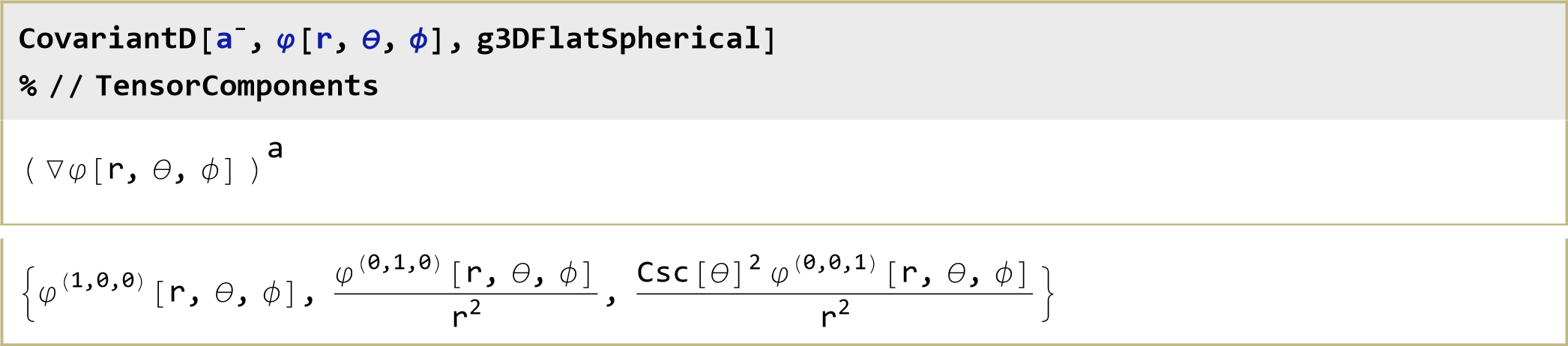}
\includecode{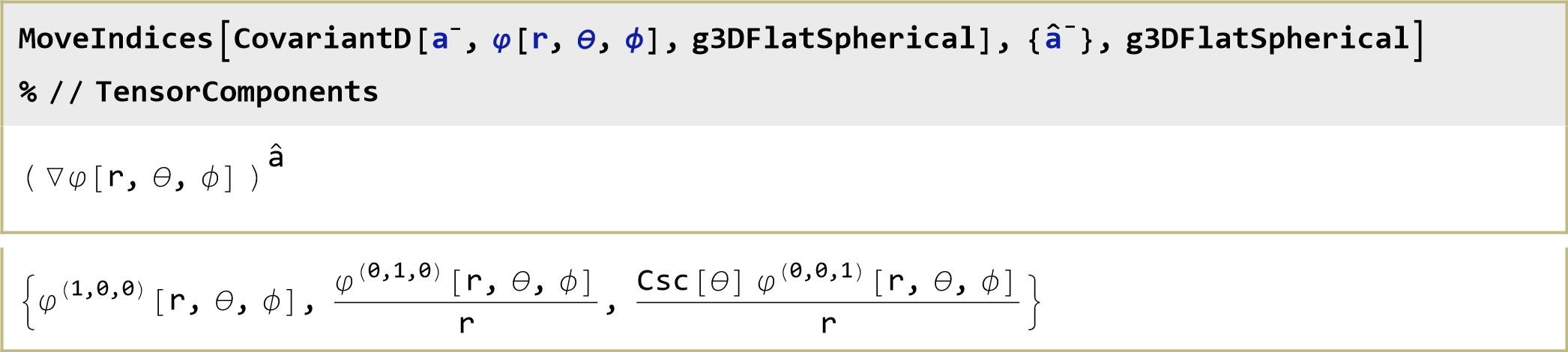}
\boldtitle{Laplacian on scalar} 
\includecode{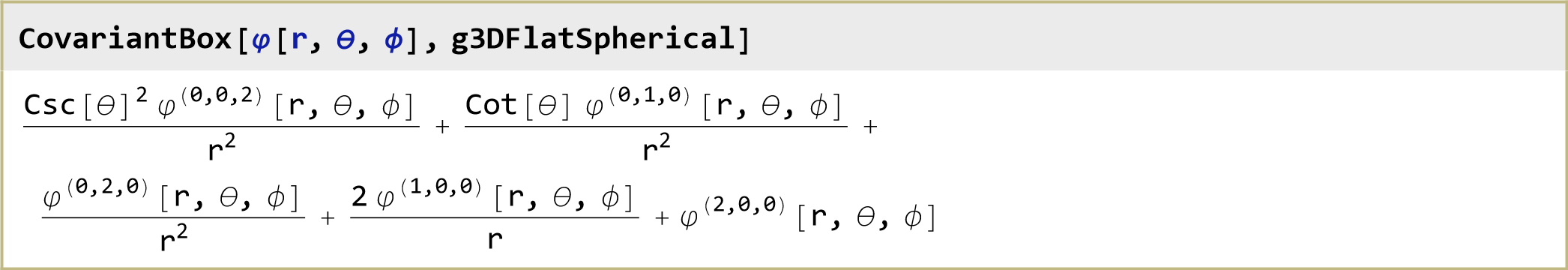}
\boldtitle{Divergence of a vector} \qquad Let us first define the vector A in a coordinate basis.
\includecode{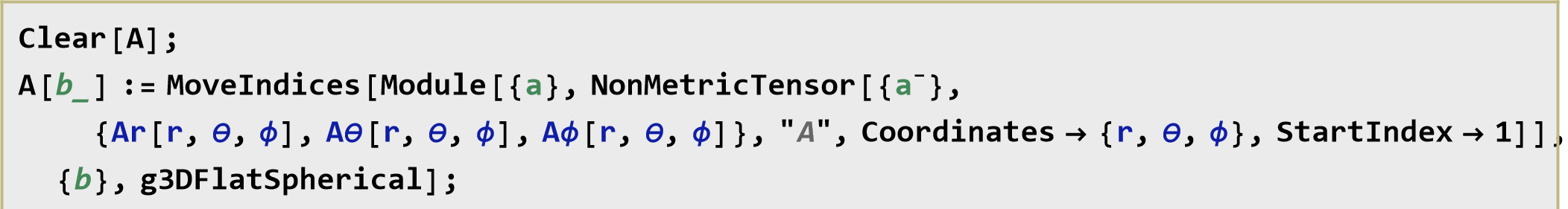}
We then use MoveIndices to convert it to an orthonormal basis, then derive from it the relations between the coordinate and orthonormal components.
\includecode{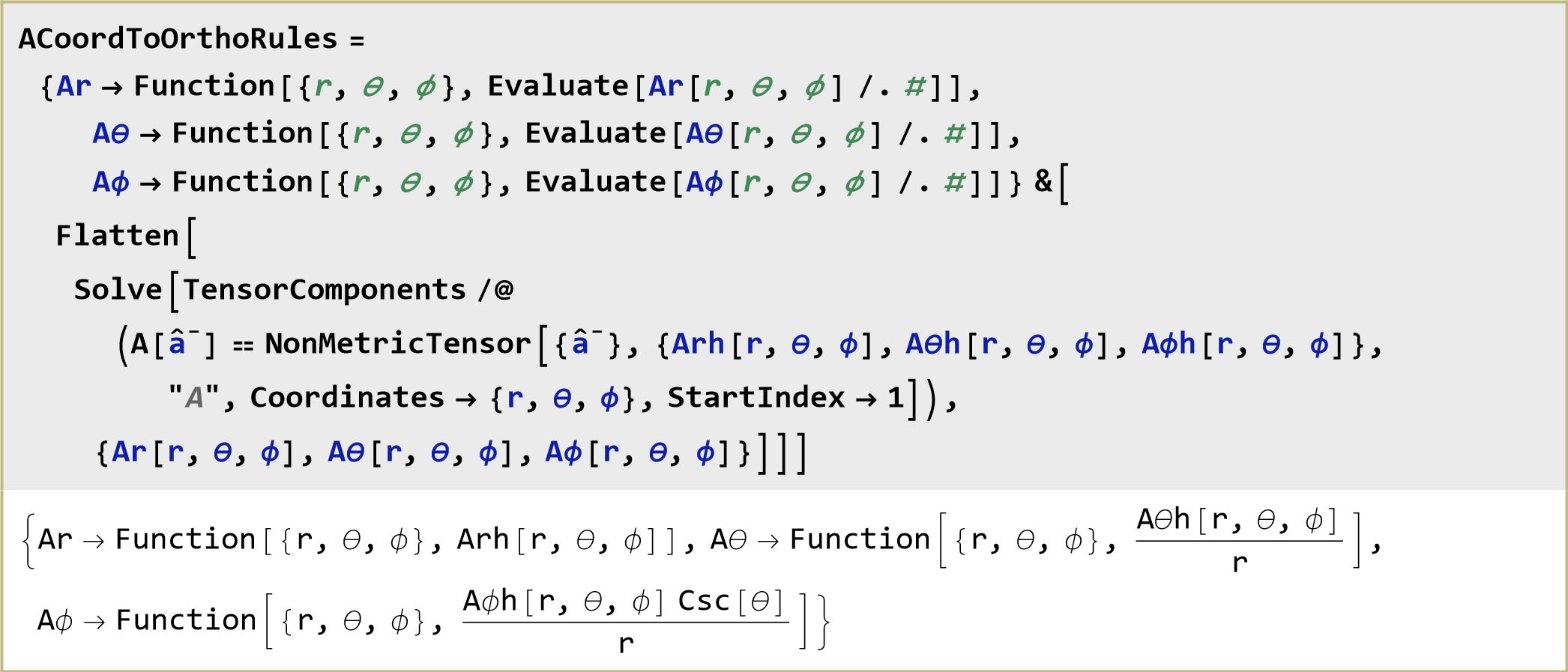}
Then we take the divergence of \textit{A}.
\includecode{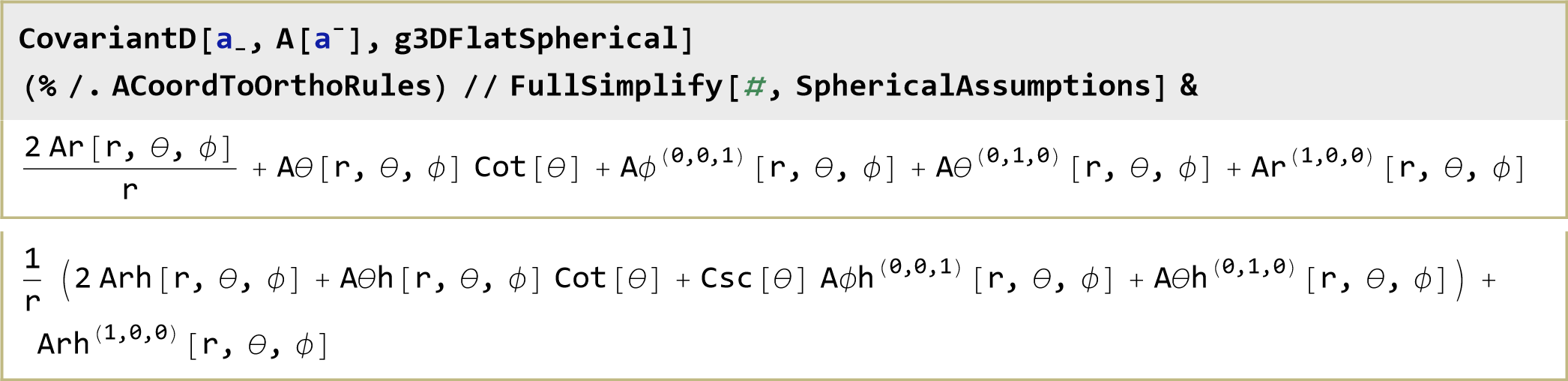}
\boldtitle{Curl of a vector} \qquad Compute the curl of \textit{A} in the coordinate basis
\includecode{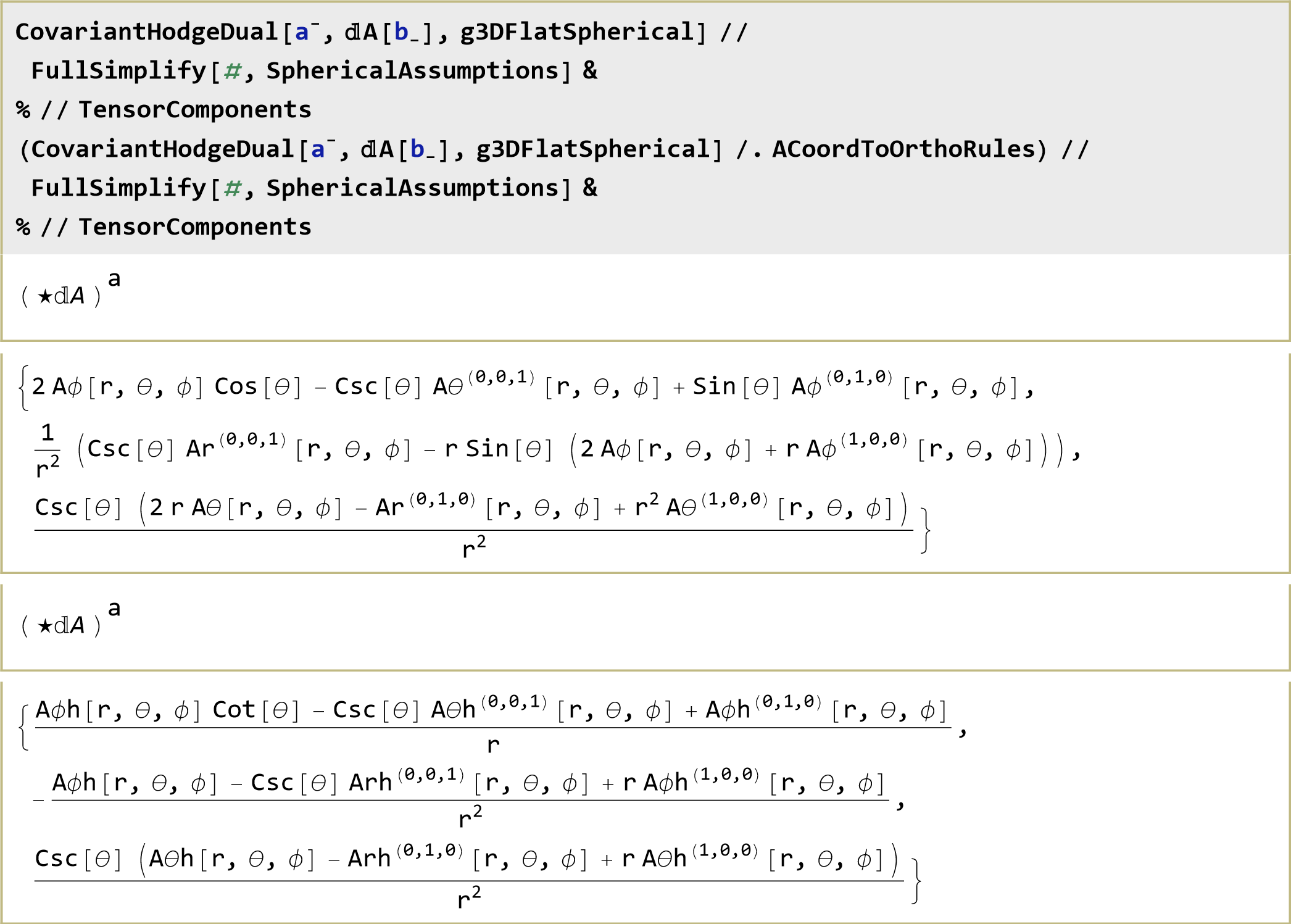}
\newpage
and in the orthonormal basis
\twopicincludecode{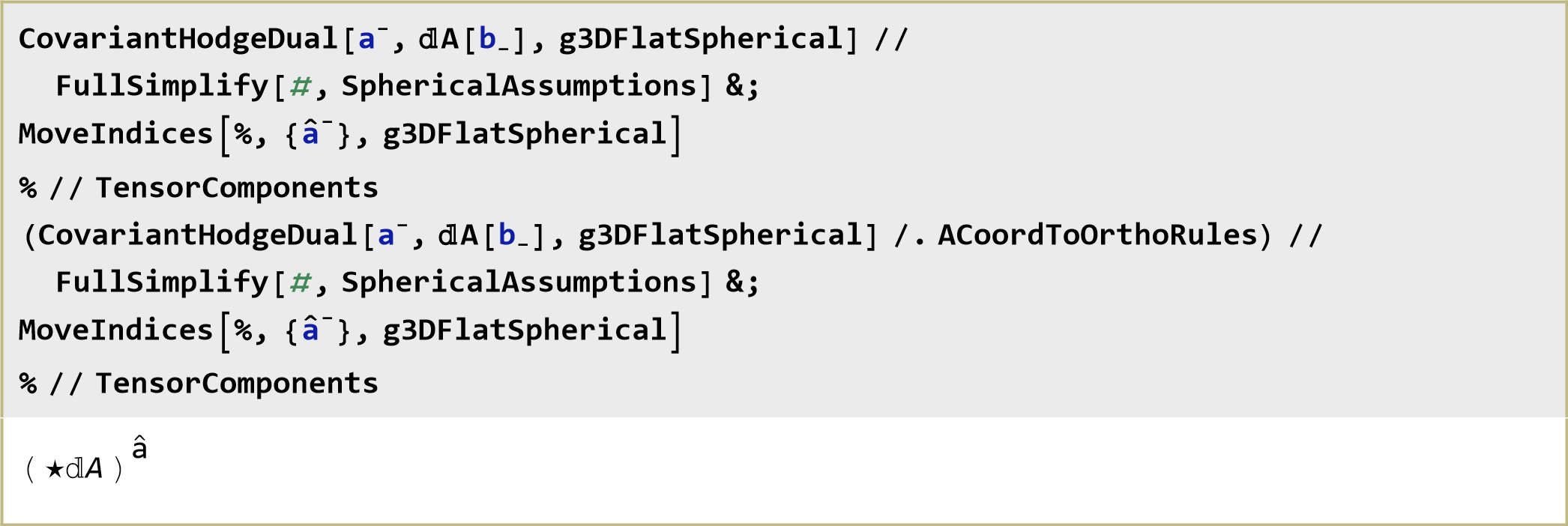}{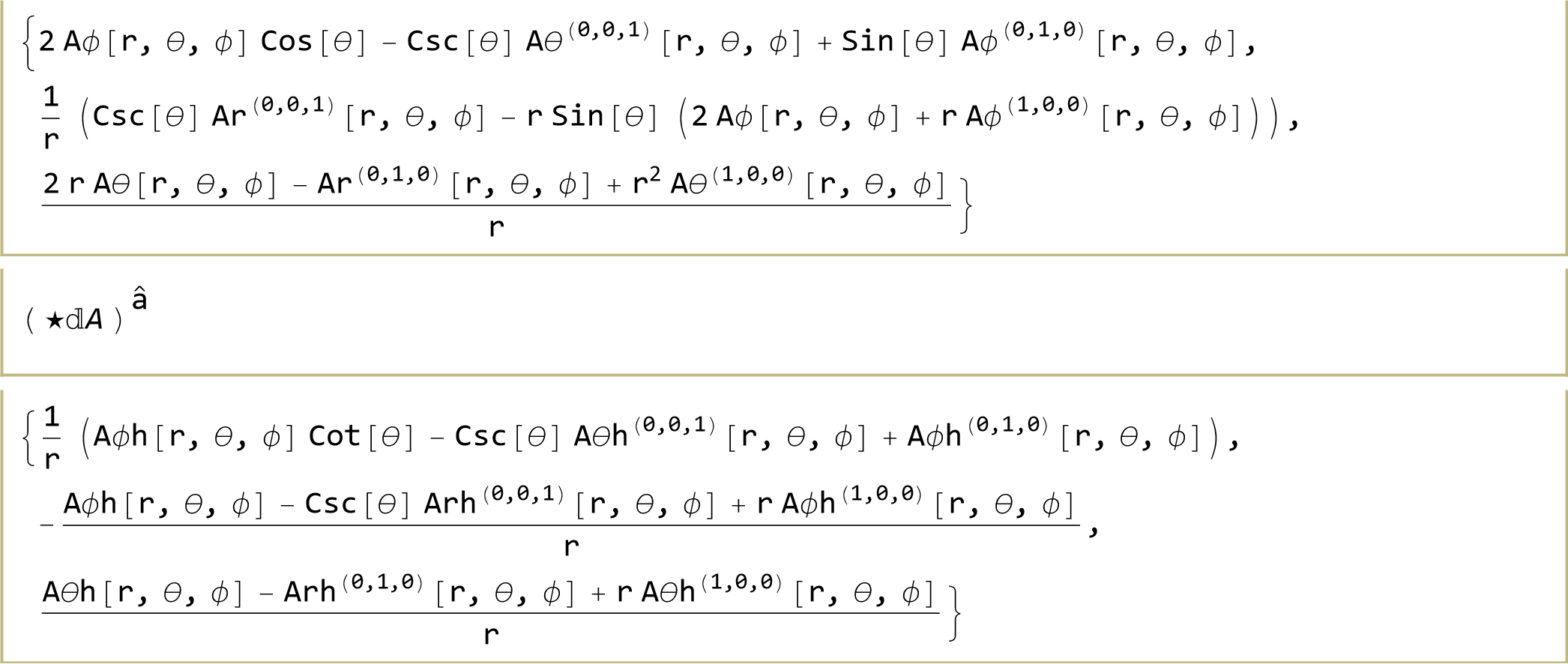}

\boldtitle{Eigensystem of Laplacian} \qquad The Laplacian on the 2-sphere (with eigenvalues $\{$-$\ell $($\ell $+1$\}$, $\ell $ $\in $ $\{$0,1,2,...$\}$) and the 3D flat space Laplacian (with eigenvalues $\{$\(-k^2\)$\}$, $k \geq 0$) are mutually commuting Hermitian operators. To find their simultaneous eigenfunctions we postulate a separation-of-variables ansatz: $R[r]$ \(Y_{\ell }{}^m\). This yields the spherical Bessel differential equation for the radial mode function $R[r]$.
\includecode{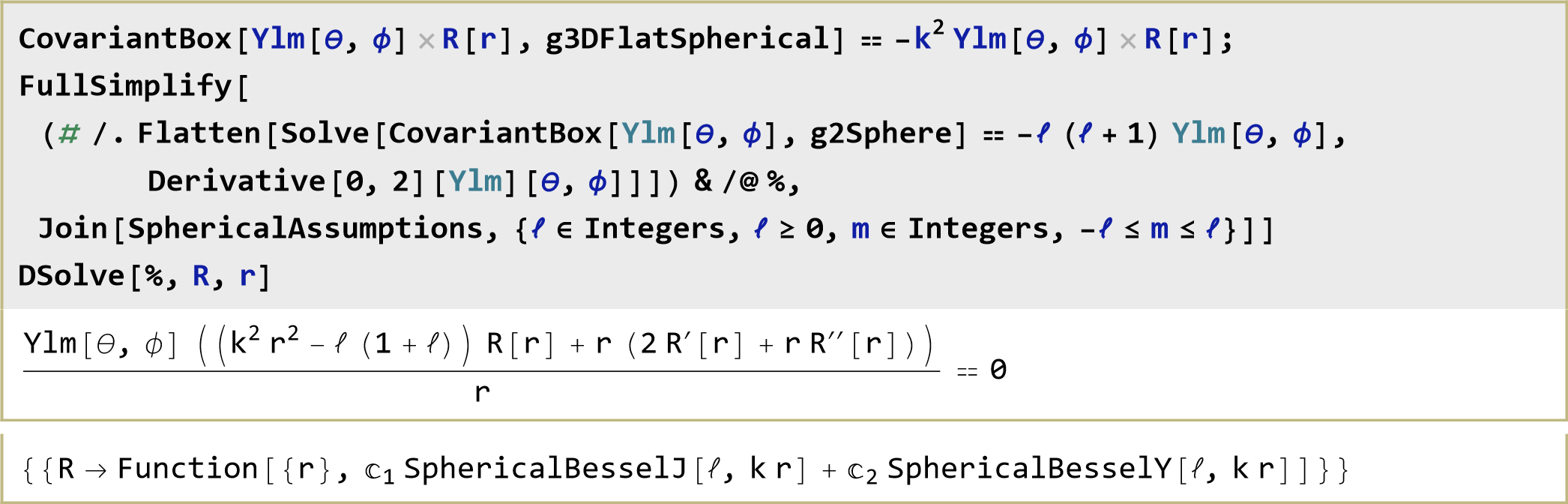}
The simultaneous eigenfunction of the Laplacian on the 2-sphere and that in 3D infinite flat space is therefore proportional to SphericalBesselJ[$\ell$, k r] \(Y_{\ell }{}^m\) because the SphericalBesselY blows up as r $\to $ 0.
\includecode{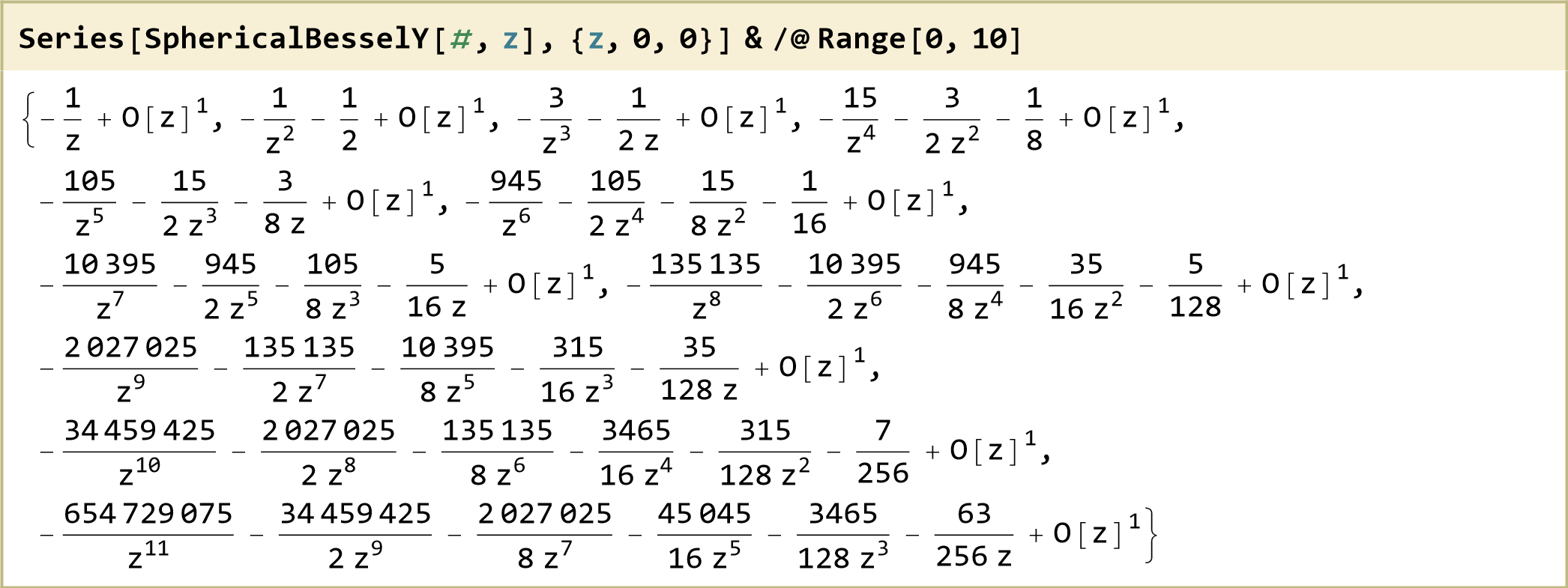}

\subsection{Exterior Derivative and the Poincare lemma}

In 3D vector calculus, and within a simply connected region of space, a vector \(\overset{\rightharpoonup}{E}\) is curl-free if and only if (iff) it is a gradient of a scalar (\(\overset{\rightharpoonup}{E}\) = -$\overset{\rightharpoonup}{\nabla} \Phi $); and a vector \(\overset{\rightharpoonup}{B}\) is divergence-free iff it is a curl of another vector (\(\overset{\rightharpoonup}{B}\) = $\overset{\rightharpoonup}{\nabla} $ $\times$ \(\overset{\rightharpoonup}{A}\)). These theorems can, respectively, be expressed using the exterior derivative: \((dE)_{\text{ab}}\) = 0 iff \(E_a=-d\Phi\) and \(\left(d\tilde{B}\right)_{\text{abc}}=0\) iff \(\tilde{B}_{\text{bc}}=(dA)_{\text{bc}}\), with \(\tilde{B}_{\text{bc}}\) $\equiv $ \(\tilde{\epsilon }_{\text{bcf}}B^f\). 

More generally, the Poincare lemma says that a fully anti-symmetric tensor \(A_{i_1\text{...}i_N}\) is the exterior derivative of a rank $N-1$ tensor (\(A_{i_1\text{...}i_N}\)= \((\dd B)_{i_1\text{...}i_N}\)) iff its own exterior derivative is zero (\((\dd A)_{i_{i_01}\text{...}i_N}\) = 0). In such a case, $B$ may be computed using $A$ via the formula
\[
B_{i_1\text{...}i_{N-1}}[\vec{x}] = \left(x^l - x'^l\right) \int_{0}^{1} A_{i_1\text{...}i_{N-1}l}\left[\vec{x}'+\lambda(\vec{x}-\vec{x}')\right] \lambda ^{N-1} \dd\lambda.
\]
We implement this formula in PotentialForm[A] = B. To illustrate its use, we shall start by defining the electric and magnetic fields.
\includecode{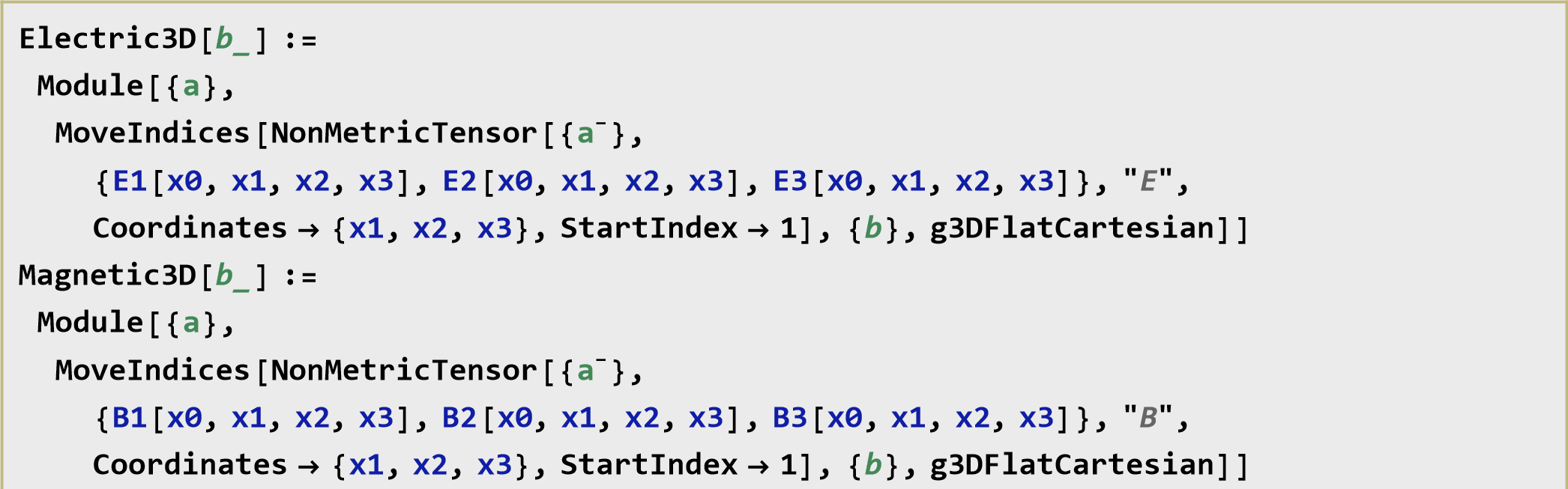}
\boldtitle{Coulomb Potential} \qquad The electric field due to a unit point charge Q is given by Q/(4$\pi $ \(r^2\)) \(\hat{r}\), where \(\hat{r}\) is the unit radial vector.
\includecode{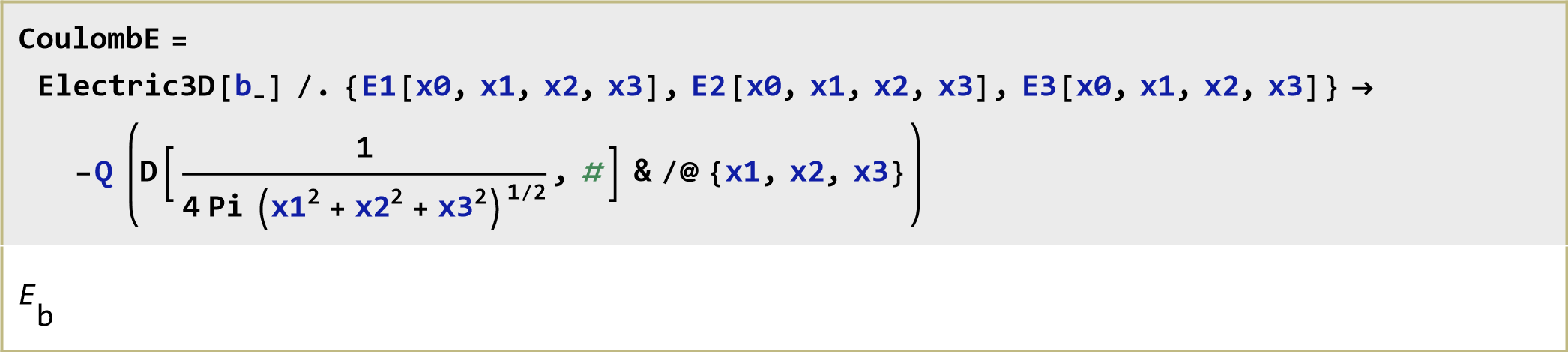}
The curl of this point charge electric field is zero; equivalently, \((\dd E)_{\text{ab}}\) = 0. 
\includecode{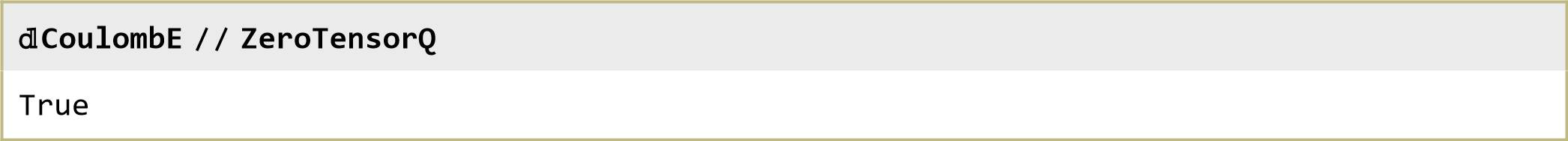}
This implies there must be some Coulomb potential $\Phi $ such that $-\dd \Phi $ = \(E_i\)d\(x^i\). To find $\Phi $ we shall use PotentialForm, and
through the optional argument StartingPoint instruct it to start its integration from an arbitrary point $\{$xp1,xp2,xp3$\}$ -- before sending
it infinitely far away at the end of the calculation.
\includecode{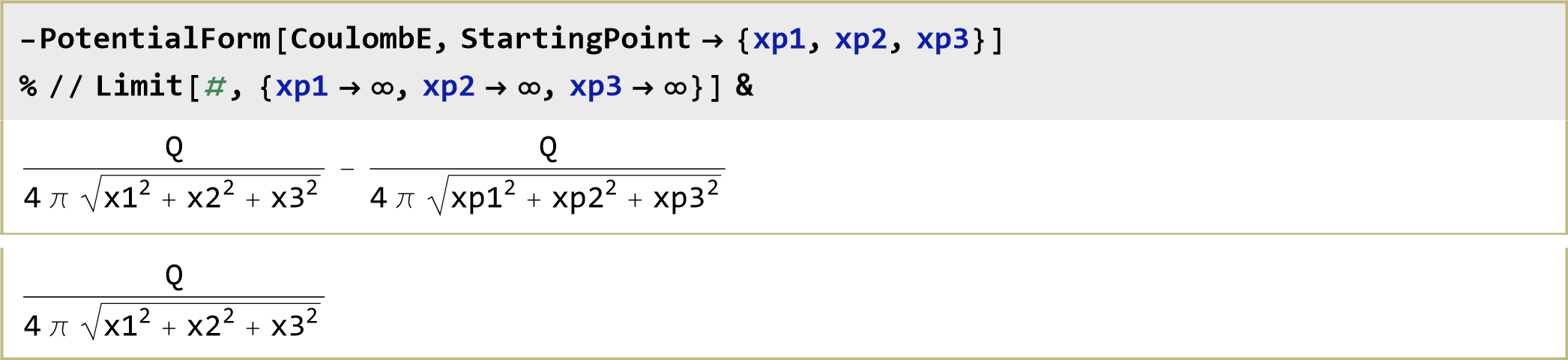}
\boldtitle{Constant Magnetic Field} \qquad Consider a constant magnetic field \(\overset{\rightharpoonup}{B}\) pointing along the 3-axis. 
\includecode{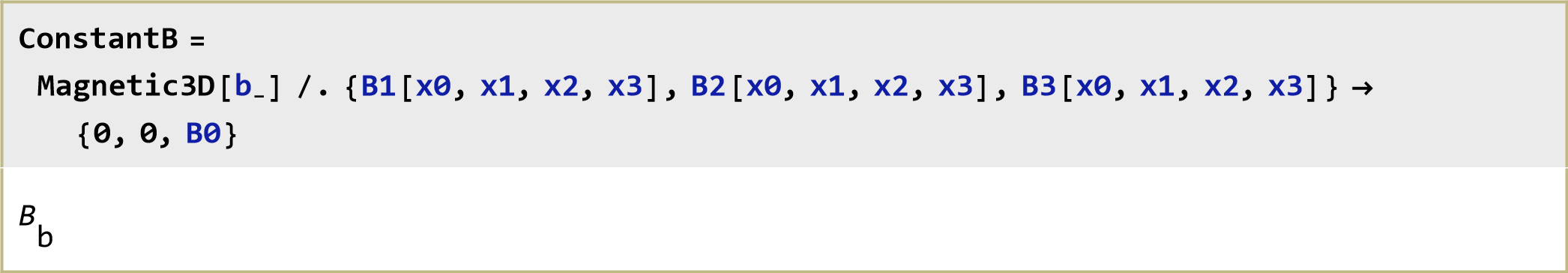}
Such a constant vector satisfies the divergence-free condition that \(\overset{\rightharpoonup}{B}\) must obey; equivalently, (d\(\tilde{B}\)\()_{\text{abc}} = 0\). Therefore, there must be some \(A_a\) such that \(\tilde{B}_a = (dA)_{\text{ab}}\). 
\includecode{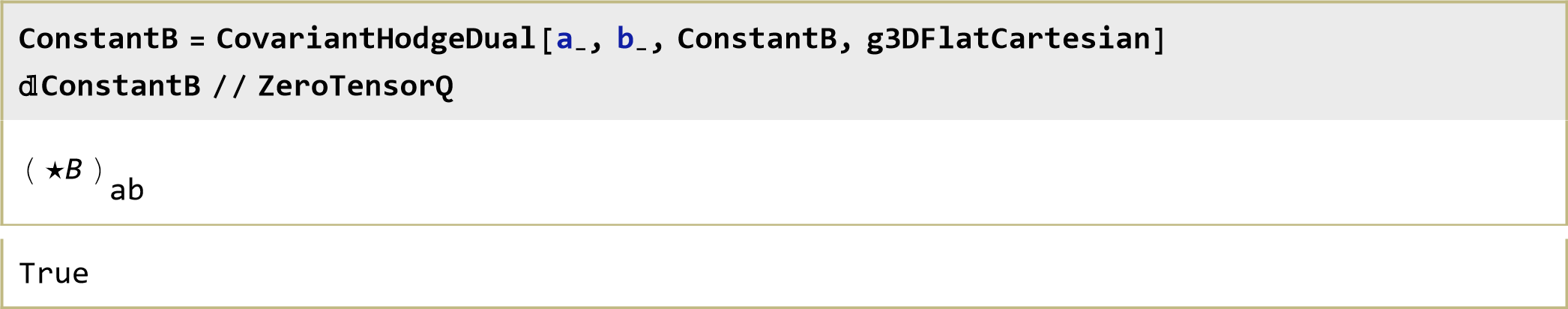}
We shall use PotentialForm to derive this vector potential \(A_a\). By default PotentialForm starts integration at the origin of the coordinate system (i.e., $\{0,...,0\}$); and specifically in 3D, we have StartingPoint $\to \{0,0,0\}$. We shall make this choice since (unlike the Coulomb case above) there are no singularities at the origin.
\includecode{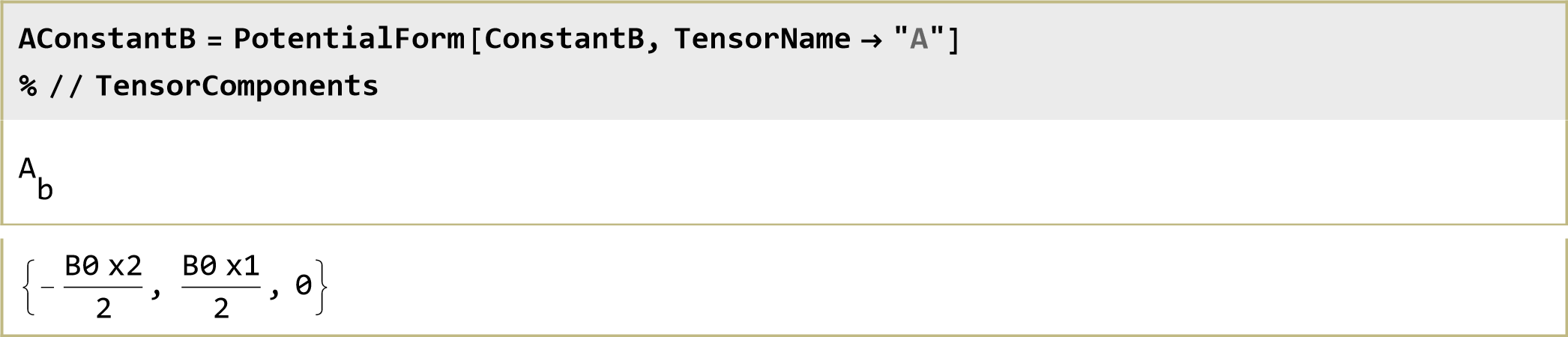}
We may readily check that the constant magnetic field along the 3-axis is recovered from its curl.
\includecode{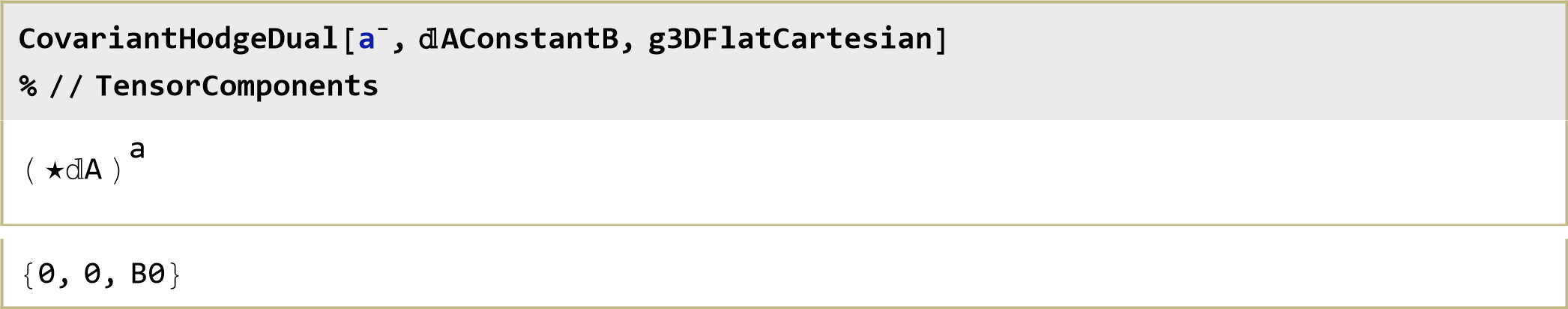}

\section{4D Minkowski Spacetime}

In this section, we shall study the Poincare group consisting of spacetime translation, Lorentz boost and spatial rotational symmetries of 4-dimensional (4D) flat spacetime (aka Minkowski spacetime) from a differential geometry point of view. In the second portion, we study the scalar wave equation and Maxwell{'}s equations in 4D flat spacetime.

\subsection{Isometries}

The most general active coordinate transformation that leaves the flat geometry
\includecode{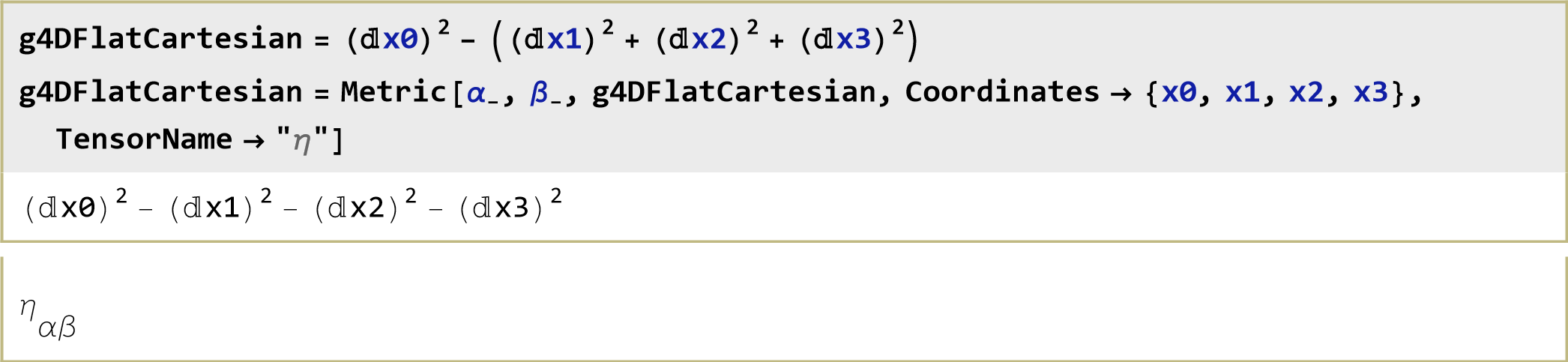}
invariant (namely, \(\eta _{\mu \nu }\) $\rightarrow $ \(\eta _{\mu \nu }\)) is given by
\[
x^{\mu }\rightarrow \Lambda ^{\mu }{}_{\nu }x^{\nu }+a^{\mu },
\]
for the spacetime-independent Lorentz transformation \(\Lambda ^{\mu }{}_{\nu }\) obeying
\[
\Lambda ^{\mu }{}_{\alpha }\Lambda ^{\nu }{}_{\beta }\eta _{\mu \nu }=\eta _{\alpha \beta };
\]
and spacetime-independent spatial translation vector \(a^{\mu }\).

\boldtitle{Isometries} \qquad Spacetime translations are generated by the four {``}linear momentum{''} operators
\[
p_{\mu } = -i\partial _{x^{\mu }}.
\]
A Lorentz transformation
\[
\Lambda = \text{Exp}\left[-\frac{I}{2} \omega _{\mu \nu }\cdot J^{\mu \nu }\right] \approx  1 -\frac{I}{2} \omega _{\mu \nu}\cdot J^{\mu \nu } + ... 
\]
acting upon the ($\mu $,$\nu $)-plane is generated by the boost-angular-momentum operators
\[
J^{\mu \nu }= +i x^{[\mu }\partial ^{\nu ]} = -x^{[\mu }p^{\nu ]}.
\]
Let us first define both the linear and boost-angular-momentum operators in \TC\ using NonMetricTensor.
\includecode{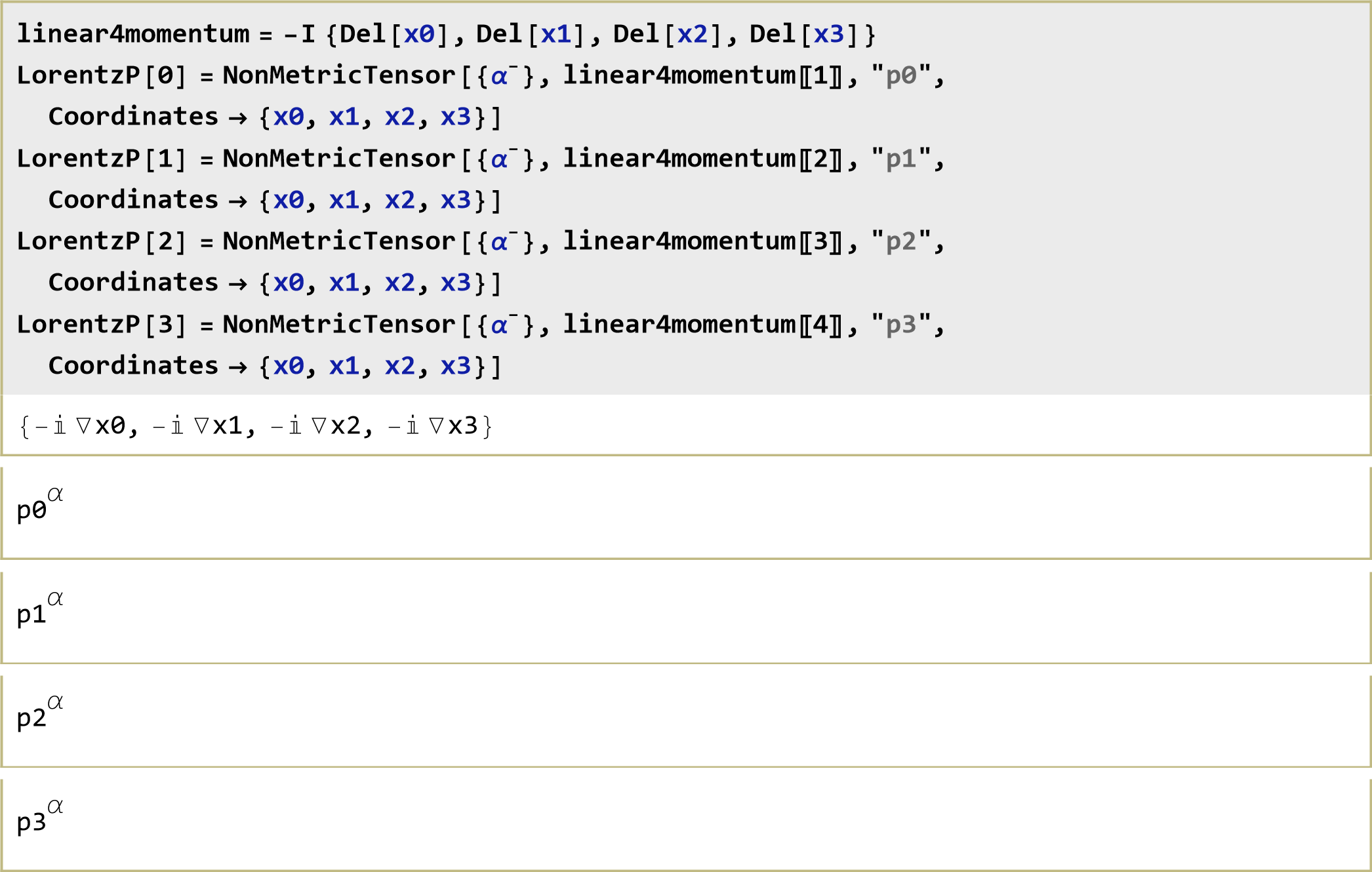}
Note that the boost generators are those involving one time component, namely $\{$ \(J^{0i}=-J^{\text{i0}}\) $\}$. Whereas the spatial angular momentum operators -- the generators of rotations -- are the $\{$ \(J^{\text{ab}}\) = \(\epsilon ^{\text{abc}}\) \(J^c\) $|$ \(\epsilon ^{123}\equiv1\) $\}$.
\includecode{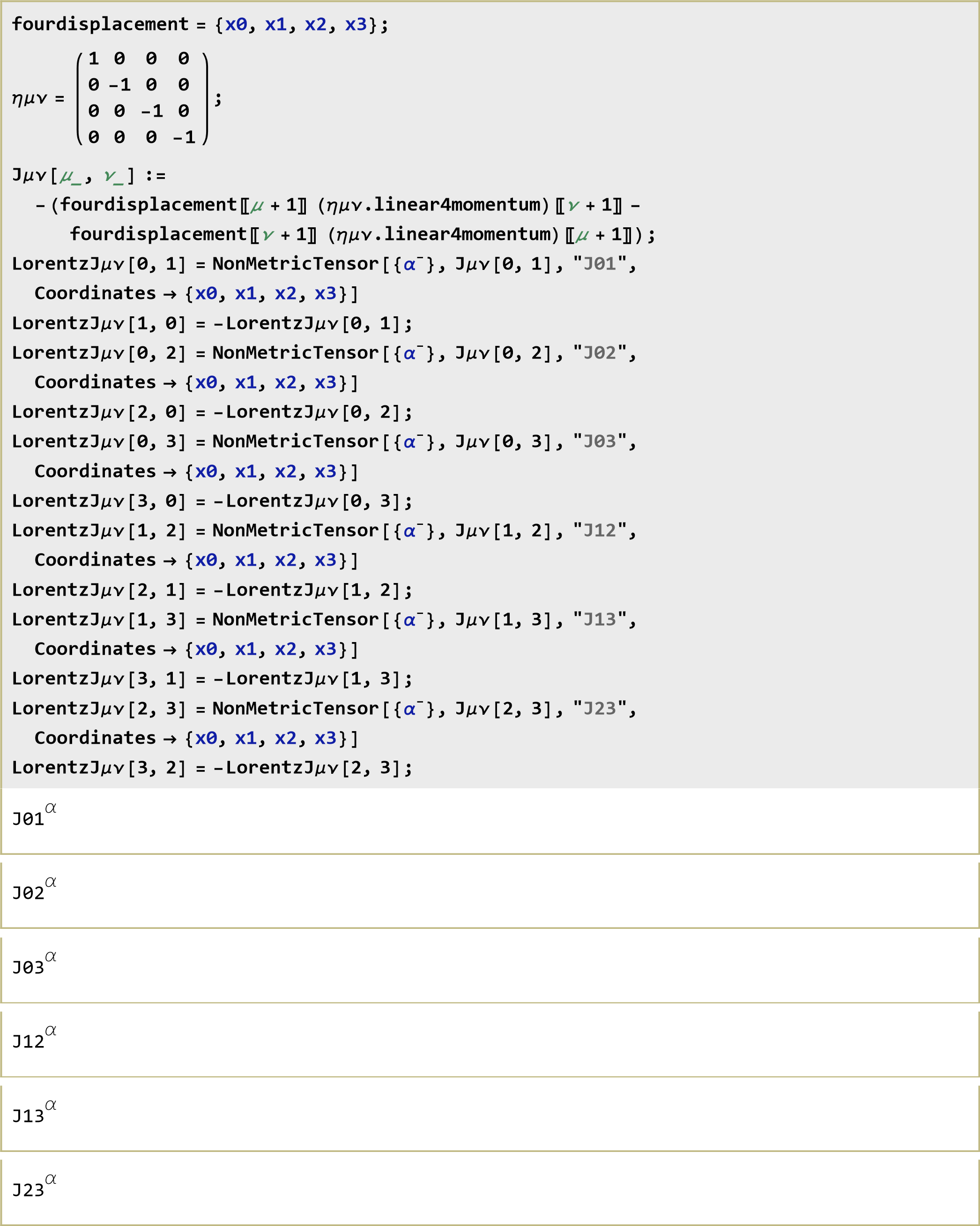}
\newpage
Under translations, $\varphi $[\(x\)] $\rightarrow $ $\varphi $[\(x+a\)] = $\varphi $[\(x\)] + \(\text{ia}\cdot p\) $\varphi $[\(x\)] + ...
\includecode{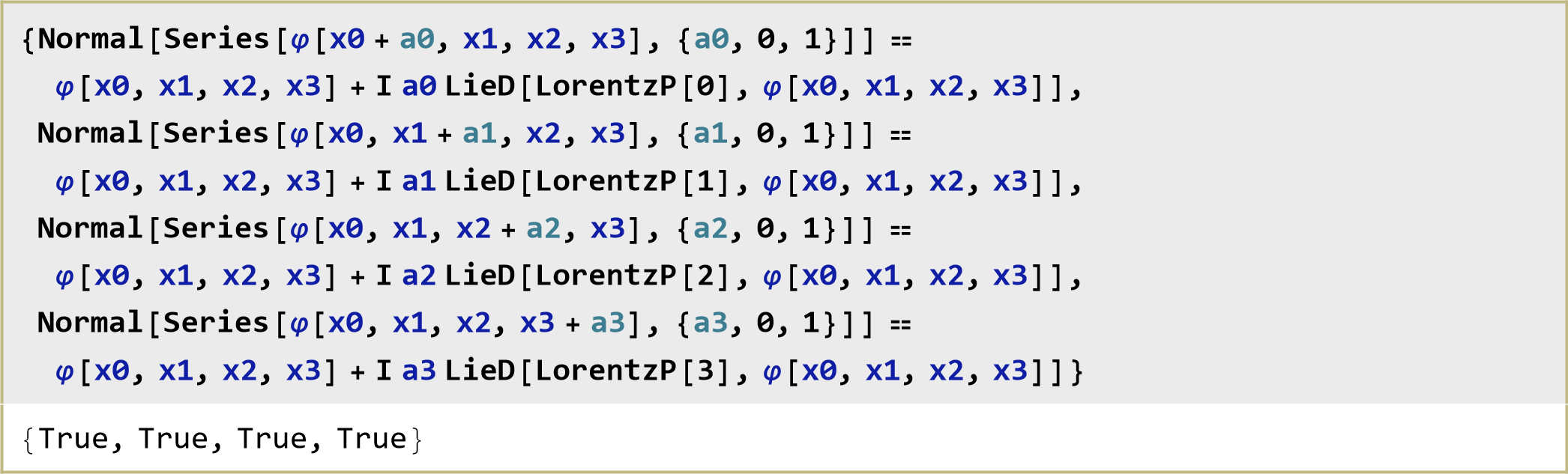}
Under Lorentz transformations, $\varphi $[\(x\)] $\rightarrow $ $\varphi $[$\Lambda \cdot $\(x\)] = $\varphi $[\(x\)] + \(\frac{i}{2} \omega _{\mu\nu } J^{\mu \nu }\) $\varphi $[\(x\)] + ... To be specific, consider a boost alone the 1-, 2-, and 3-axes.
\includecode{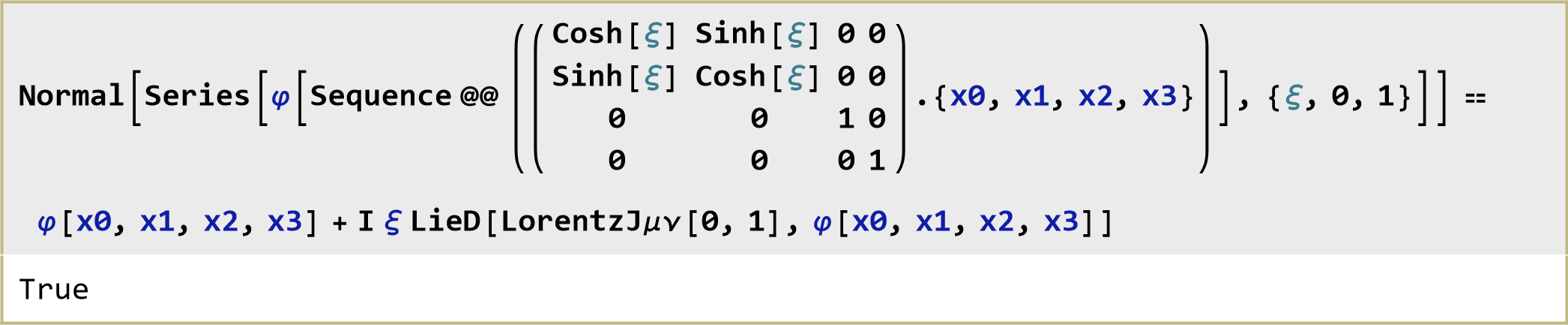}
\includecode{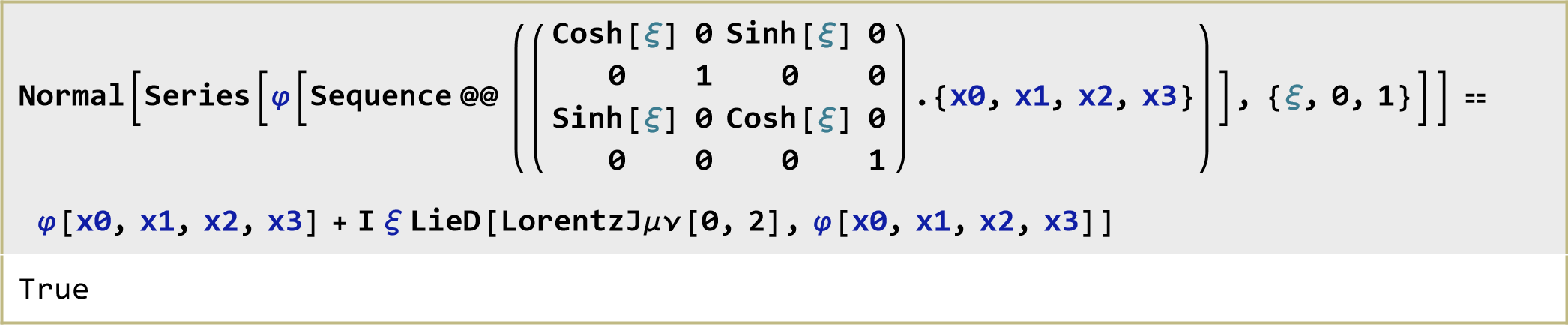}
\includecode{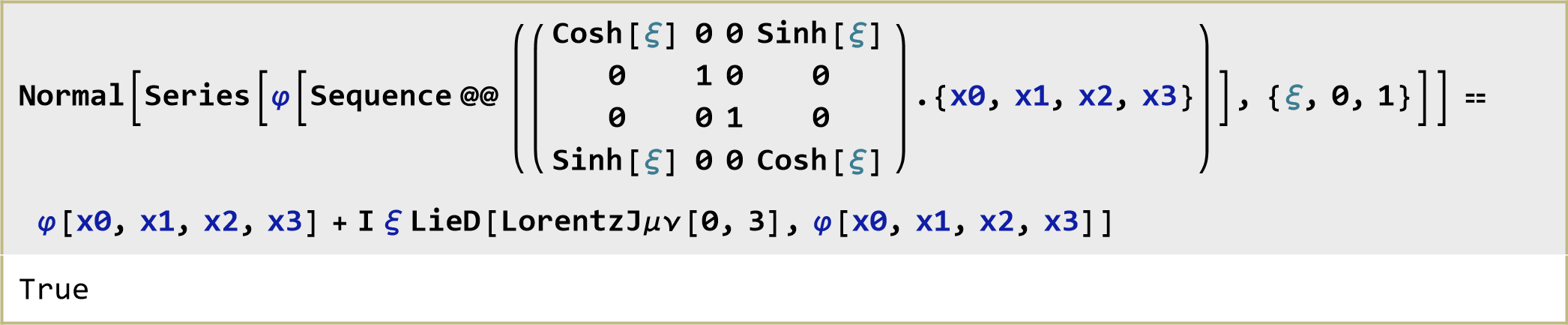}
Consider, too, rotations of the (1,2)-, (1,3)-, (2,3)-plane. These are equivalent to the above rotations along, respectively, the 3-, 2-, and 1-axis.
\includecode{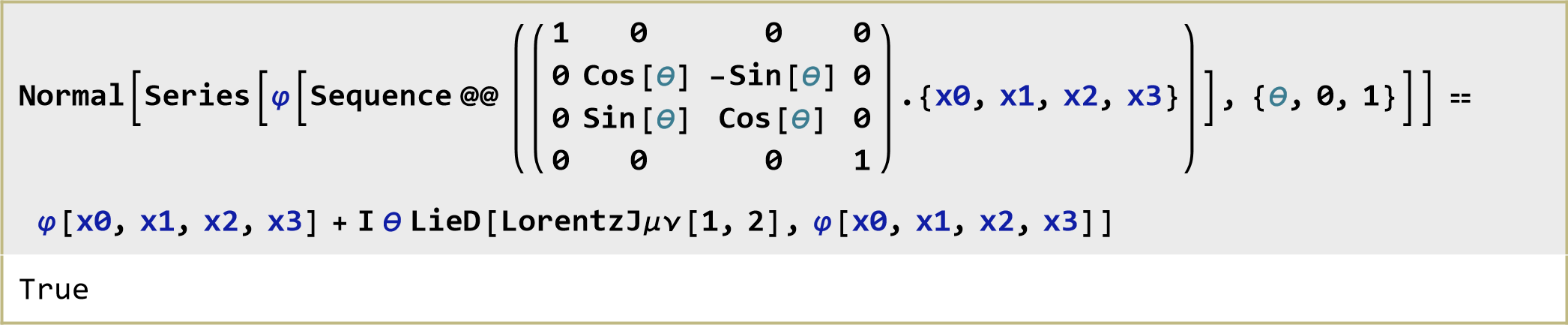}
\includecode{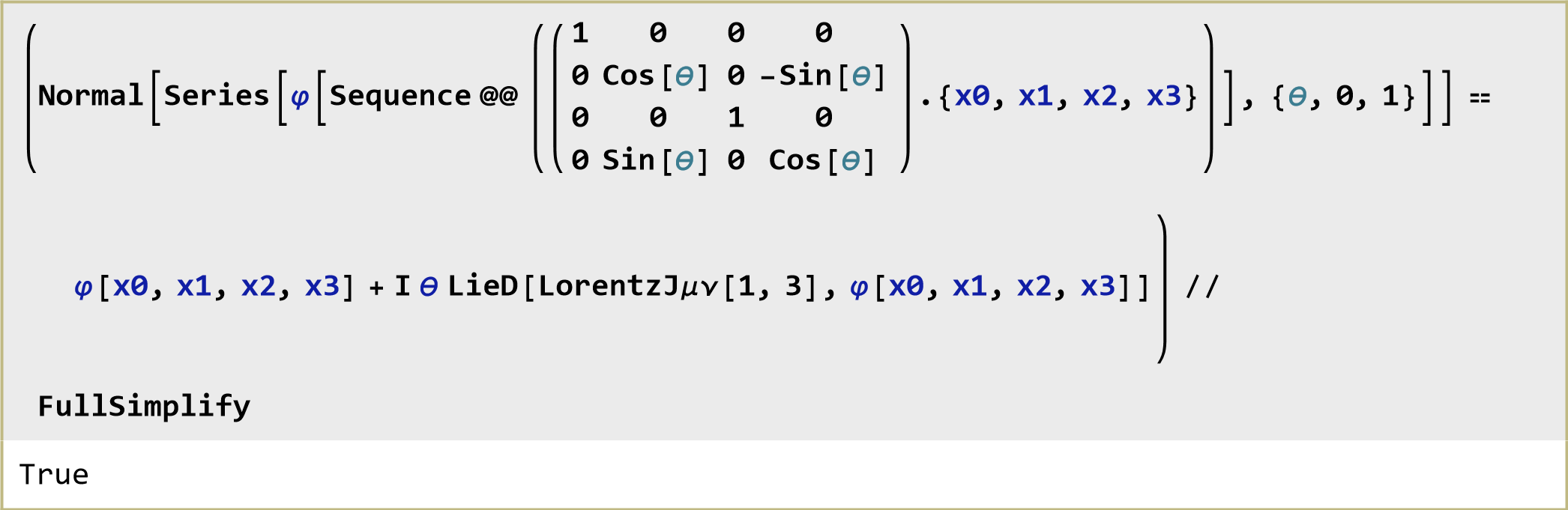}
\includecode{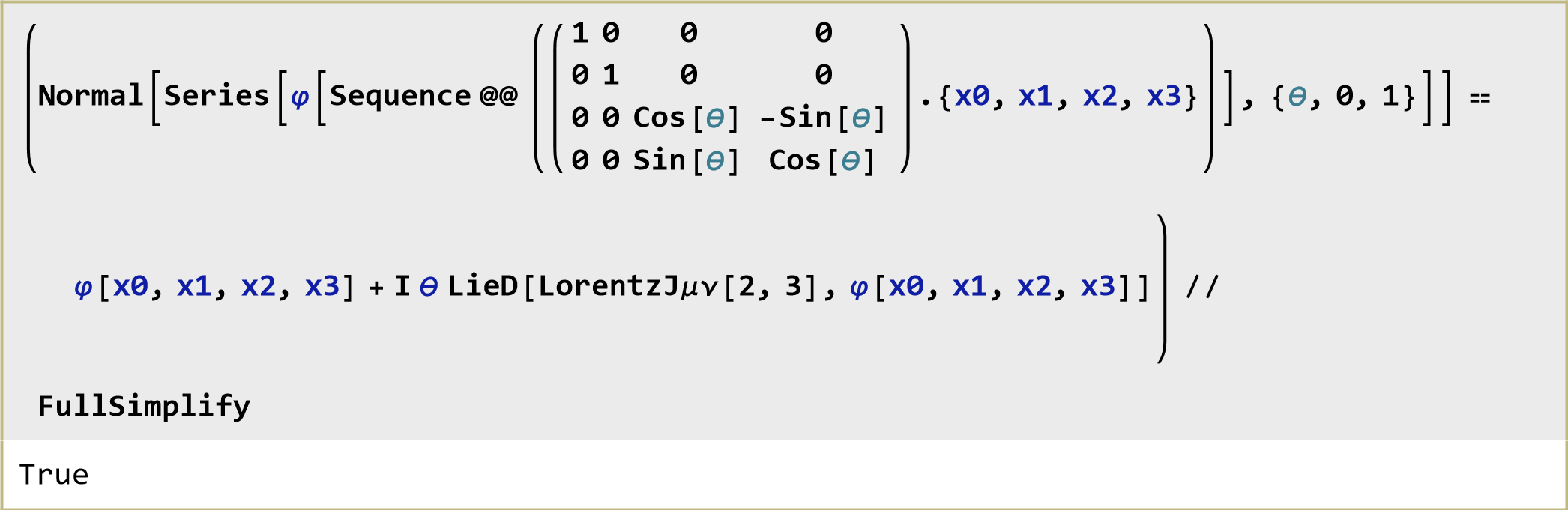}
That the infinitesimal coordinate transformations generated by these vector fields leave the flat 4D Minkowski metric invariant means the Lie derivative of the metric along them is zero -- these vectors satisfy Killing{'}s equations.
\includecode{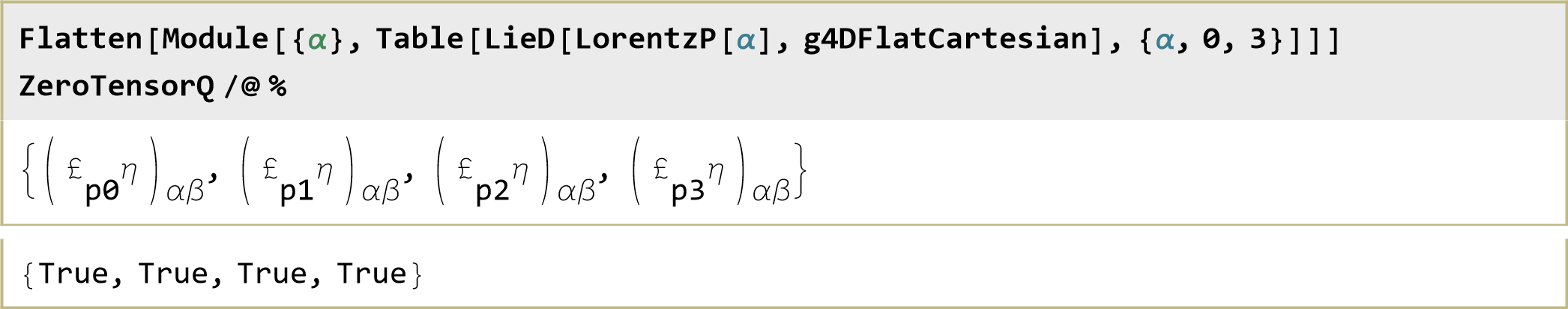}
\includecode{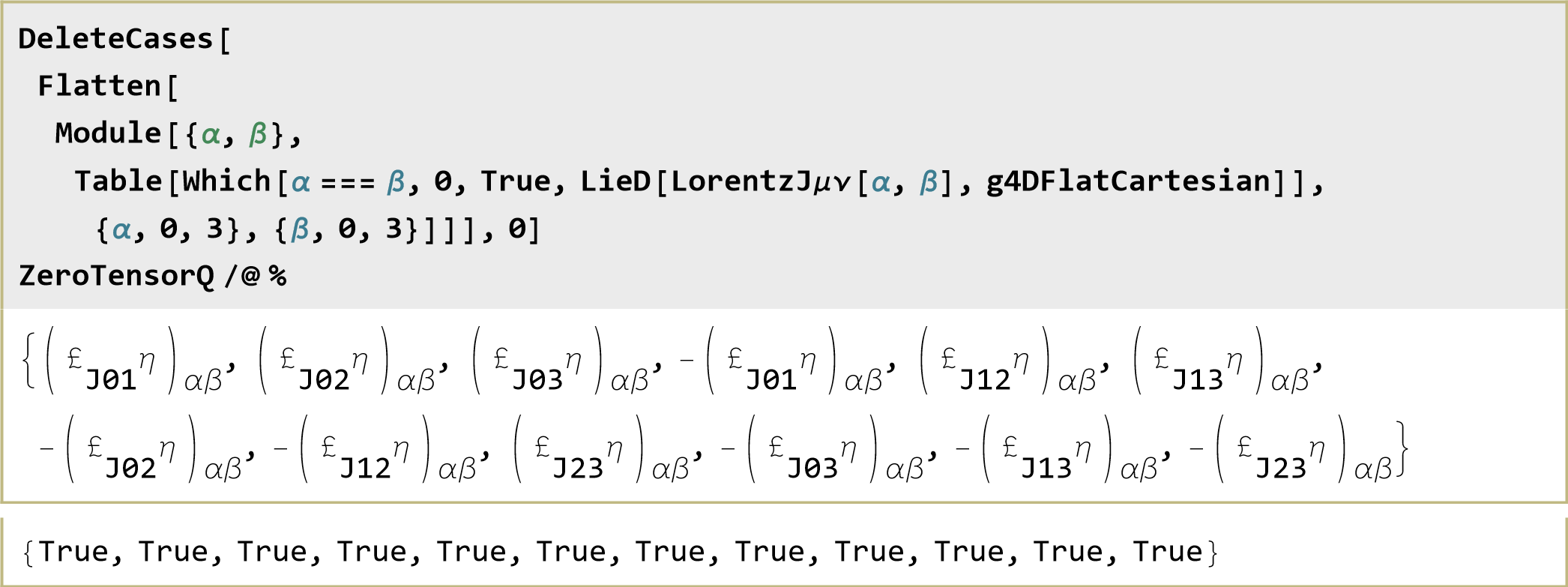}
Defining
\[
J^a\equiv  (1/2)\epsilon ^{\text{abc}}J^{\text{bc}}\quad \text{and}\quad K^a\equiv J^{0a},
\]
as well as
\[
J_{\pm }{}^a \equiv  (1/2)(J^a \pm  i K^a) .
\]
\includecode{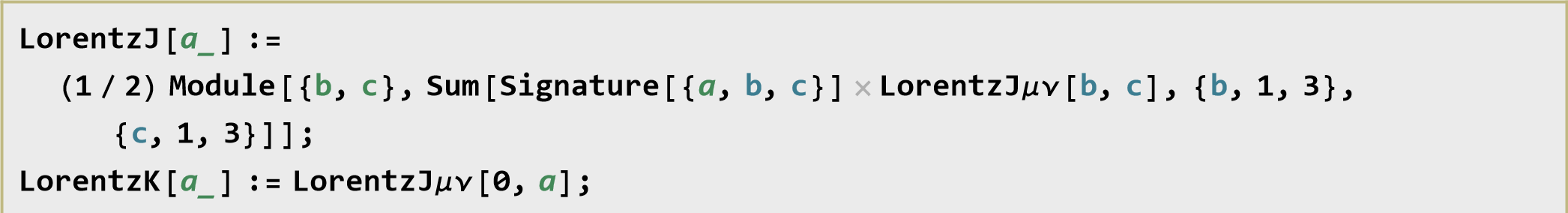}
-- let us now turn to verifying the 4D Lorentz group Lie algebra.
It is a pair of independent \(\text{so}_3\) Lie algebras.
\[
[J_{\pm }{}^a, J_{\pm }{}^b] = i \epsilon ^{\text{abc}} J_{\pm }{}^c\quad \text{and}\quad \quad \left[J_+{}^a,J_-{}^b\right]=0
\]
In quantum field theory these relations allow for the existence of {``}spin-1/2{''} fermions.
\newpage
Lie algebra of the \(J_+\){'}s:
\includecode{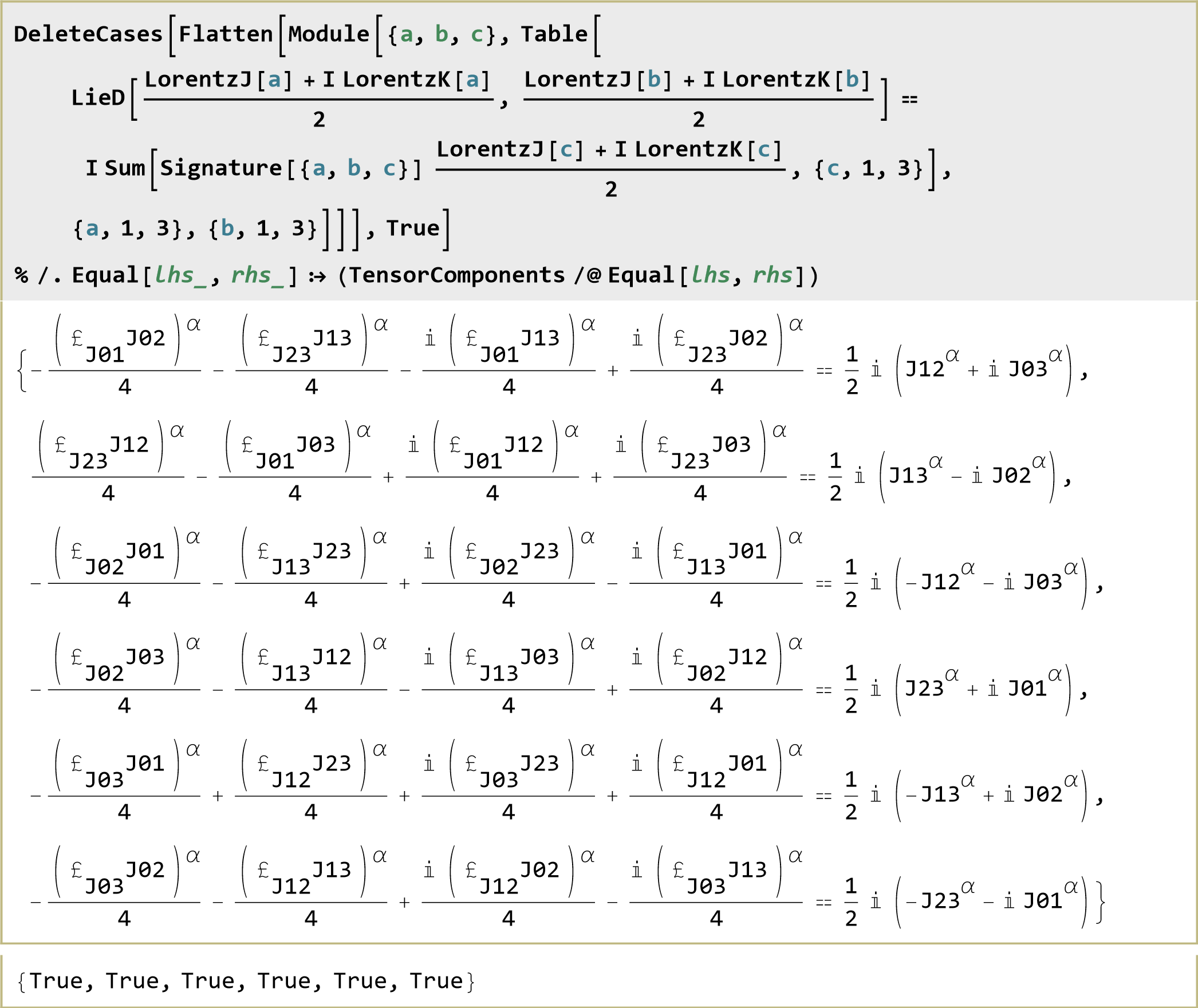}
\newpage
Lie algebra of the \(J_-\){'}s:
\includecode{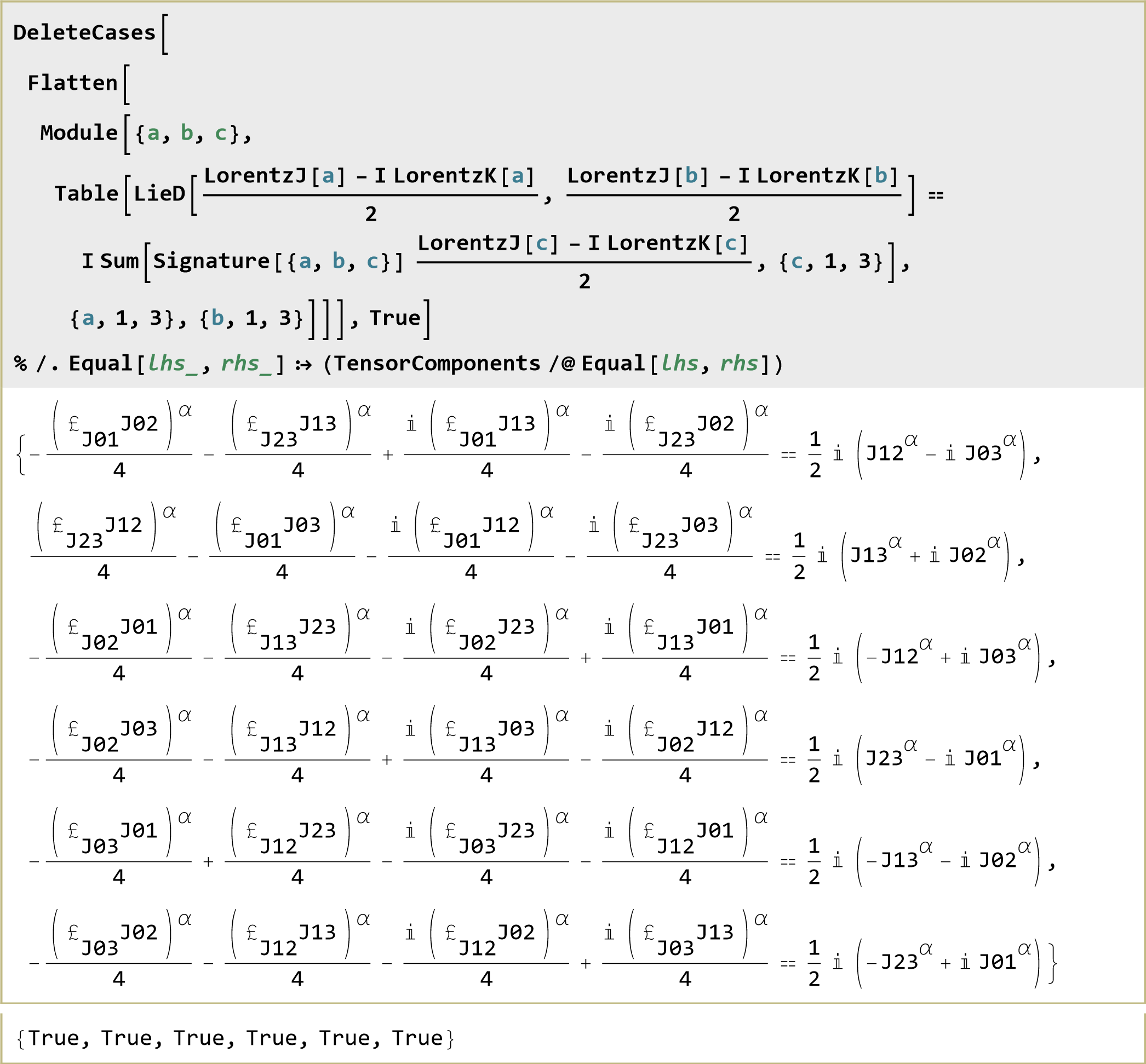}
\newpage
The \(J_+\){'}s and \(J_-\){'}s commute.
\includecode{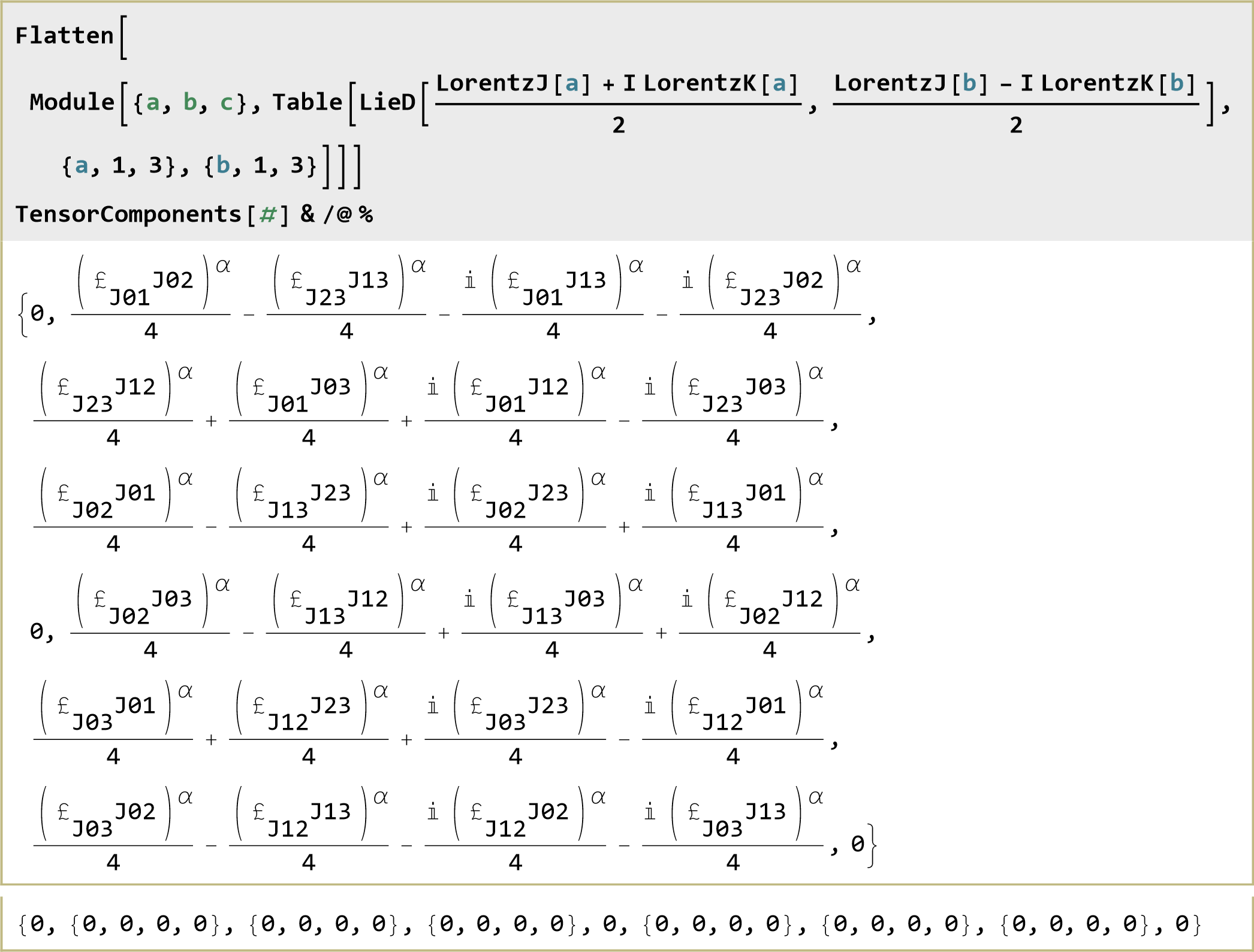}
Notice LieD obeys linearity.

While the 4-momentum \(p_{\mu }\) commutes with itself, [\(p_{\alpha }\), \(p_{\beta }\)] = 0, its commutator with the Lorentz generators are non-trivial because it transforms as a Lorentz 1-form:
\[
[J^{\alpha \beta }, p_{\mu }] = -i \delta _{\mu }{}^{[\alpha }\eta ^{\beta ]\gamma }p_{\gamma }
\]
\includecode{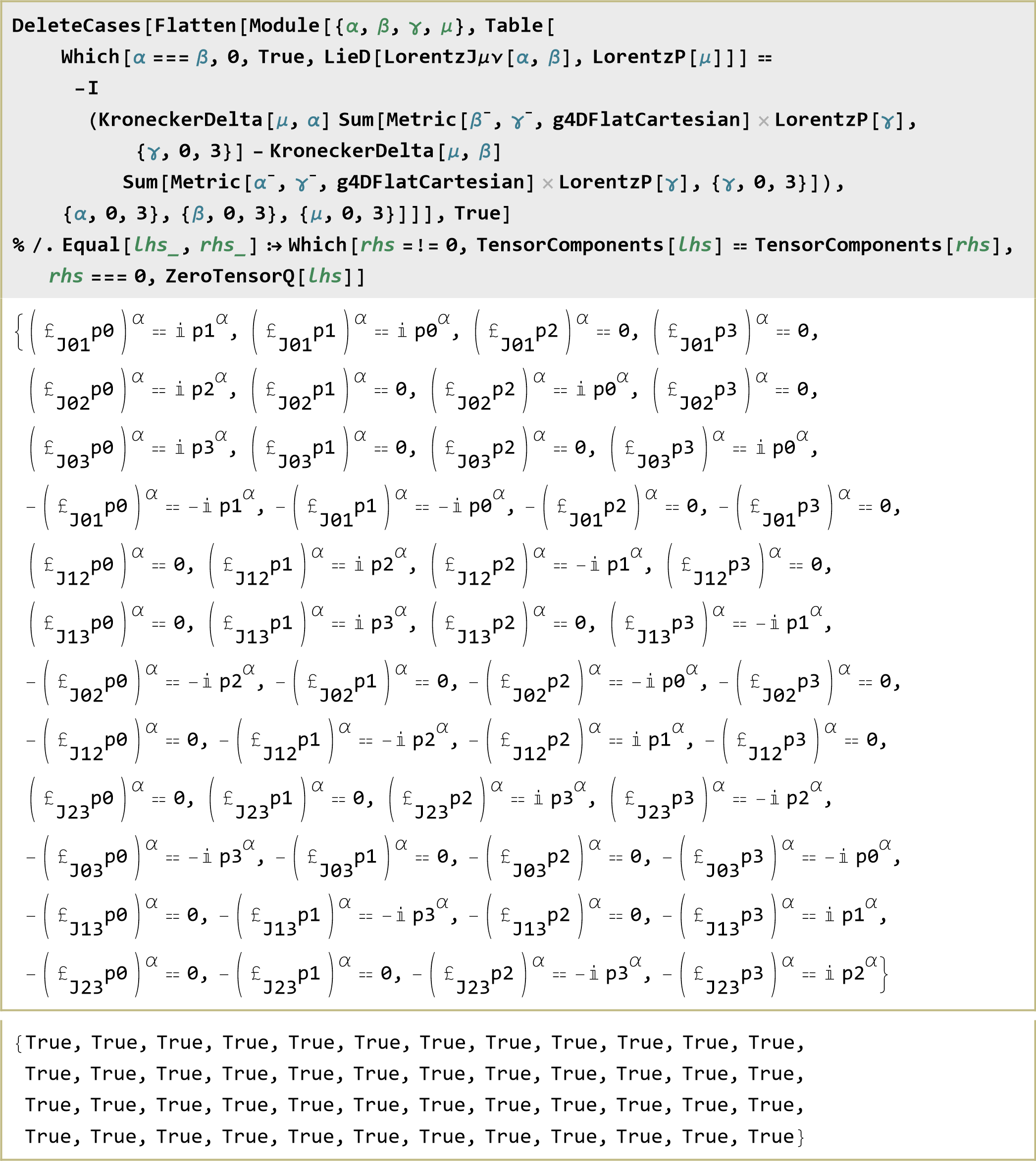}
Altogether, we have successfully verified the Lie algebra of the 4D Poincare group consisting of Lorentz boosts, spatial rotations, and spacetime translations.

\boldtitle{Finite Lorentz Boosts} \qquad The most general Lorentz boost \(\Lambda ^{\mu }{}_{\nu }\) corresponding to a change of inertial frames can be parametrized by a spatial velocity \(\overset{\rightharpoonup}{v}\) = d\(\overset{\rightharpoonup}{x}\)/dt. The interpretation is that the boost changes the given inertial frame \(\left\{x^{\mu }\right.\)$\}$ to one \(\left\{x'^{\mu }\equiv \Lambda ^{\mu }{}_{\nu }x^{\nu }\right.\)$\}$ moving at \(\overset{\rightharpoonup}{v}\) relative to it. Let us define it as a Tensor object.
\includecode{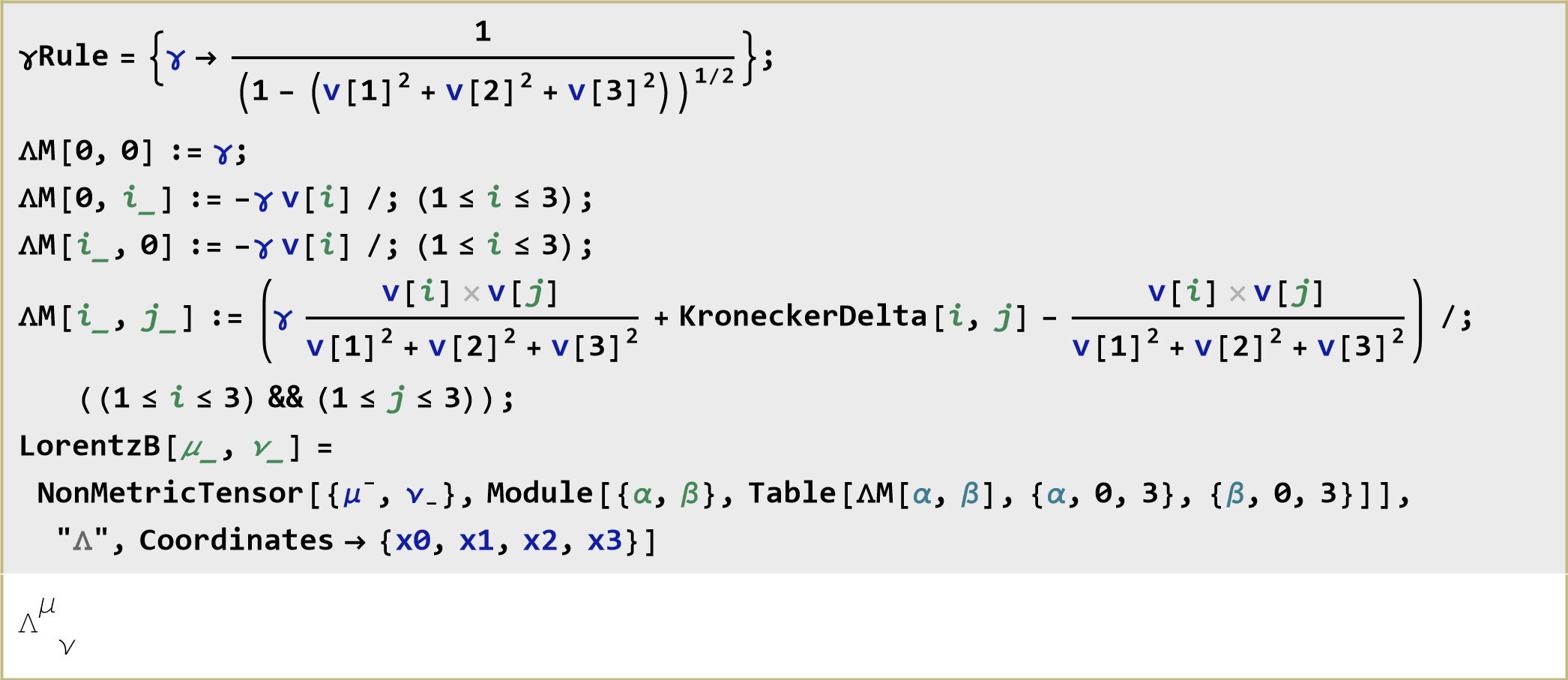}
Let us verify the defining property of the Lorentz group.
\includecode{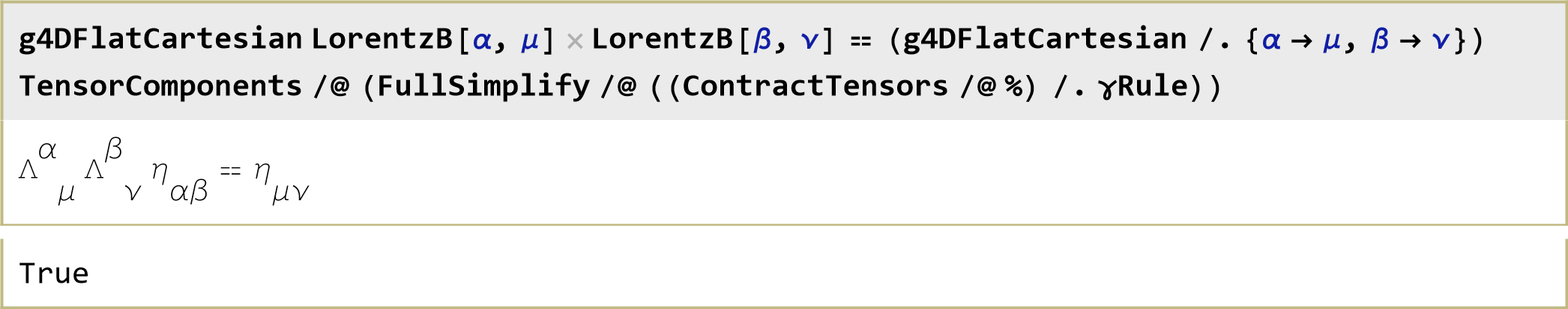}
Defining a 4-vector through NonMetricTensor --
\includecode{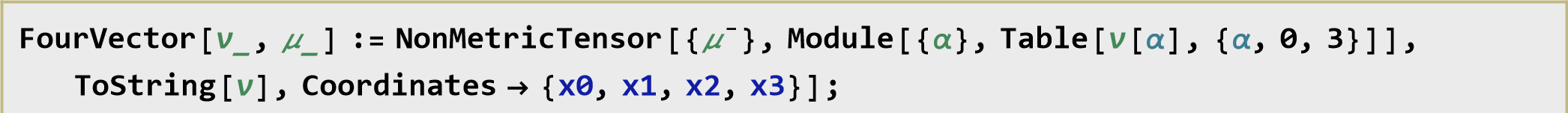}
-- we may witness how the Lorentz boost brings \(U^{\mu }\) $\equiv $ $\gamma $(1,\(\overset{\rightharpoonup}{v}\)) to its rest frame.
\includecode{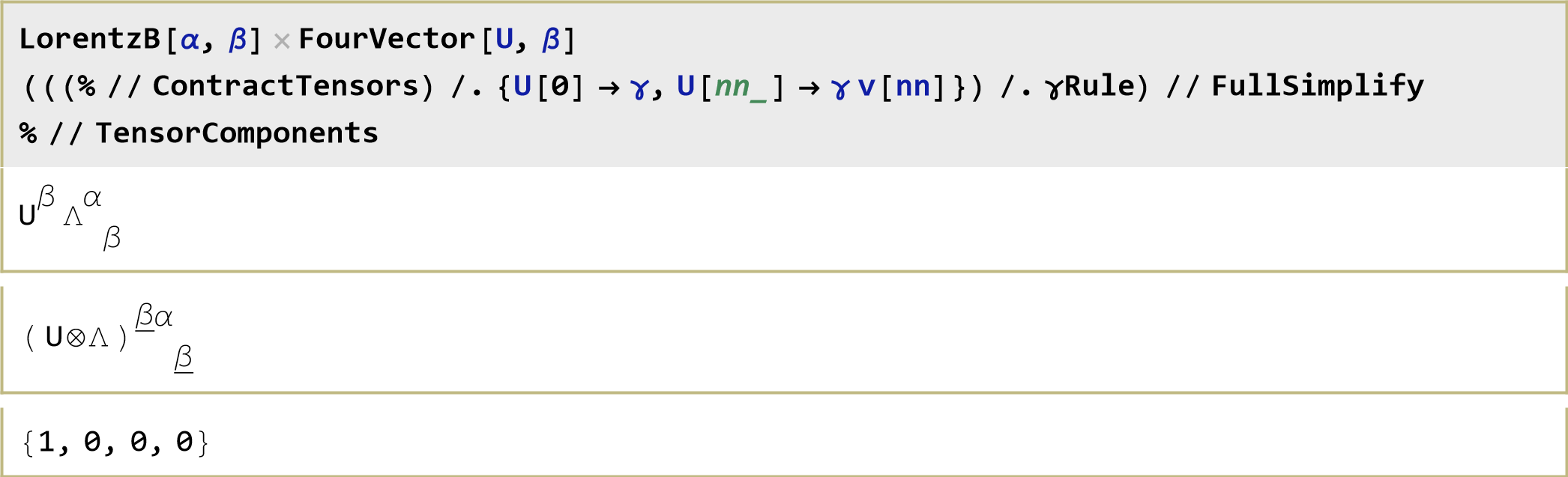}
Two events $x_1$ and $x_2$ become, in the boosted frame, $x_1'$ and $x_2'$.
\includecode{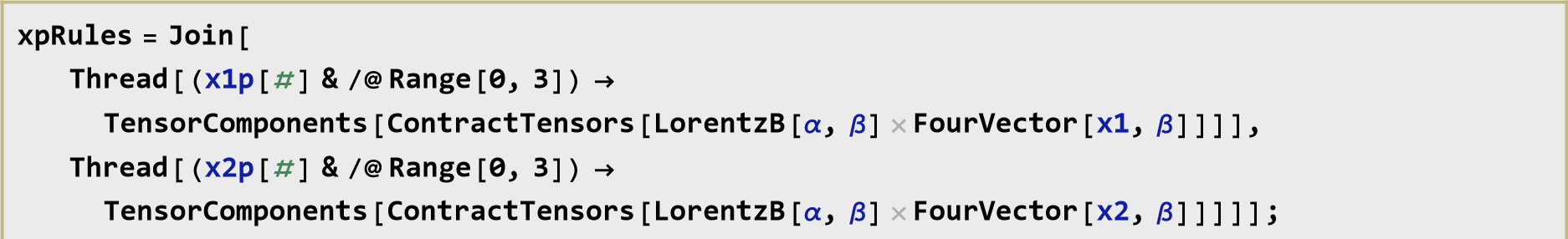}
In the boosted frame, two events occurring at the same time obeys $x'^0_1 = x'^0_2$.
\includecode{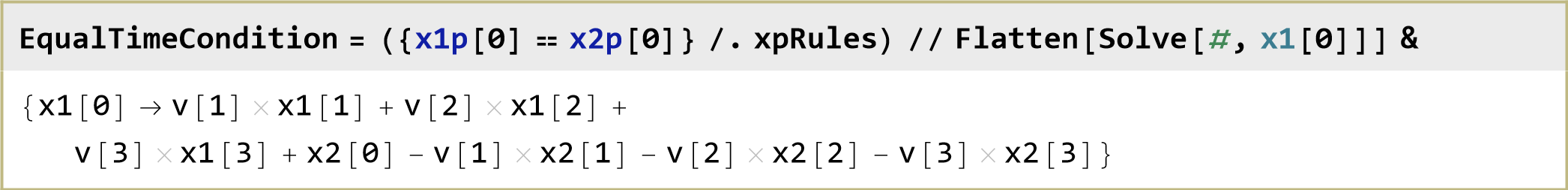}
Using this condition to solve for the relative displacement
\begin{align}
x_1'^\mu - x_2'^\mu
= \left(0,\vec{x}'_1 - \vec{x}'_2\right) .
\end{align}
and denoting $\Delta \vec{x} \equiv \vec{x}_1 - \vec{x}_2$ -- for example, $(\vec{x}_1 - \vec{x}_2) \cdot \vec{v} \equiv \vec{\Delta} \cdot \vec{v}$,
\includecode{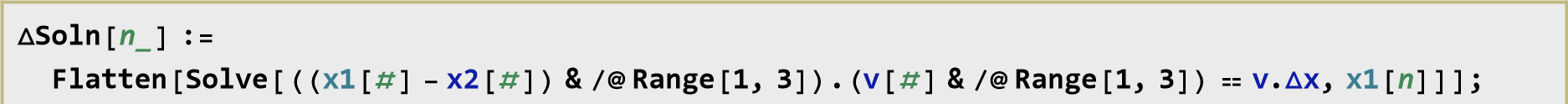}
-- we obtain the following length contraction formula:
\[
\Delta \overset{\rightharpoonup}{x}' = \Delta \overset{\to}{x}-\hat{v}\left(\hat{v}.\Delta\overset{\to}{x}\right)(1-\left(1-v^2\right)^{1/2})
\]
\includecode{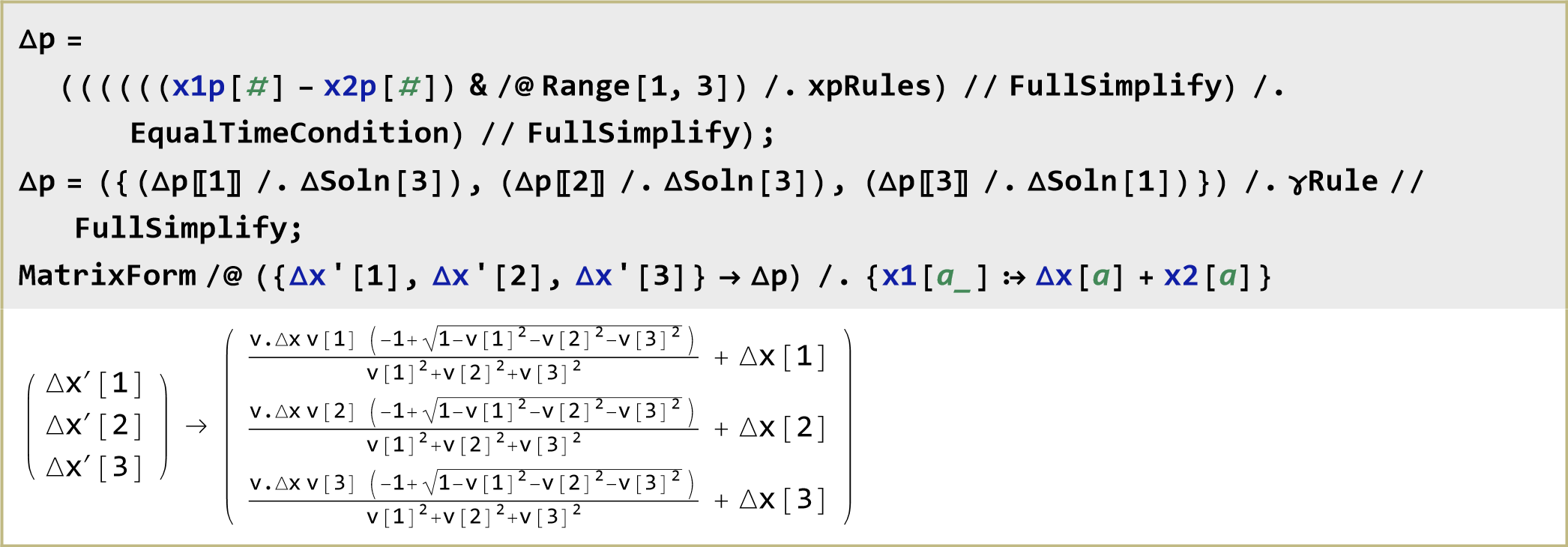}
Displacements perpendicular to \(\overset{\rightharpoonup}{v}\) are not affected by Lorentz boosts, as can be seen by dotting both sides with any vector perpendicular to \(\overset{\rightharpoonup}{v}\). Whereas, those parallel to \(\overset{\rightharpoonup}{v}\) are length contracted. In particular the component of $\vec{x}'_1 	- \vec{x}'_2$ parallel to \(\overset{\rightharpoonup}{v}\) reads as follows.
\includecode{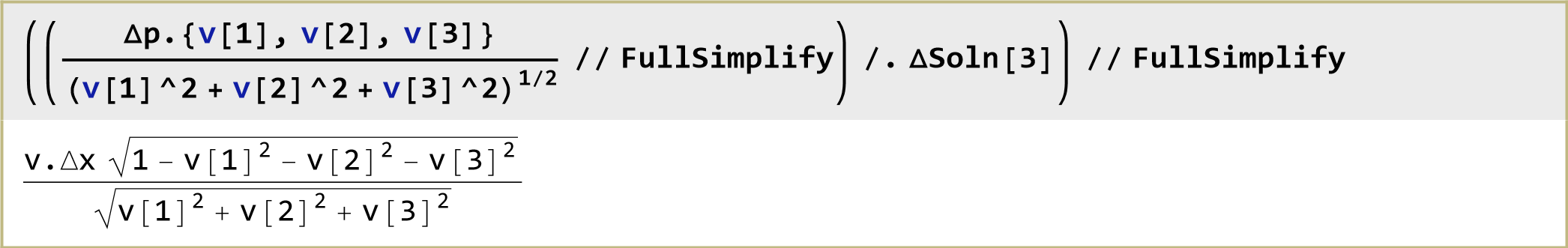}

\subsection{Wave Equations}

\boldtitle{Scalar Wave Equation} \qquad We say a scalar field $\varphi $ obeys the homogeneous wave equation in Minkowski spacetime if the following holds.
\includecode{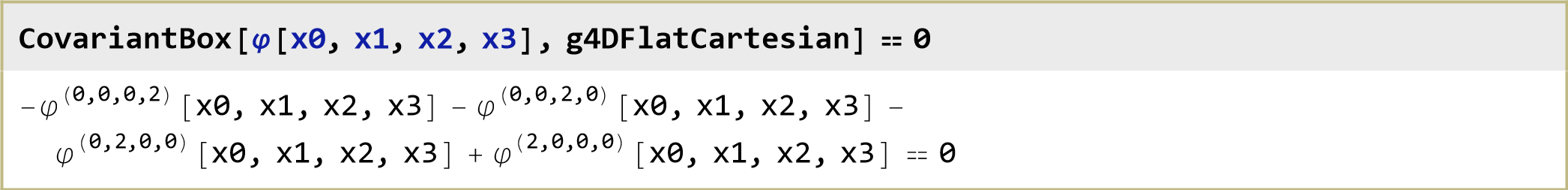}
Suppose we Fourier decompose $\varphi $.
\includecode{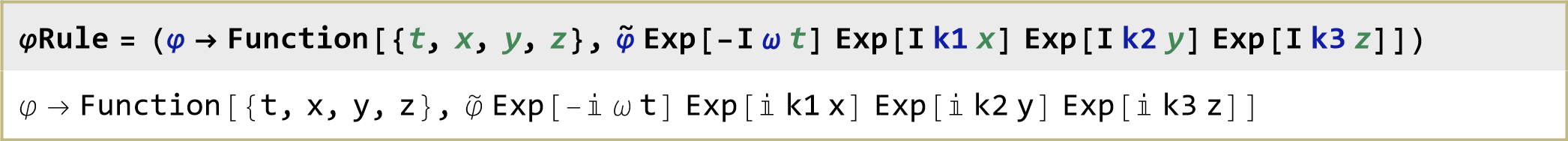}
We then recover the null dispersion relations \(k_{\mu }k^{\mu }=0\).
\includecode{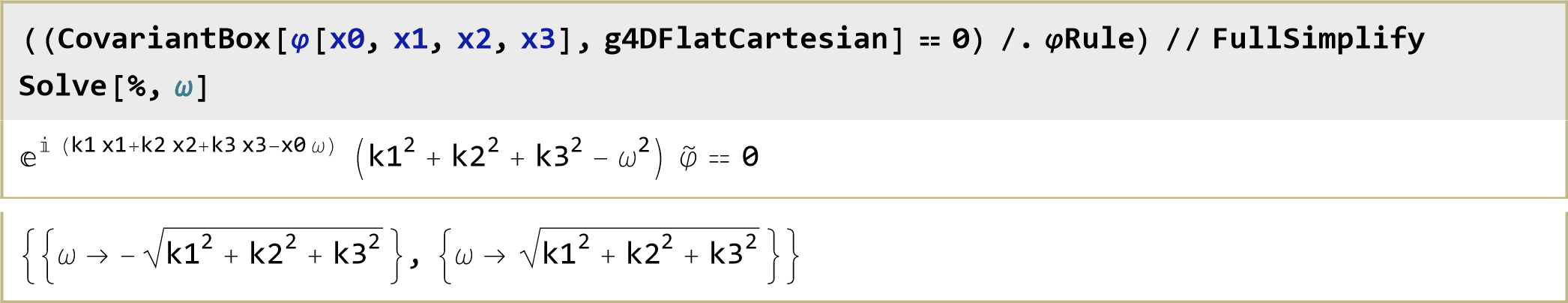}
Therefore any homogeneous real scalar solution to the wave equation may be decomposed as
\[
\varphi = \int _{R^3} \frac{\dd^3 \vec{k}}{(2\pi)^3} \textit{a}[\vec{k}] E^{-\text{I} |\vec{k}| x^0}E^{\text{I}\overset{\rightharpoonup}{k}.\overset{\rightharpoonup}{x}}+c.c.
\]
where \textit{a} is some appropriate $\vec{k}$-dependent (but spacetime-independent) complex coefficient and {``}c.c.{''} means {``}complex conjugation of the preceding terms{''}.

\boldtitle{Maxwell{'}s Equations} \qquad Next, we turn to a brief discussion of Maxwell{'}s equations. To write Maxwell{'}s equations in a Lorentz covariant form, we need to re-package the electric and magnetic fields into the anti-symmetric Faraday tensor \(F^{\mu \nu }=-F^{\nu \mu }\).
\includecode{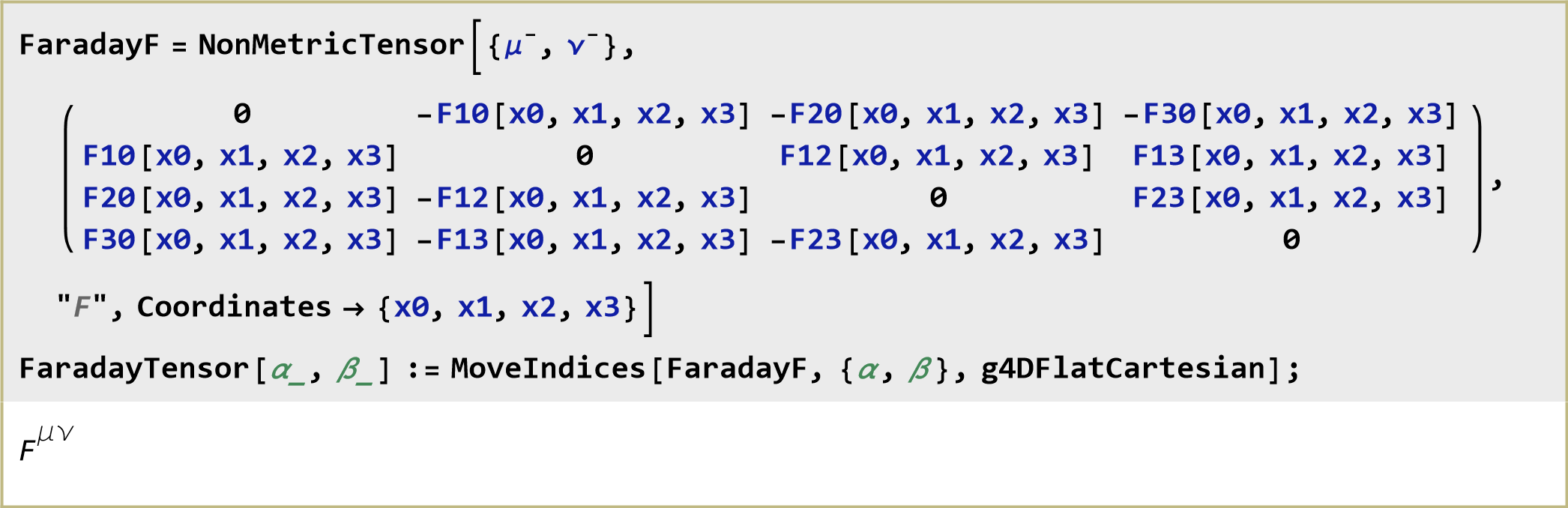}
We may verify the anti-symmetric character of this electromagnetic field strength tensor in two different ways.
\includecode{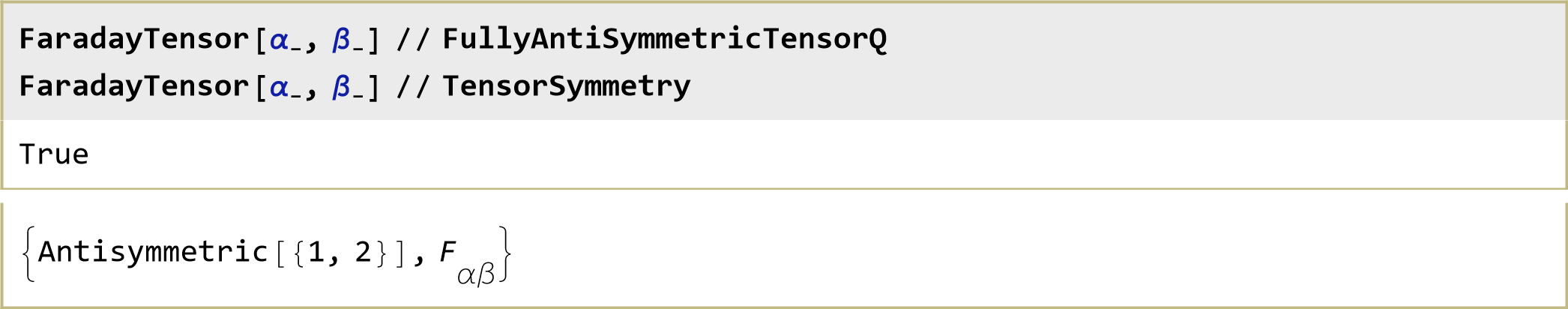}
Notice, TensorSymmetry is a \MMA\ built-in function. Its functionality has been extended in \TC\ to extract the (anti)symmetric properties
of a given Tensor object. There are other \MMA\ built-in functions, such as Series, Dimensions, Normal, Simplify, FullSimplify, Conjugate,
etc. that works on Tensor objects too.

To interpret the components of the Faraday tensor we examine the Lorentz force law; namely, the Lorentz covariant version of \(\overset{\rightharpoonup}{f}\)
$\equiv $ \(\overset{\rightharpoonup }{E}\) + \(\overset{\rightharpoonup}{v}\) $\times$ \(\overset{\rightharpoonup}{B}\). We postulate -- since it is a 4-vector -- it is simply the Faraday tensor contracted into the point charge{'}s 4-velocity: \(f^{\mu }\) $\propto $ \(F^{\mu }{}_{\nu }U^{\nu }\).
\includecode{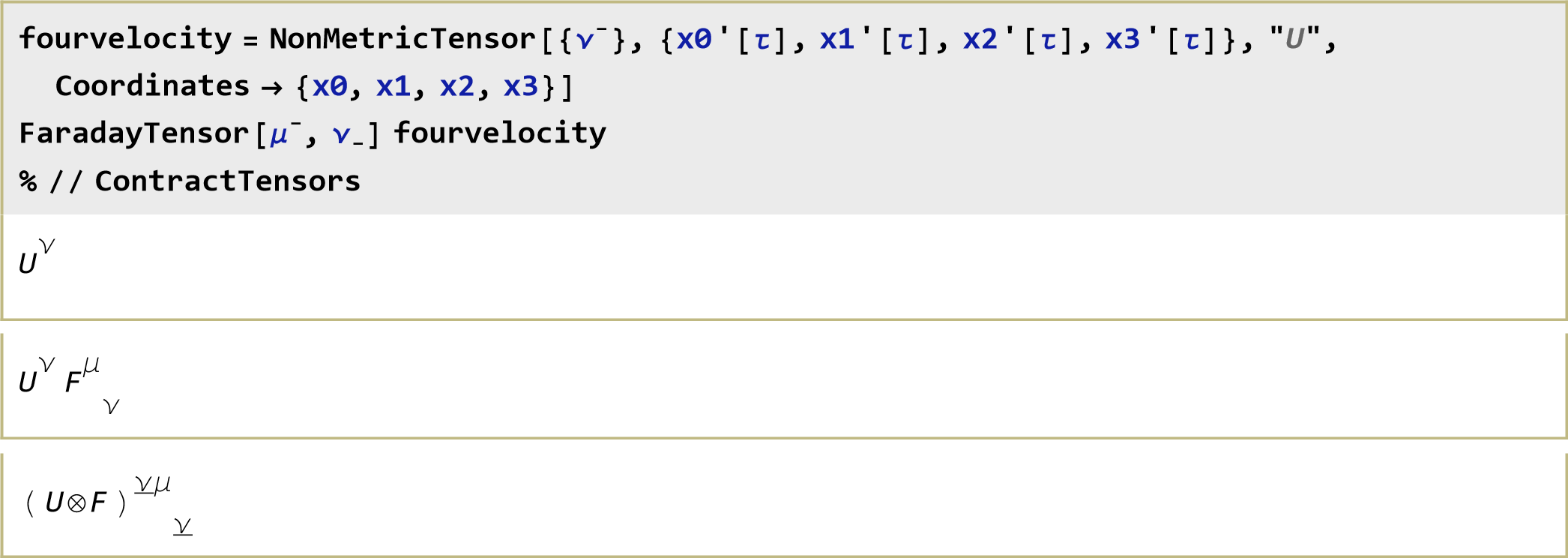}
Focus on the spatial components, and proceed to identify \(E^i=F^{\text{i0}}=-F^{0i}=-F_{\text{i0}}=F_{0i}\) and \(F_{\text{ab}}=F^{\text{ab}}=-\epsilon^{\text{abc}}B^c\) with \(\epsilon ^{123}\equiv 1\). Denoting, \(\overset{\rightharpoonup}{v}\) $\times$ \(\overset{\rightharpoonup}{B}\) as v $\wedge$ B, and recognizing d\(x^0\)/d$\tau $ $\cdot $ \(\overset{\rightharpoonup}{v}\) = d\(x^0\)/d$\tau $ $\cdot $ d\(\overset{\rightharpoonup}{x}\)/dt = d\(\overset{\rightharpoonup}{x}\)/d$\tau$, we recover the Lorentz force law up to an overall (relativistic) d\(x^0\)/d$\tau $ $\equiv $ dt/d$\tau $ factor, where $\tau $ is proper time.
\includecode{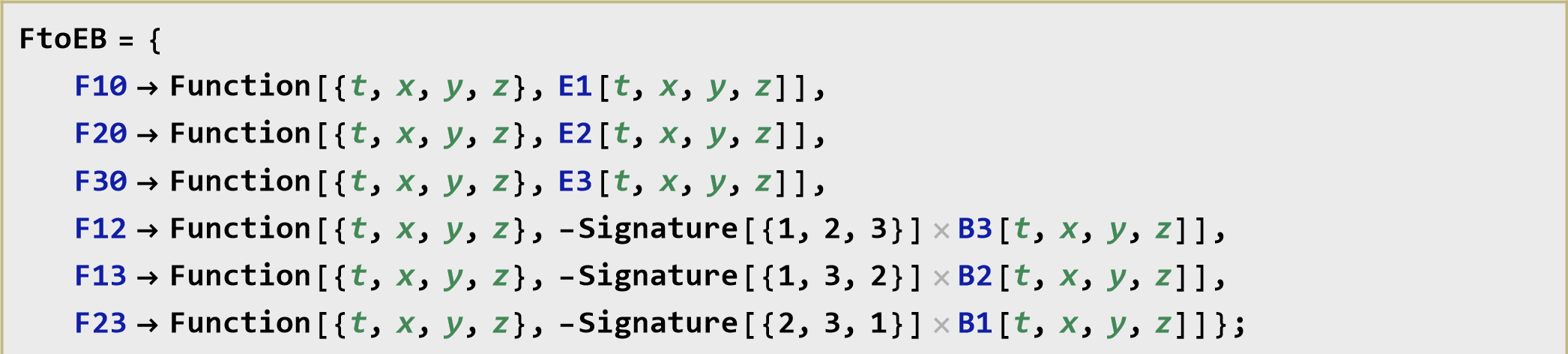}
\includecode{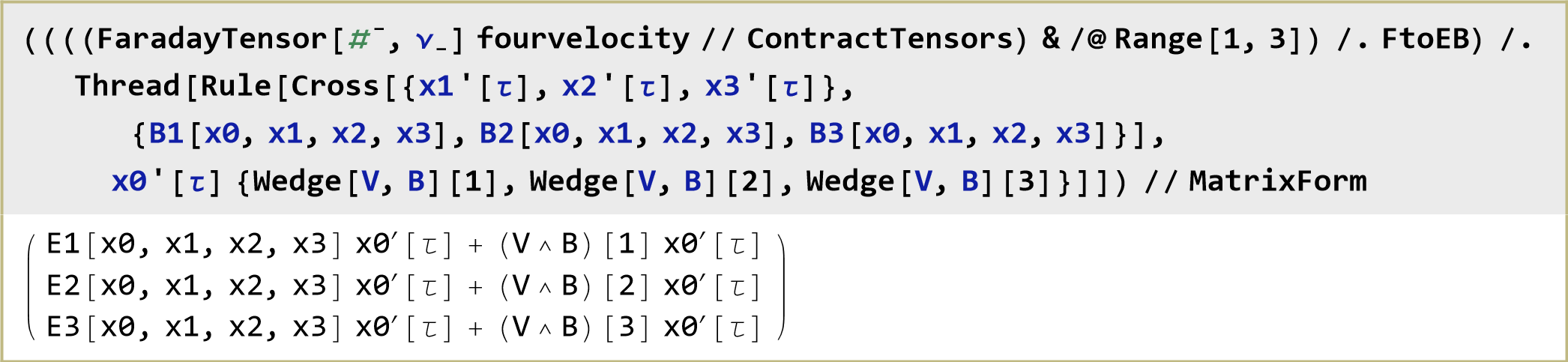}
Because \(F_{\mu \nu }\) is anti-symmetric, ToExpressionForm converts FaradayTensor into a sum of basis wedge products of infinitesimal displacements.
\includecode{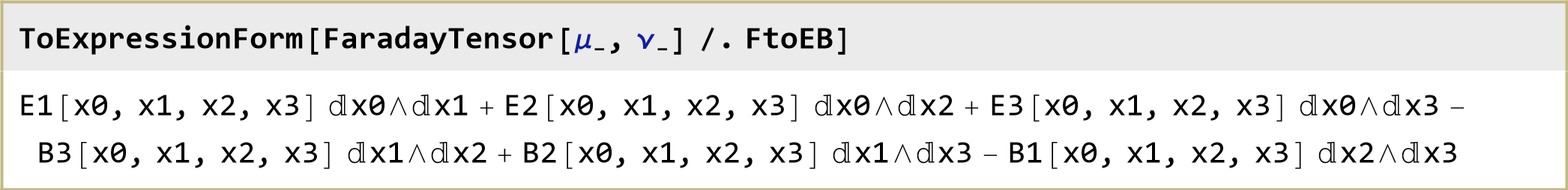}
That \(F_{\mu \nu }\) is a rank-2 Lorentz tensor allows us to work out the Lorentz transformation rules for the electric and magnetic fields under the boost $x = \Lambda x'$. The latter itself may first be constructed as follows.
\includecode{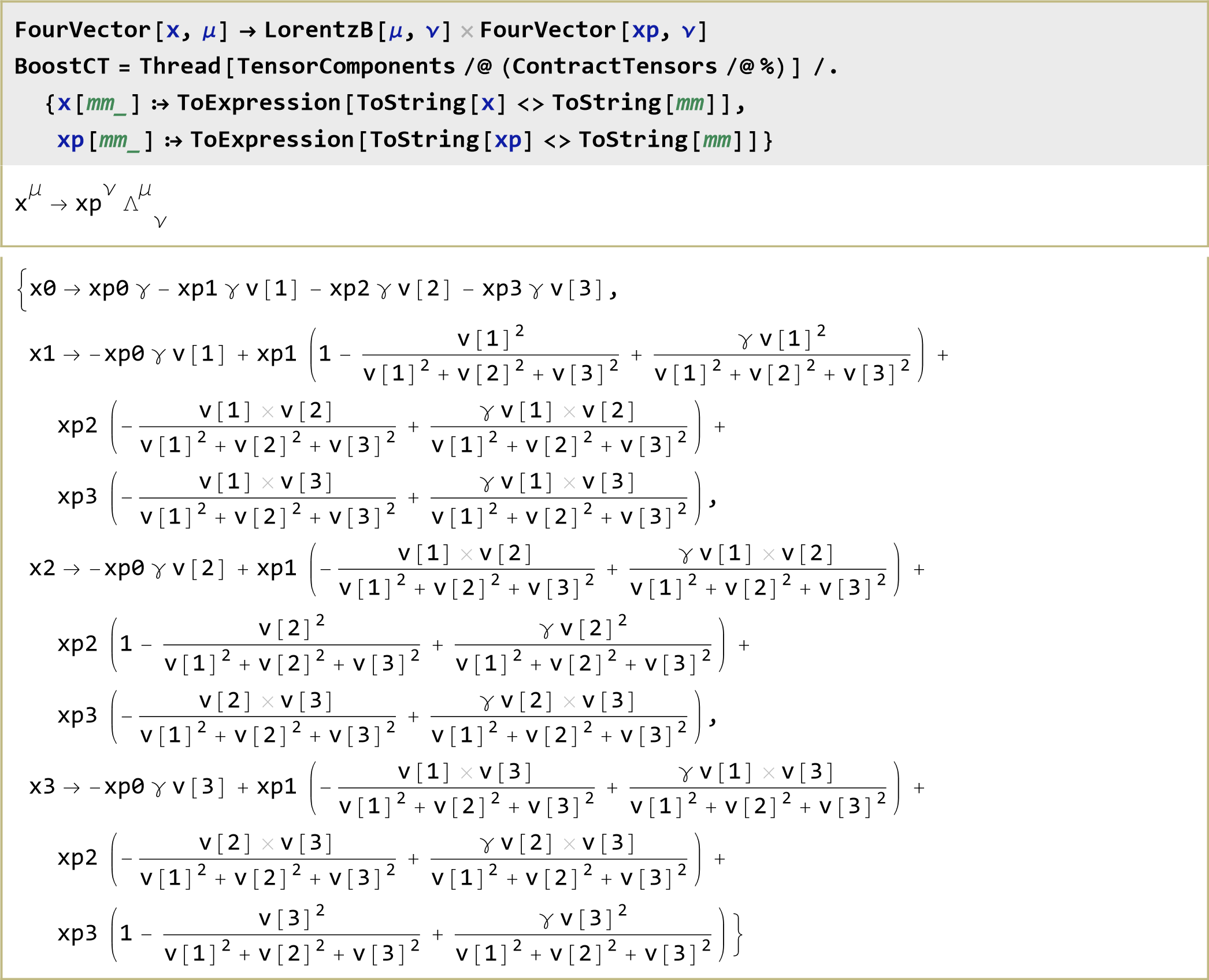}
Apply this coordinate transformation to the Faraday tensor.
\includecode{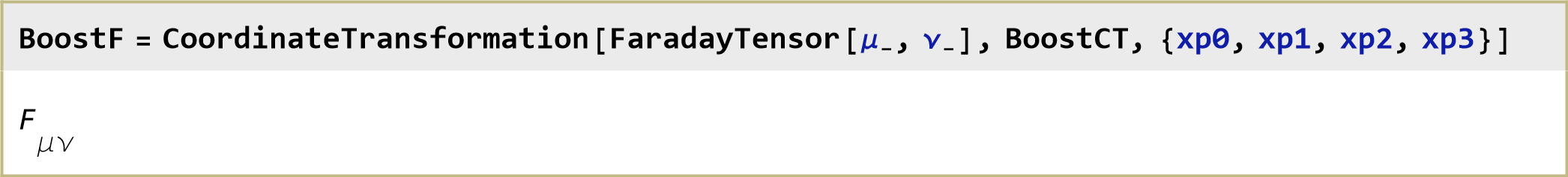}
Let us remind ourselves, the following components of the Faraday tensor form, respectively, the electric and magnetic fields in a given inertial
frame.
\includecode{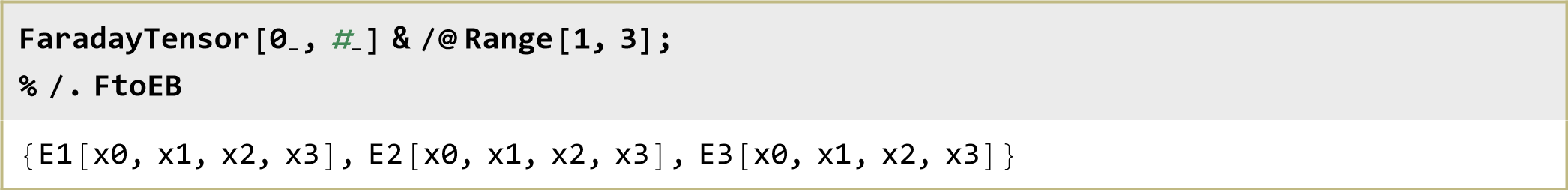}
\includecode{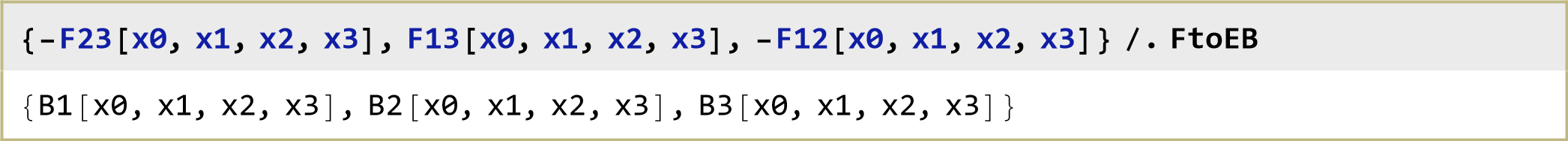}
\newpage
Equipping ourselves with the following notation-related Rule{'}s --
\sizeincludecode{1}{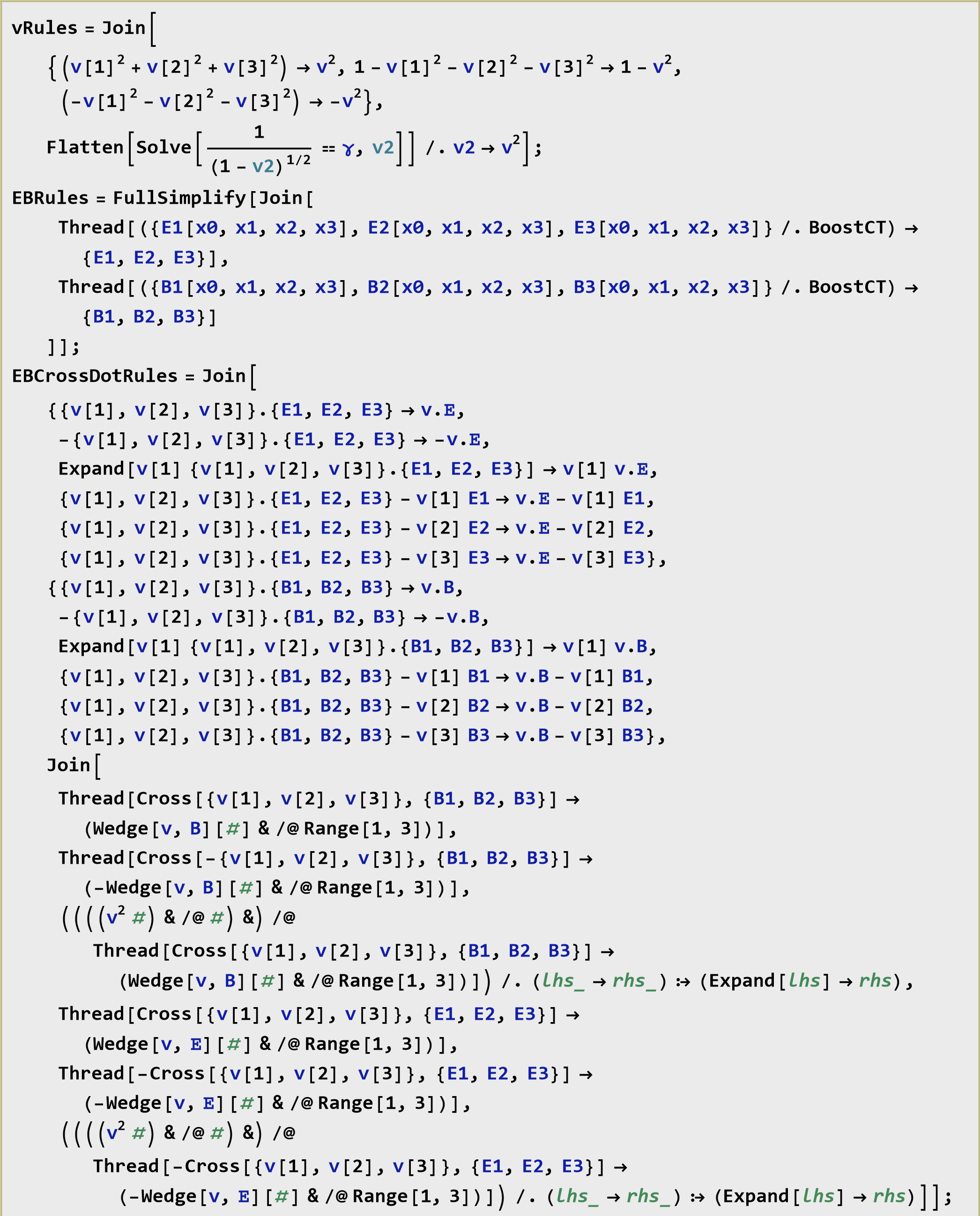}
-- we may turn to the electric field components in the primed frame, but in terms of the electric and magnetic fields in the un-primed one. Here and below, we denote Eperp = \(\overset{\rightharpoonup}{E}\) - \(\hat{v}\) \(\hat{v}\).\(\overset{\rightharpoonup}{E}\) and \\
Bperp = \(\overset{\rightharpoonup}{B}\) - \(\hat{v}\) \(\hat{v}\).\(\overset{\rightharpoonup}{B}\) { }to be the components of, respectively, the electric and magnetic fields perpendicular to the velocity \(\overset{\to}{v}\), with the unit norm version \(\hat{v}\) = \(\overset{\rightharpoonup}{v}\)/$v$, $v$ $\equiv $ $|$\(\overset{\rightharpoonup}{v}\)$|$, 
\includecode{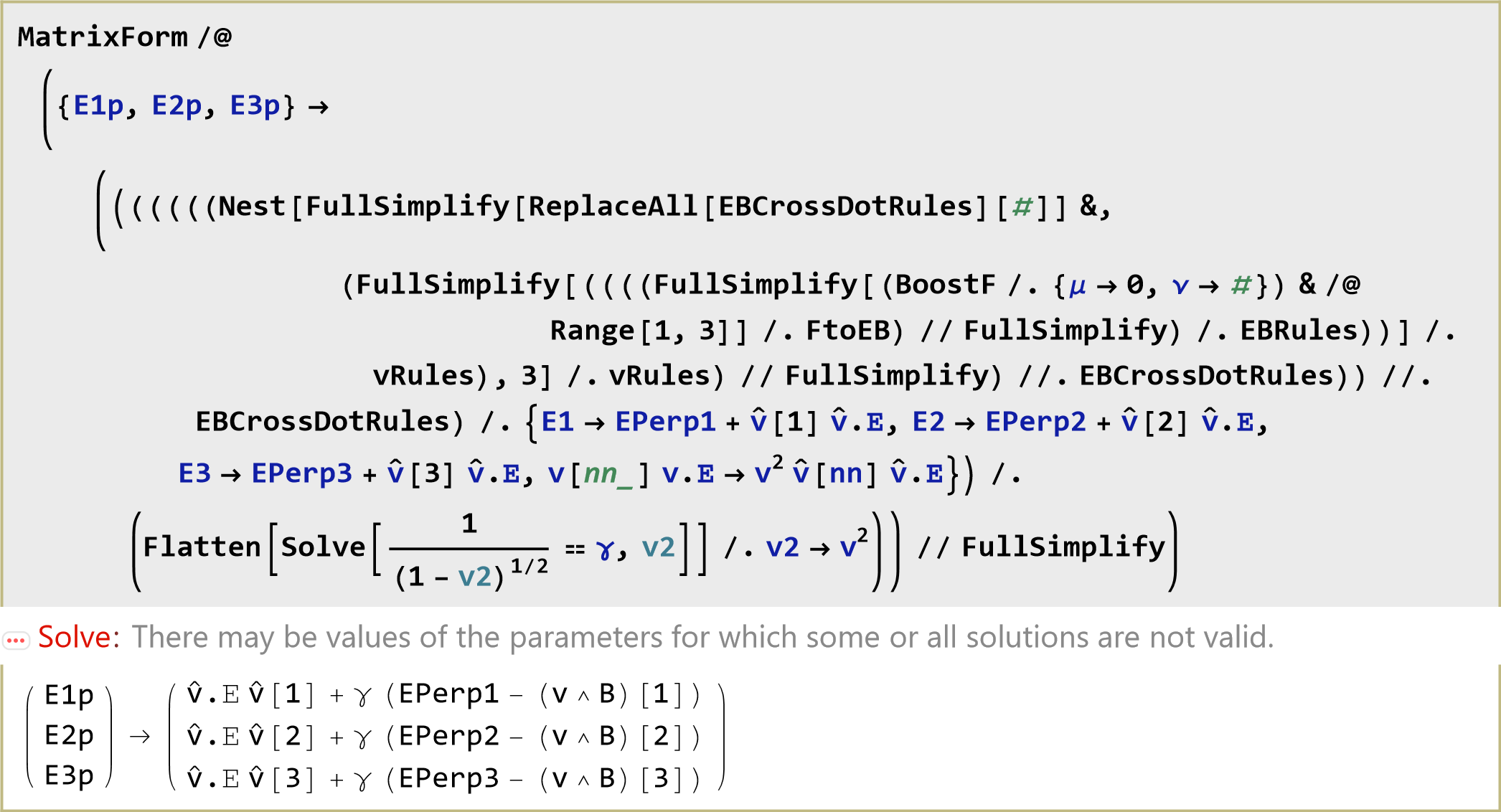}
\newpage
Similarly, the magnetic field components in the primed frame, but in terms of the electric and magnetic fields in the un-primed one are as follows.
\includecode{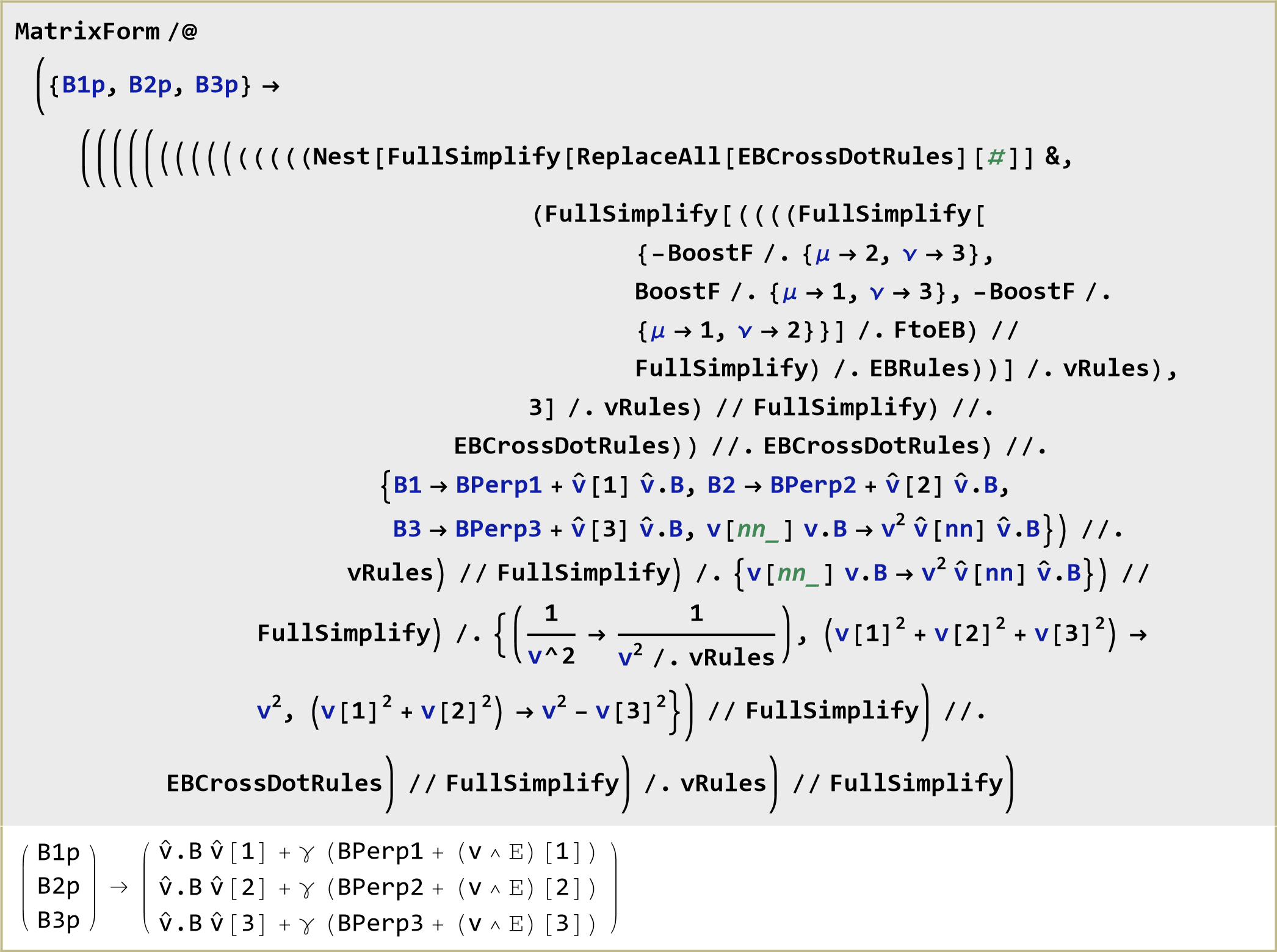}
With the electric and magnetic fields properly identified we may now recover Maxwell{'}s equations. 
\[
\partial _{\mu }F^{\mu \nu }=J^{\nu }\quad \text{and} \quad \quad \partial _{[\alpha }F_{\mu \nu ]}=0\quad (\text{or, equivalently, }\tilde{\epsilon }^{\alpha \beta \mu \nu } \partial _{[\beta }F_{\mu \nu ]}=0).
\]
\newpage
First we define the electric 4-current and the electric and magnetic 3-vectors.
\includecode{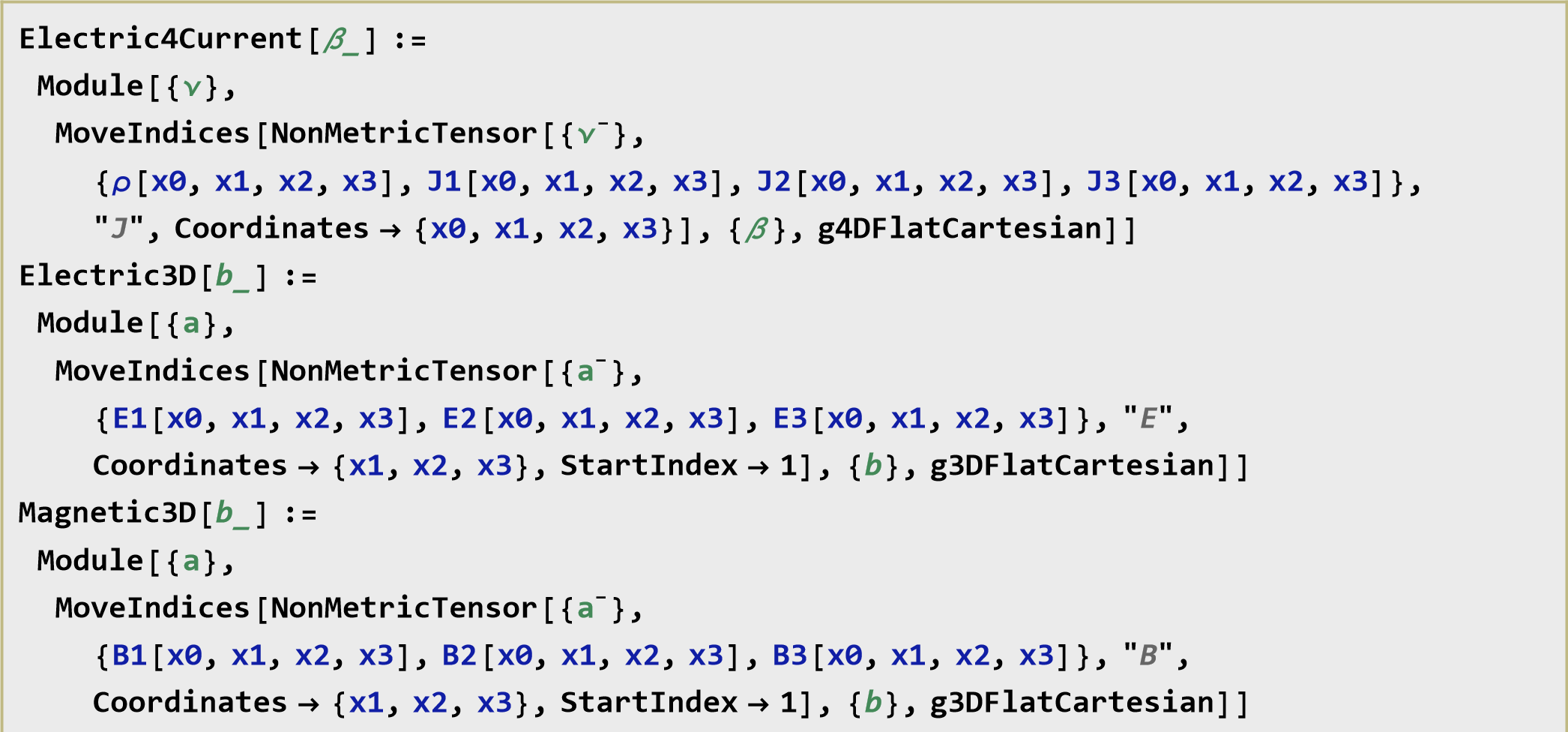}
The zeroth equation \(\partial _{\mu }F^{\text{$\mu $0}}=J^0\) is Gauss{'} law: divergence of \(\overset{\rightharpoonup}{E}\) = charge density $\equiv$ $\rho $.
\includecode{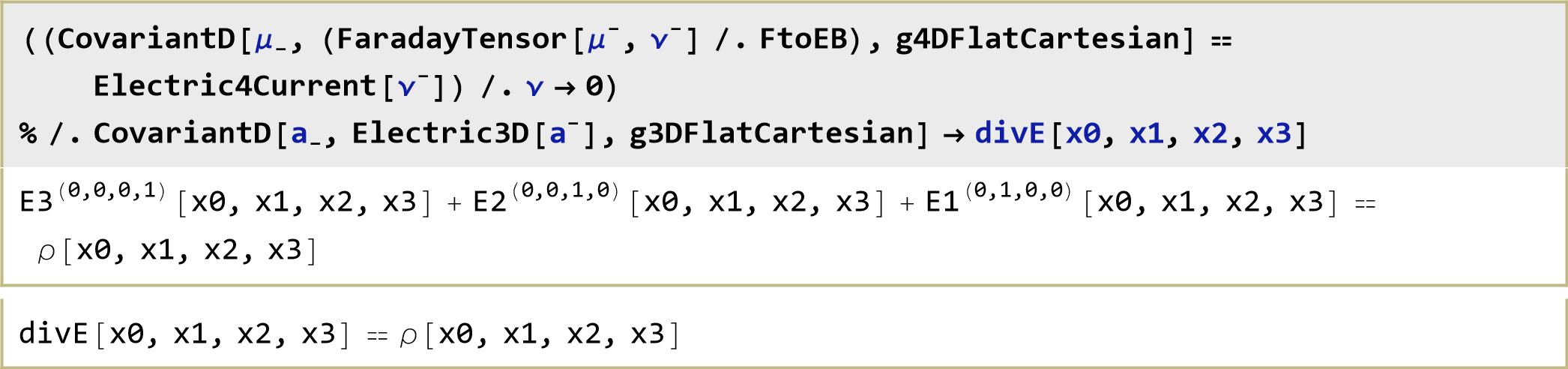}
Next, to help us recover the vector calculus form of Maxwell{'}s equations, we create Rule{'}s identifying the curl of $B$ and curl of $E$.
\includecode{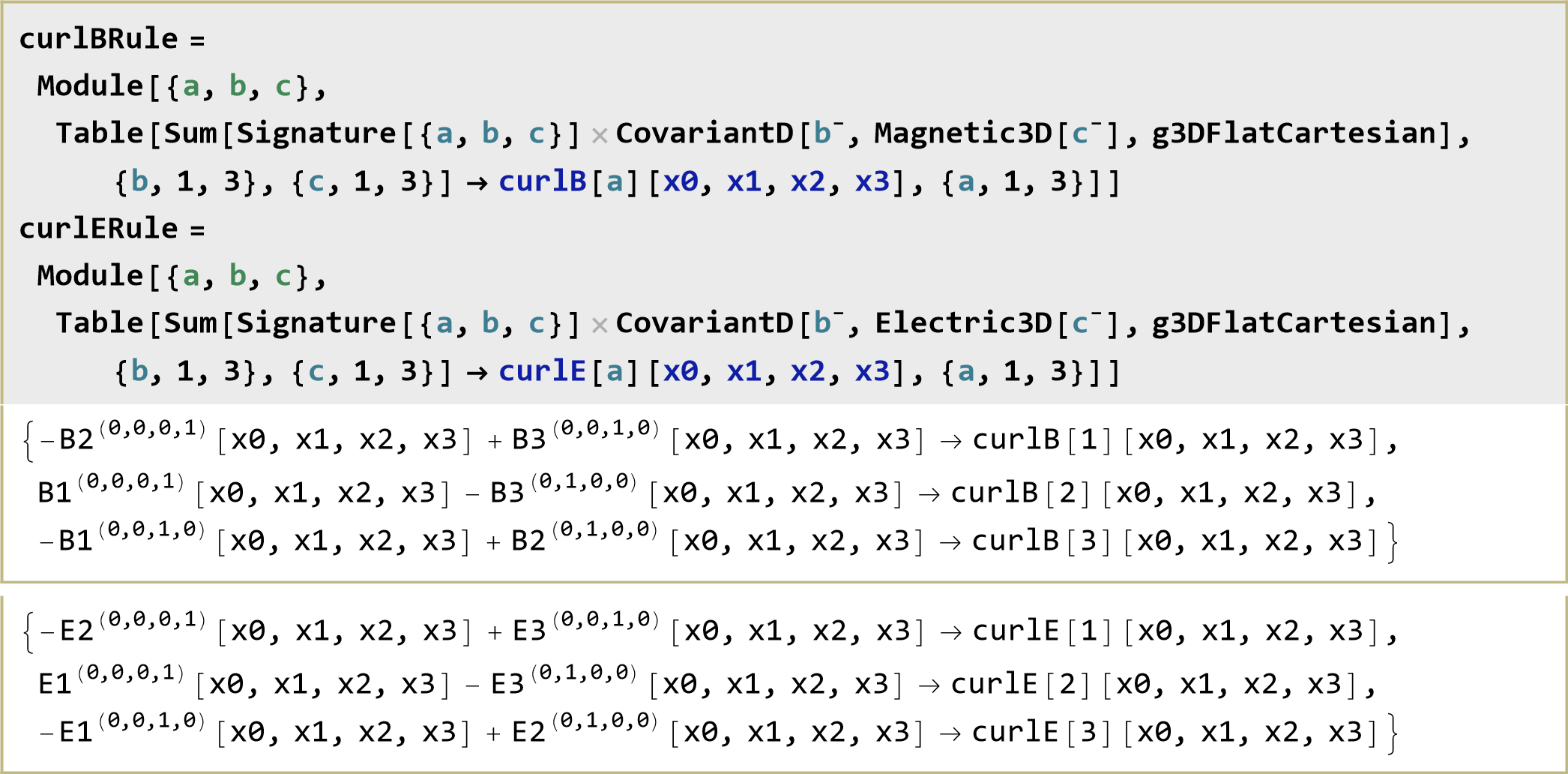}
The spatial equations \(\partial _{\mu }F^{\text{$\mu $i}}=J^i\) involves the time derivative of \(\overset{\rightharpoonup}{E}\), the curl of \(\overset{\rightharpoonup}{B}\),
and the spatial electric current -- Ampere{'}s law.
\includecode{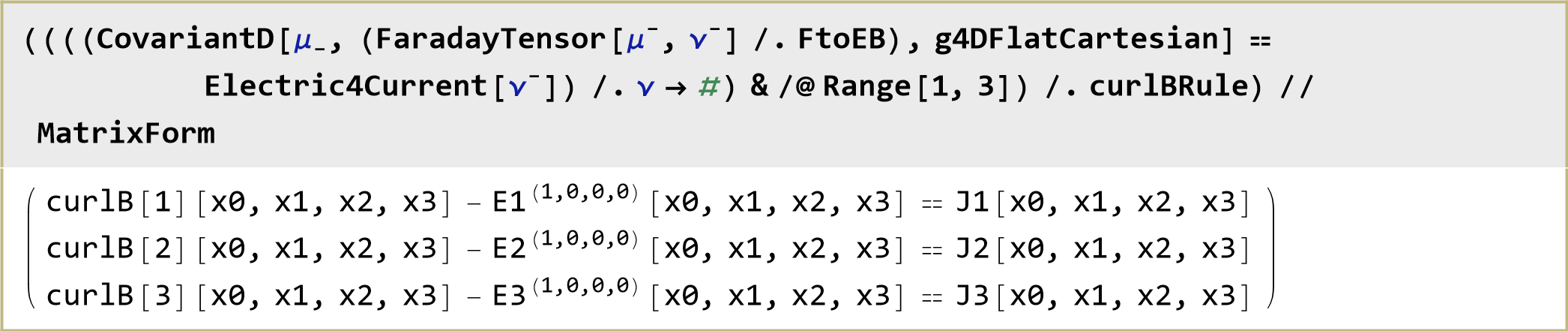}
The \(\partial _{[\alpha }F_{\mu \nu ]}=0\) equations can be denoted as d$F = 0$, where \((\dd F)_{\alpha \mu \nu }\) itself can be computed as follows.
\includecode{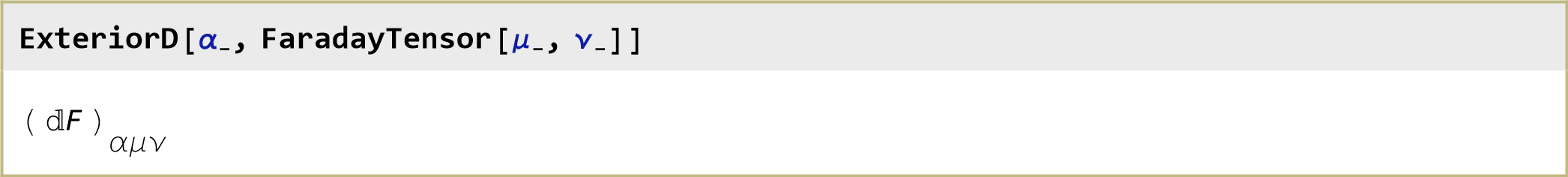}
Let us work out the equivalent form \(\tilde{\epsilon }^{\alpha \beta \mu \nu }\) \(\partial _{[\beta }F_{\mu \nu ]}=0\) = \((\star dF)^{\alpha }\). The zeroth equation yields divergence of \(\overset{\rightharpoonup}{B}\) = 0.
\includecode{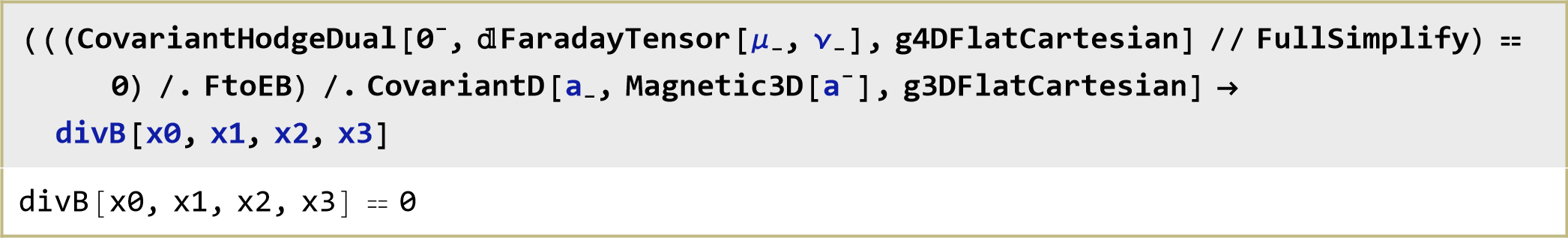}
The spatial ones return curl $E$ = -\(\partial _tB\) -- Faraday{'}s law.
\includecode{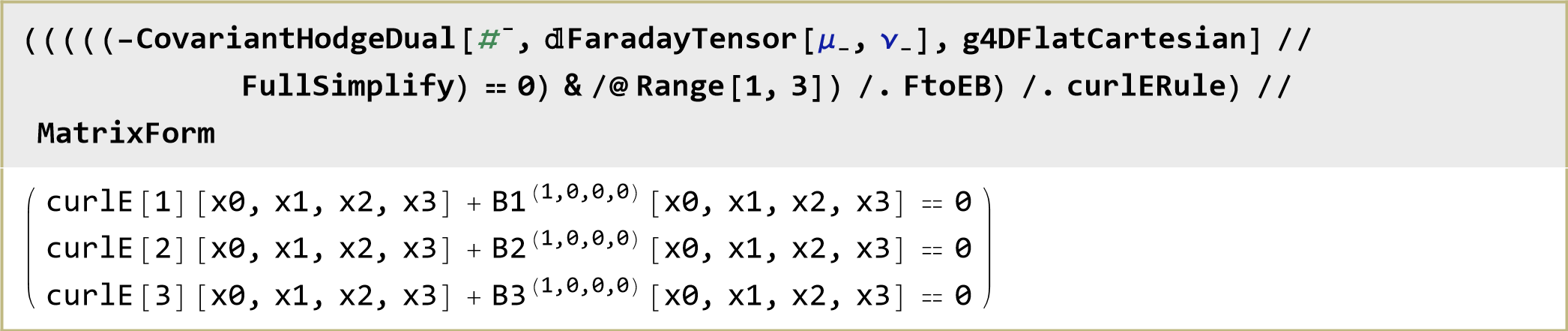}
\boldtitle{Poincare lemma} \qquad Actually, the Poincare lemma tells us, d$F=0$ implies $F=\dd A$ for some 1-form $A$. Hence, we may first define A; followed by taking its exterior derivative; which then allows us to define $F$ in terms of $A$.
\includecode{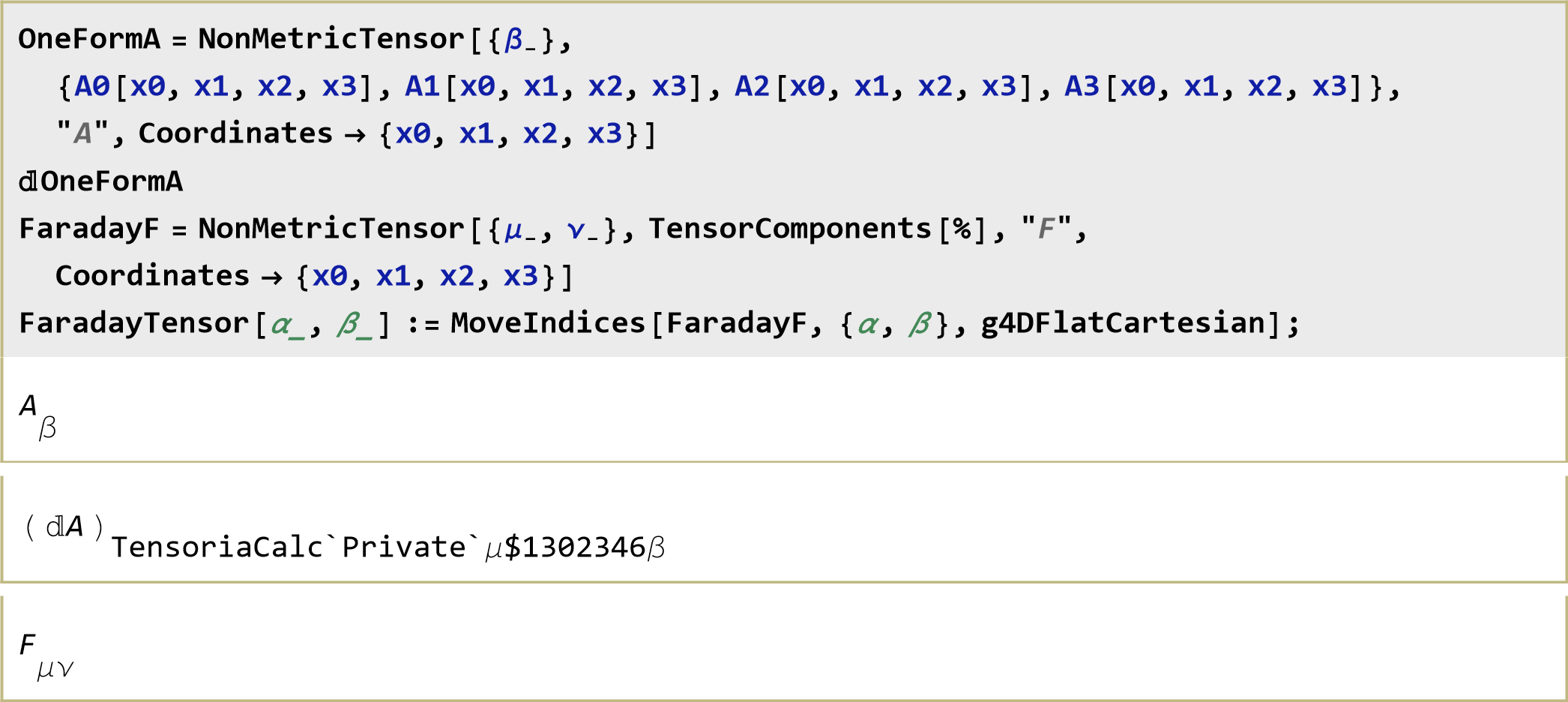}
One may readily check that the electromagnetic field strength (i.e., Faraday) tensor is invariant under \(A_{\mu }\rightarrow A_{\mu }+\partial _{\mu}\Lambda\), where $\Lambda $ is an arbitrary scalar function.
\includecode{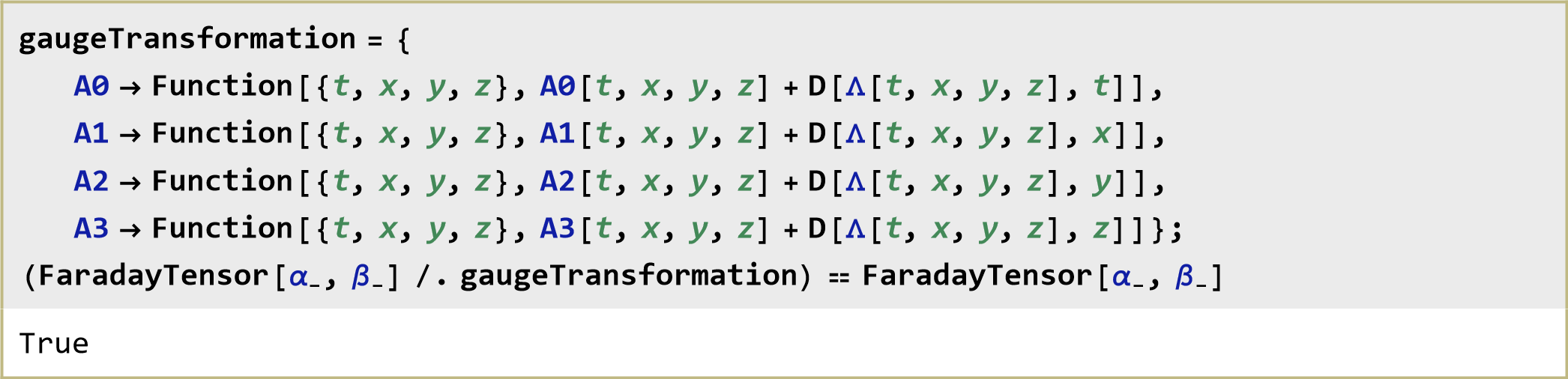}
In Fourier spacetime \(\partial _{\mu }F^{\mu \nu }\rightarrow (-i)^2 \left(k_{\mu }k^{\mu } \delta ^{\nu }{}_{\gamma }-k^{\nu }k_{\gamma }\right)\tilde{A}^{\gamma}\equiv M^{\nu }{}_{\gamma }\tilde{A}^{\gamma }\). That \(F^{\mu \nu }\) is invariant gauge transformations is also why the Fourier wave operator \(M^{\nu }{}_{\gamma }\) is non-invertible due to the existence of a null eigenvector: \(M^{\nu }{}_{\gamma }\)\(k^{\gamma }=0\). Hence, to solve \(\partial _{\mu }F^{\mu \nu }\) = \(J^{\nu }\) systematically, one needs to {``}fix a gauge{''}. We shall choose the Lorentz gauge \(\partial ^{\mu}A_{\mu }=0\).
\includecode{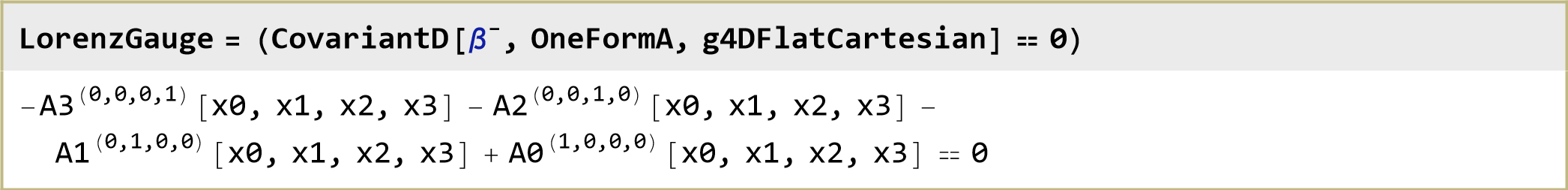}
Utilizing the Lorenz gauge, Maxwell{'}s equations reduces to four wave equations \(\partial _{\mu }\partial ^{\mu }A_{\nu }=J_{\nu }\).
\includecode{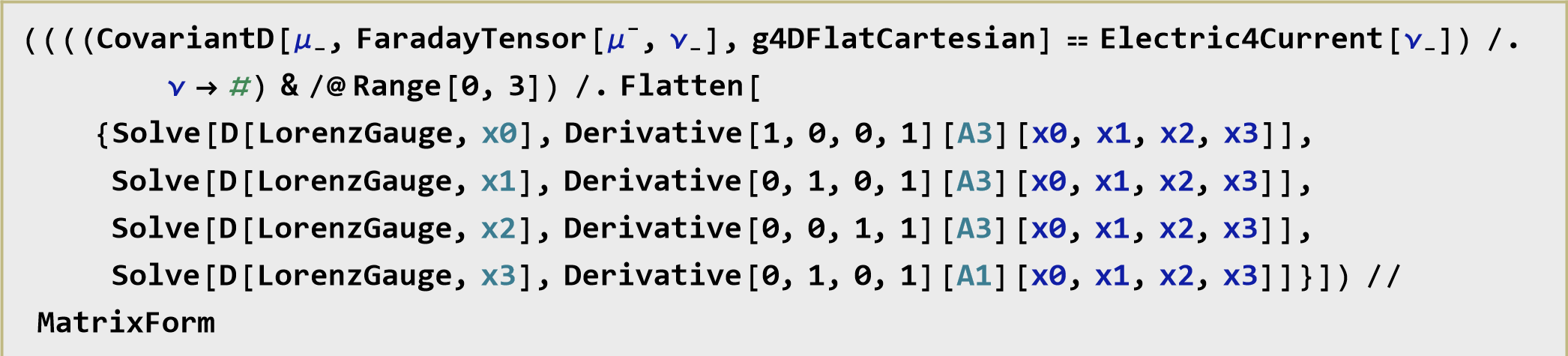}
\includecode{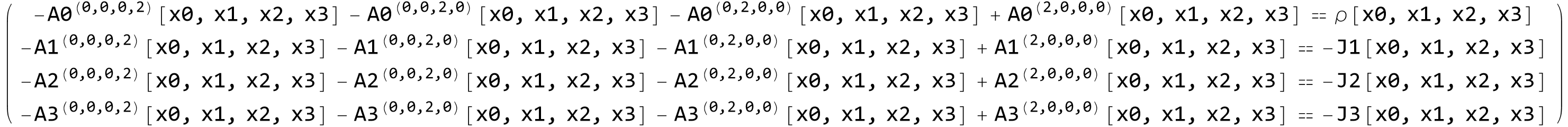}

\section{Weakly Curved Static Spacetimes}

In this section we shall study weakly curved static (time-independent) spacetimes, where geometric curvature is very small. 

Let us begin with the following ansatz for the spacetime interval.
\includecode{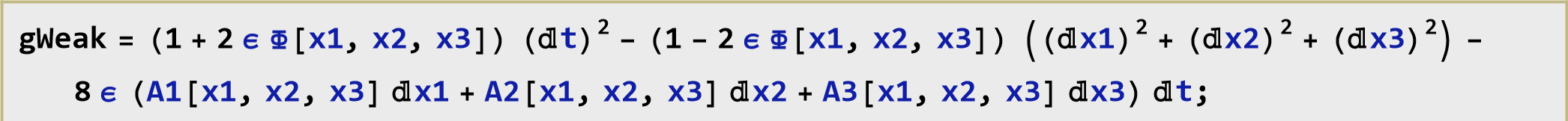}
We shall impose Einstein{'}s equations
\[
G^{\alpha \beta }=8\text{$\pi $G}_NT^{\alpha \beta }
\]
with the space-space of the stress tensor set to zero,
\[
T^{\text{ab}}=0;
\]
but since $\Phi $ and \(A_i\) are supposed to be small, we will expand the Einstein tensor up to only first order in $\Phi $ and \(A_i\). This is why we have inserted a fictitious parameter $\epsilon $; it allows for easy power-counting via Series. Specifically, Series[$\tau \_$Tensor,$\{\epsilon$,0,n$\}$] would return $\tau $ but with Series[$\#$,$\{\epsilon $,0,n$\}$]$\&$ applied to its TensorComponents.
\includecode{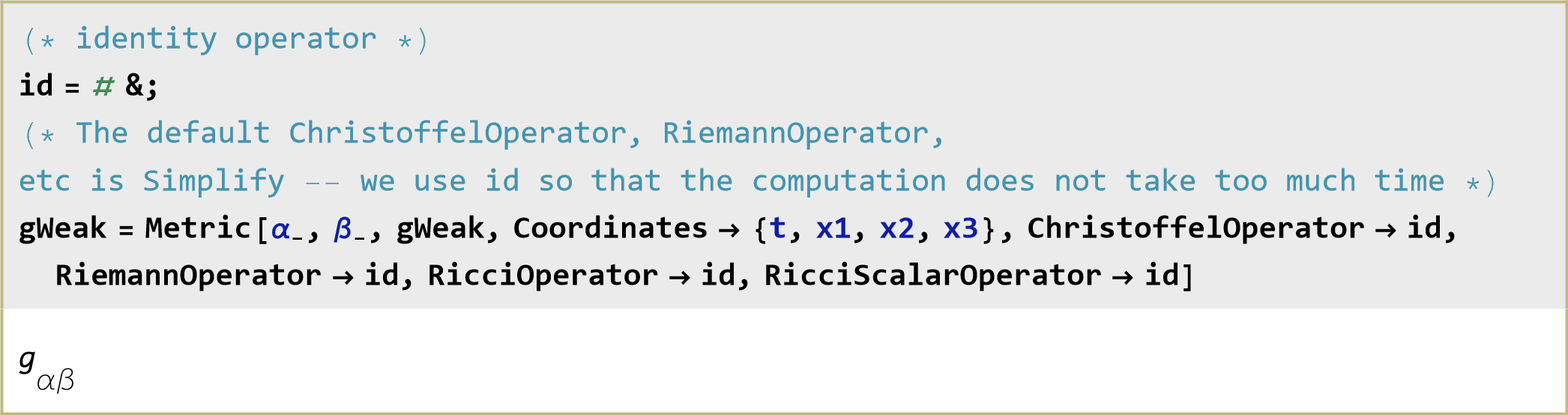}
The linearized Einstein tensor is as follows.
\includecode{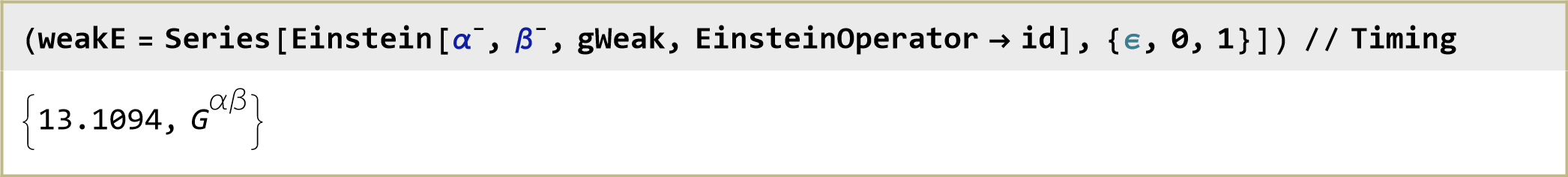}
We see the metric ansatz is consistent with the assumption that the space-space components of the matter stress tensor are zero. Defining the matter stress tensor as a Tensor object
\includecode{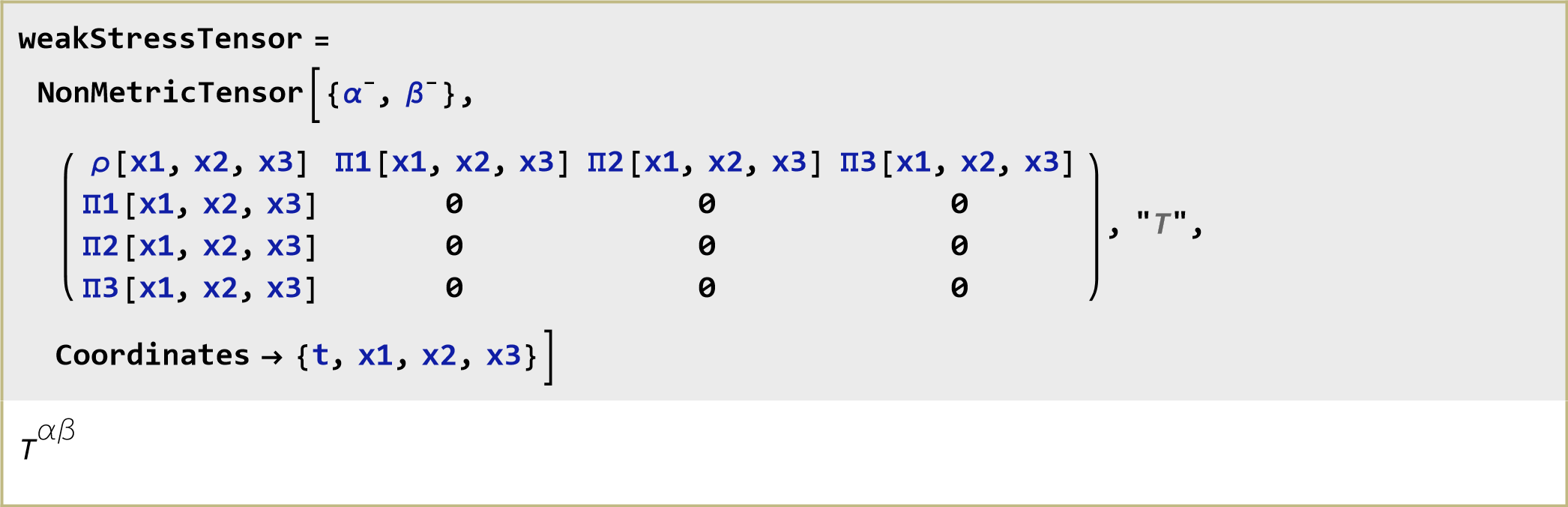}
we may impose its flat spacetime conservation (divergence-free) condition, because the Einstein tensor itself is already order \(\epsilon ^1\). This in turn informs us, the spatial momentum is divergence-free.
\includecode{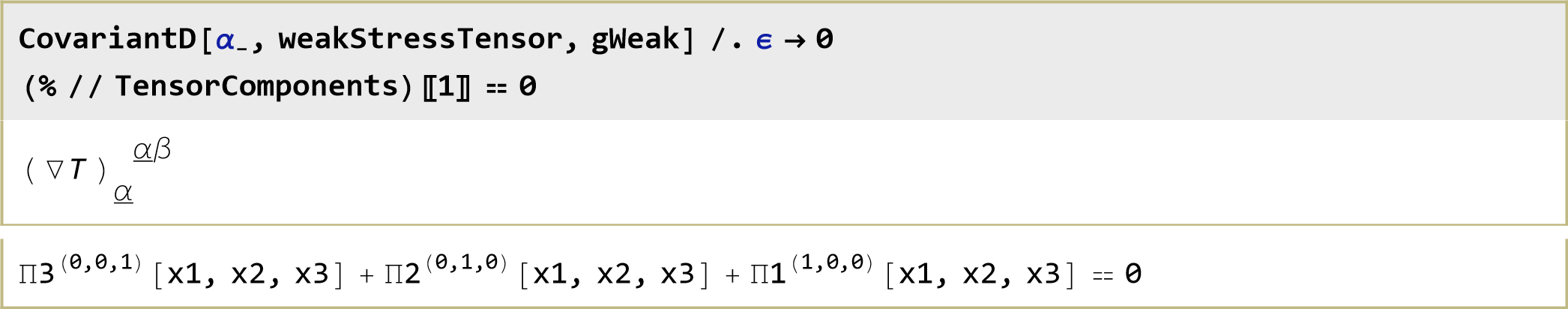}
\newpage
Let us introduce div $\vec{A}$ = \(\partial _jA_j\) to denote the spatial divergence of the 1-form potential \(A_i\).
\includecode{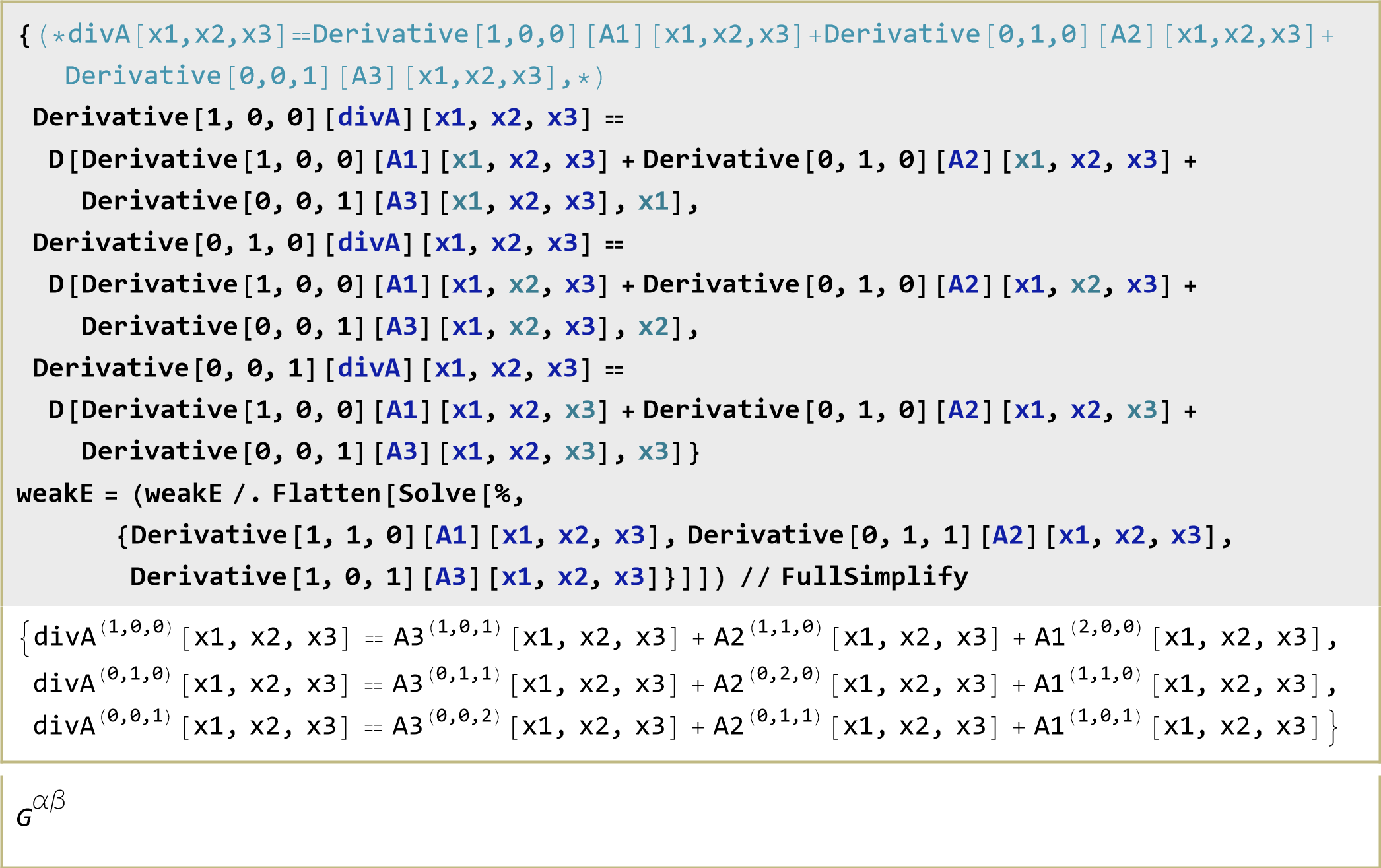}
The linearized Einstein{'}s equations now reduces to the following.
\begin{align}
\label{Weak1}
	\nabla^2\Phi =4\text{$\pi $G}_N\rho 
	\quad \text{and} \qquad
	\nabla ^2A_i-\partial_i\partial_jA_j=4\text{$\pi $G}_N\Pi^i .
\end{align}
\includecode{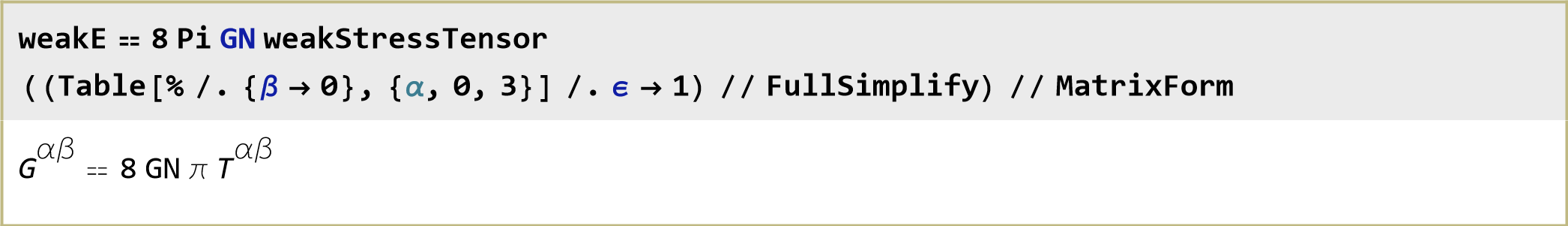}
\includecode{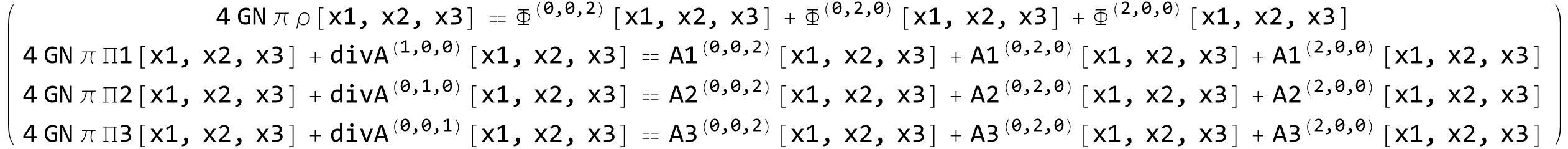}
Since the conservation of the matter stress tensor requires \(\Pi ^i\) to be divergence free, by decomposing \(A_i\) into a gradient plus a divergence-free piece, \(A_i=\partial _iA+A_i{}^T\) (where \(\partial _iA_i{}^T=0\)), we see that the gradient portion drops out; and the divergence-free spatial momentum sources the divergence-free portion of \(A_i\): \(\nabla ^2\partial _iA+\nabla ^2A_i{}^T-\partial _i\nabla ^2A=4\text{$\pi $G}_N\Pi ^i\).
\begin{align}
\label{Weak2}
\nabla ^2A_i{}^T=4\text{$\pi $G}_N\Pi ^i .
\end{align}
To sum, whenever the matter stress tensor is consists solely of some time-independent energy $\rho $ and spatial momentum \(\Pi ^i\), it sources
the Newtonian potential $\Phi $ and the spatial-divergence-free {``}1-form potential{''} \(A_i\) with the Poisson equations of equations \eqref{Weak1} and \eqref{Weak2}.

\section{Cosmology}

The Friedmann-Lema{\^ \i}tre-Robertson-Walker (FLRW) geometry describing the universe at the largest scales is based on the assumption of spatial homogeneity and isotropy. This, in turn, amounts to the assumption that the 3-space at constant observer time \textit{t} is a maximally
symmetric one. Let us step through this construction for positive, negative and zero spatial curvatures. 

\subsection{Positive Spatial Curvature}

A 3D maximally symmetric positively curved space can be constructed by viewing it as a 3-sphere of some radius \textit{L} embedded in a 4D flat space; namely, \(\left(x^1\right)^2+\left(x^2\right)^2+\left(x^3\right)^2+\left(x^4\right)^2=L^2\). Solving for \(x^4\) and inserting it back to the 4D flat space metric then returns the induced metric on the 3-sphere.
\includecode{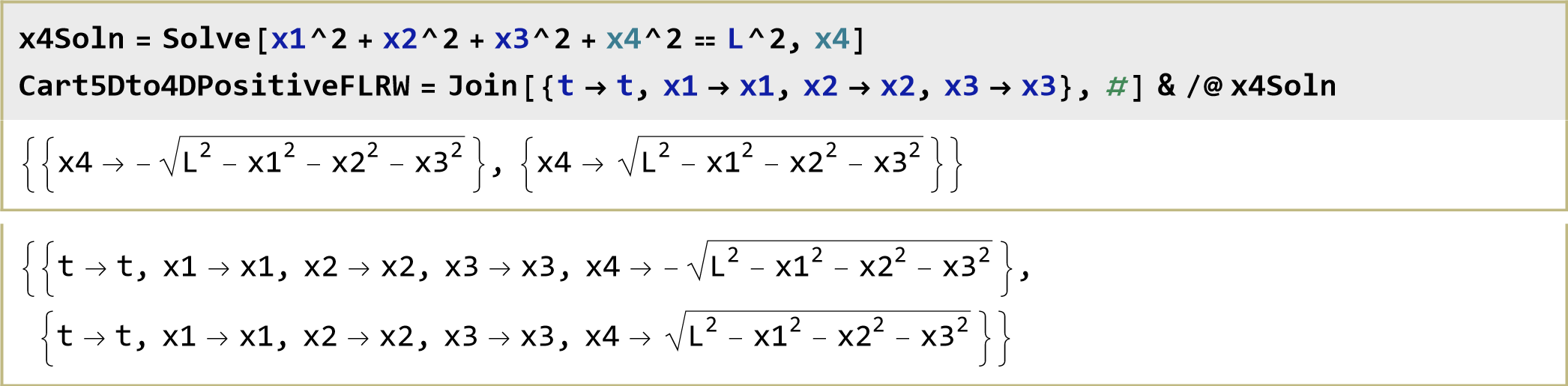}
\includecode{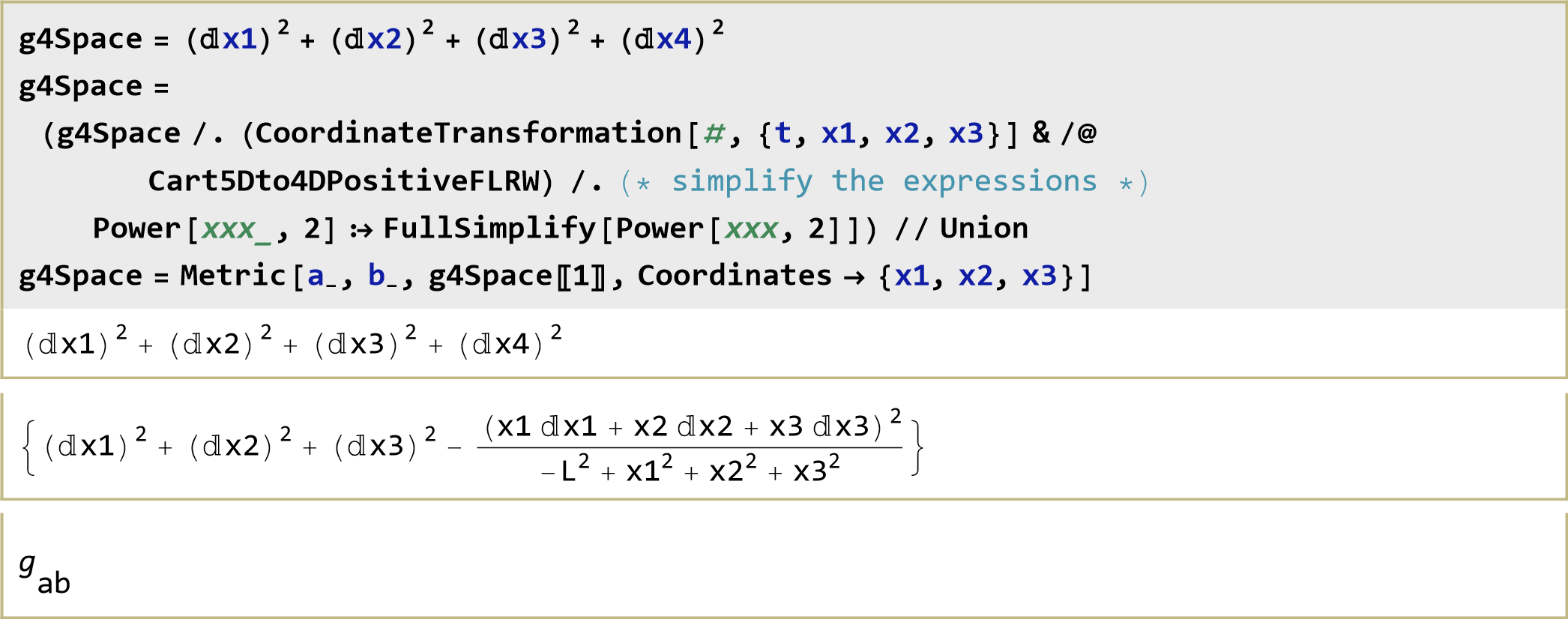}
The following eigenvalues of the induced metric are positive, since
\begin{align}
\left(x^1\right)^2+\left(x^2\right)^2+\left(x^3\right)^2 
\leq \left(x^1\right)^2+\left(x^2\right)^2+\left(x^3\right)^2+\left(x^4\right)^2 
= L^2 .
\end{align}
-- as must be the case for spatial metrics.
\includecode{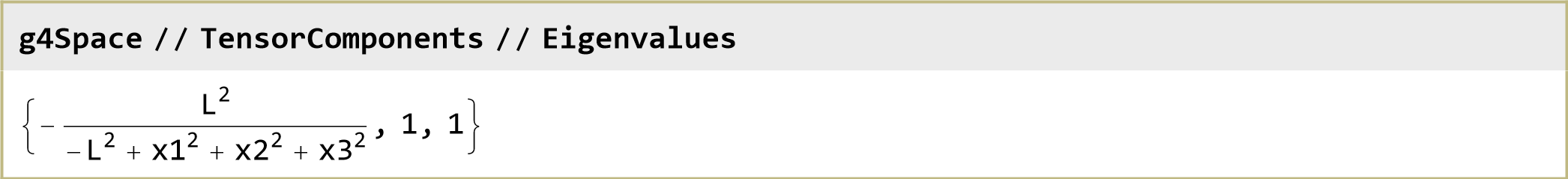}
The Ricci scalar is a positive constant.
\includecode{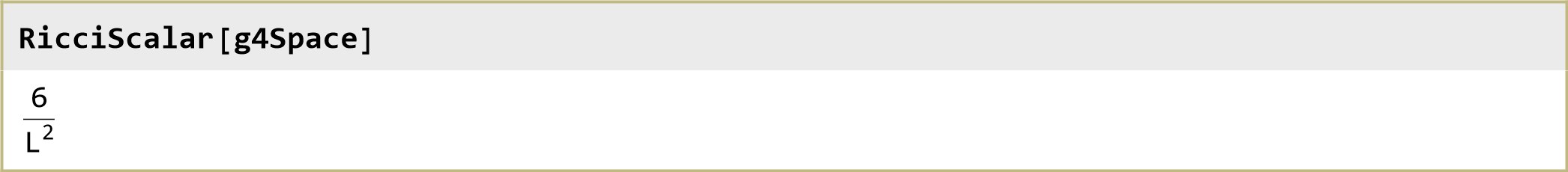}
Viewing the 3D maximally symmetric positively curved space as a 3-sphere allows us to identify its isometries as that of spatial rotations in the ambient 4D flat space. The associated generators (i.e., Killing vectors) are \(x^{[A}\partial ^{B]}\). We first do so without inducing the tensors on the 3-sphere.
\includecode{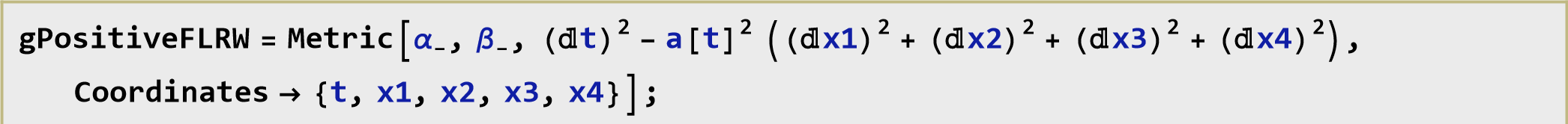}
\includecode{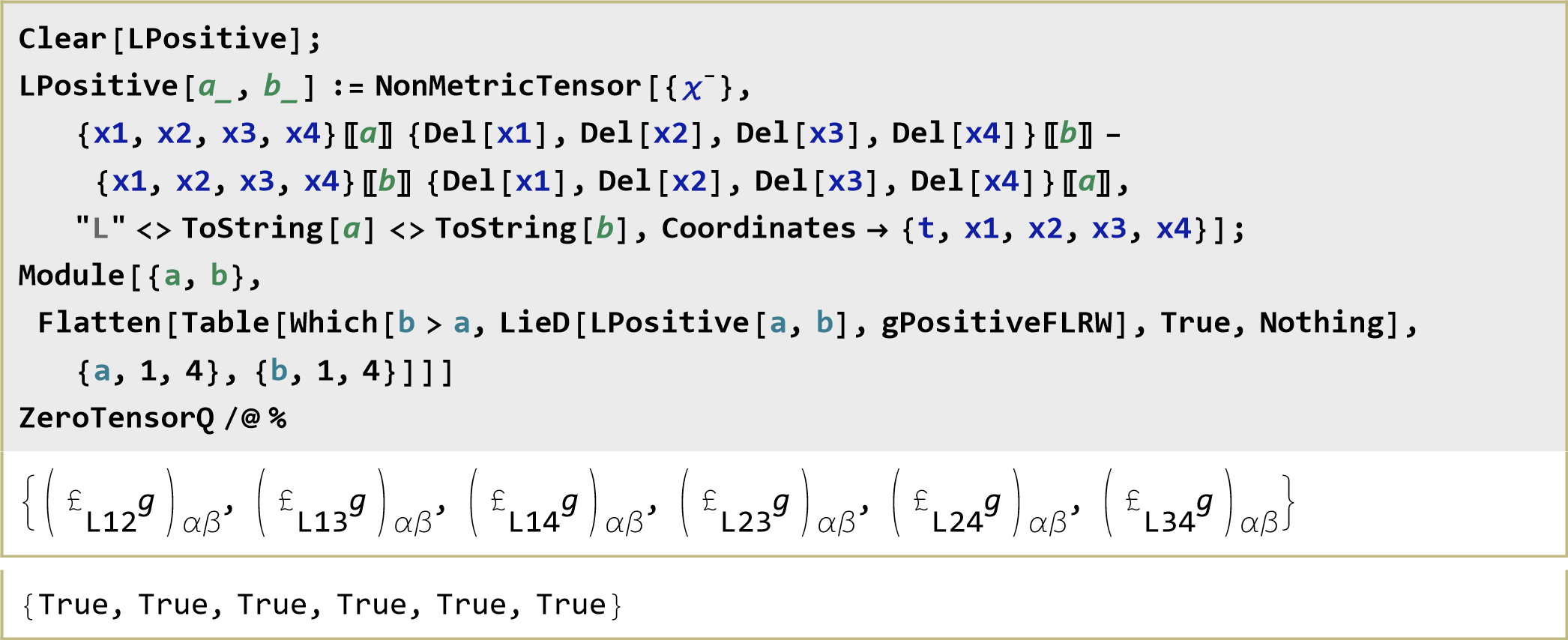}
\includecode{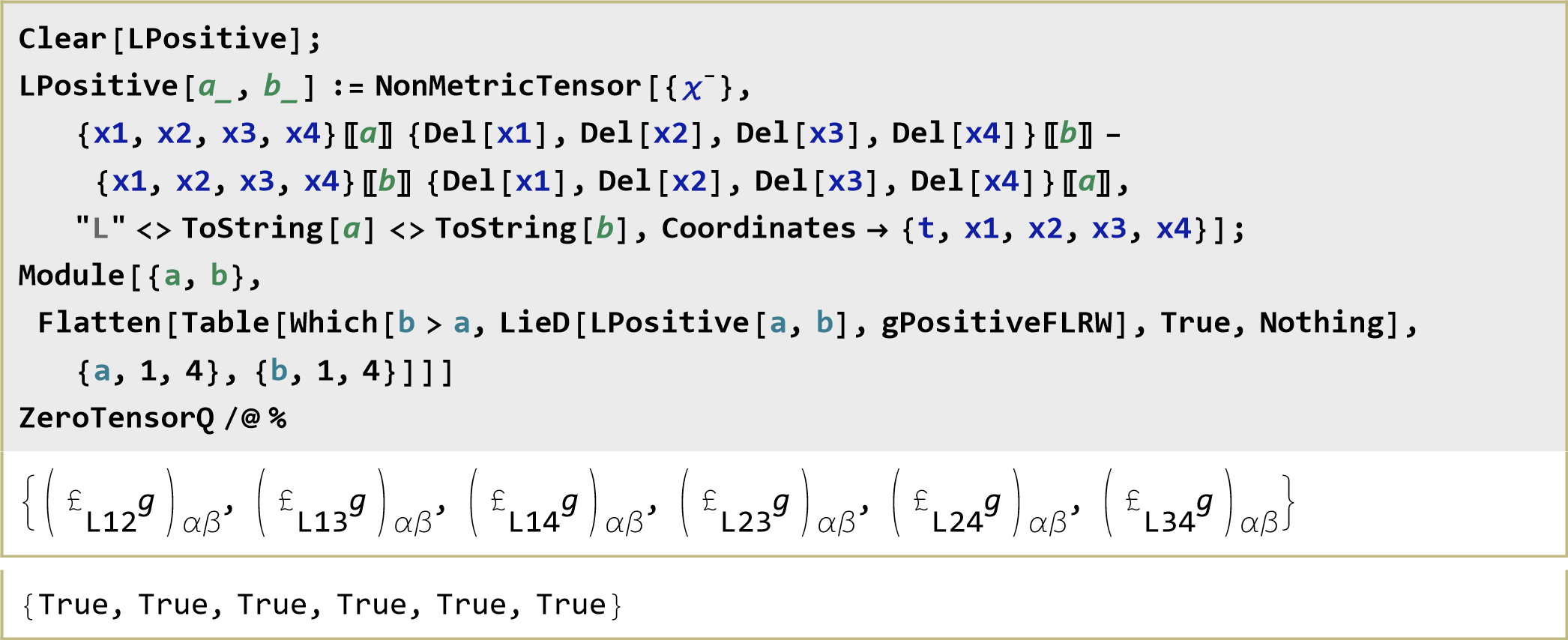}
Next, we induce the Killing vectors and metric onto the 3-sphere.
\includecode{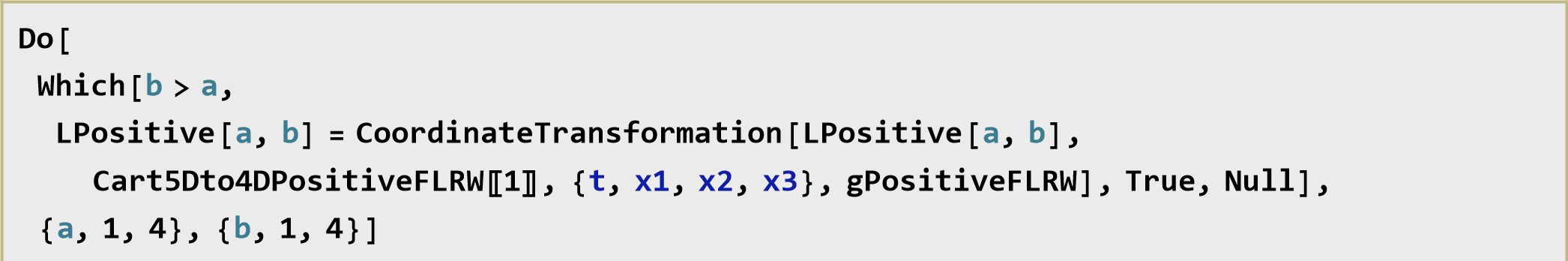}
\includecode{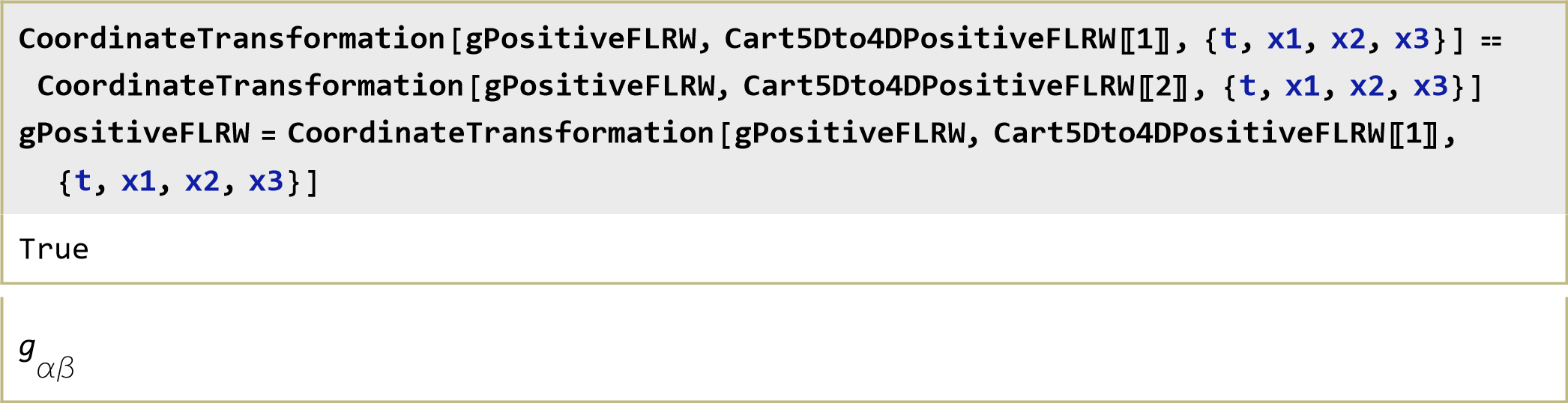}
The off-diagonal character of the metric in the quasi-Cartesian $\{$x1, x2, x3$\}$ system can be cumbersome for practical calculations. Hence, we now switch to trigonometric ones, defined as follows.
\includecode{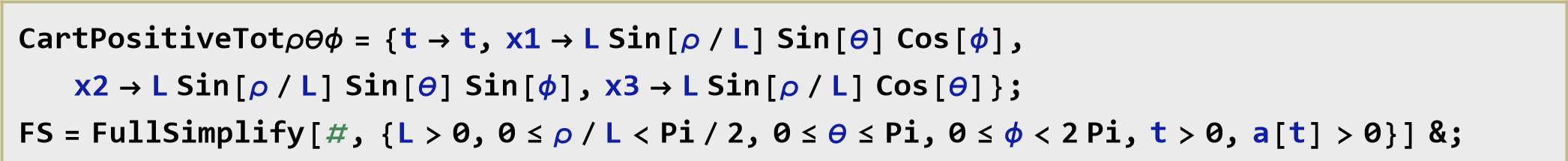}
Apply this to the metric and Killing vectors.
\includecode{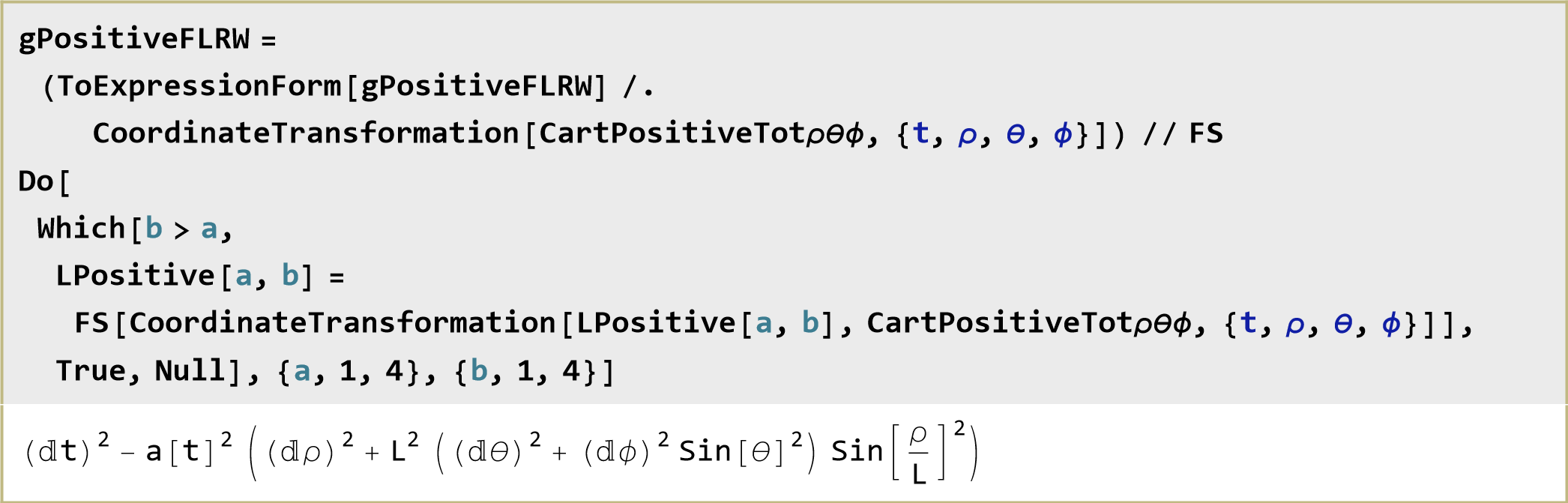}
In this new $\{\rho $, $\theta $, $\phi \}$ system, let us compute the metric together with its orthonormal frame fields. The FlatMetric $\to $ $\{$1,-1,-1,-1$\}$ here indicates the non-zero components of the orthonormal frame fields are \(\varepsilon ^{\hat{t}}{}_t=\left(+g_{\text{tt}}\right){}^{1/2}\), \(\varepsilon^{\hat{\rho }}{}_{\rho }=\left(-g_{\rho \rho }\right){}^{1/2}\), \(\varepsilon ^{\hat{\theta }}{}_{\theta }=\left(-g_{\theta \theta }\right){}^{1/2}\) and \(\varepsilon ^{\hat{\phi }}{}_{\phi }=\left(-g_{\phi \phi }\right){}^{1/2}\).
\includecode{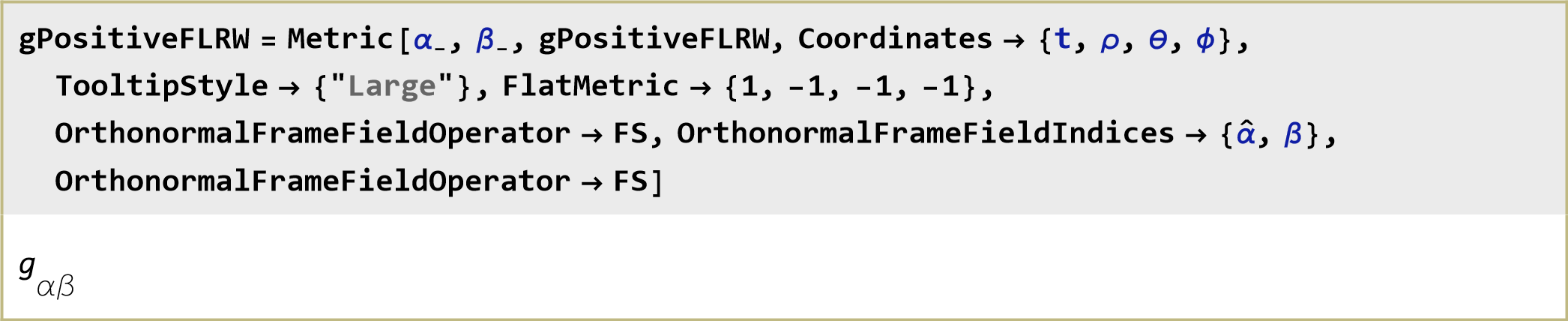}
Killing{'}s equations must still hold.
\includecode{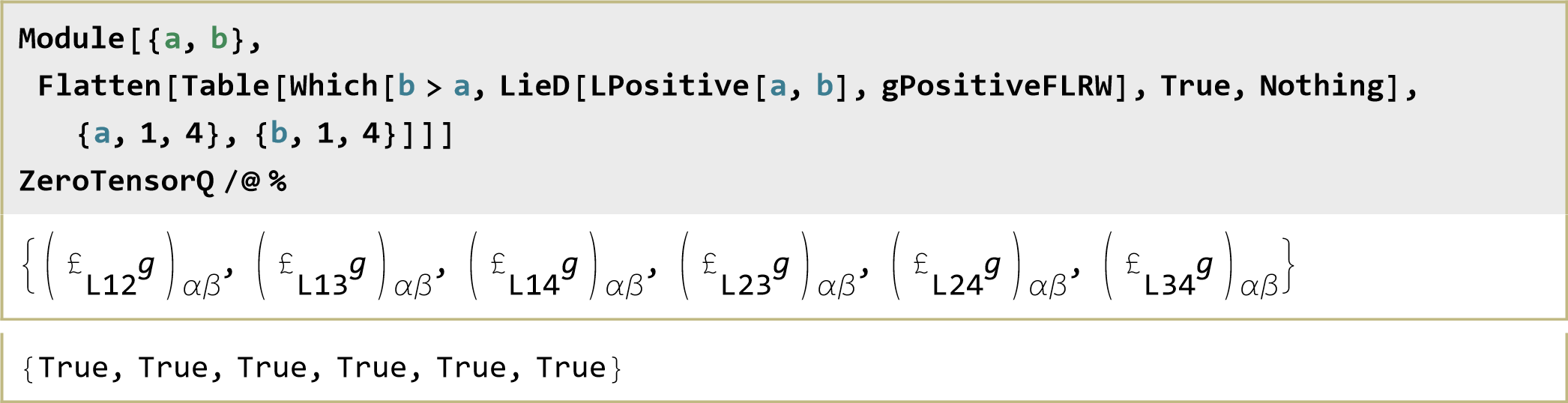}
\includecode{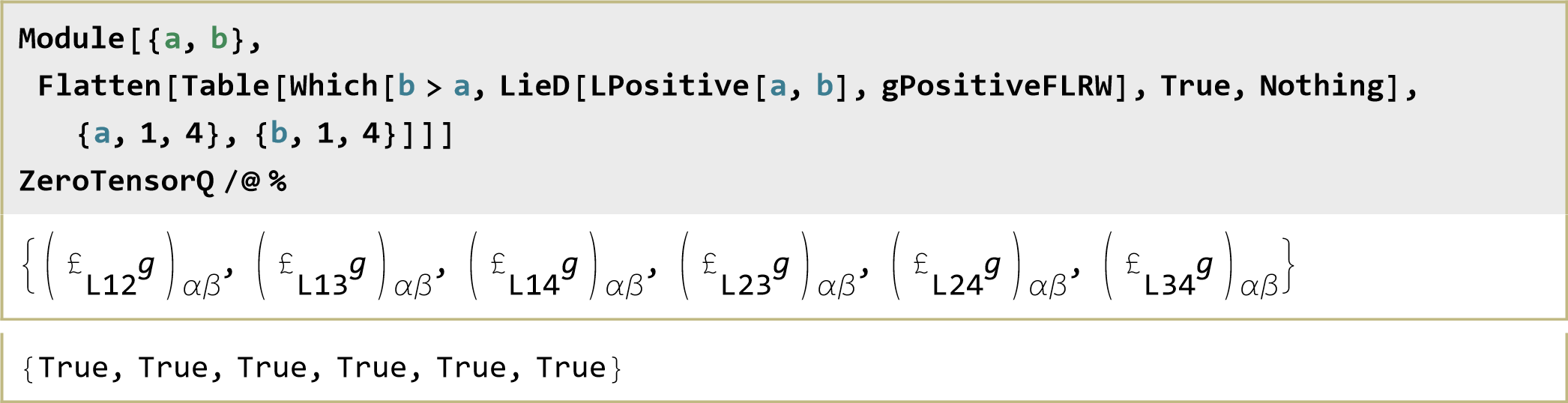}
Let \(U^{\mu }\) = \(\delta ^{\mu }{}_0\) be the timelike vector tangent to the worldline of a particle at rest with the universe. We may verify
its unit length and geodesic nature.
\includecode{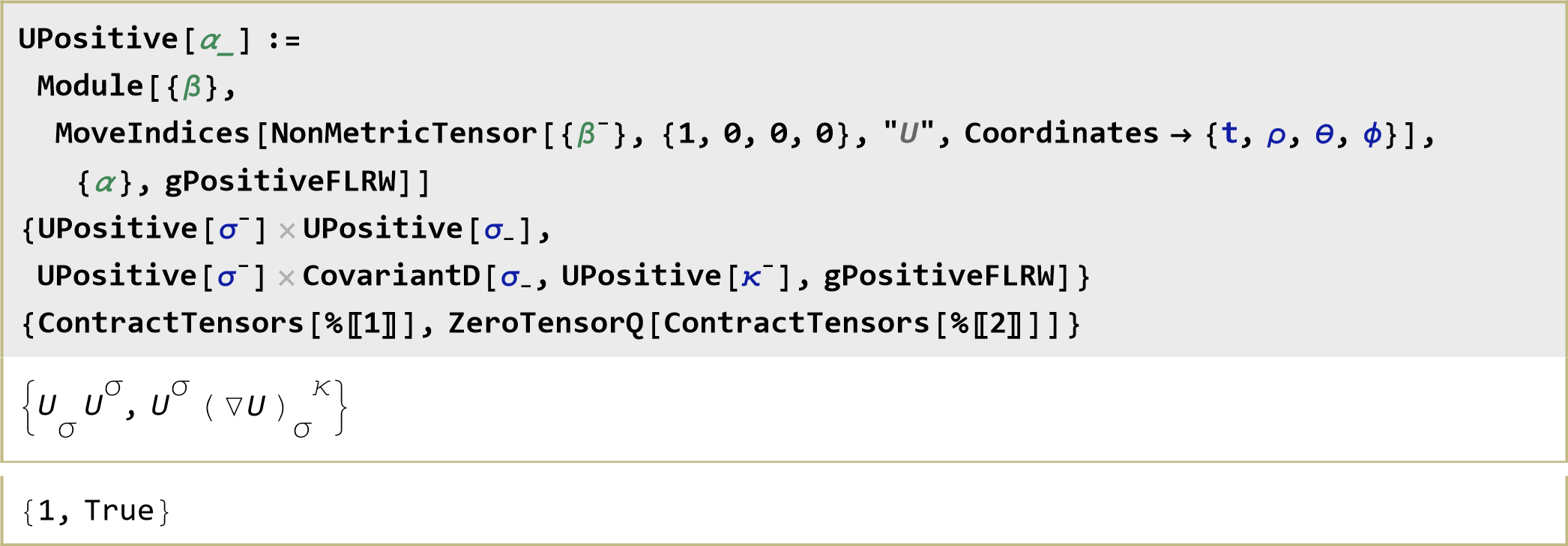}
With this \textit{U} we may also construct the Killing tensor of the FLRW geometry.
\includecode{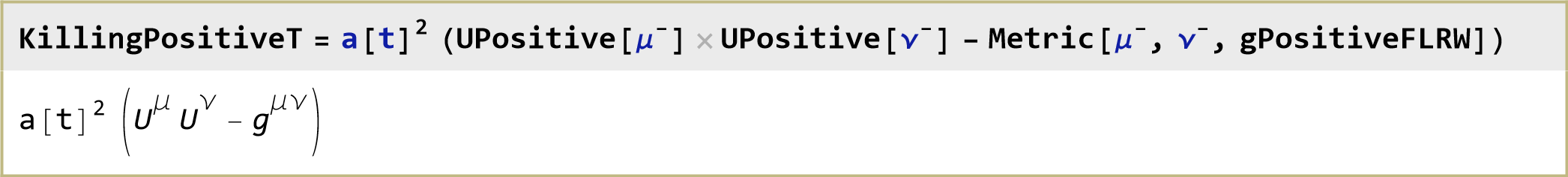}
The key property is that its fully symmetrized first covariant derivative is zero: \(\nabla _{\{\alpha }K_{\mu \nu \}}\) = 0. If \(q^{\mu }\) were tangent to some geodesic worldline, obeying \(q^{\sigma }\nabla _{\sigma }q^{\alpha }=0\), the quantity \(K_{\mu \nu }q^{\mu }q^{\nu }\) is then constant along the geodesic itself. We verify this Killing tensor equation by taking the first covariant derivative of K, then extract its TensorComponents and use \MMA{'}s Symmetrize to fully symmetrize the array. (SymmetrizeIndices and AntiSymmetrizeIndices are being developed in \TC, that would accomplish index (anti-)symmetrization.)
\includecode{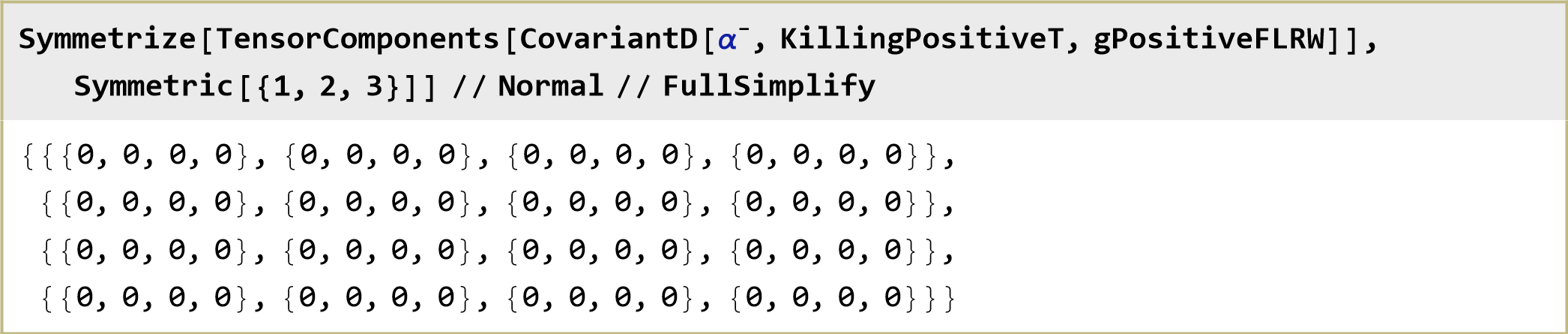}
We turn to verify that the Weyl tensor is zero. This implies there exists a coordinate system whereby the positively spatially curved FLRW geometry is proportional to the flat Minkowski metric -- i.e., gPositiveFLRW is conformally flat.
\includecode{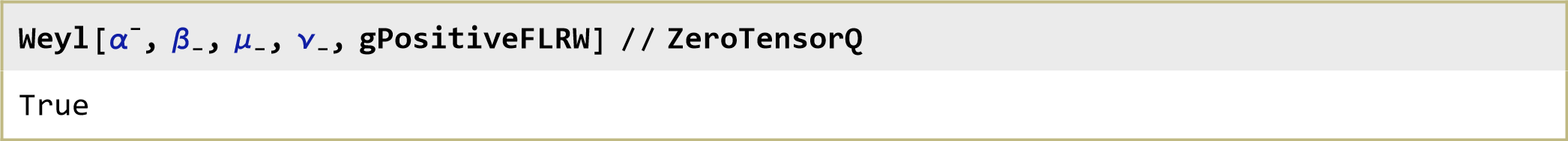}
Turning to physics, let us ask: what sort of matter would source such a positively spatially curved spacetime like gPositiveFLRW? That is, what sort of stress tensor \(T^{\mu \nu }\) would satisfy \(G^{\mu \nu }\) - $\Lambda $ \(g^{\mu \nu }\) = 8$\pi $\(G_N\) \(T^{\mu \nu }\) ($\Lambda $ is the cosmological constant)? We first observe that the Einstein tensor \(G^{\mu \nu }\) of such a geometry enjoys the same Killing symmetries.
\includecode{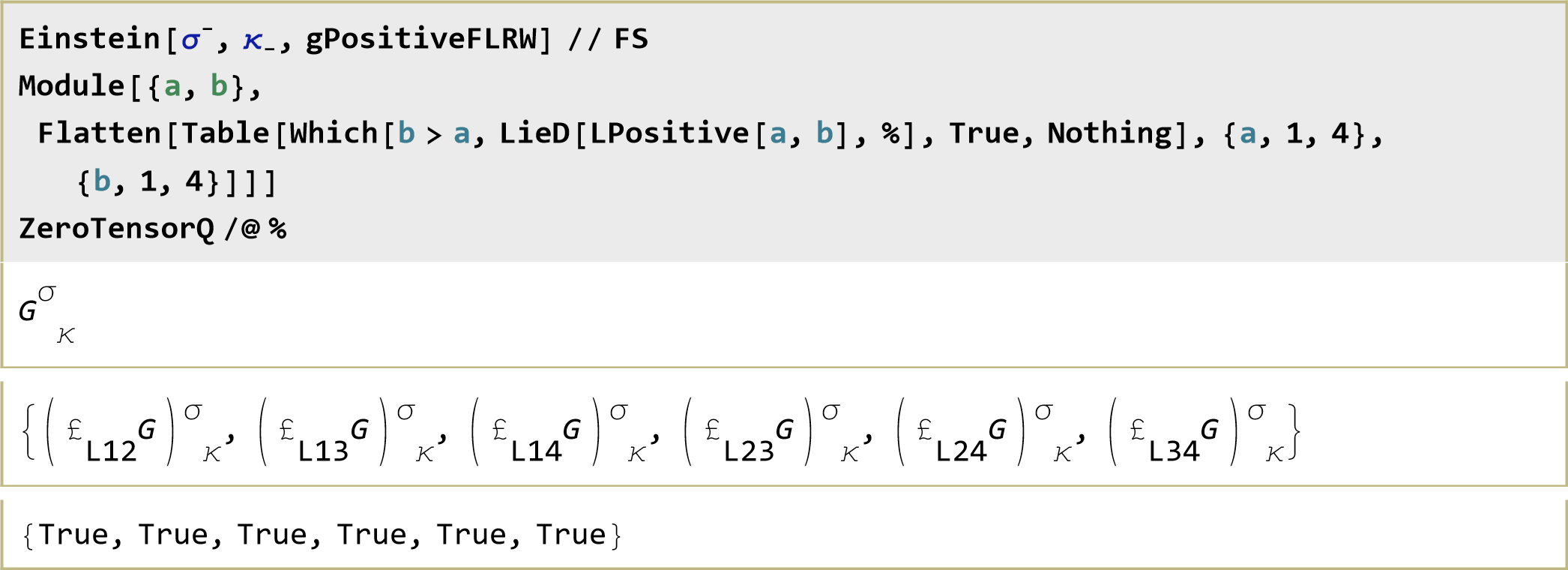}
Therefore we may constrain the form of the stress tensor \(T^{\mu }{}_{\nu }\) by demanding that it satisfy \(\pounds _{\xi }\)\(T^{\mu }{}_{\nu
}\)= 0, where $\xi $ is one of the above Killing vectors. Let us define the stress tensor in the form \(T^{\mu }{}_{\nu }\). 
\includecode{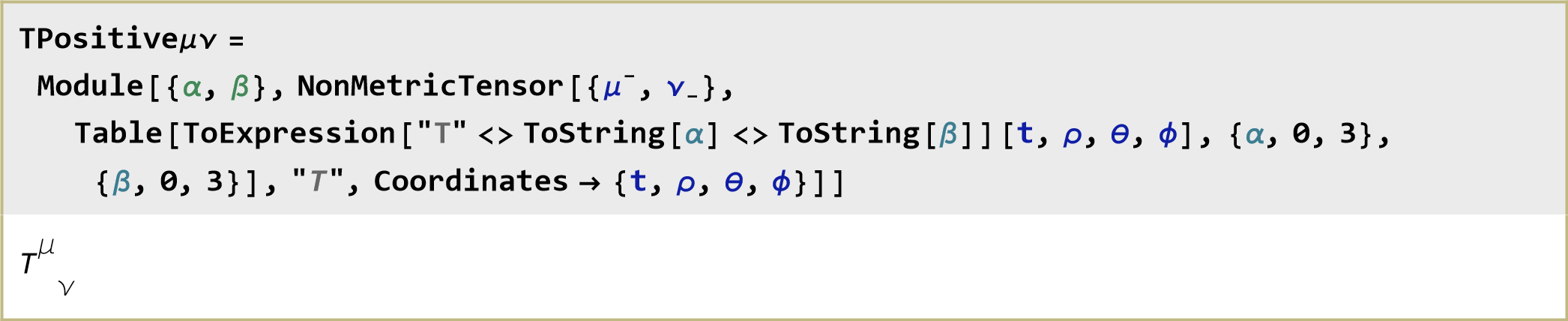}
That its Lie derivative along L12 must be zero tells us its components must be independent of $\phi $.
\includecode{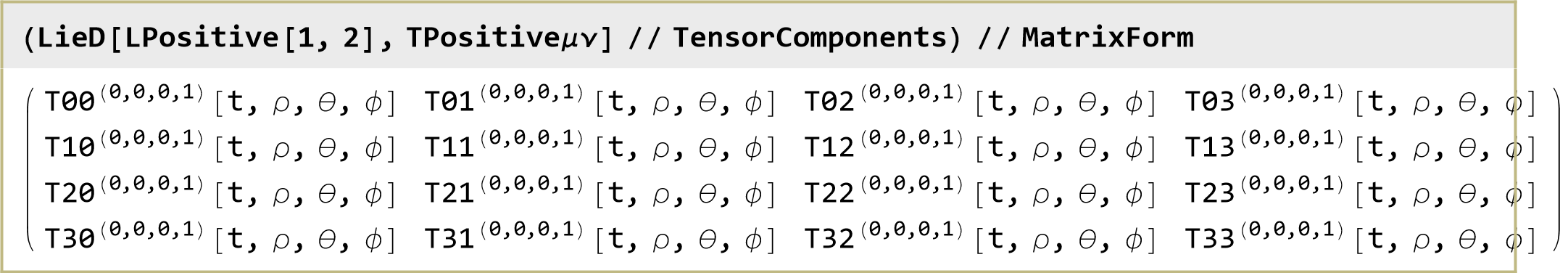}
\includecode{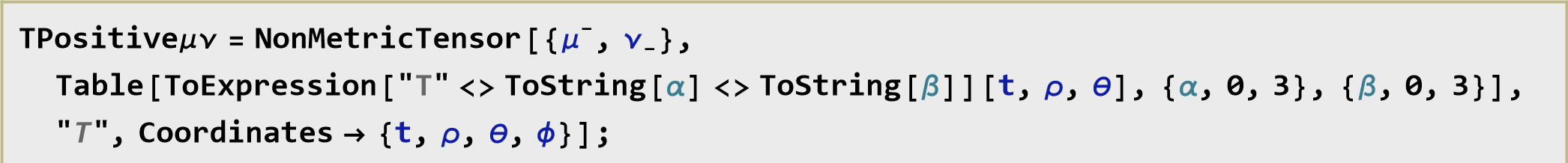}
\includecode{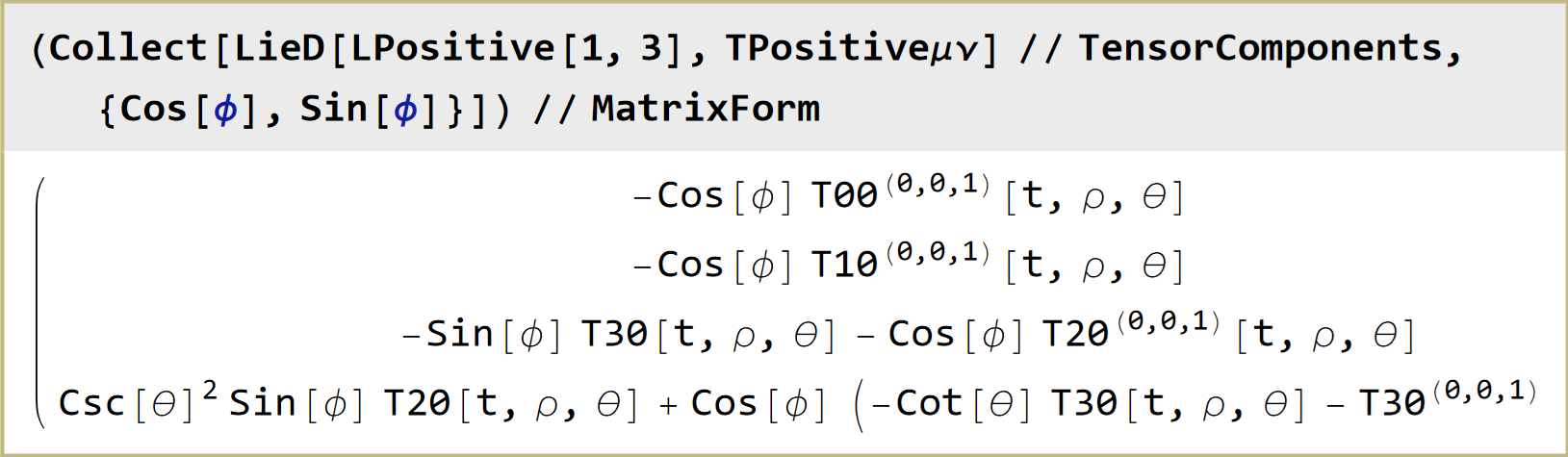}
That the stress tensor{'}s Lie derivative along L13 must be zero allows us to read off the relevant constraints component-by-component. Note that, since cos $\phi $ and sin $\phi $ are independent functions, their coefficients must separately vanish. Additionally, we also recognize the following.
\includecode{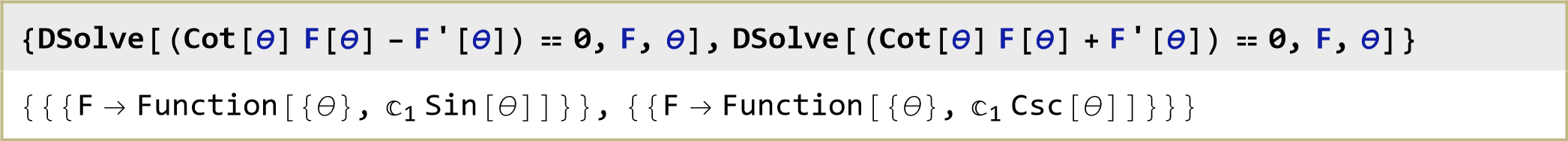}
We gather
\includecode{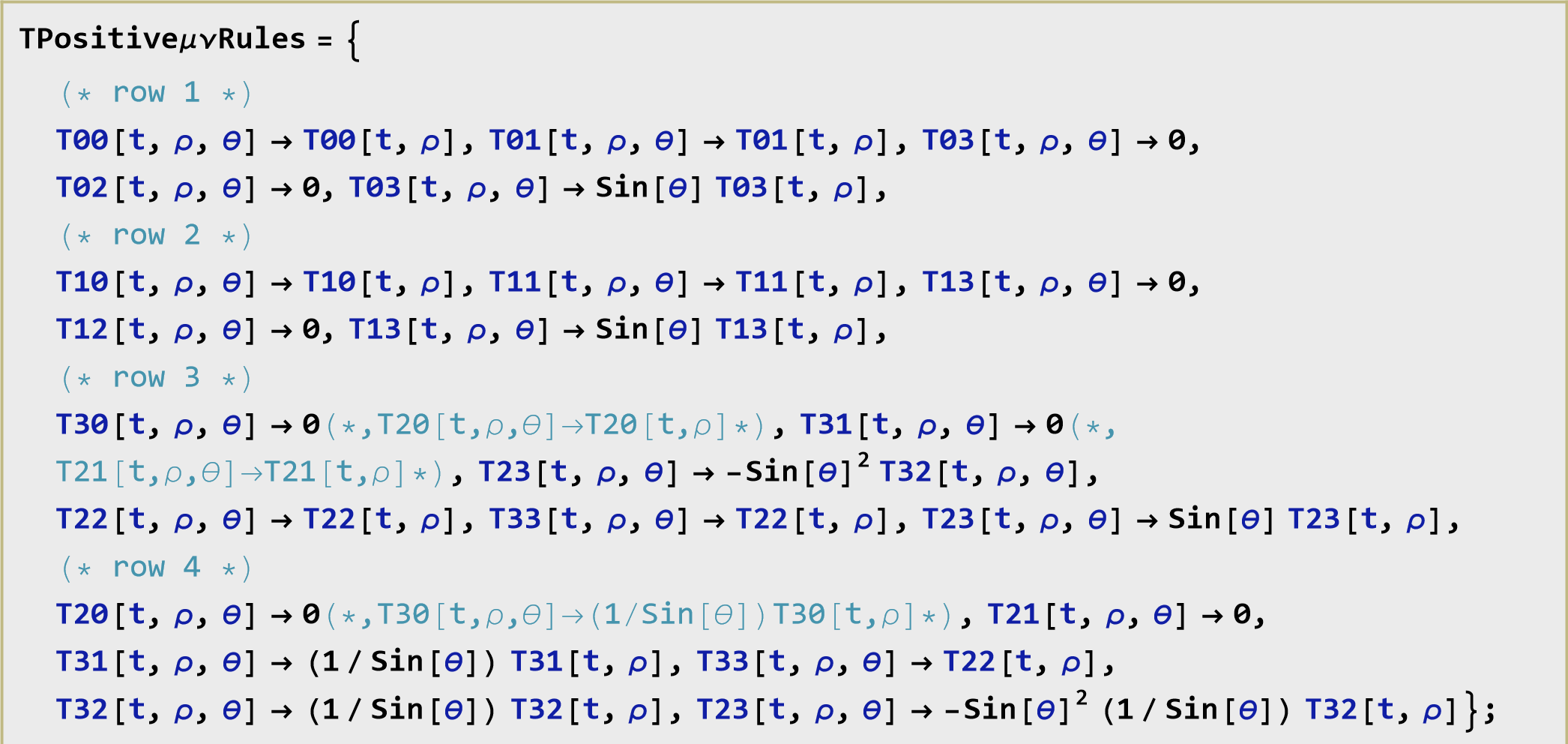}
Next we turn to the stress tensor{'}s Lie derivative along L14.
\includecode{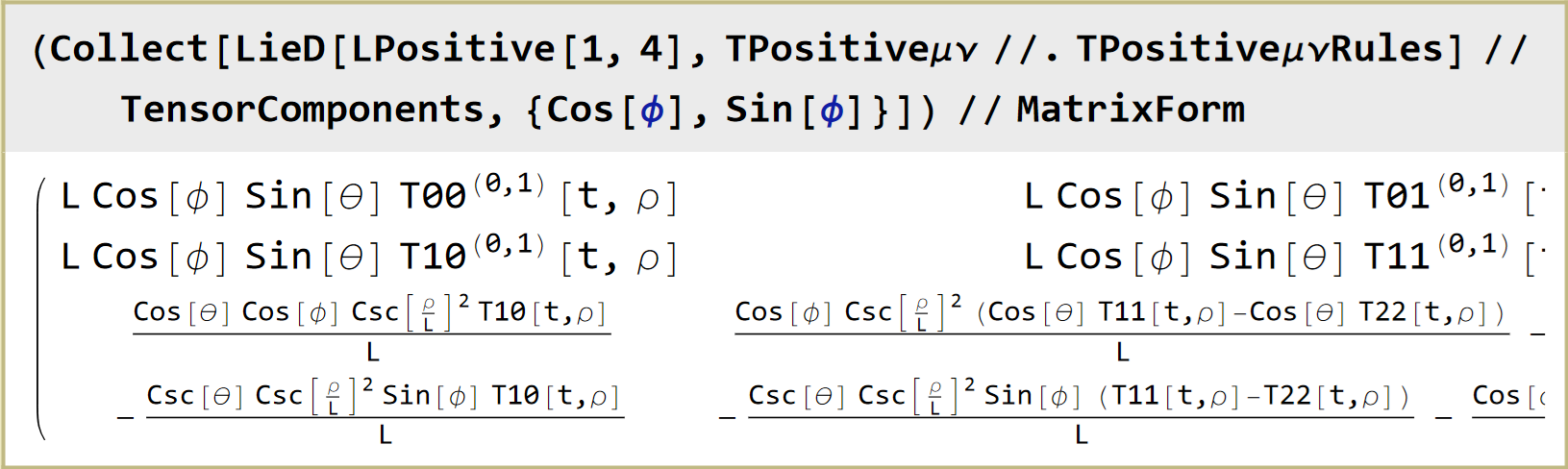}
Requiring the above to be zero yields the following.
\includecode{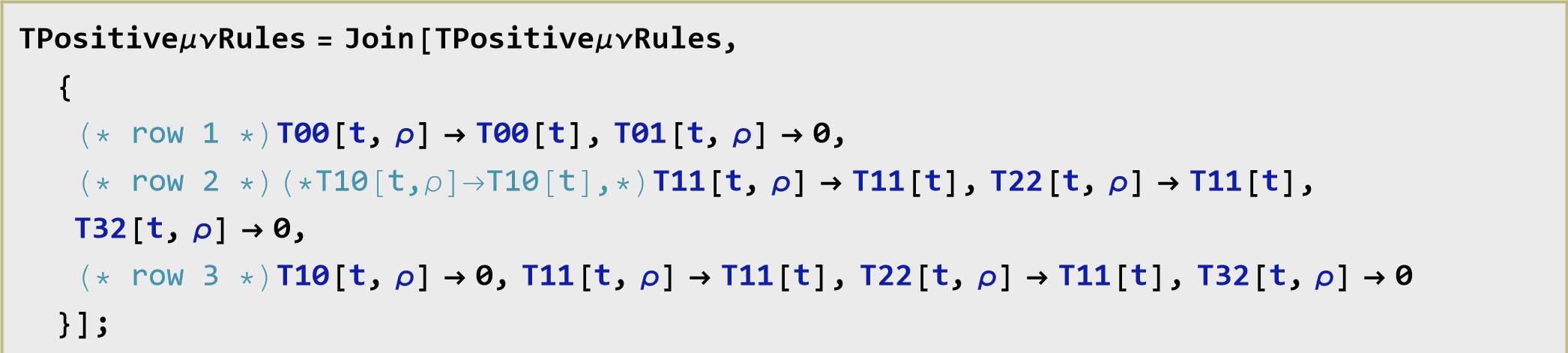}
Taking into account all the constraints obtained thus far, we see the stress tensor{'}s Lie derivatives along the remaining Killing vectors are zero.
\includecode{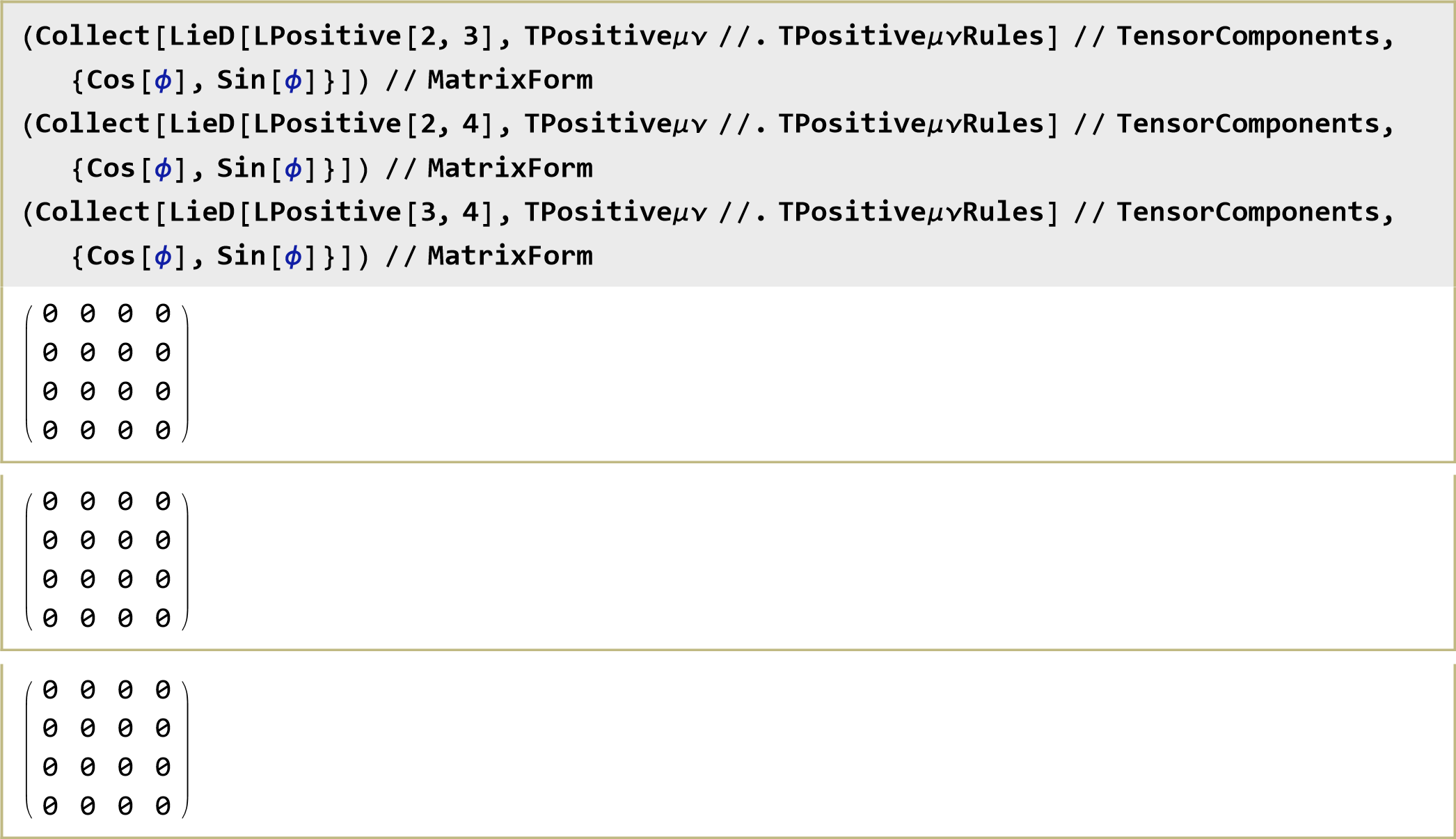}
At this point, the stress tensor takes a diagonal form -- where all the space-space components are equal.
\includecode{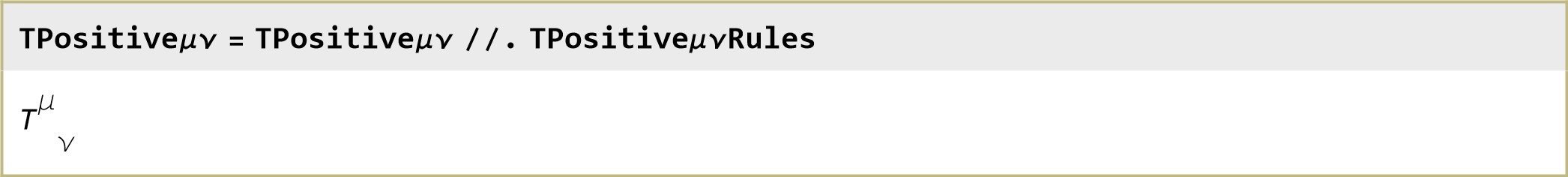}
To interpret the components we express the stress tensor in an orthonormal basis, where \(T^{\hat{0}\hat{0}}\) is the energy density and \(T^{\hat{a}\hat{a}}\) (no sum over \textit{a}) is the pressure along the \textit{a}-th spatial direction. 
\includecode{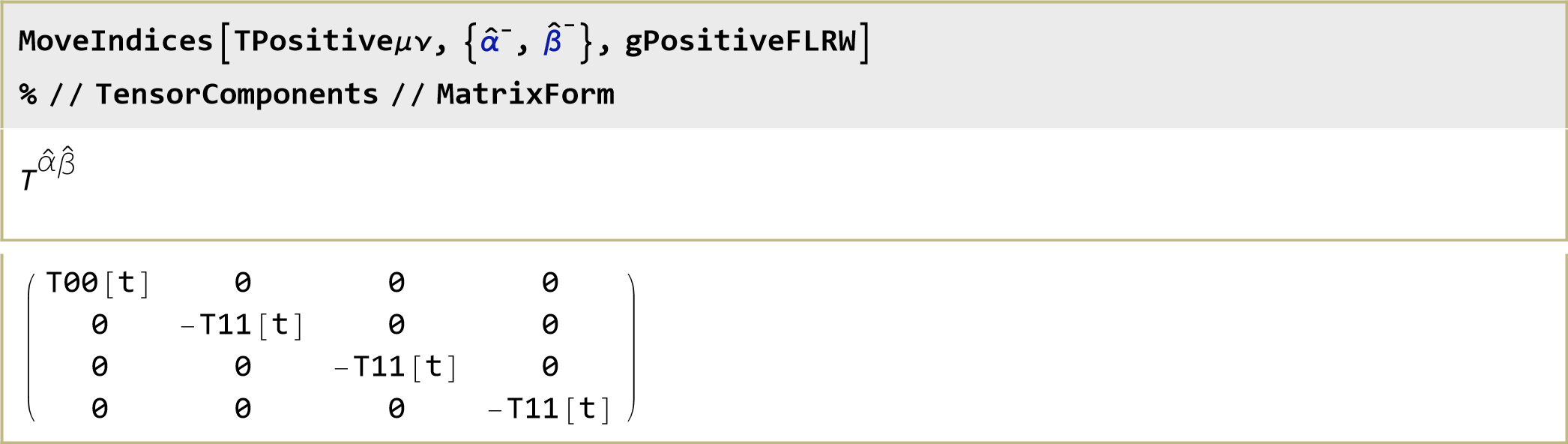}
This implies \(T^0{}_0\equiv \rho\), the energy density; and \(-T^1{}_1\equiv p\), the pressure density.
\includecode{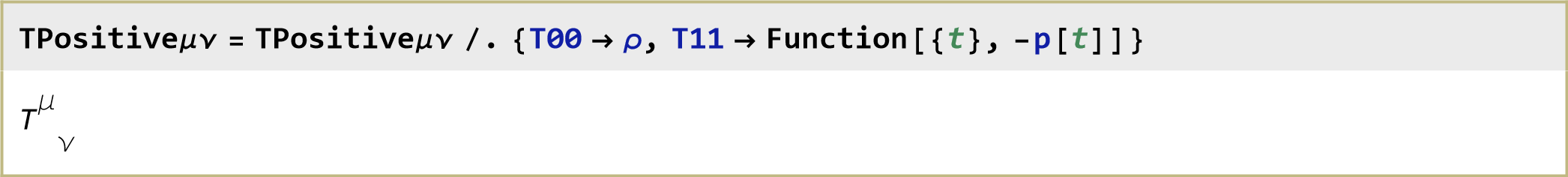}
That Einstein{'}s tensor is divergence-free means the stress tensor must be too.
\includecode{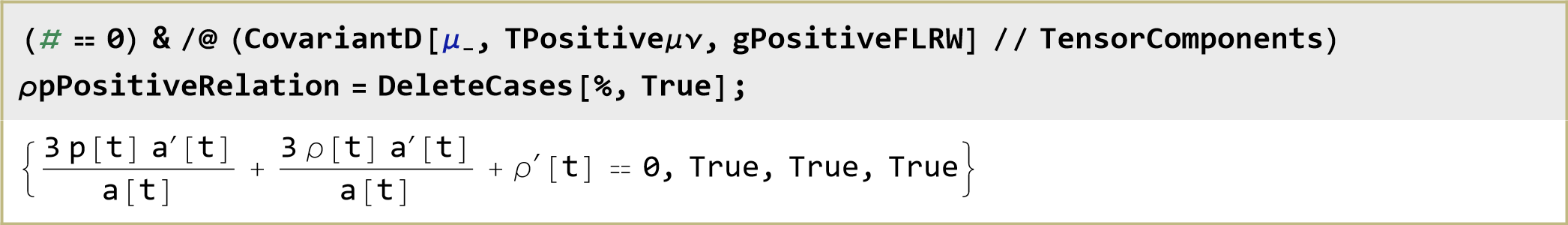}
If p/$\rho $ $\equiv $ \textit{w} were a constant -- for instance, in a radiation, matter and Dark Energy dominated universe, $$w = 1/3, 0, -1$$
respectively -- we may show that $\rho $[t] = $\rho $[\(t_0\)] (\(\left.\left.a\left[t_0\right]\right/a[t]\right){}^{3(1+w)}\) for some {``}initial time{''} \(t_0\).
\includecode{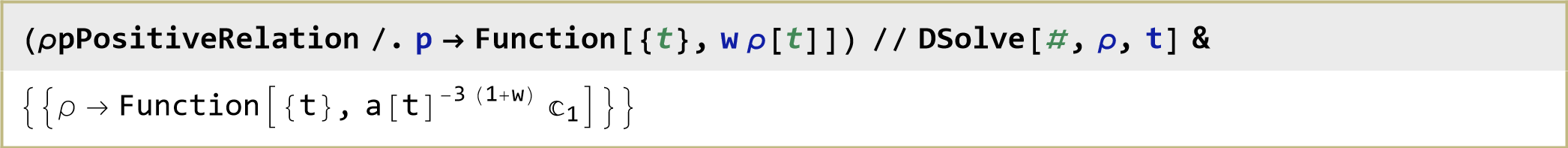}
Einstein{'}s equations governing the large scale dynamics of the positively spatially curved universe read as follows.
\twopicincludecode{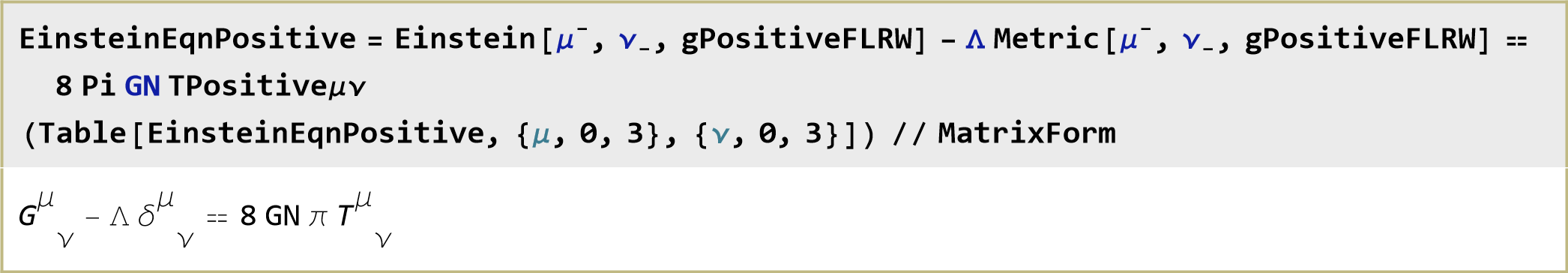}{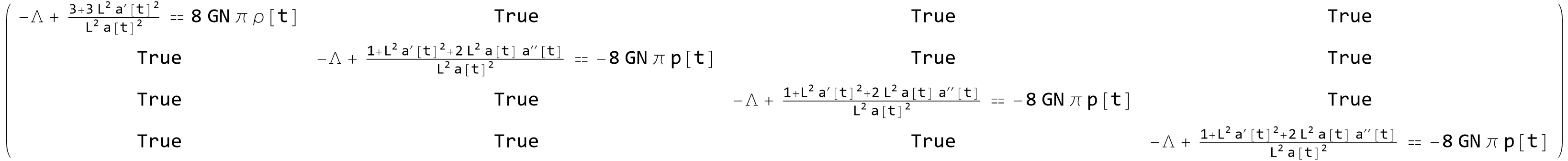}
The diagonal space-space components yield the same equations; hence the distinct equations arise from the 00 and 11 (or 22, or 33) components. We use them to solve for the scale factor{'}s velocity and acceleration.
\includecode{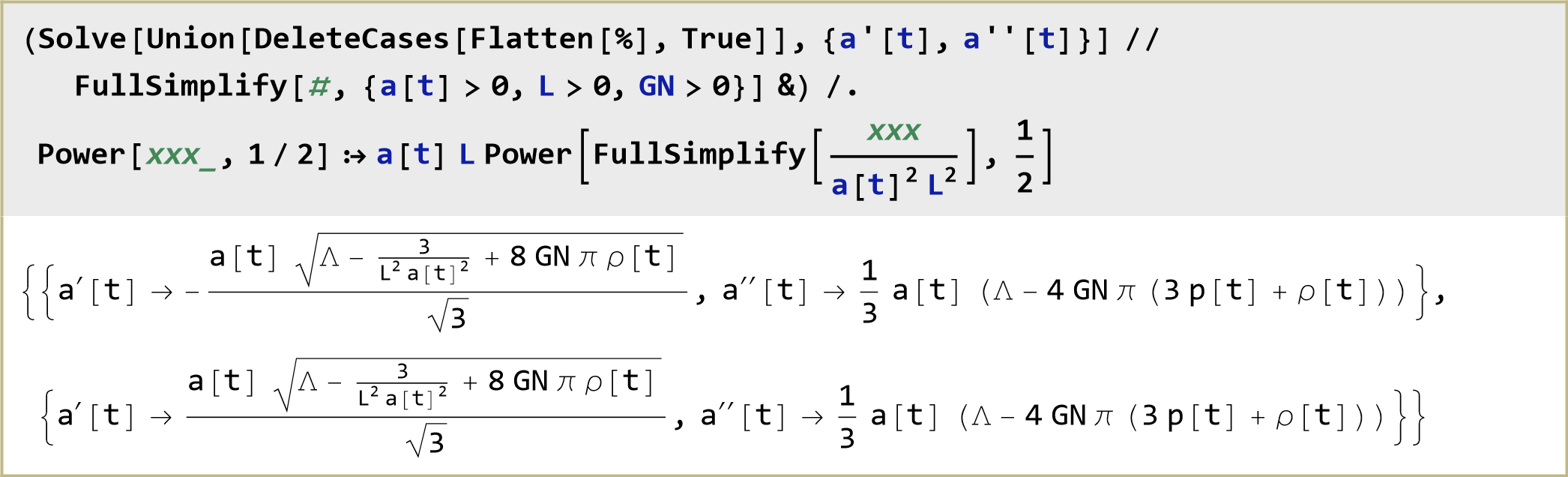}
The 11 component of Einstein{'}s equations is not independent of the 00 component, in that the former follows from the latter -- provided the
divergence-free property of the stress tensor holds. To see this, we differentiate the 00 component with respect to time, and proceed to utilize
both the \(\nabla _{\mu }T^{\mu \nu }=0\) and 00 component to reduce the result to that of the 11 component.
\includecode{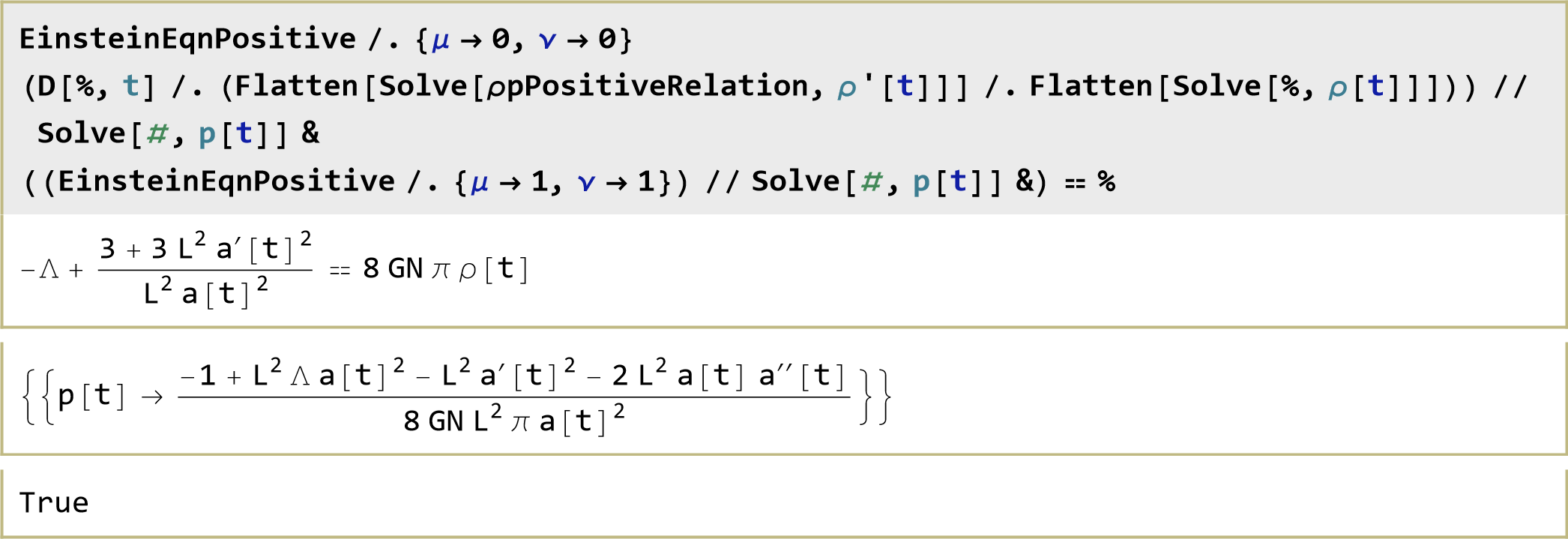}
Finally, let us observe that the stress tensor may be written a perfect fluid form, \(T^{\mu \nu }=U^{\mu }U^{\nu }(\rho +p) - g^{\mu \nu }p\).
\includecode{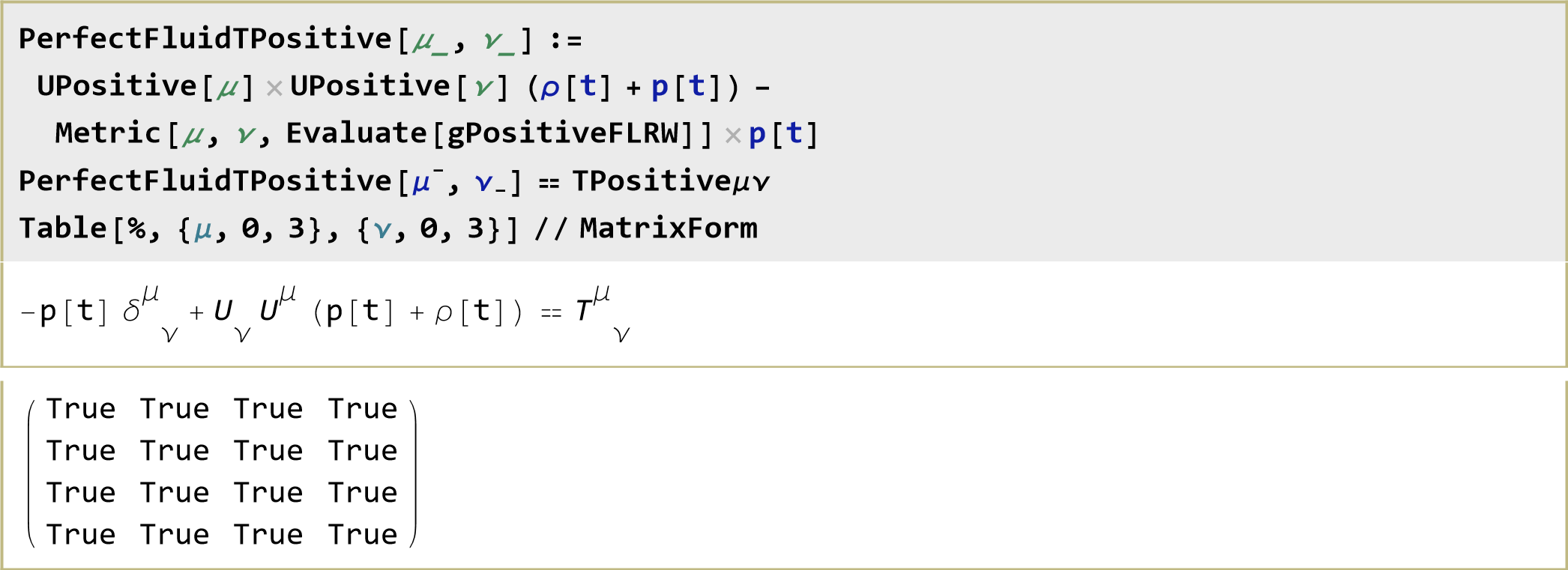}
CovariantD obeys linearity and the product rule, allowing us to recover the $\rho $pPositiveRelation above.
\includecode{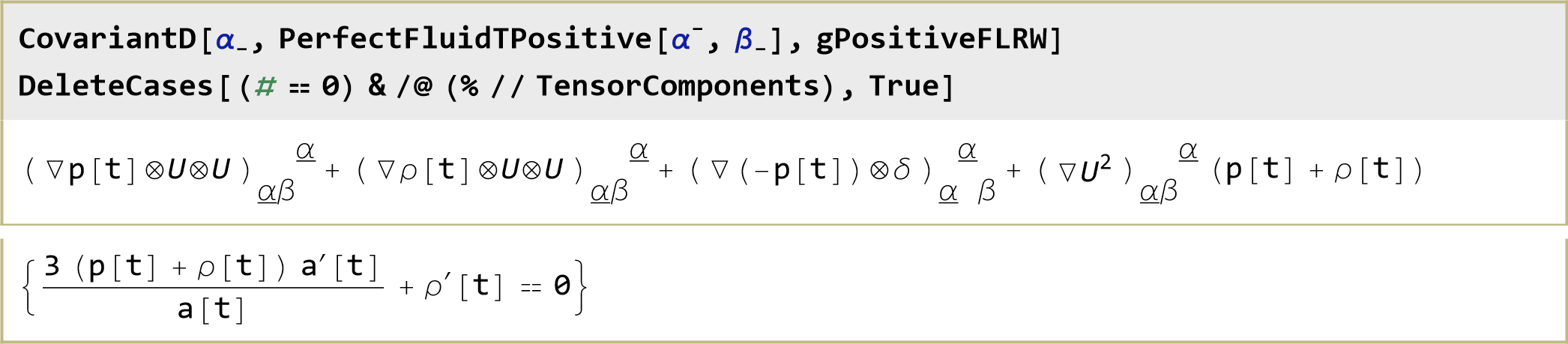}

\subsection{Negative Spatial Curvature} 

Next, we tackle the negative spatial curvature case. A 3D maximally symmetric negatively curved space can be constructed by viewing it as a 3-hyperboloid embedded in a 4D flat spacetime; specifically, \(\left(x^0\right)^2-\left(x^1\right)^2-\left(x^2\right)^2-\left(x^3\right)^2=L^2\). 
\includecode{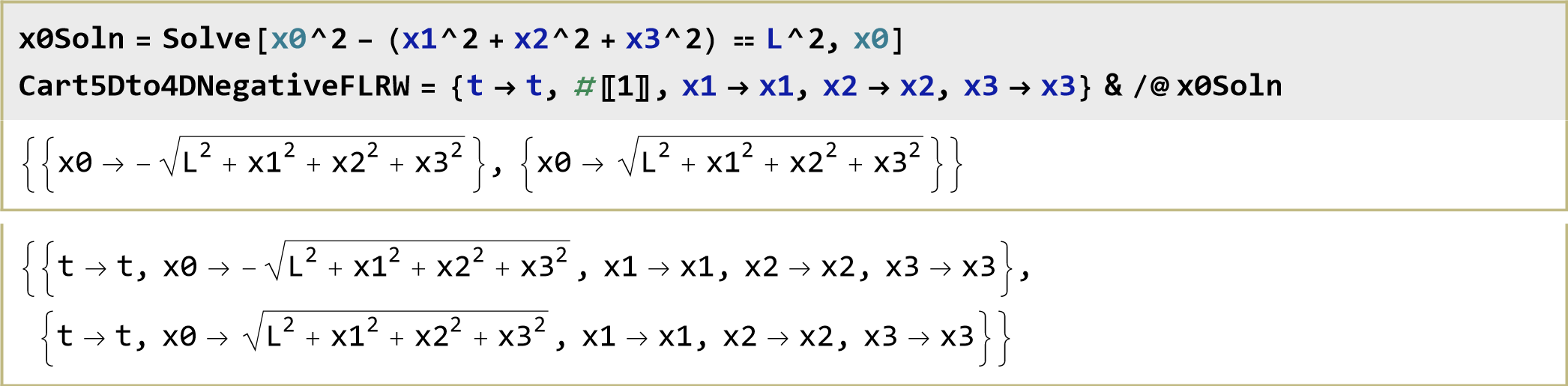}
We choose the {``}mostly plus{''} convention here so that the induced metric on the 3-hyperboloid would have 3 positive eigenvalues; the {``}mostly minus{''} convention would return 3 negative eigenvalues.
\includecode{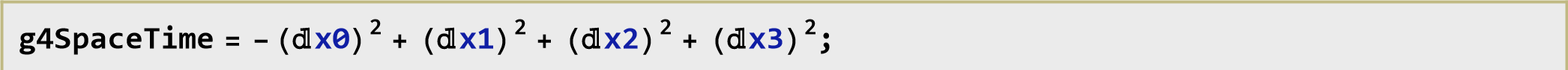}
Solving for \(x^0\) and inserting it back to the 4D Minkowski metric then returns the induced metric on the 3-hyperboloid.
\includecode{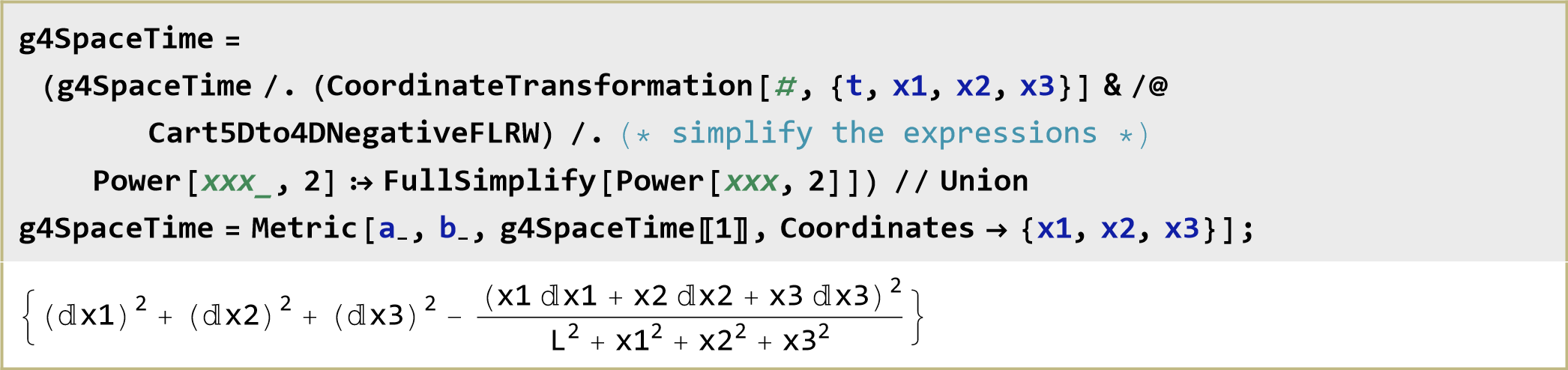}
The eigenvalues of the induced metric are all positive, as must be the case for spatial metrics.
\includecode{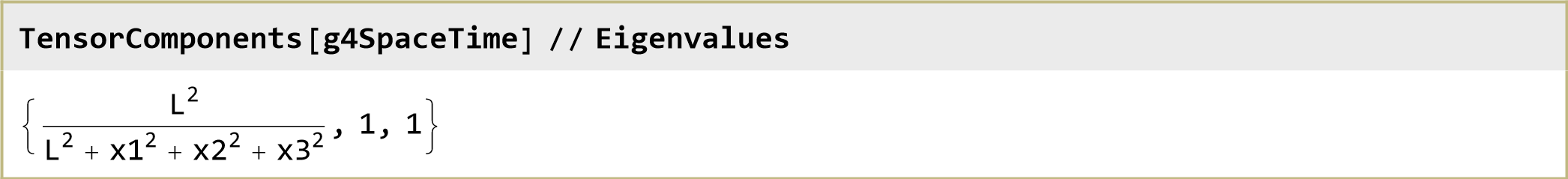}
The Ricci scalar is a negative constant.
\includecode{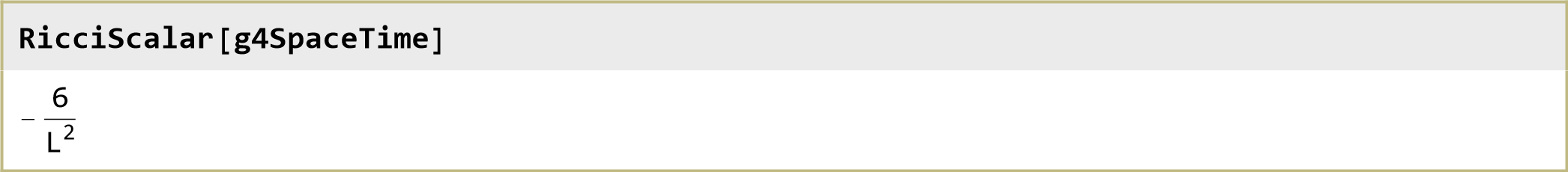}
Viewing the 3D maximally symmetric negatively curved space as a 3-hyperboloid allows us to identify its isometries as that of spatial rotations in the (1,2,3)-space and Lorentz boosts of the (0,1,2,3)-spacetime. The associated generators (i.e., Killing vectors) are \(x^{[\mu }\)\(\partial ^{\nu]}\). We first do so without inducing the tensors on the 3-hyperboloid.
\includecode{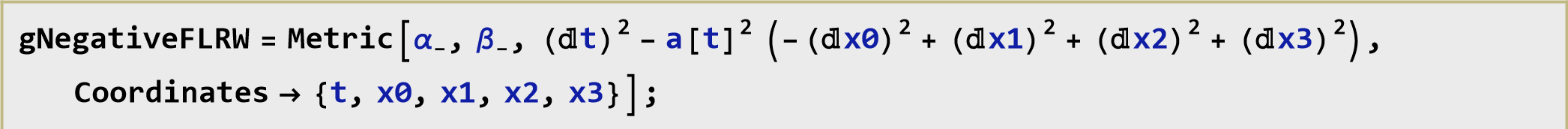}
\includecode{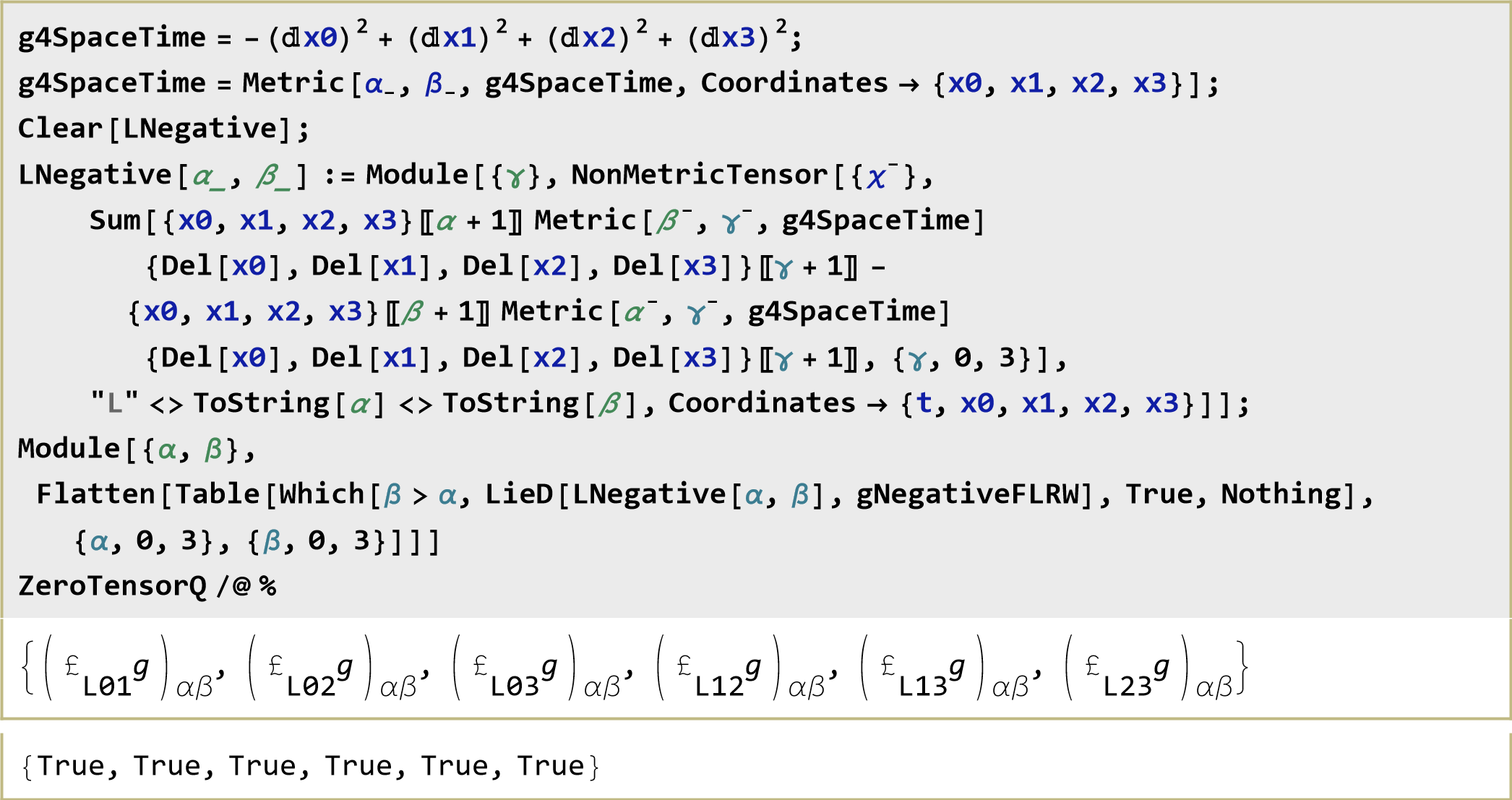}
Next, we induce the Killing vectors and metric onto the 3-hyperboloid.
\includecode{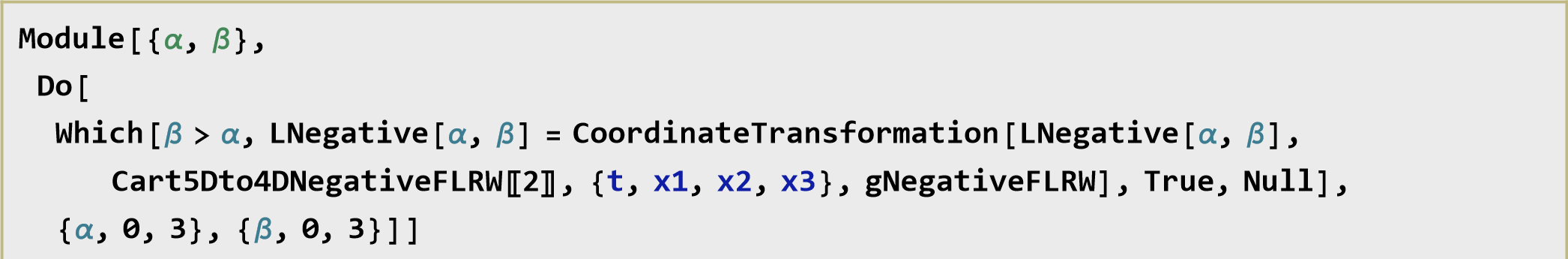}
\includecode{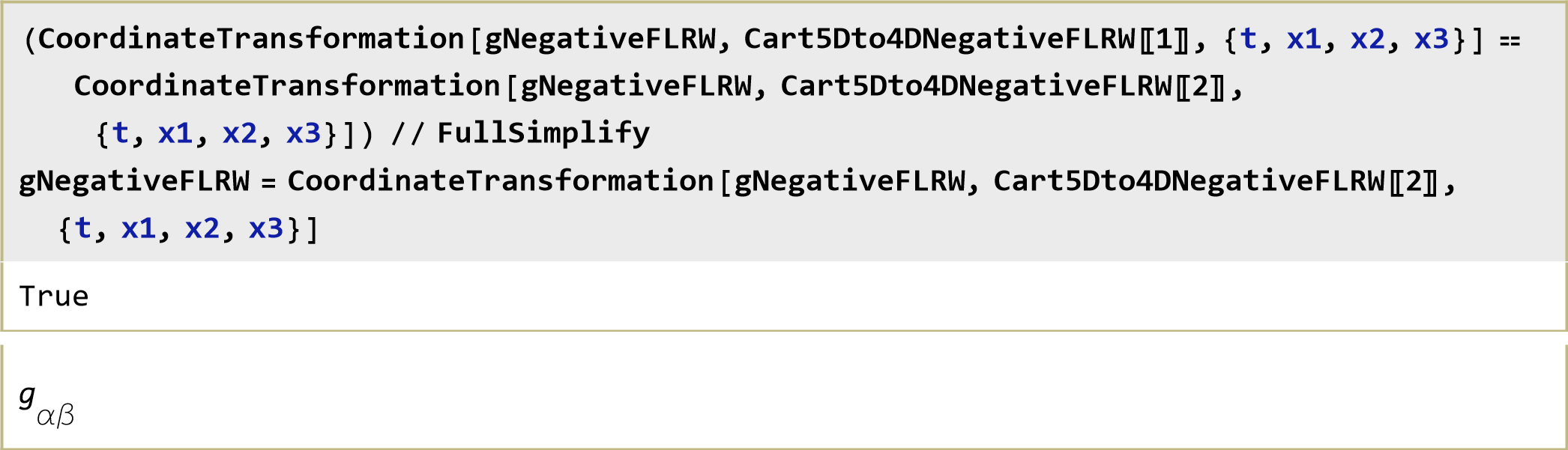}
Following the positively curved case, we now switch to hyperbolic-trigonometric spatial coordinates -- defined as 
\includecode{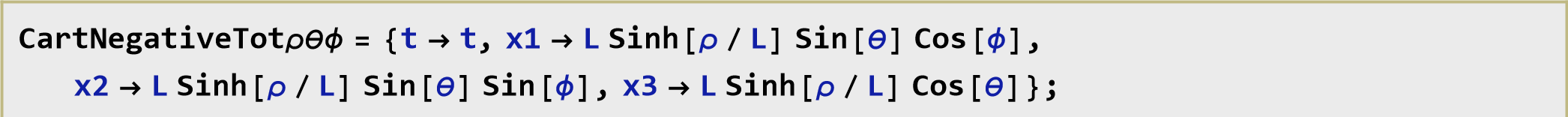}
-- so as to render the metric diagonal:
\includecode{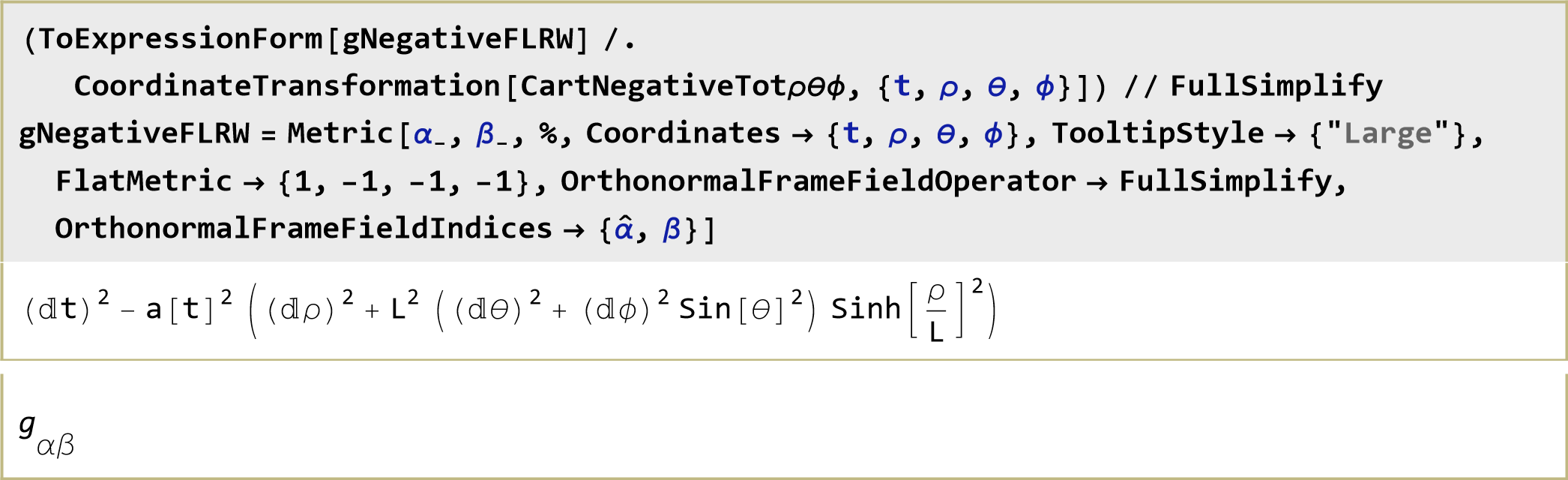}
\includecode{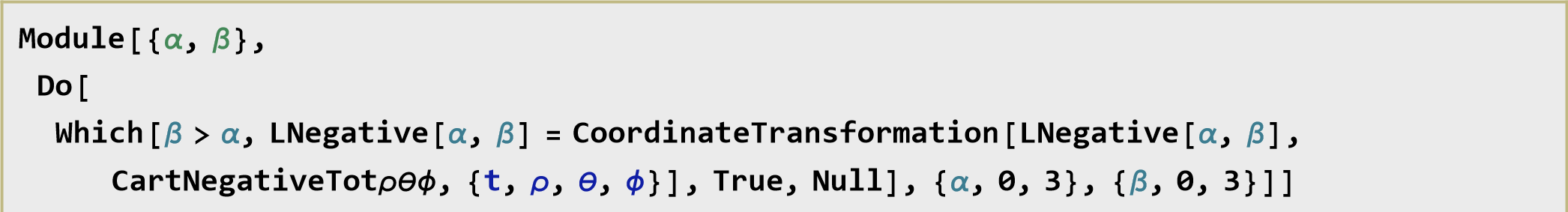}
We may readily verify, in this $\{$\textit{t}, $\rho $, $\theta $, $\phi \}$ coordinate system, Killing{'}s equations are still satisfied.
\includecode{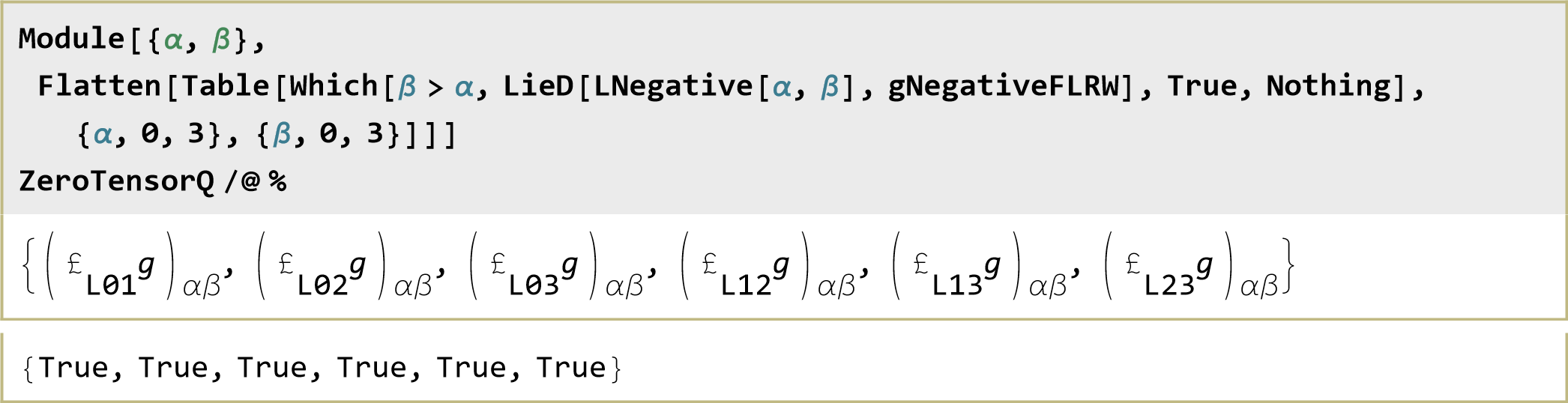}
Let \(U^{\mu }\) = \(\delta ^{\mu }{}_0\) be the timelike vector tangent to the worldline of a particle at rest with the universe. We may verify
its unit length and geodesic nature.
\includecode{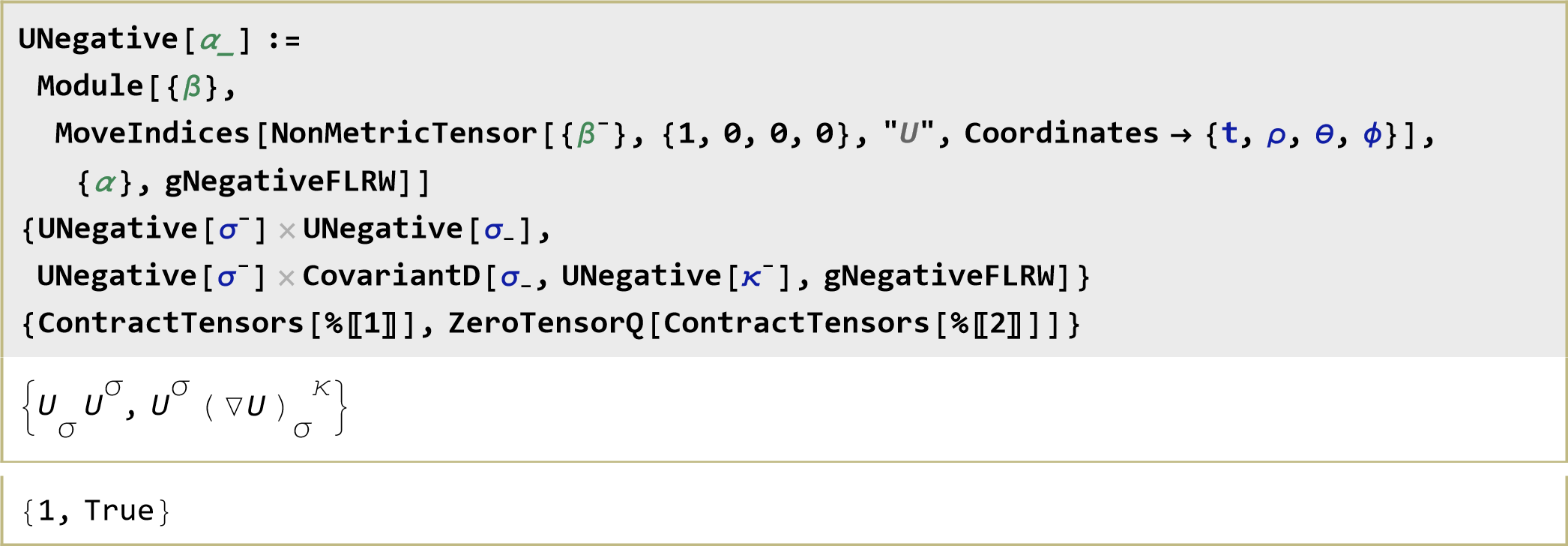}
With this \textit{U} we may also construct the Killing tensor of the FLRW geometry.
\includecode{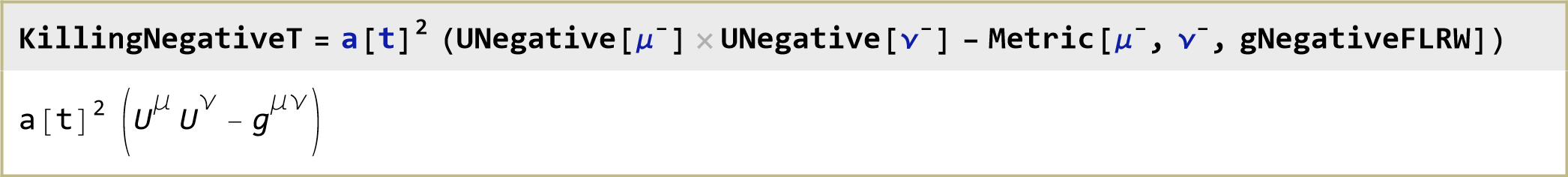}
The key property is that its fully symmetrized first covariant derivative is zero: \(\nabla _{\{\alpha }K_{\mu \nu \}}\) = 0. If \(q^{\mu }\) were tangent to some geodesic worldline, obeying \(q^{\sigma }\nabla _{\sigma }q^{\alpha }=0\), the quantity \(K_{\mu \nu }q^{\mu }q^{\nu }\) is then constant along the geodesic itself.
\includecode{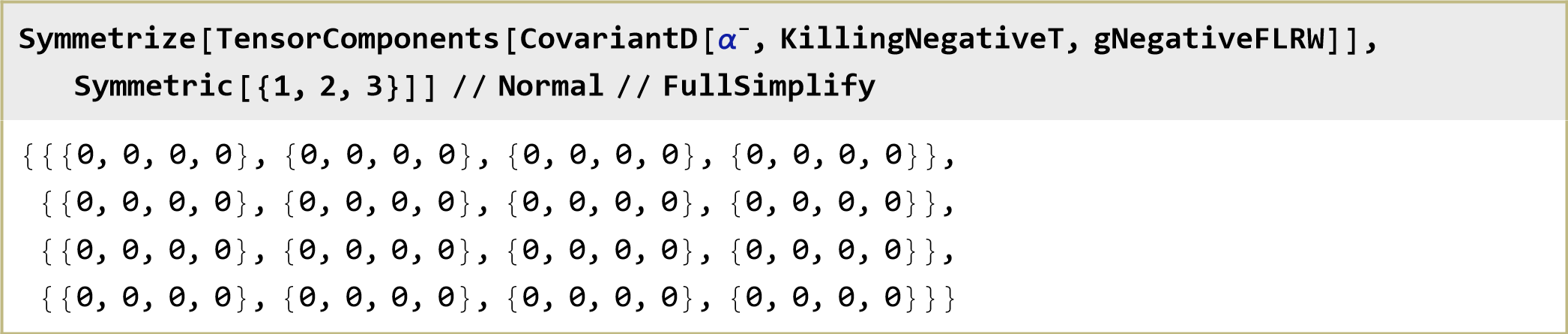}
Moreover, just like the positively curved case, the negatively curved FLRW geometry is conformally flat because its Weyl tensor is also zero.
\includecode{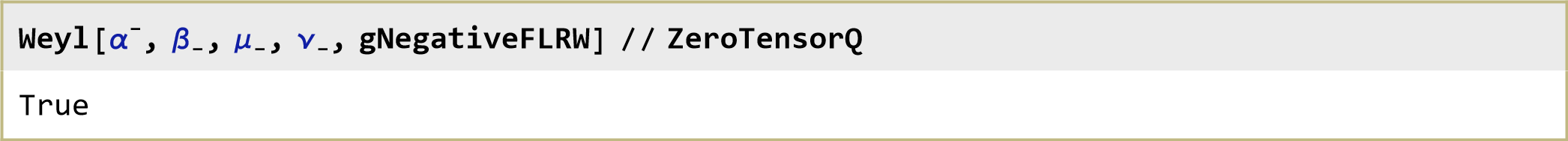}
Turning to physics, let us ask: what sort of matter would source such a negatively spatially curved spacetime like gNegativeFLRW? That is, what sort of stress tensor \(T^{\mu \nu }\) would satisfy \(G^{\mu \nu }\) - $\Lambda $ \(g^{\mu \nu }\) = 8$\pi $\(G_N\) \(T^{\mu \nu }\) ($\Lambda $ is the cosmological constant)? We first observe that the Einstein tensor \(G^{\mu \nu }\) of such a geometry enjoys the same Killing symmetries.
\includecode{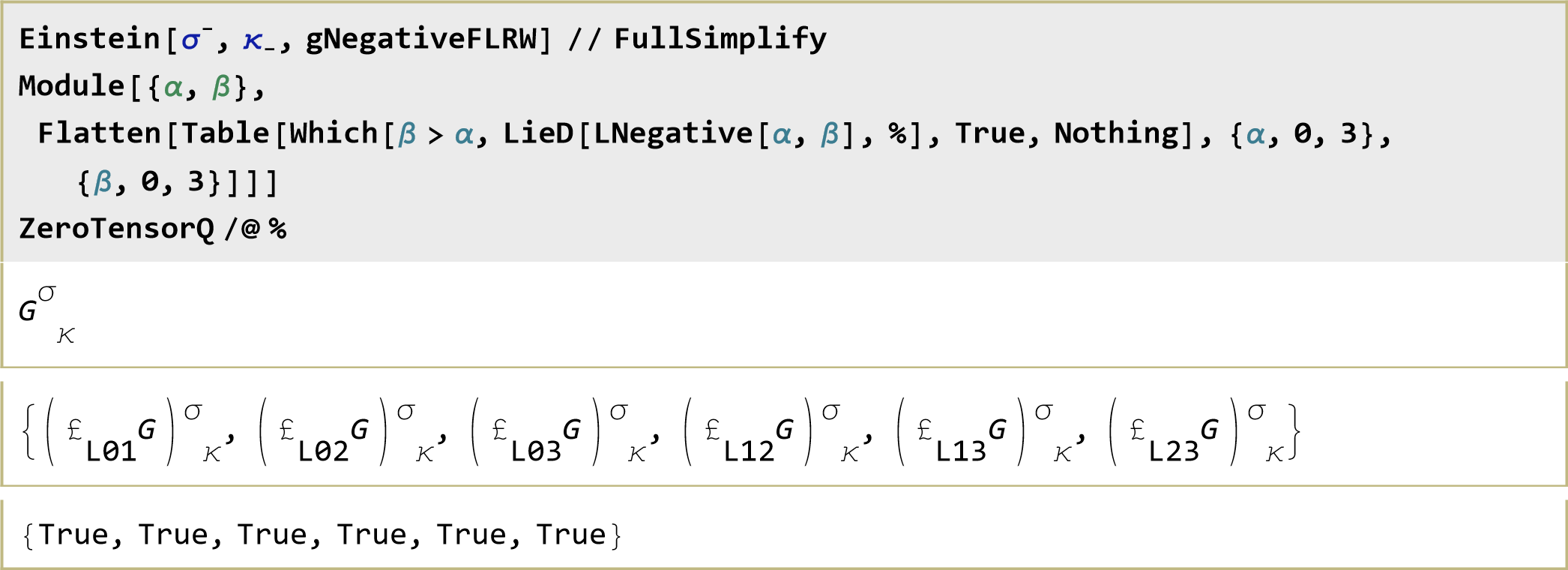}
Therefore we may constrain the form of the stress tensor \(T^{\mu }{}_{\nu }\) by demanding that it satisfy \(\pounds _{\xi }\)\(T^{\mu }{}_{\nu}\) = 0, where $\xi $ is one of the above Killing vectors. Moreover, recall from the discussion in the positively curved case that { }\(\pounds _{\text{L12}}\)\(T^{\mu}{}_{\nu }\) = 0 and { }\(\pounds _{\text{L13}}\)\(T^{\mu }{}_{\nu }\) = 0, were enough to completely determine its diagonal form. If we recognize the \(\text{L12}^{\chi }\) and \(\text{L13}^{\chi }\) are the same in both the positive and negative spatial curvature cases --
\includecode{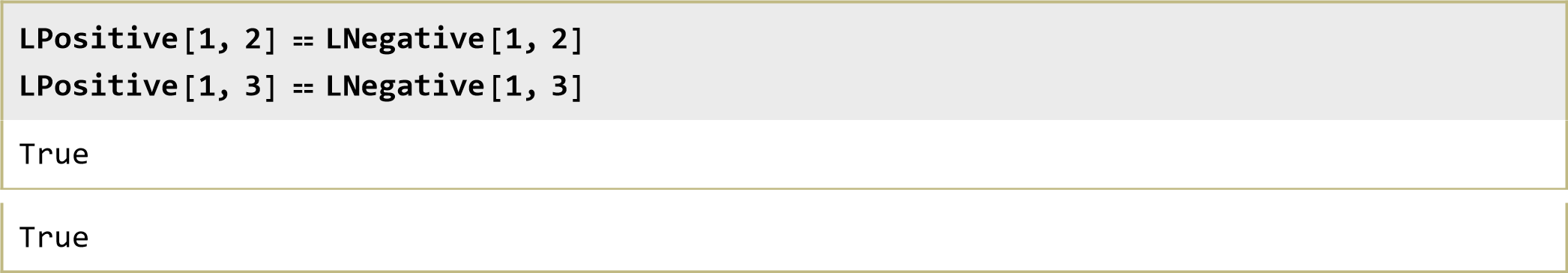}
-- we may therefore immediately write down the stress tensor \(T^{\mu }{}_{\nu }\), since it must take the same form as the positive curvature
case.
\includecode{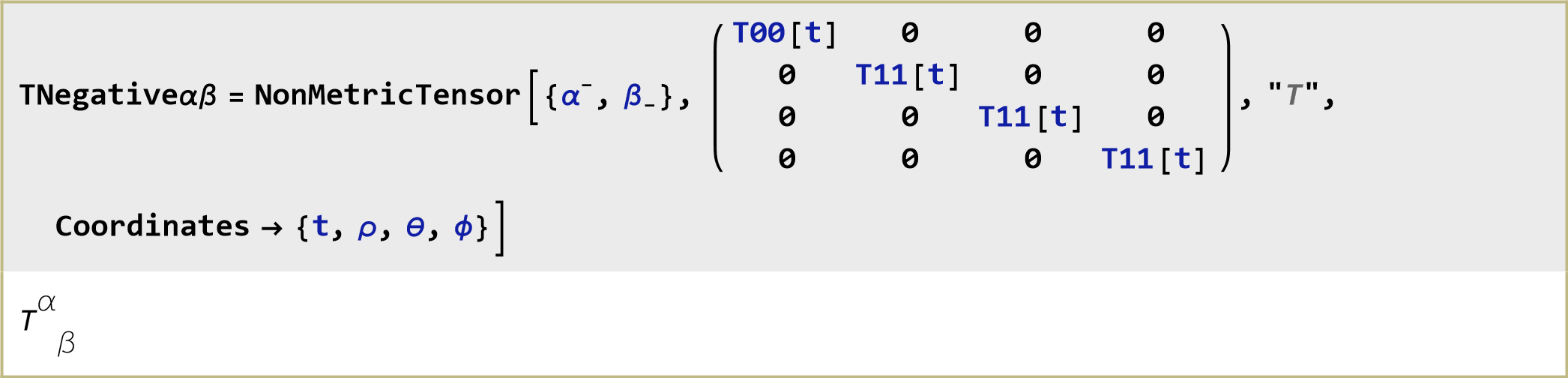}
Let us verify its Lie derivative along all the Killing vectors are indeed zero. 
\includecode{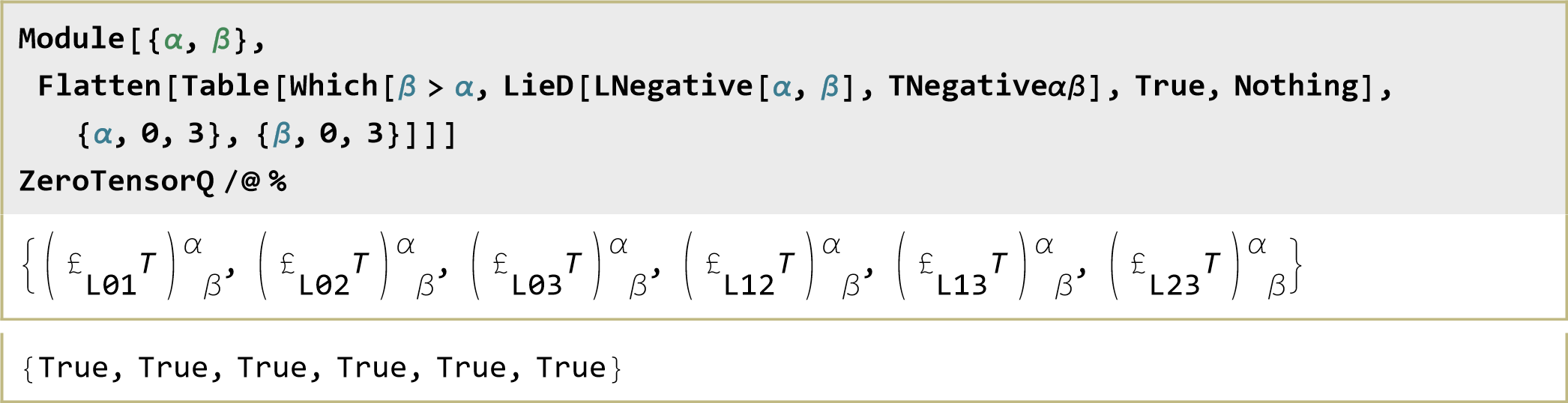}
To interpret the components we express the stress tensor in an orthonormal basis, where \(T^{\hat{0}\hat{0}}\) is the energy density and \(T^{\hat{a}\hat{a}}\) (no sum over \textit{a}) is the pressure along the \textit{a}-th spatial direction. 
\includecode{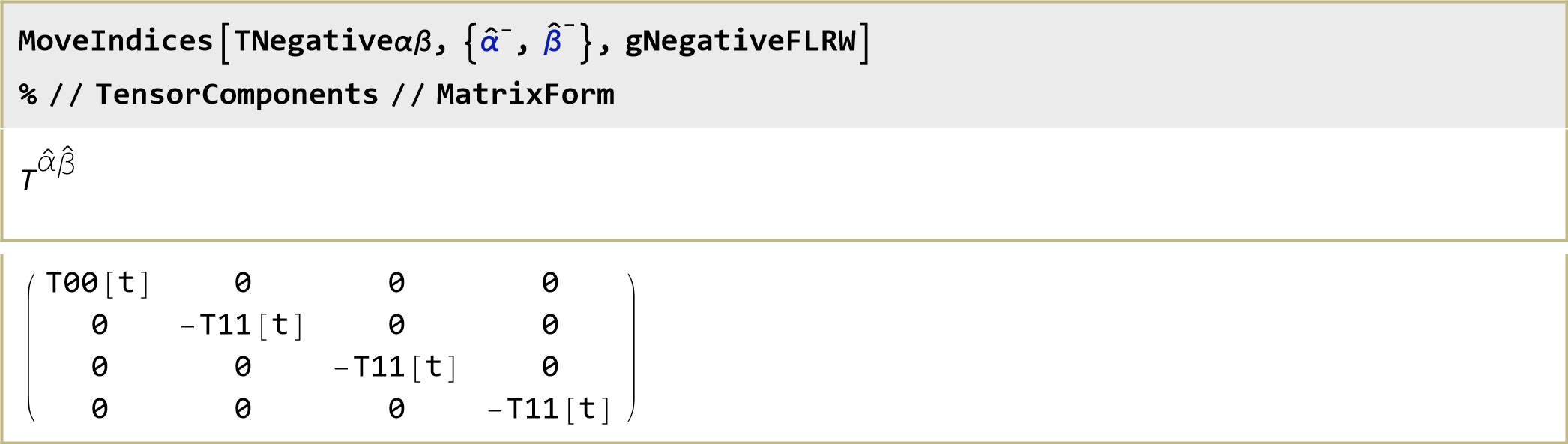}
This implies \(T^0{}_0\equiv \rho\), the energy density; and \(-T^1{}_1\equiv p\), the pressure density.
\includecode{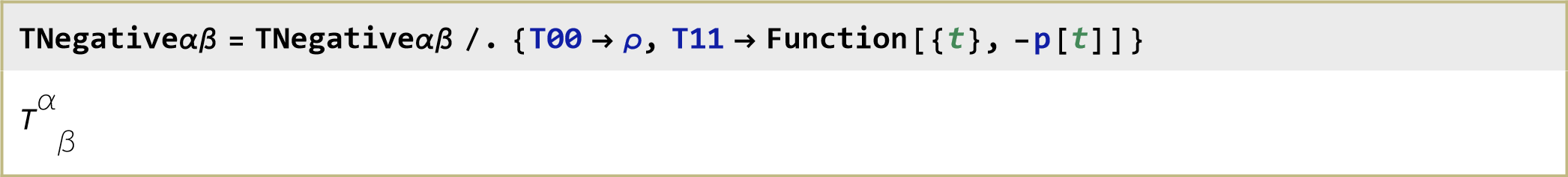}
That Einstein{'}s tensor is divergence-free means the stress tensor must be too.
\includecode{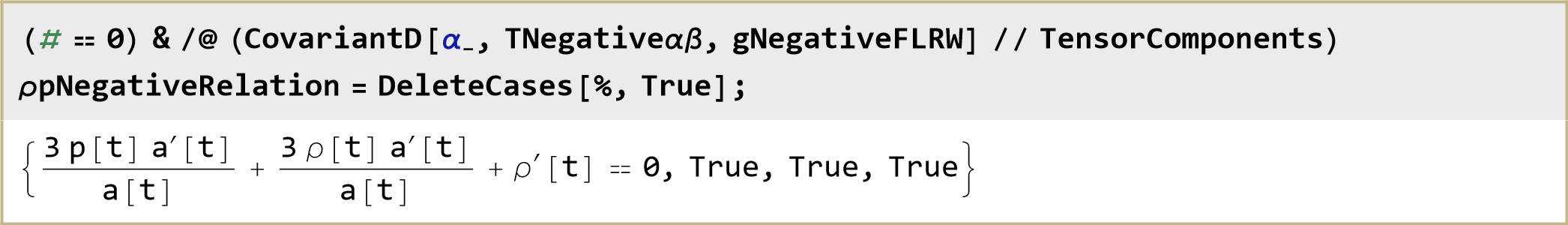}
If p/$\rho $ $\equiv $ \textit{w} were a constant -- for instance, in a radiation, matter and Dark Energy dominated universe, $$w = 1/3, 0, -1$$
respectively -- we may show that $\rho $[t] = $\rho $[\(t_0\)] (\(\left.\left.a\left[t_0\right]\right/a[t]\right){}^{3(1+w)}\) for some {``}initial time{''} \(t_0\).
\includecode{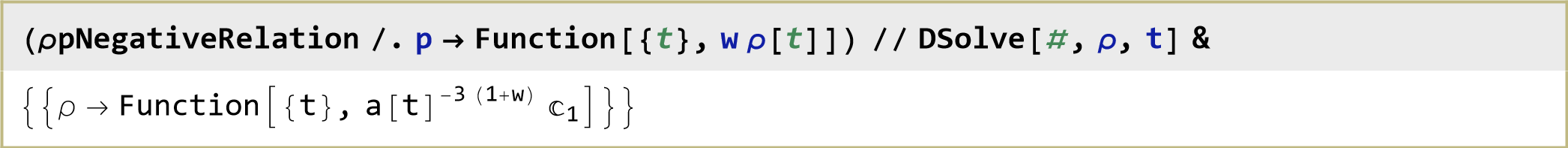}
Einstein{'}s equations governing the large scale dynamics of the spatially negatively curved universe read as follows.
\twopicincludecode{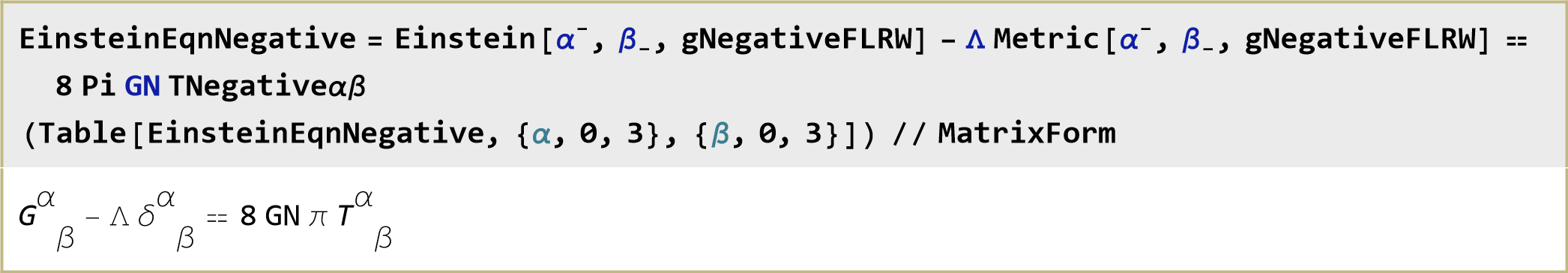}{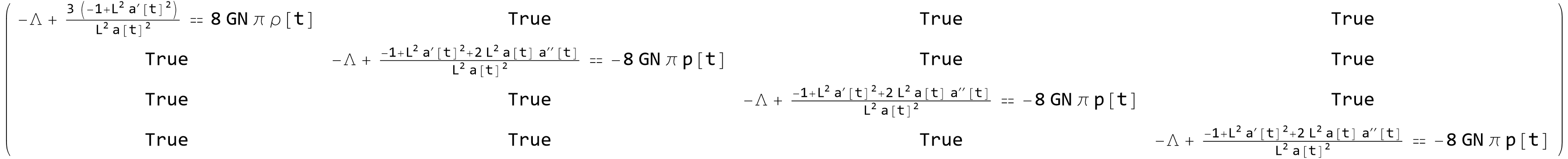}
Like the positively curved case, there are only two distinct equations. We use them to solve for the scale factor{'}s velocity and acceleration.
\includecode{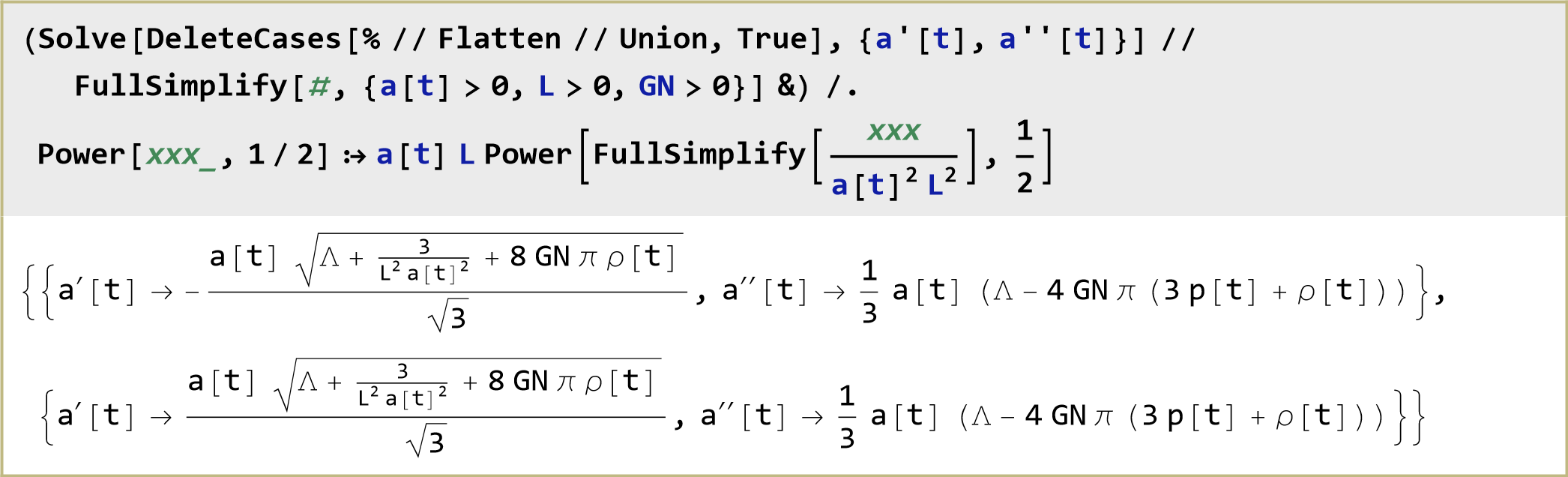}
The 11 component of Einstein{'}s equations is not independent of the 00 component, in that the former follows from the latter -- provided the
divergence-free property of the stress tensor holds. To see this, we differentiate the 00 component with respect to time, and proceed to utilize
both the \(\nabla _{\mu }T^{\mu \nu }=0\) and 00 component to reduce the result to that of the 11 component.
\includecode{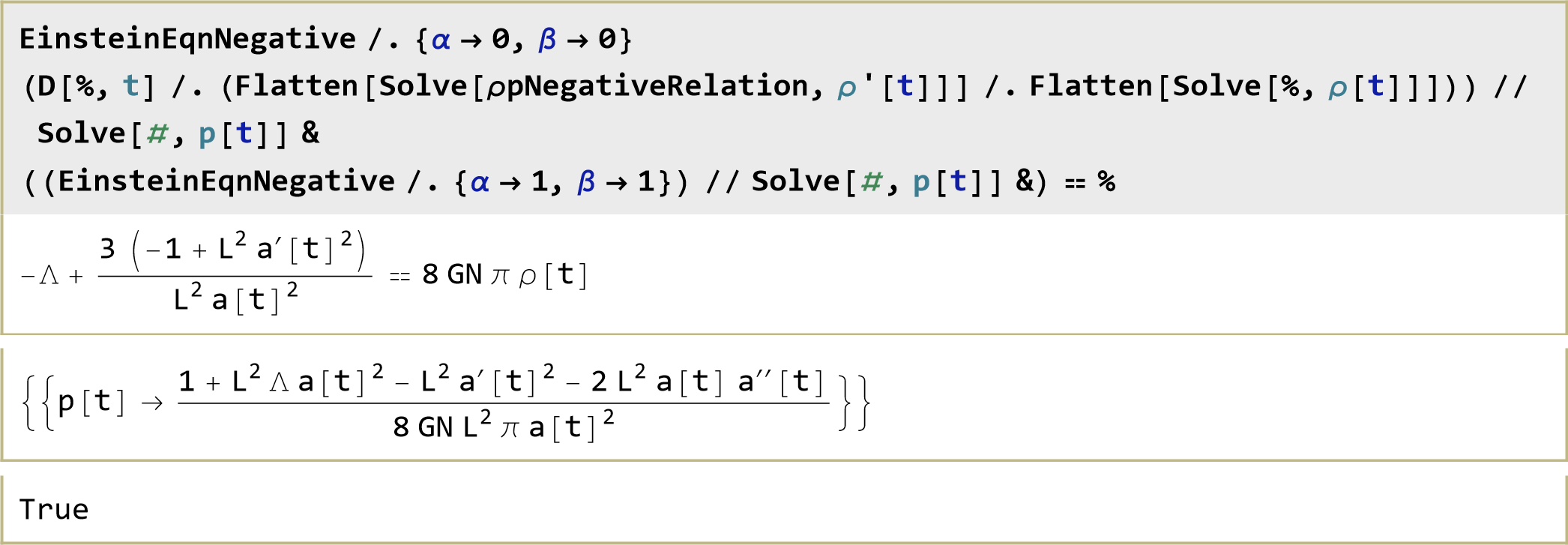}
Finally, let us observe that the stress tensor may be written a perfect fluid form, { }\(T^{\mu \nu }=U^{\mu }U^{\nu }(\rho +p) - g^{\mu \nu }p\).
\includecode{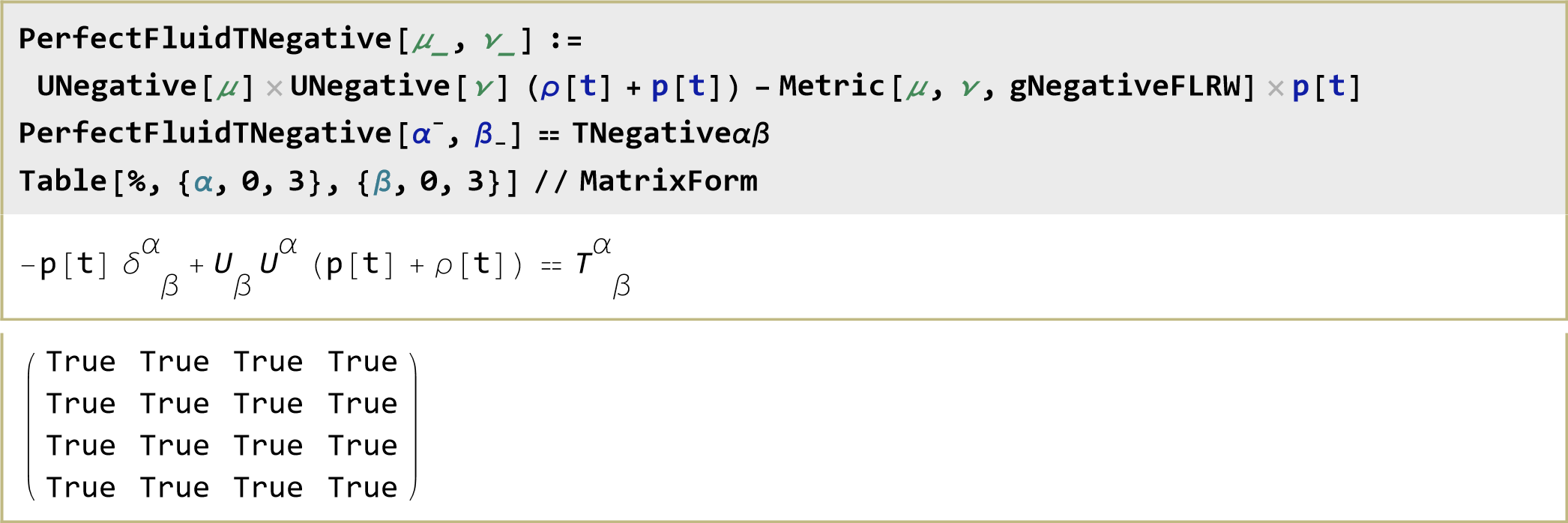}
CovariantD obeys linearity and the product rule, allowing us to recover the $\rho $pNegativeRelation above. 
\includecode{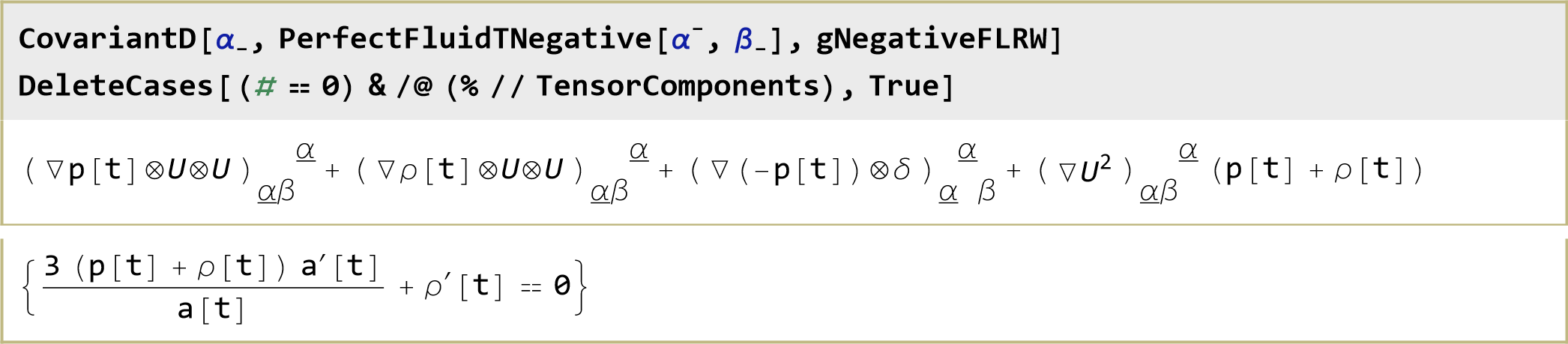}

\subsection{Zero Spatial Curvature}

Finally, we turn to the spatially flat case.
\includecode{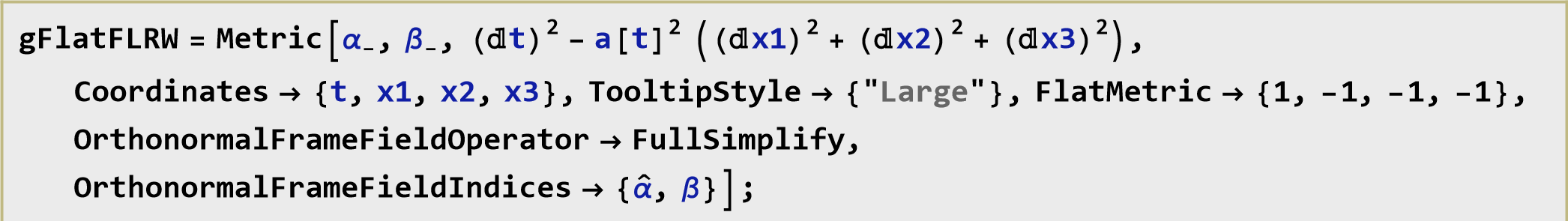}
At a fixed time, the geometry becomes the flat space \(a^2\left(\left(dx^1\right)^2+\left(dx^2\right)^2+\left(dx^3\right)^2\right)\). This suggests spatial translations and rotations are its isometries. The corresponding linear and angular momentum generators are defined as follows.
\includecode{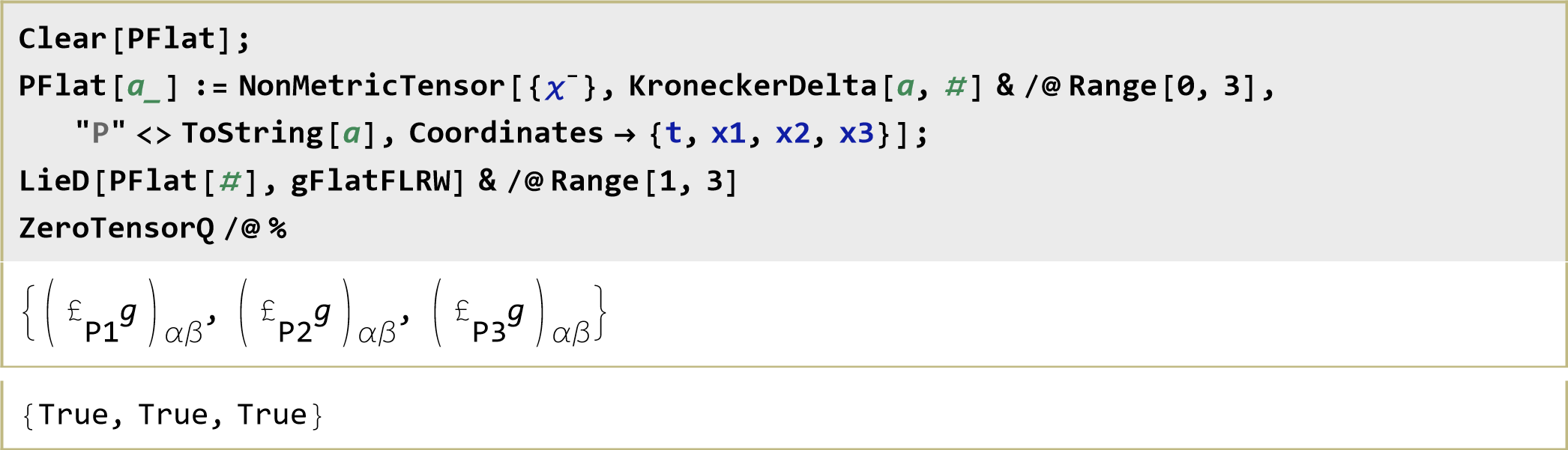}
\includecode{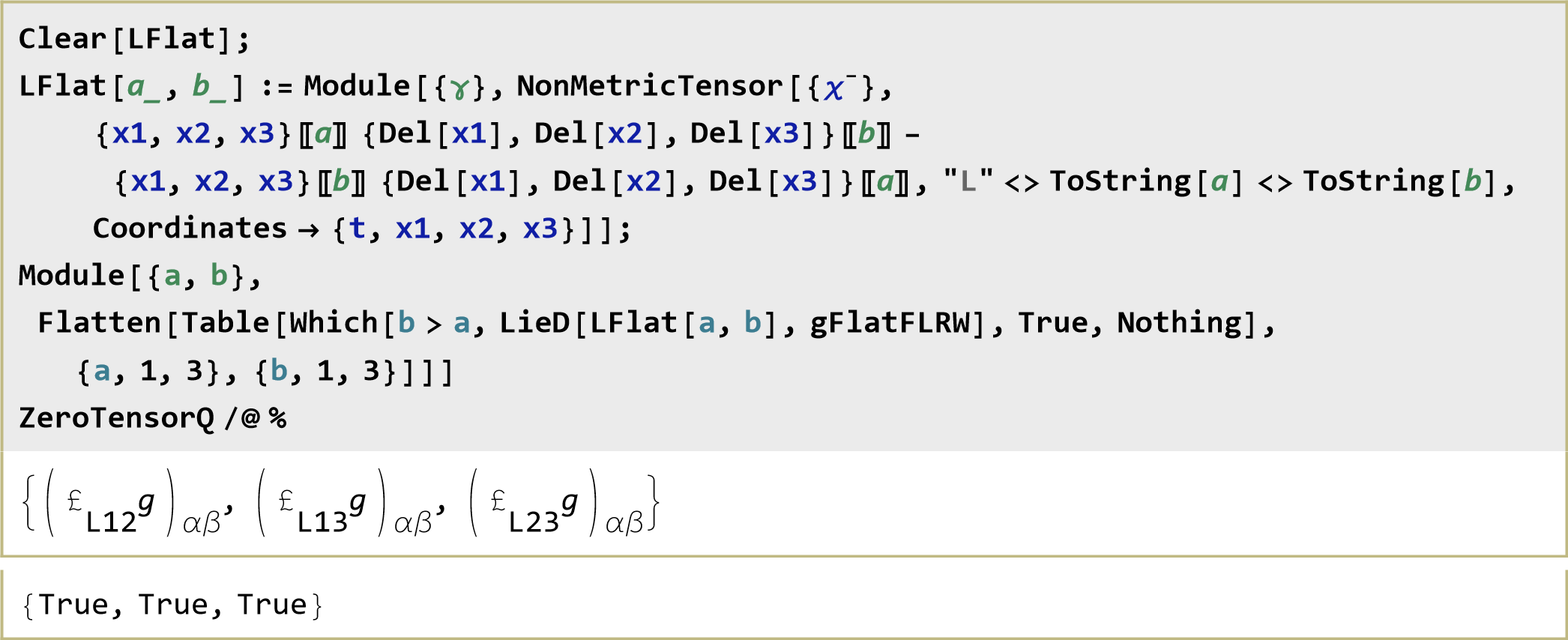}
Let \(U^{\mu }\) = \(\delta ^{\mu }{}_0\) be the timelike vector tangent to the worldline of a particle at rest with the universe. We may verify
its unit length and geodesic nature.
\includecode{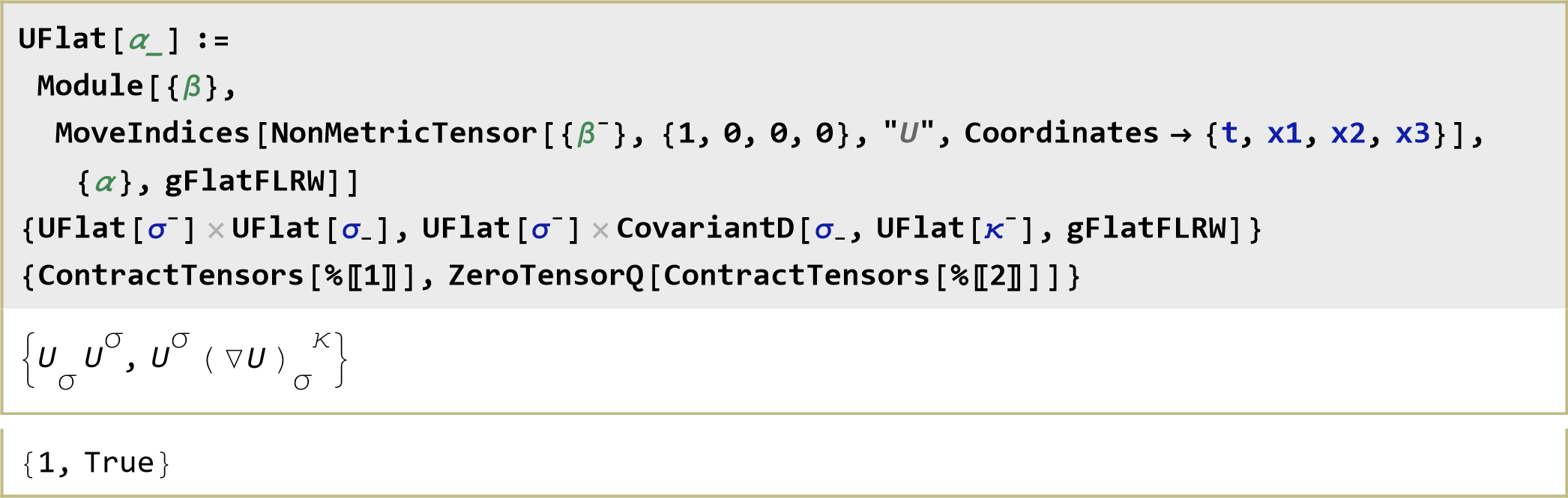}
With this \textit{U} we may also construct the Killing tensor of the FLRW geometry.
\includecode{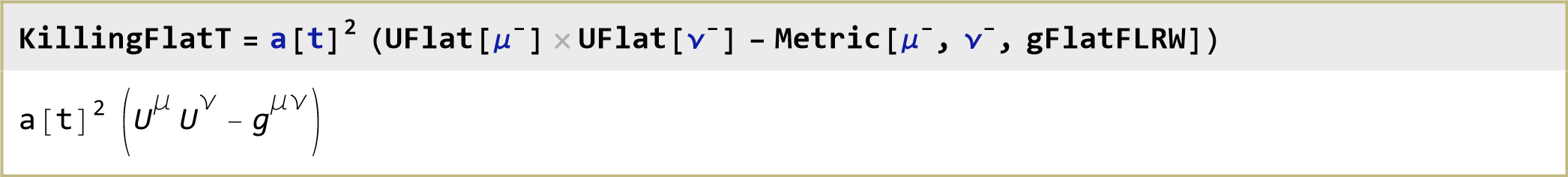}
The key property is that its fully symmetrized first covariant derivative is zero: \(\nabla _{\{\alpha }K_{\mu \nu \}}\) = 0. If \(q^{\mu }\) were tangent to some geodesic worldline, obeying \(q^{\sigma }\nabla _{\sigma }q^{\alpha }=0\), the quantity \(K_{\mu \nu }q^{\mu }q^{\nu }\) is then constant along the geodesic itself.
\includecode{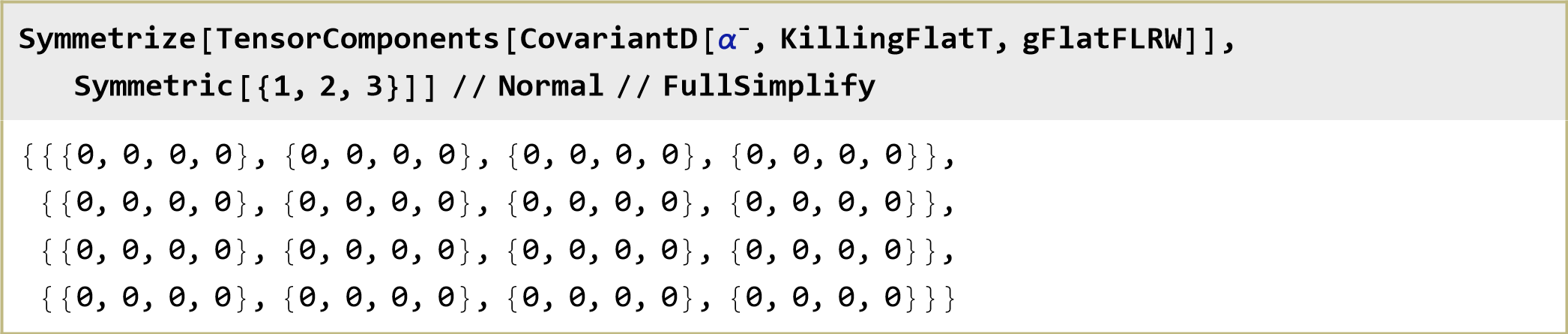}
Just like both the positively and negatively curved cases, the flat FLRW geometry is conformally flat because its Weyl tensor is also zero.
\includecode{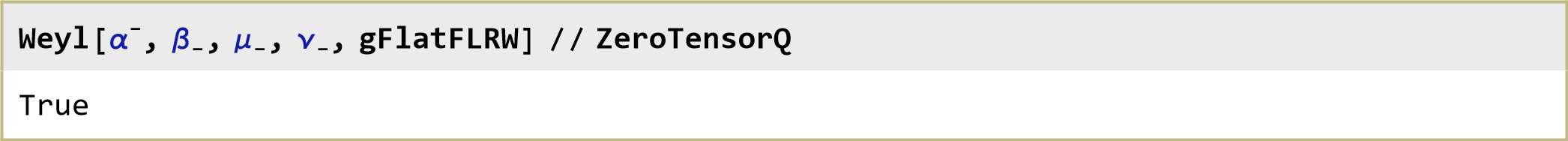}
In this case we may switch to conformal time $\eta $, which is related to observer time \textit{t} through the relation d$t \equiv $ a[$\eta $]
d$\eta $, to render the metric explicitly proportional to Minkowski spacetime \((\dd\eta)^2-\dd\overset{\rightharpoonup}{x}\).\(\dd\overset{\rightharpoonup}{x}\).
\includecode{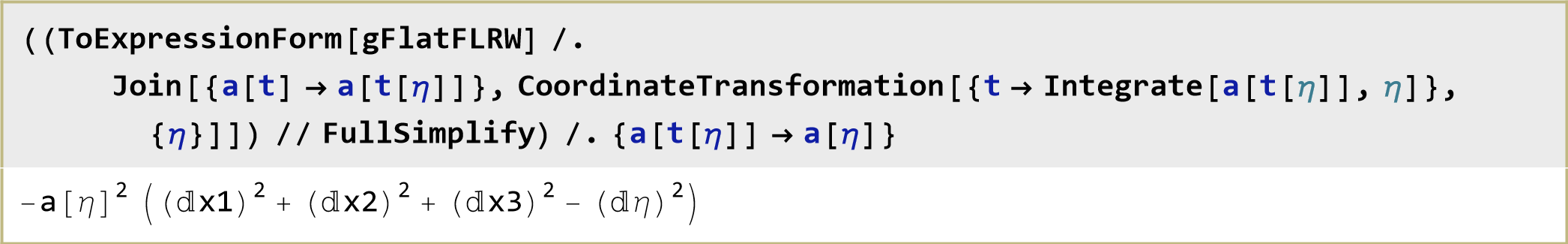}
Turning to physics, let us ask: what sort of matter would source such a spatially flat curved spacetime like gFlatFLRW? That is, what sort of stress tensor \(T^{\mu \nu }\) would satisfy \(G^{\mu \nu }\) - $\Lambda $ \(g^{\mu \nu }\) = 8$\pi $\(G_N\) \(T^{\mu \nu }\) ($\Lambda $ is the cosmological constant)? We first observe that the Einstein tensor \(G^{\mu \nu }\) of such a geometry enjoys the same Killing symmetries.
\includecode{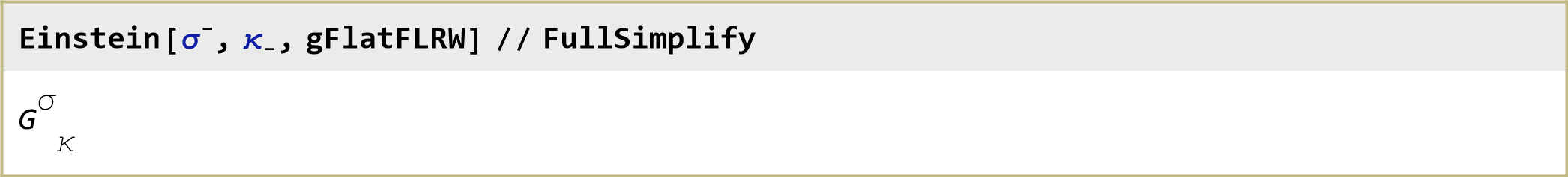}
\includecode{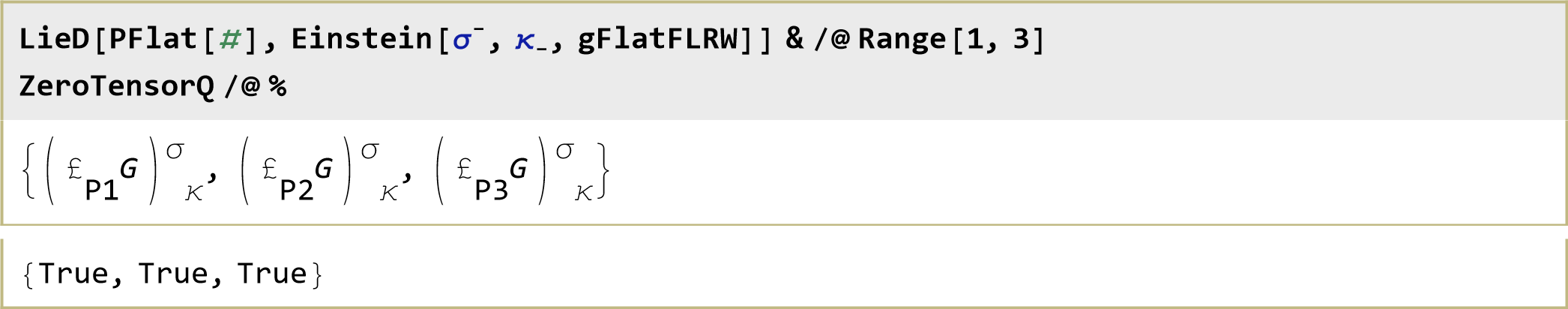}
\includecode{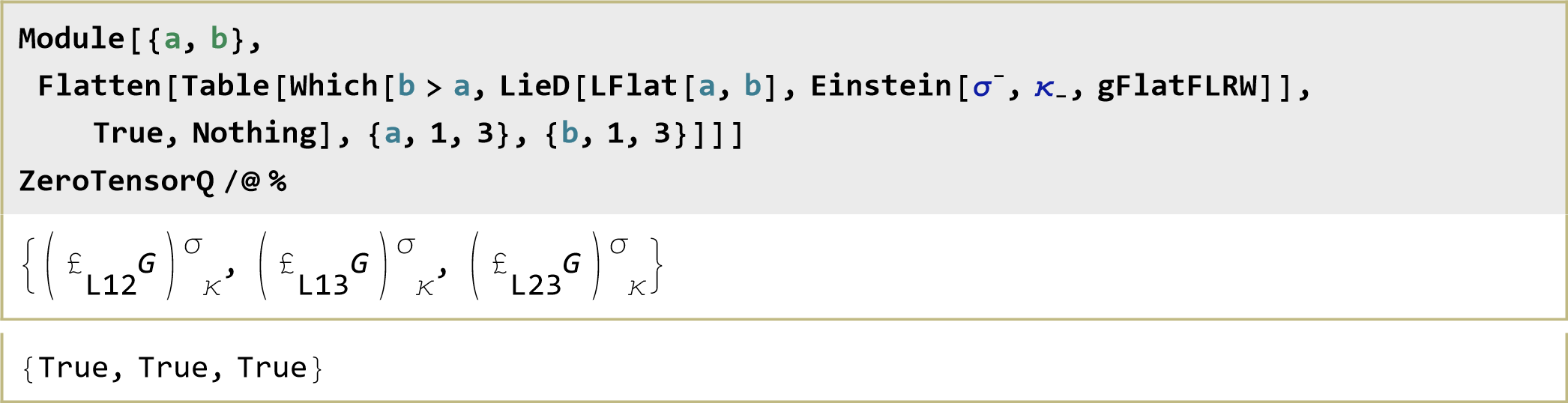}
Therefore we may constrain the form of the stress tensor \(T^{\mu }{}_{\nu }\) by demanding that it satisfy \(\pounds _{\xi }\)\(T^{\mu }{}_{\nu}\) = 0, where $\xi $ is one of the above Killing vectors.
\includecode{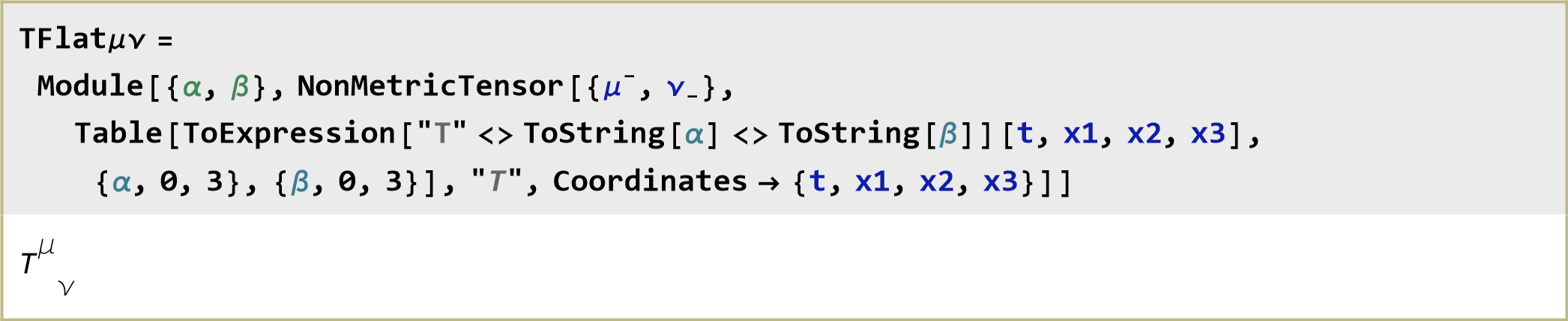}
That the Lie derivative of the stress tensor along the {``}linear momentum{''} vectors must be zero implies every component must be in fact space-independent.
\twopicincludecode{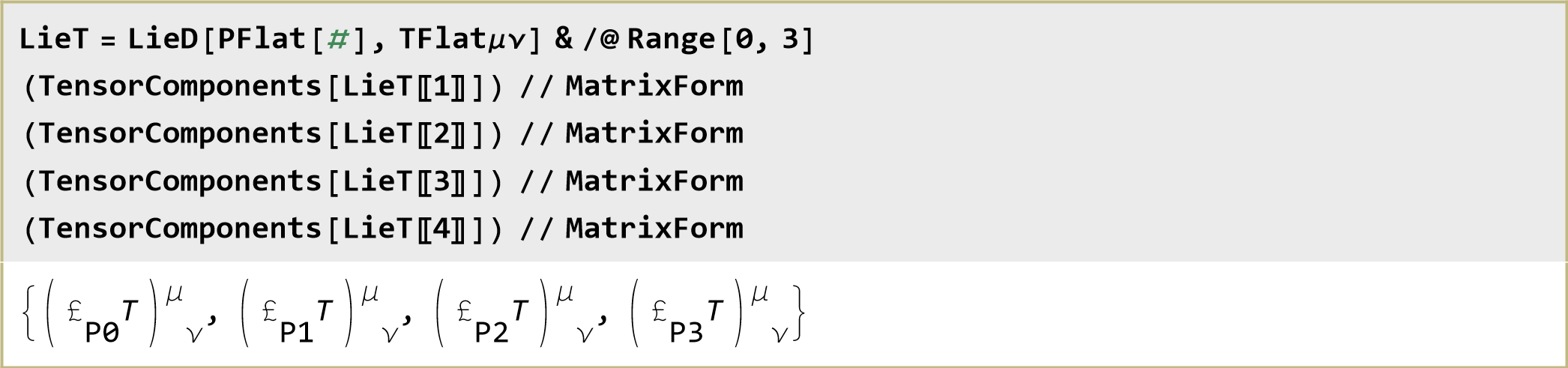}{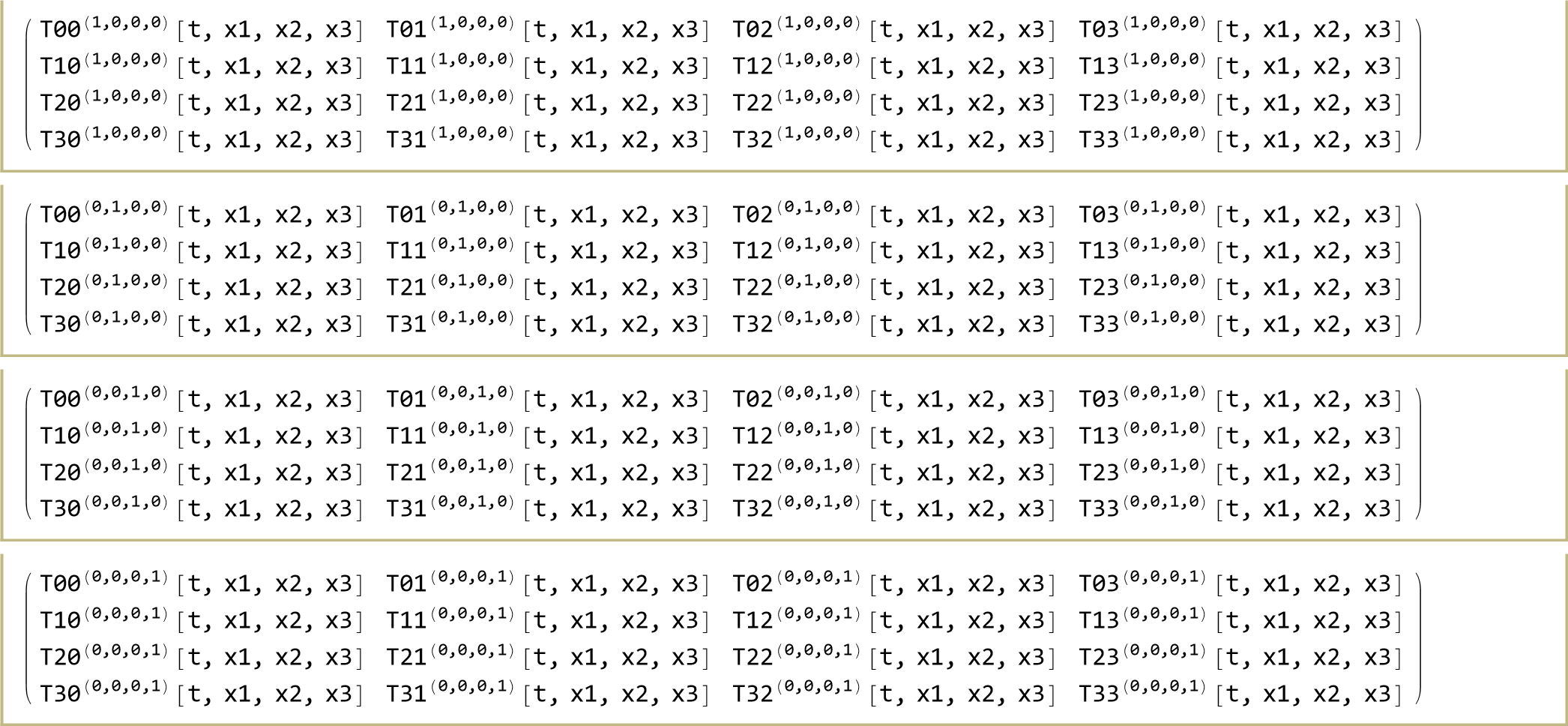}
\includecode{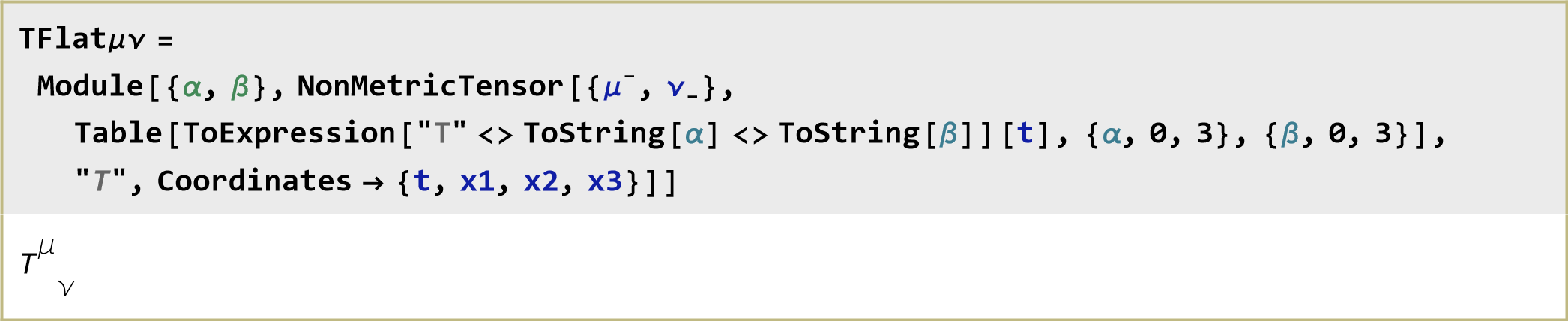}
That the Lie derivative of the stress tensor along the {``}angular momentum{''} vectors must be zero implies not only is it diagonal, the space-space components must be equal.
\twopicincludecode{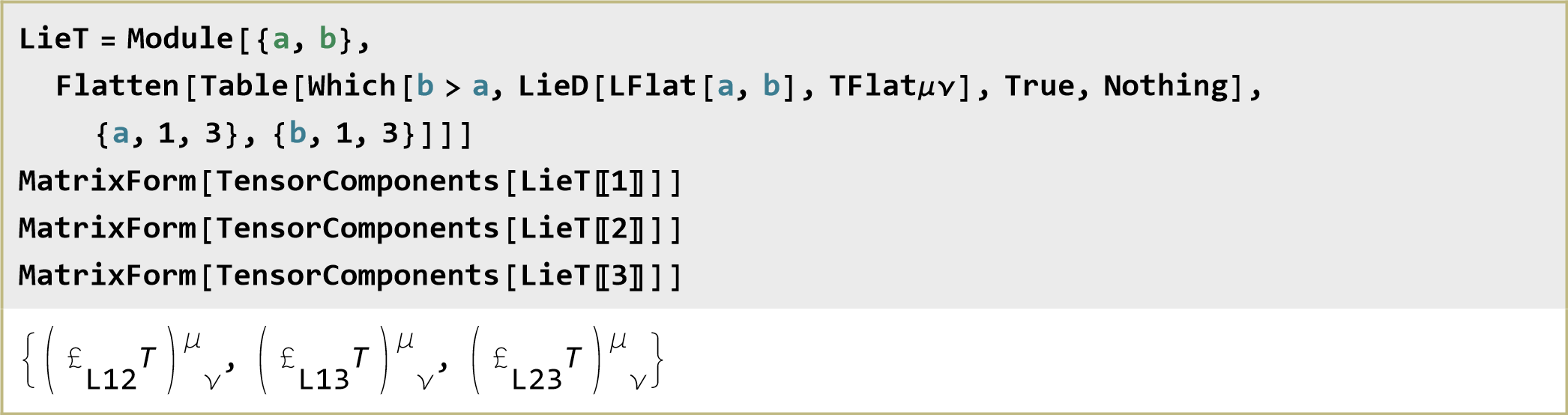}{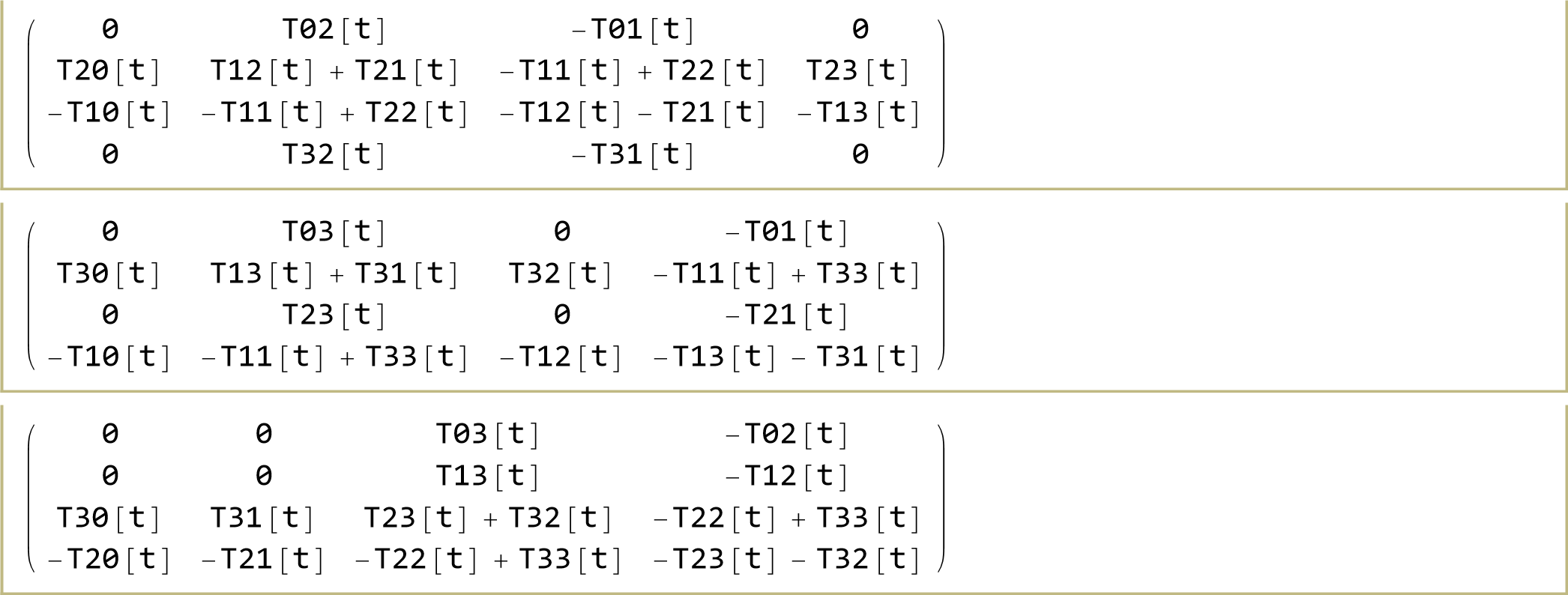}
\includecode{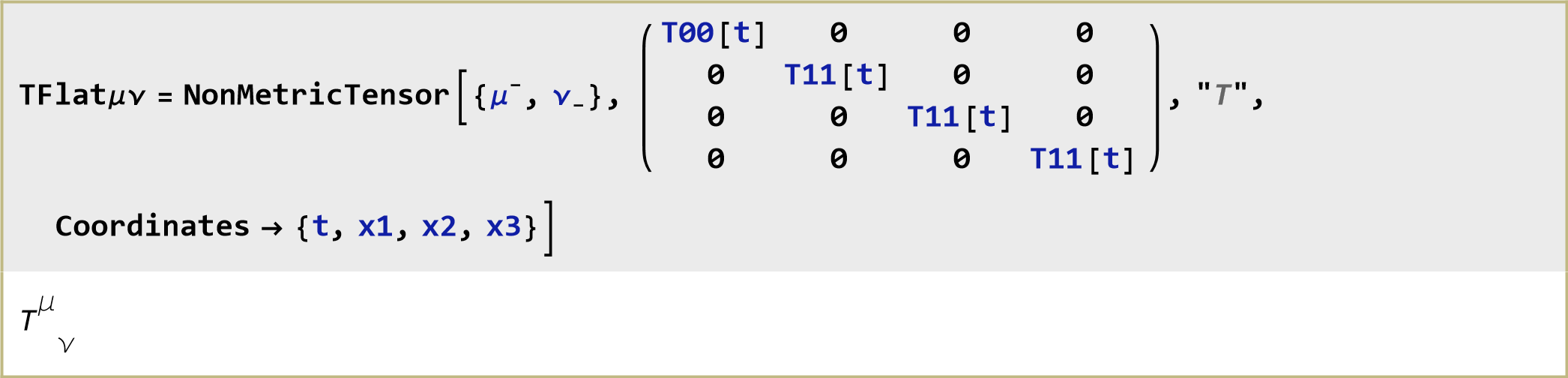}
To interpret the components we express the stress tensor in an orthonormal basis, where \(T^{\hat{0}\hat{0}}\) is the energy density and \(T^{\hat{a}\hat{a}}\) (no sum over \textit{a}) is the pressure along the \textit{a}-th spatial direction. 
\includecode{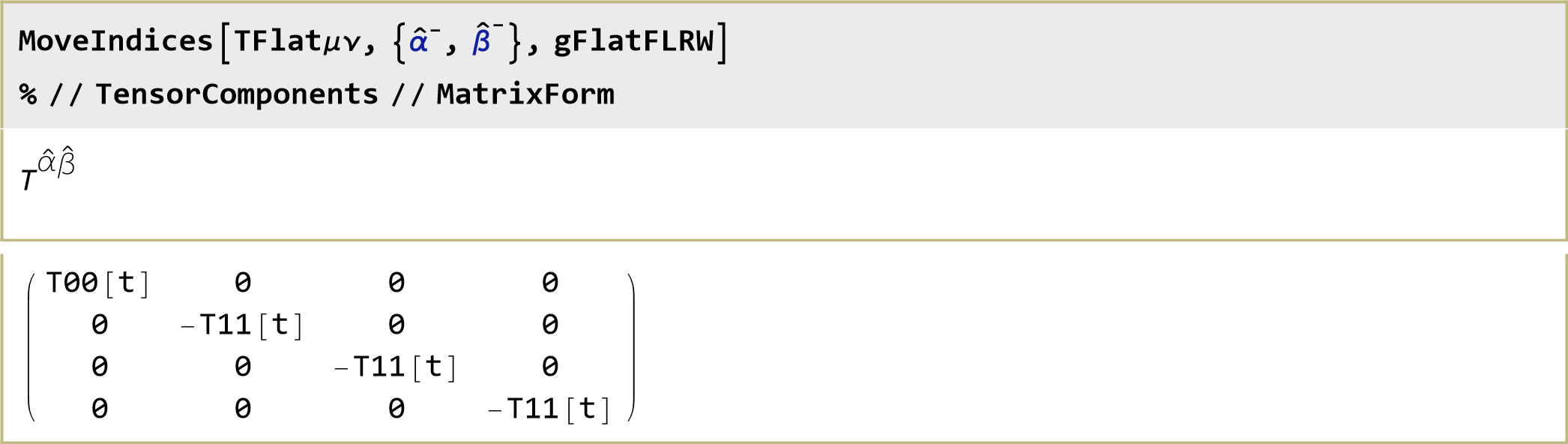}
This implies \(T^0{}_0\equiv \rho\), the energy density; and \(-T^1{}_1\equiv p\), the pressure density.
\includecode{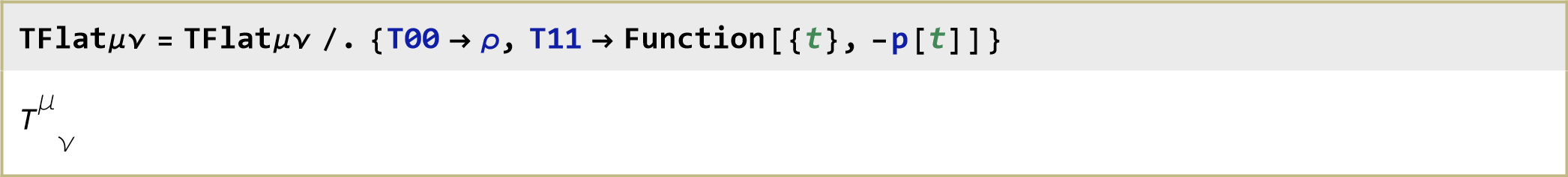}
That Einstein{'}s tensor is divergence-free means the stress tensor must be too.
\includecode{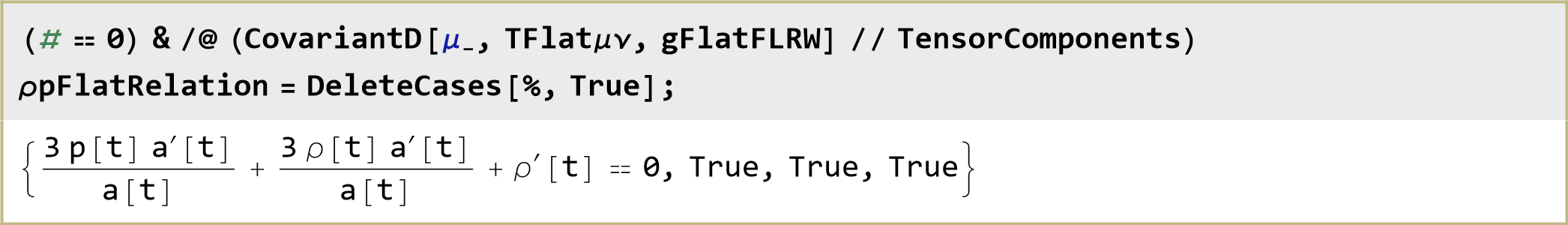}
If p/$\rho $ $\equiv $ \textit{w} were a constant -- for instance, in a radiation, matter and Dark Energy dominated universe, $$w = 1/3, 0, -1$$
respectively -- we may show that $\rho $[t] = $\rho $[\(t_0\)] (\(\left.\left.a\left[t_0\right]\right/a[t]\right){}^{3(1+w)}\) for some {``}initial time{''} \(t_0\).
\includecode{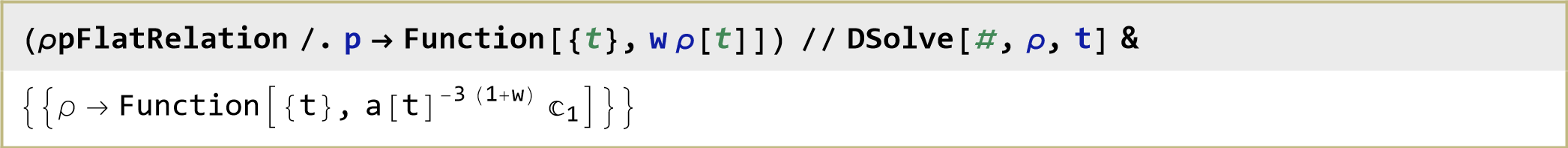}
Einstein{'}s equations governing the large scale dynamics of the spatially flat universe read as follows.
\twopicincludecode{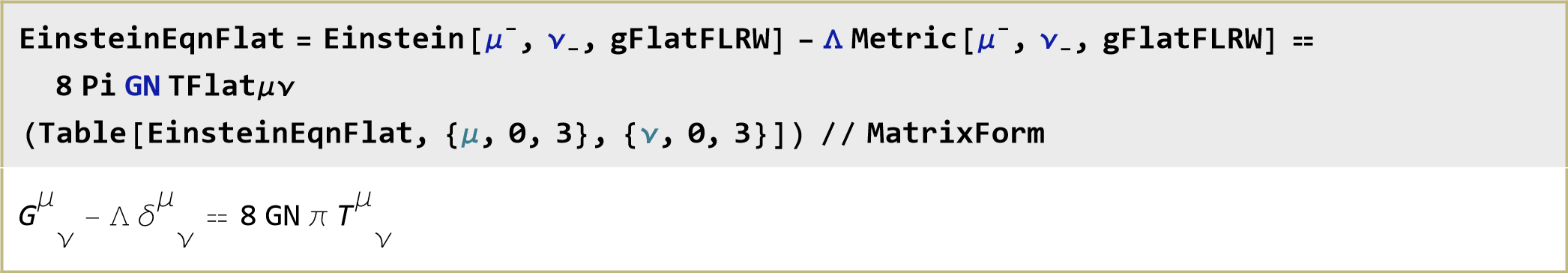}{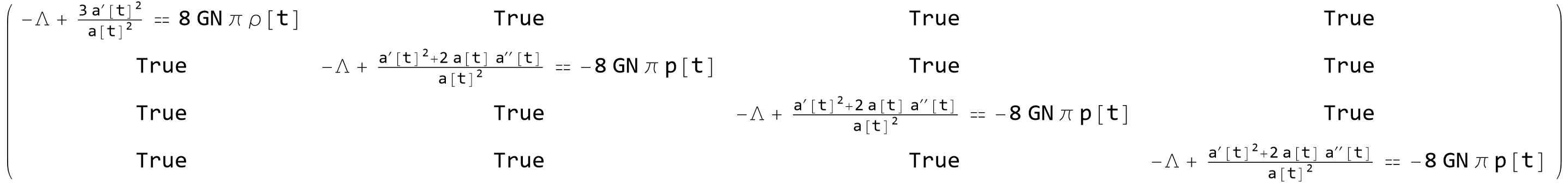}
Like the positively and negatively curved cases, there are only two distinct equations. We use them to solve for the scale factor{'}s velocity and acceleration.
\includecode{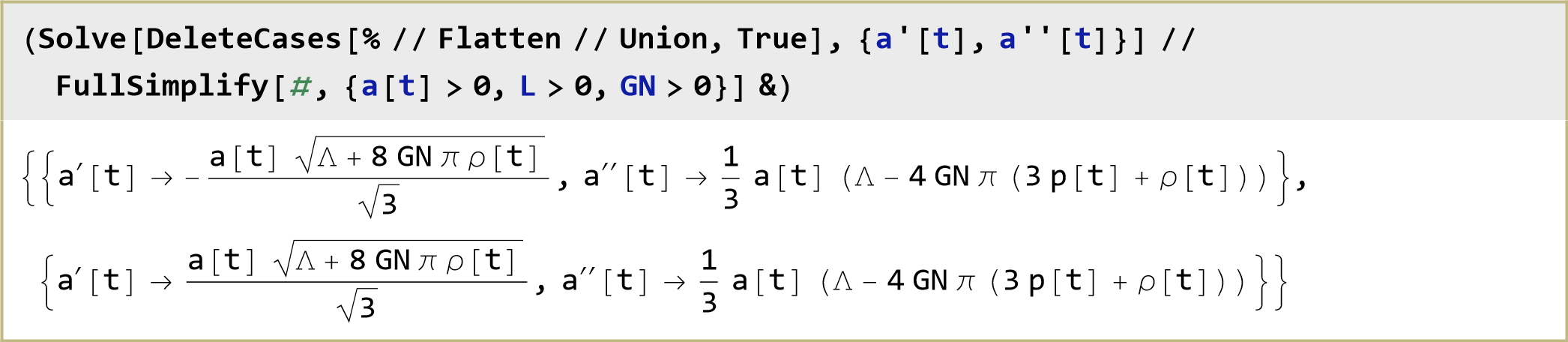}
The 11 component of Einstein{'}s equations is not independent of the 00 component, in that the former follows from the latter -- provided the
divergence-free property of the stress tensor holds. To see this, we differentiate the 00 component with respect to time, and proceed to utilize
both the \(\nabla _{\mu }T^{\mu \nu }=0\) and 00 component to reduce the result to that of the 11 component.
\includecode{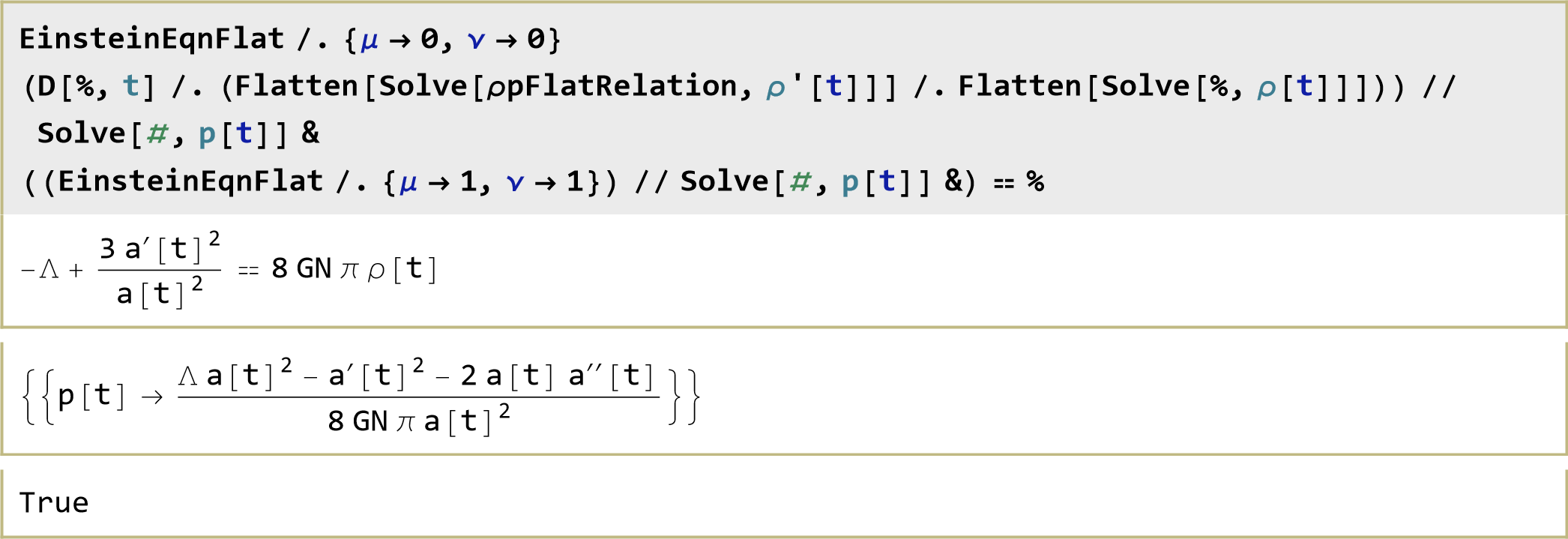}
Finally, let us observe that the stress tensor may be written a perfect fluid form, \(T^{\mu \nu }=U^{\mu }U^{\nu }(\rho +p) - g^{\mu \nu }p\).
\includecode{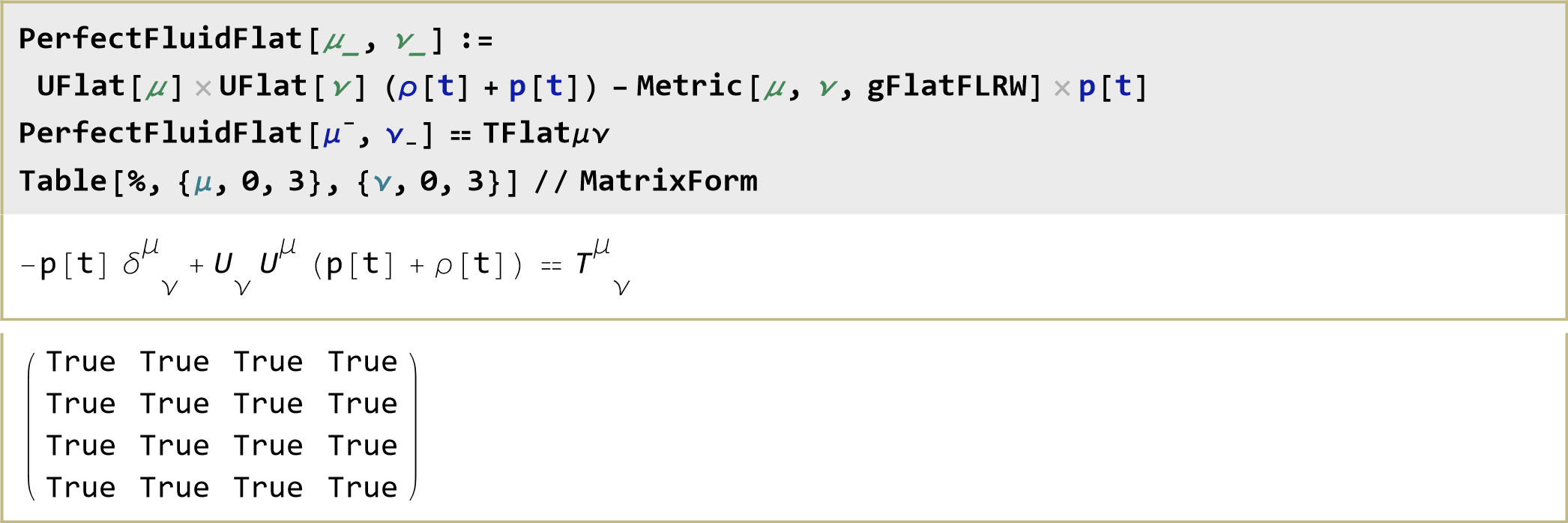}
CovariantD obeys linearity and the product rule, allowing us to recover the $\rho $pNegativeRelation above.
\includecode{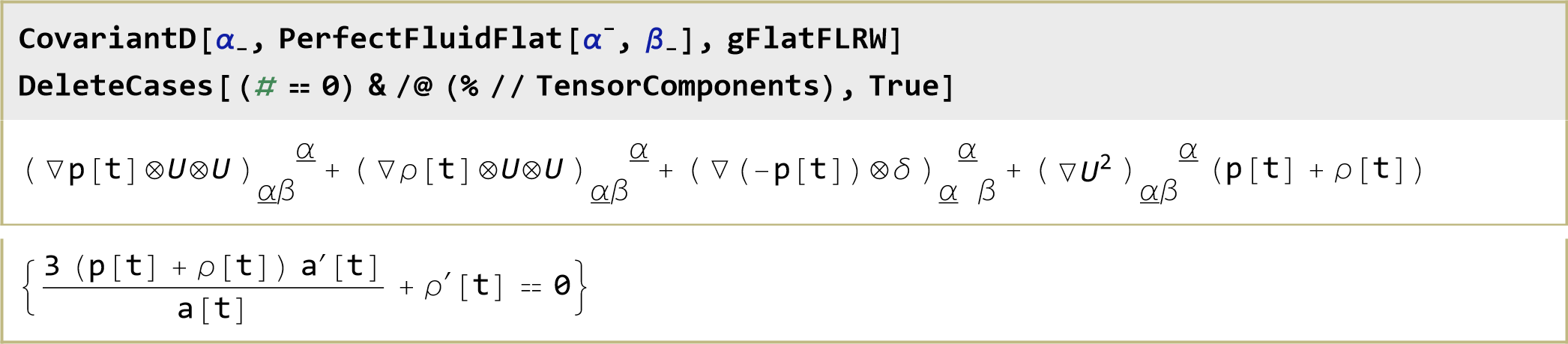}
\boldtitle{JWKB and Cosmological Redshift} \qquad For a generic scale factor a[t] it may be impossible to obtain exact analytic solutions to the wave equation \(\nabla _{\alpha }\)\(\nabla ^{\alpha}\)$\varphi $ = 0 in cosmological spacetimes. However, we may study the problem in the high frequency or short wavelength approximation, where the
wavelength of the solution is much shorter than the curvature scale of the universe at large.

Specifically, we suppose $\varphi $ = Re $A$ exp[I \(S_{\text{JWKB}}\)], where $A$ is some slowly varying amplitude and the exponential involving \(S_{\text{JWKB}}\) is the rapidly evolving phase. Furthermore, by the spatial translation symmetry at hand, we expect to be able to exploit Fourier decomposition -- doing so then tells us
\[
S_{\text{JWKB}} = \overset{\rightharpoonup}{k}\cdot \overset{\rightharpoonup}{x} + \Sigma [t]. 
\]
To leading order in the JWKB approximation, the key result is that the gradient of this \(S_{\text{JWKB}}\) is a null vector.
\includecode{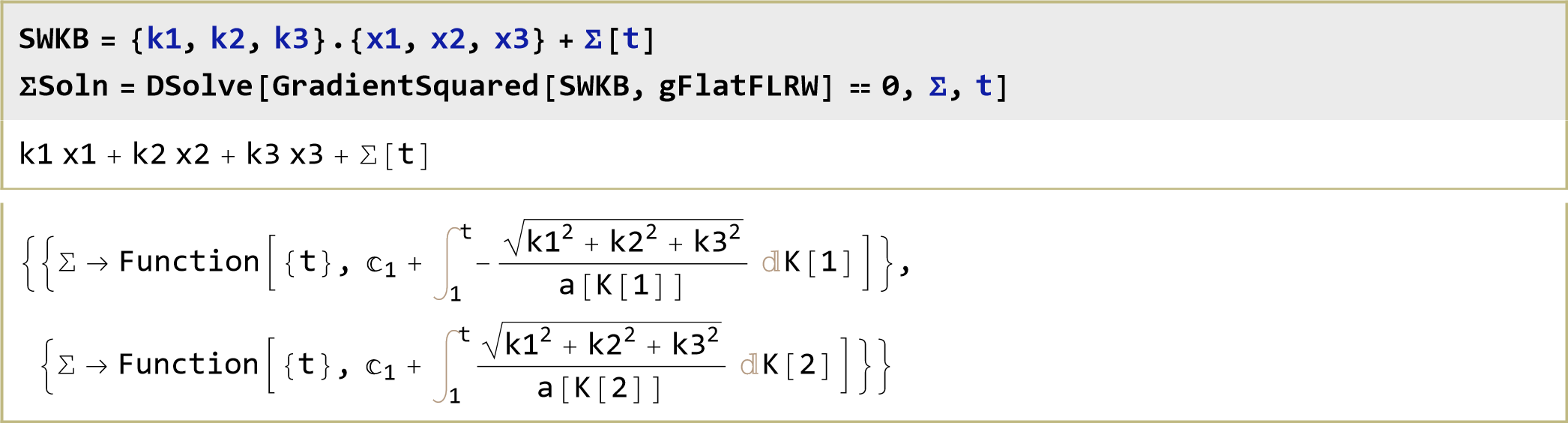}
The observer at rest with the universe detects a frequency given by \(\omega\) = \(U^{\alpha }\partial _{\alpha }S_{\text{JWKB}}\)., where \textit{U} is above the timelike geodesic tangent vector UFlat. By comparing the signal detected by the observer at \(t_{\text{now}}\) to that at \(t_{\text{emission}}\), we may obtain the formula for cosmological redshift 1+z $\equiv $ $\omega $[now]/$\omega $[emission] = a[\(t_{\text{emission}}\)]/a[\(t_{\text{now}}\)].
\includecode{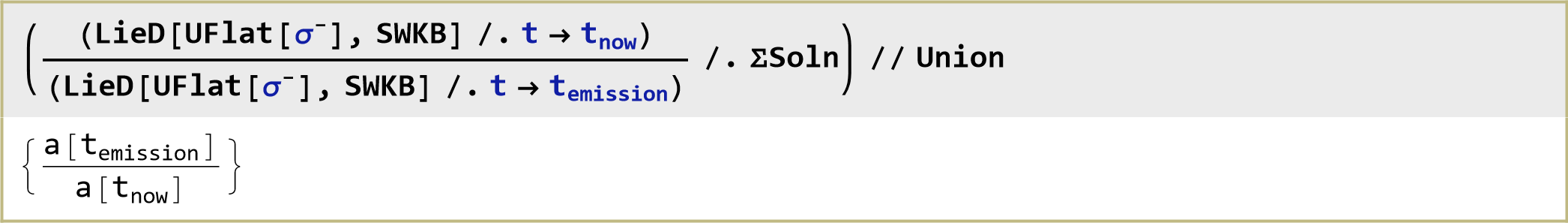}

\section{Black Holes}

In this section, we shall develop a brief study of vacuum Black Hole solutions of Einstein{'}s equations with zero cosmological constant.
\[
G^{\alpha \beta }=8\text{$\pi $G}_NT^{\alpha \beta } = 0
\]
\boldtitle{Rotationally Invariant Vacuum Solution} \qquad We begin with the fact that the most general rotationally symmetric 4D spacetime geometry may be put in the following form.
\includecode{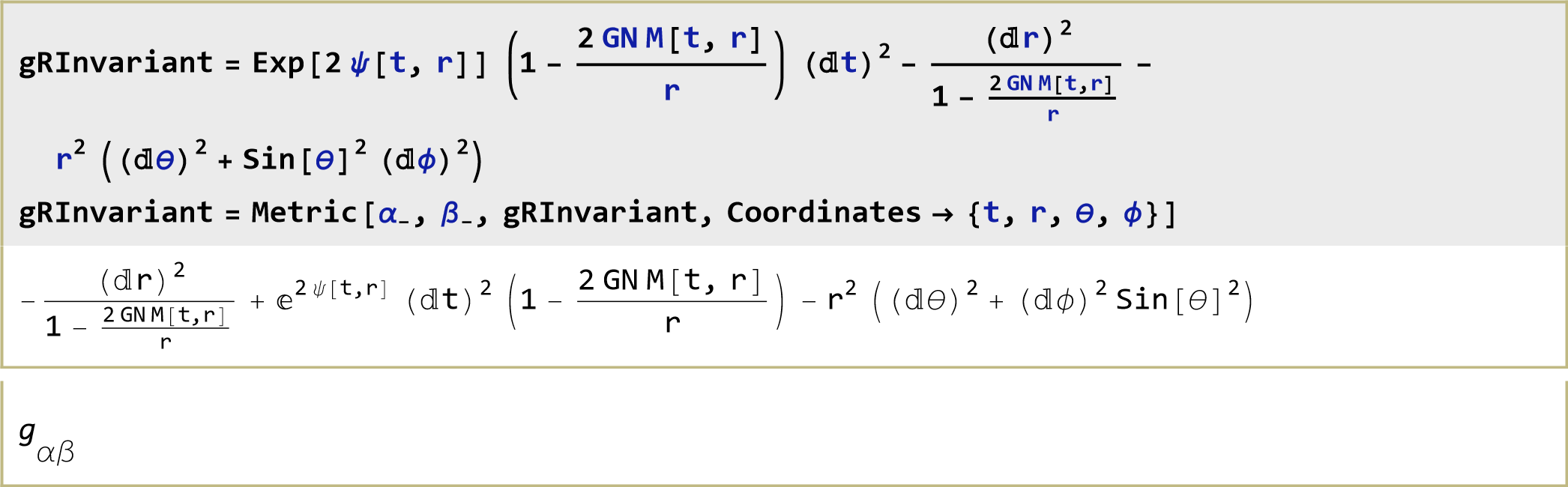}
In vacuum the Einstein tensor is zero. The (\textit{t}, \textit{t}) and (\textit{t}, \textit{r}) components tell us the mass function \textit{M} must be constant.
\includecode{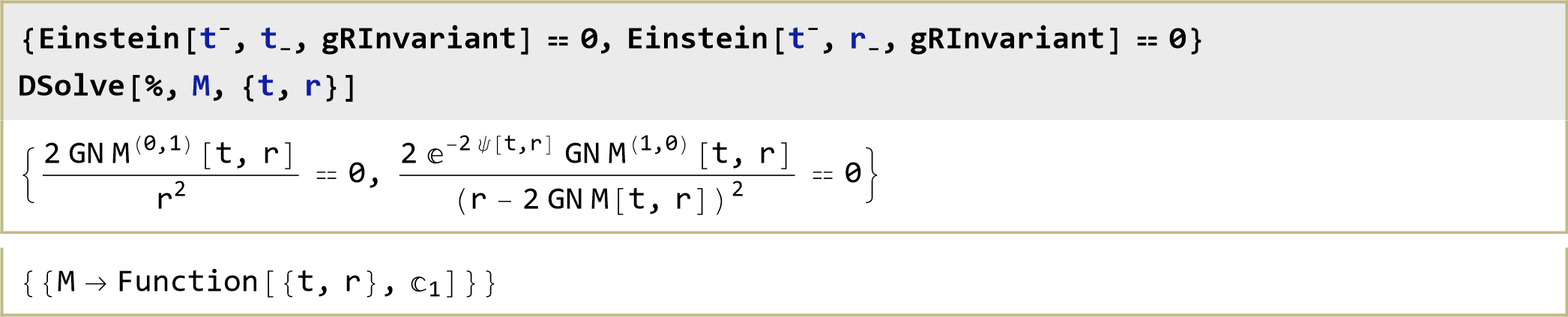}
Whereas the (\textit{t}, \textit{t}) minus (\textit{r}, \textit{r}) components of Einstein's equations tell us $\psi $ must be independent
of \textit{r}. 
\includecode{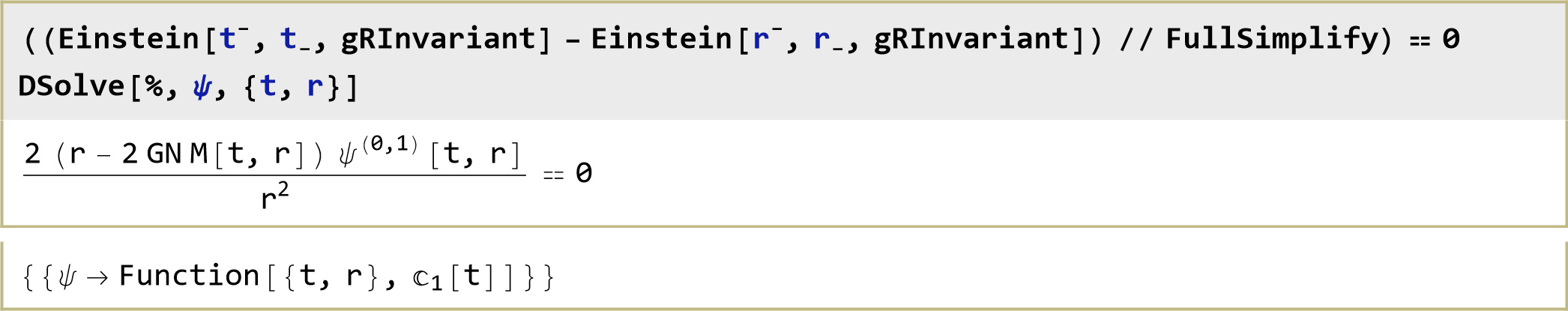}
Since the only occurrence of $\psi $ is in the combination \(\left(E^{\psi }dt\right)^2\), and since $\psi $ does not depend on \textit{r}, we may then re-define the time coordinate \textit{t} to set $\psi $ $\equiv $ 1; namely, \(E^{\psi }dt\) $\rightarrow $ d\textit{t}.

\subsection{Schwarzschild}

We have therefore derived Birkhoff{'}s theorem: the vacuum spherically symmetric region of spacetime obeying Einstein{'}s equations (with zero cosmological constant) is the following Schwarzschild geometry.
\includecode{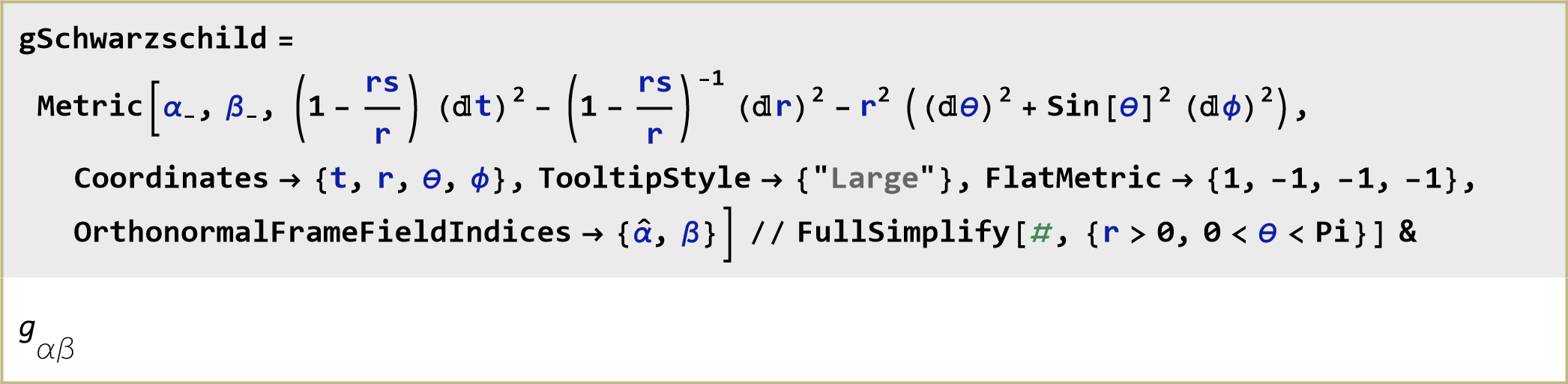}
where we have defined the Schwarzschild radius to be \(r_s\equiv\) 2 \(G_NM\).

By choosing FlatMetric $\to $ $\{$1,-1,-1,-1$\}$, and therefore \(\varepsilon ^{\hat{0}}{}_{\mu }=\delta _{\mu } {}^0g_{00}{}^{1/2}\) and \(\varepsilon^{\hat{1}}{}_{\mu }=\delta _{\mu } {}^1\left(-g_{\text{rr}}\right){}^{1/2}\) we have assumed \textit{r} $>$ \(r_s\) as far as the construction of the orthonormal frame fields is concerned, otherwise the square roots would be purely imaginary. To assume \textit{r} $<$ \(r_s\), we would choose FlatMetric $\to $ $\{$-1,+1,-1,-1$\}$ instead.
\includecode{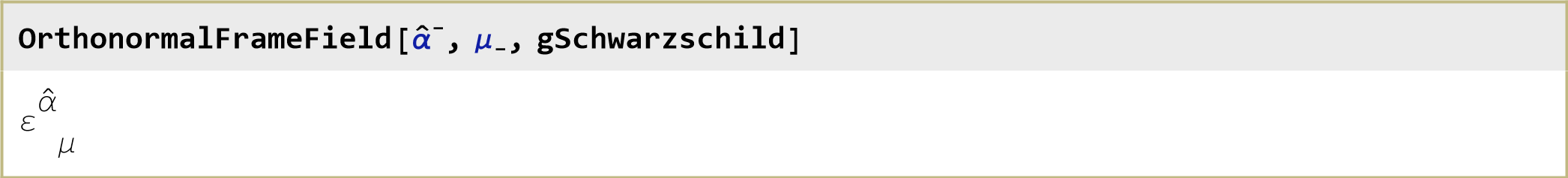}
\boldtitle{Schwarzschild Isometries} \qquad The Schwarzschild geometry is not only rotationally symmetric it is also invariant under time translation. To construct the rotation generators (Killing vectors) in the Schwarzschild geometry, we borrow \hyperref[sent:LIn3C]{our results} from the 3D flat space section.
\includecode{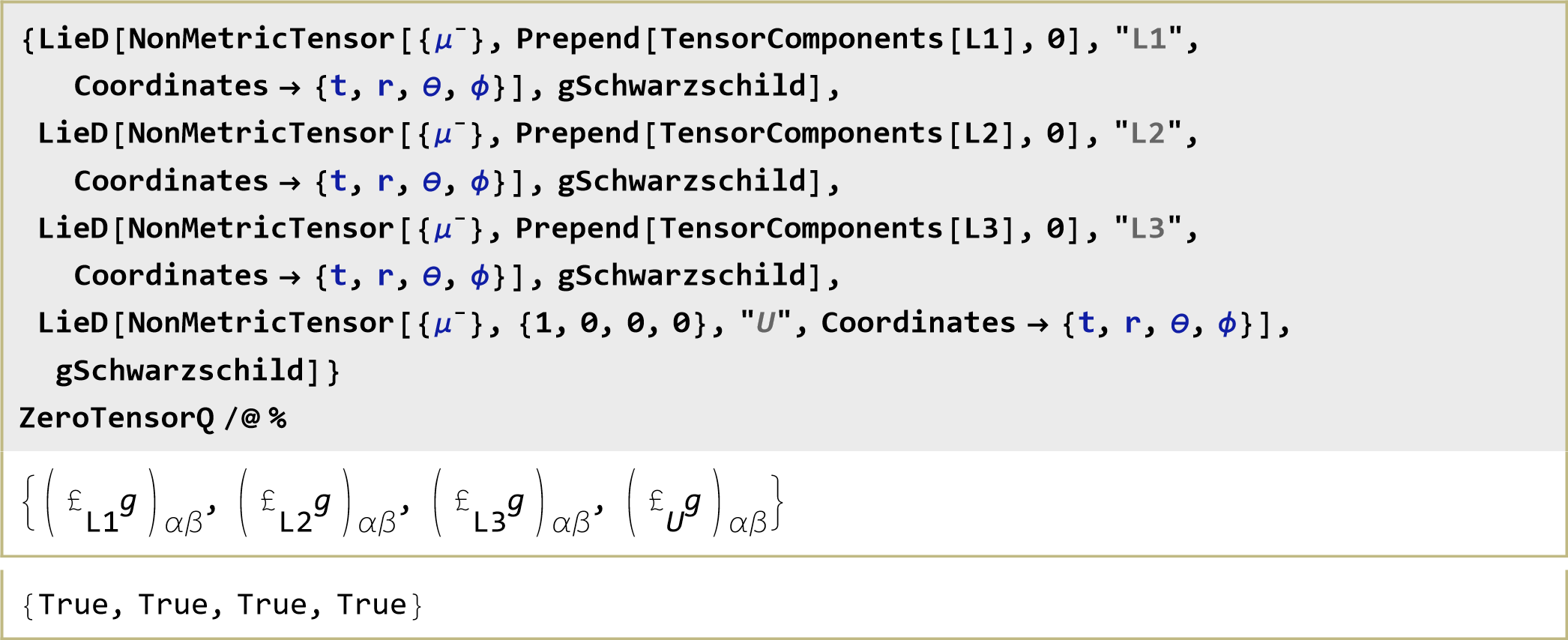}
\boldtitle{Coordinate Singularity vs Spacetime Curvature} \qquad As $r \to r_s$, notice \(g_{00}\to 0\) and \(g_{\text{rr}}\to \infty\). There is a coordinate singularity there. However, spacetime curvature is in fact finite everywhere expect at \textit{r} = 0. For instance, we may compute the {``}square{''} of the Riemann tensor, a scalar invariant.
\includecode{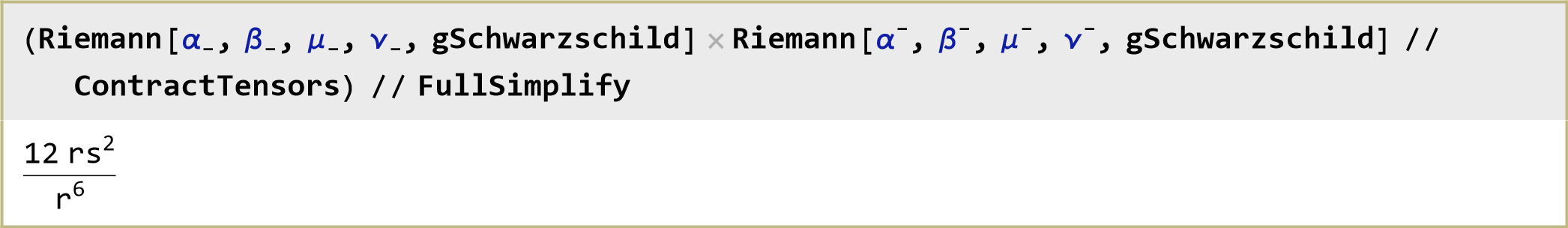}
Alternatively, we may simply compute the Riemann tensor components in an orthonormal frame. First we use MoveIndices to allow us to convert the coordinate basis Riemann into one where the indices can take arbitrary positions or basis (i.e., either orthonormal or coordinate).
\includecode{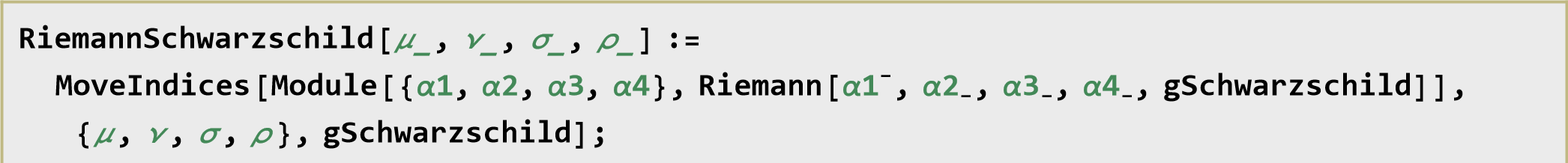}
Let us extract only the non-zero components. We see that, for \textit{r} $\geq $ \(r_s\), the geometric curvature is finite. In fact, the more massive the black hole, the smaller the geometric curvature is at the horizon \textit{r} = \(r_s\), since it goes as \(1\left/r_s{}^2\right.\).
\includecode{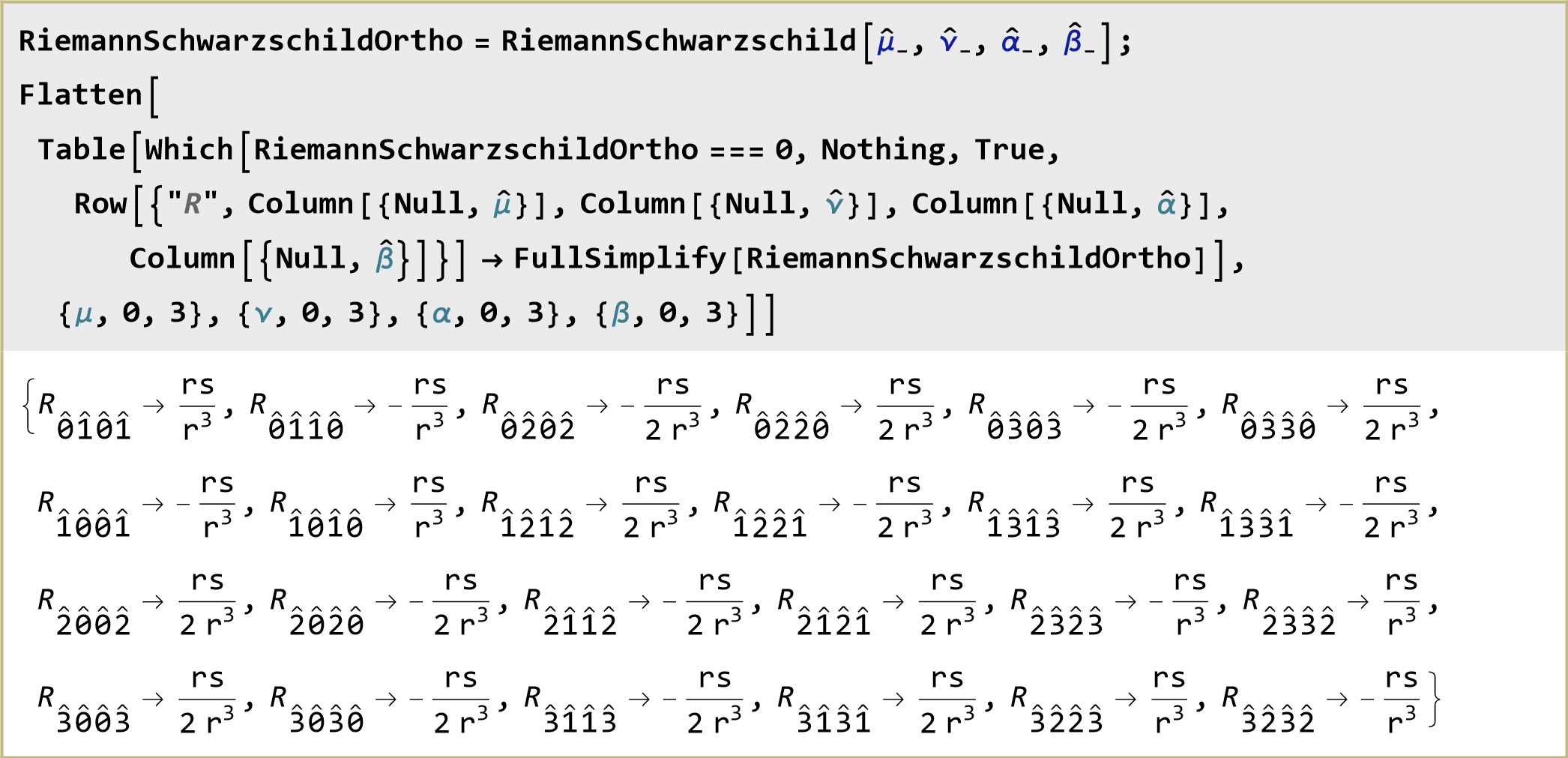}
\boldtitle{Gravitational Redshift} \qquad Consider a JWKB spherical wave $\varphi $ = Re \textit{A} \(E^{-I (\omega  t-\Omega [r,\theta ,\phi ])}\) propagating on the Schwarzschild geometry -- the time-translation invariance of the geometry allows us to do a frequency transform of the wave; hence the \(E^{-I \omega  t}\) factor. Suppose there are two observers remaining at rest at different spatial coordinates: (\(r_{1,2}\), \(\theta _{1,2}\), \(\phi _{1,2}\)). The unit timelike vector \textit{U} tangent to their worldlines must be \(U^{\mu }\) = \(\delta ^{\mu }{}_0\)/\(g_{00}{}^{1/2}\), since their spatial velocities are zero.
\includecode{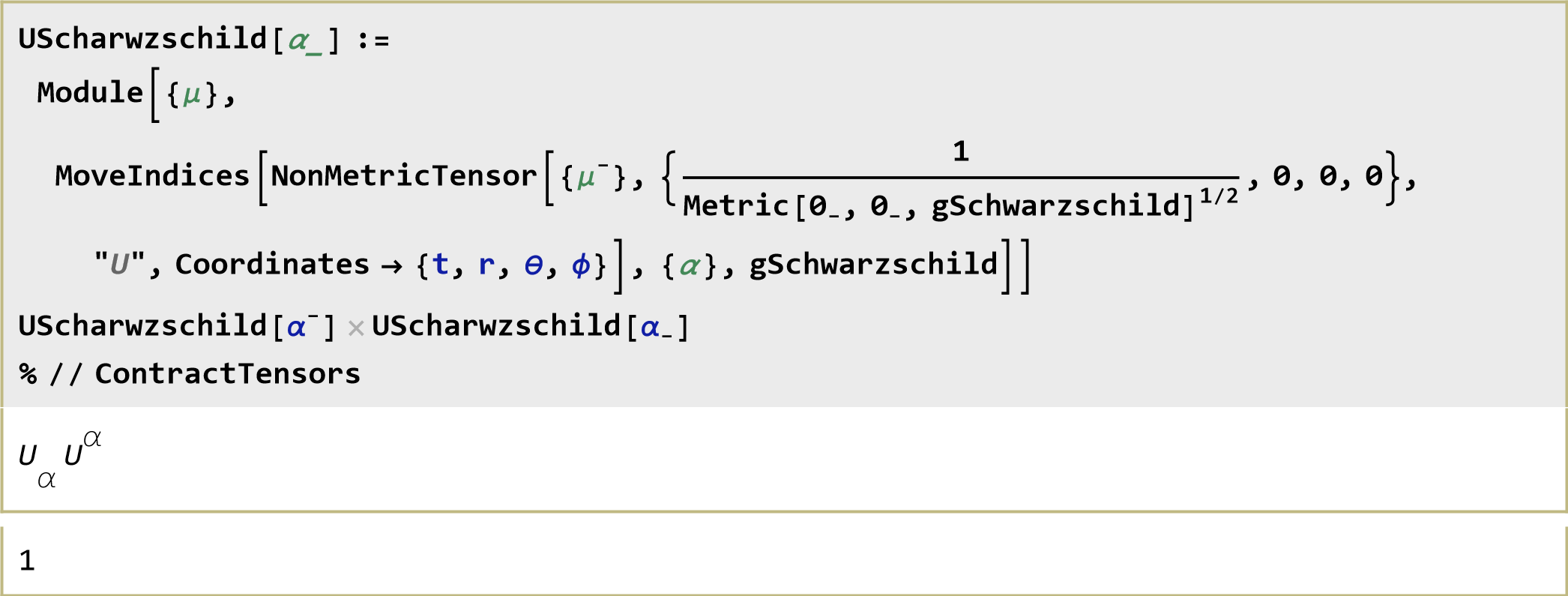}
JWKB also says, the ratio of the observed frequency at \(r_1\) to that at \(r_2\) is given by \(\left(U^{\alpha }\partial _{\alpha }S\right){}_{r_1}/\left(U^{\alpha}\partial _{\alpha }S\right){}_{r_2}\). We see that is it given by the ratios of 1/\(g_{00}{}^{1/2}\).
\includecode{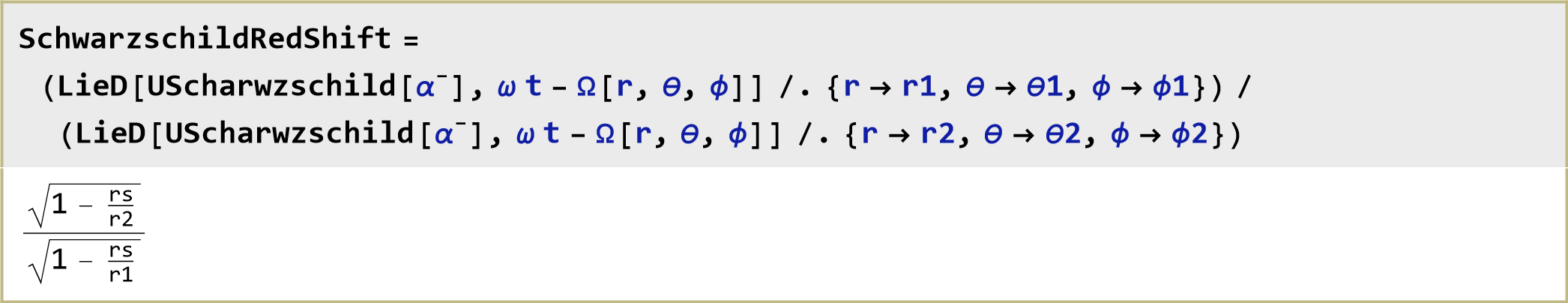}
Let us work out the order-of-magnitude of this redshift effect for a pair of observers, the first one based at the bottom and the second on the top of a 500 meter tall structure on Earth. Since \(r_s\)/$r$ is very small for the Earth -- specifically, using the speed of light = 299,792,458 meters/second = 1 to convert \(\left.r_s\right/(\text{Earth } \text{radius})\) $\to $ 2 \(G_N\) \(M_{\text{Earth}}/R_{\text{radius}}\) to a pure number,
\includecode{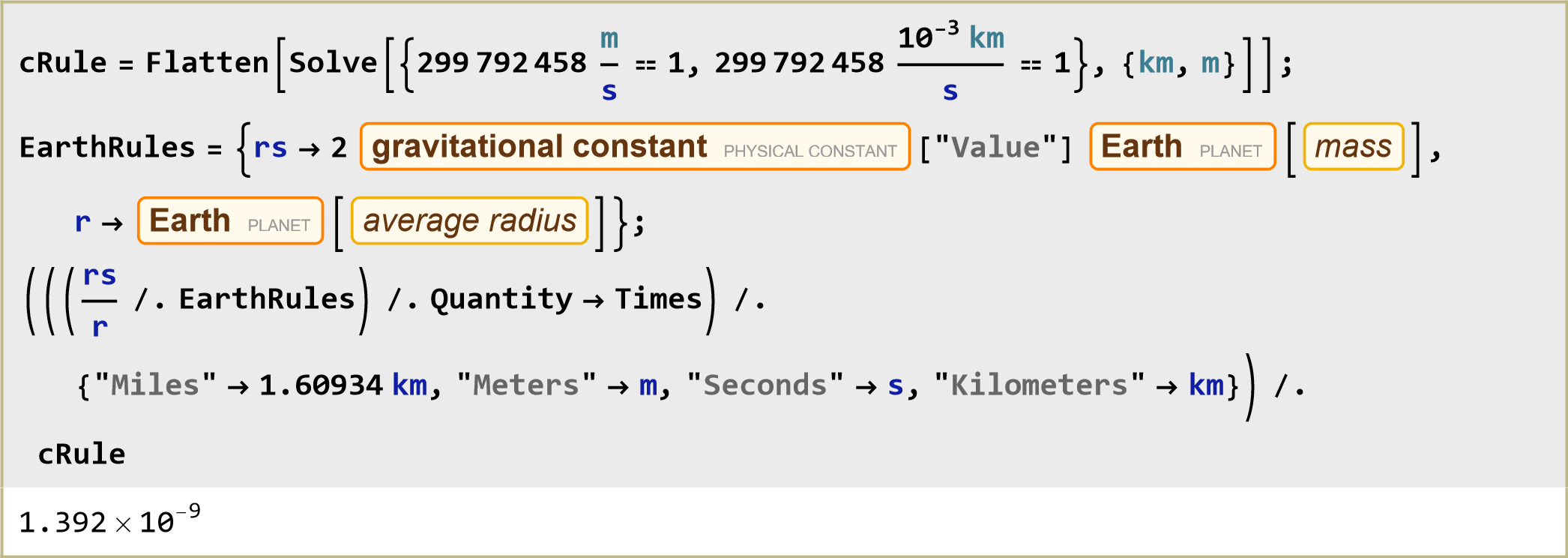}
-- we may proceed do a Taylor expansion in \(r_s\), followed by an expansion in the ratio 500m/\(R_{\text{radius}}\). If we define 1+z $\equiv
$ \(\left(U^{\alpha }\partial _{\alpha }S\right){}_{r_1}/\left(U^{\alpha }\partial _{\alpha }S\right){}_{r_2}\) then the redshift z itself is given by the following. The result is essentially the difference between the gravitational Newtonian potential at 2 different radii.
\includecode{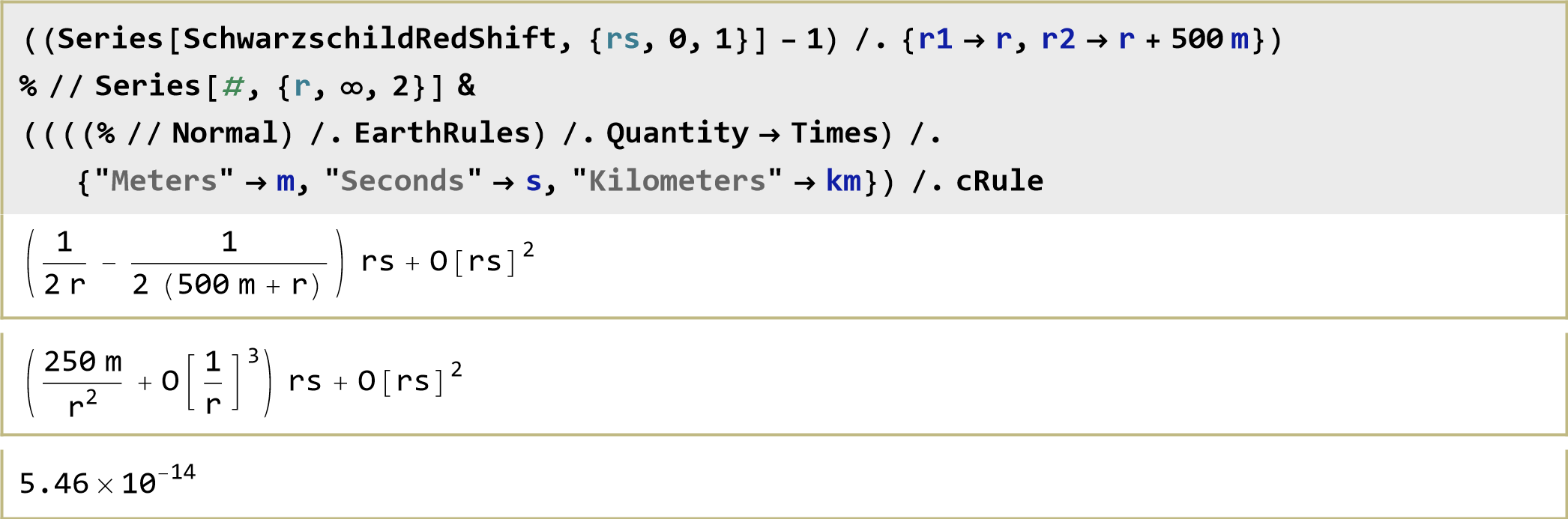}

\boldtitle{Schwarzschild Geodesics} \qquad We denote affinely parametrized geodesic 4-velocity as \(u^{\alpha } [\lambda ]\equiv  \left.\text{dz}^{\alpha }\right/\text{d$\lambda $}\) and non-affinely parametrized one as \(u^{\alpha } [s]\equiv \left.\text{dz}^{\alpha }\right/\text{ds}\). Given an input metric Tensor \textit{M}, GeodesicSystem[M, AffineParameter $\to $ $\lambda $, NonAffineParameter $\to $ s] returns a List consisting of the affinely parameterized geodesic Lagrangian.
\[
\frac{1}{2}g_{\alpha \beta } u^{\alpha } [\lambda ]u^{\beta } [\lambda ],
\]
the non-affinely parametrized geodesic Lagrangian
\[
|g_{\alpha \beta } u^{\alpha } [s]u^{\beta } [s]|^{1/2},
\]
List of affinely parametrized geodesic equations
\[
\frac{d}{\text{d$\lambda $}}u^{\alpha } [\lambda ] + \Gamma ^{\alpha }{}_{\mu \nu }u^{\mu } [\lambda ]u^{\nu } [\lambda ]
= 0,
\]
and a List of non-affinely parametrized geodesic equations
\[
\frac{d}{\text{ds}}u^{\alpha } [s] + \Gamma ^{\alpha }{}_{\mu \nu }u^{\mu } [s]u^{\nu } [s]= u^{\alpha } [s] \frac{d}{\text{ds}}\text{ln}\left|g_{\alpha \beta }u^{\alpha } [s]u^{\beta } [s]\right|^{\frac{1}{2}}.
\]
\sizeincludecode{1}{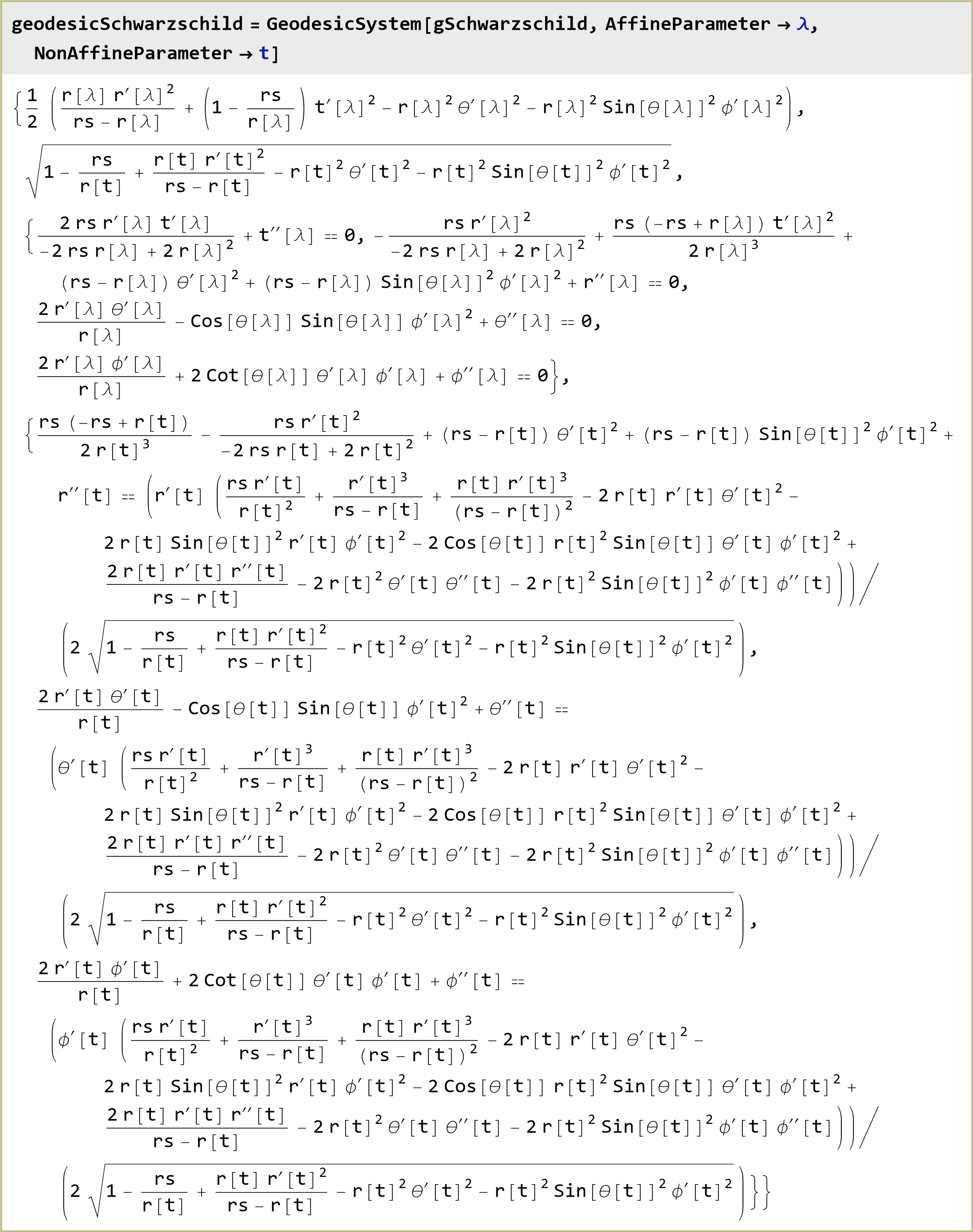}
Due to the spherically symmetric nature of the Schwarzschild geometry, we may restrict geodesic motion on the $\theta $ = $\pi $/2 equatorial plane and exploit conservation of angular momentum $\ell \equiv -\partial L/\partial \phi'$. Time translation symmetry implies conservation of energy $E \equiv \partial L/\partial t'$. We also note that, as $ r\rightarrow $ $\infty $, dt/d$\lambda $ tends to E; hence, for future pointing geodesics we may demand that E $>$ 0.
\includecode{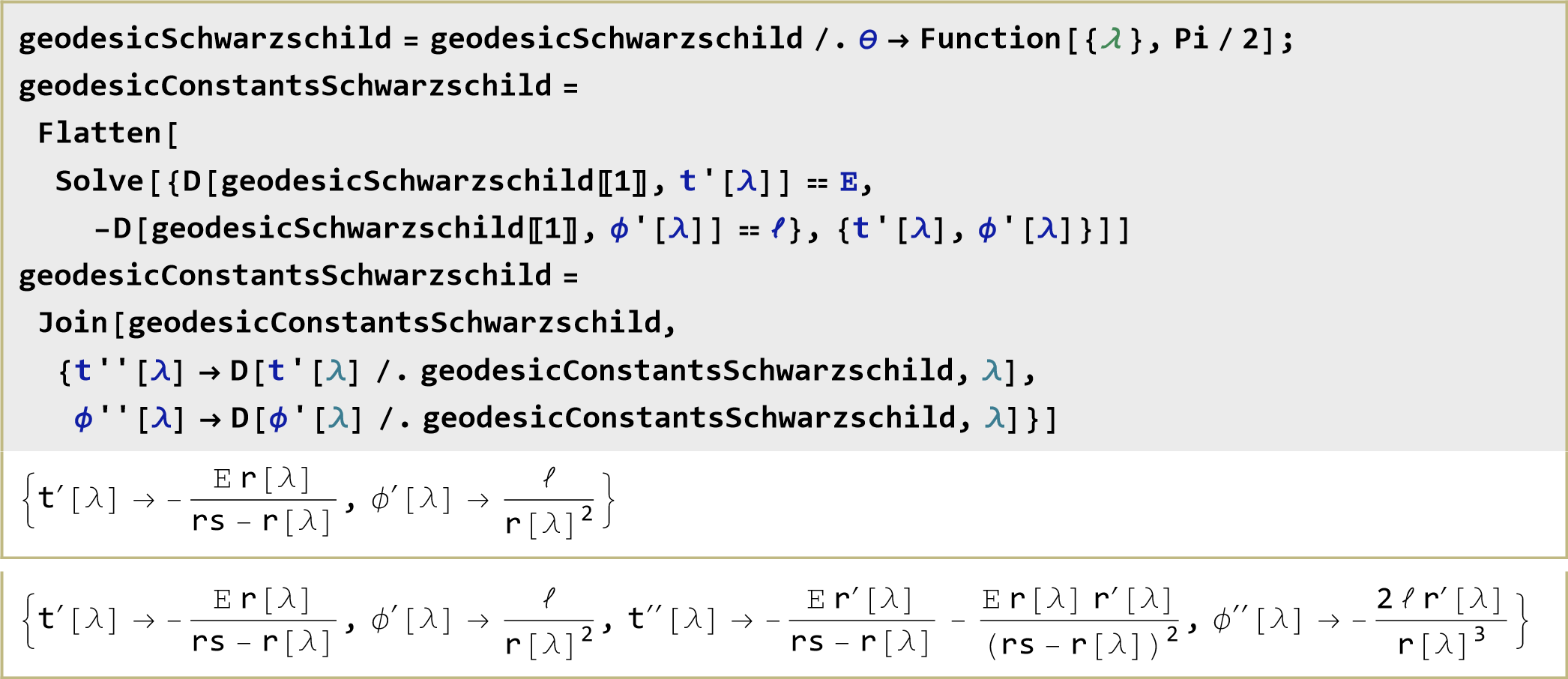}
These conservation laws reduce the set of four affinely parametrized geodesic equations to the radial one.
\includecode{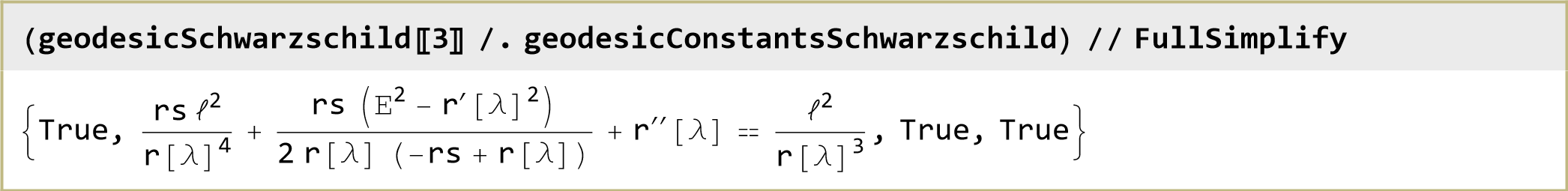}
\boldtitle{Schwarzschild Geodesics -- Massive particles} \qquad A massive particle undergoing geodesic motion would yield an affinely parametrized Lagrangian of 1/2 when we choose $\lambda $ to be its proper time. This allows us to derive a conservation of {``}kinetic \(r'[\lambda ]^2\)/2 plus potential energy{''} of sorts.
\includecode{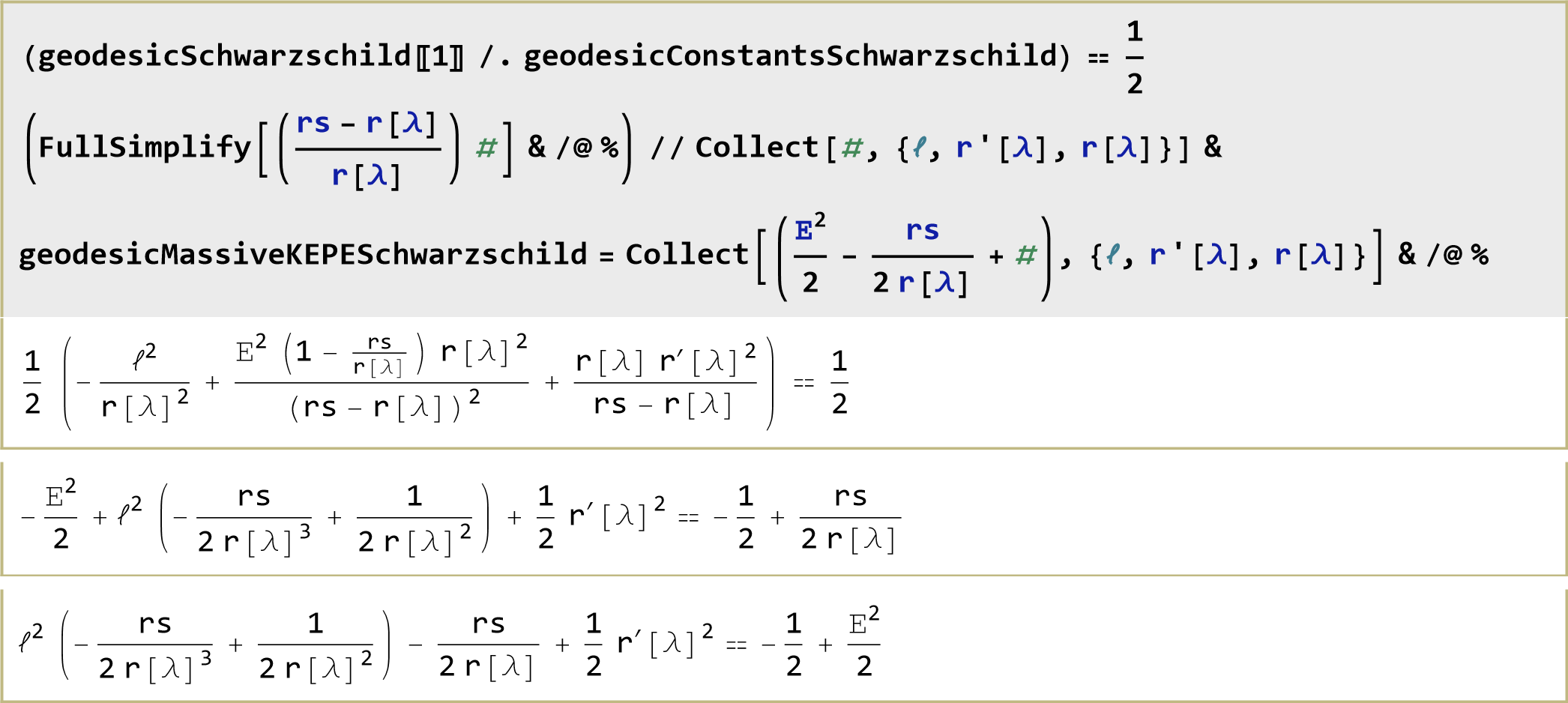}
For instance, we may extract the potential energy as a function of $\rho $ $\equiv $ \textit{r}/ \(r_s\). so as to compute its two turning points and study the qualitative behavior of the radial motion for a fixed E.
\includecode{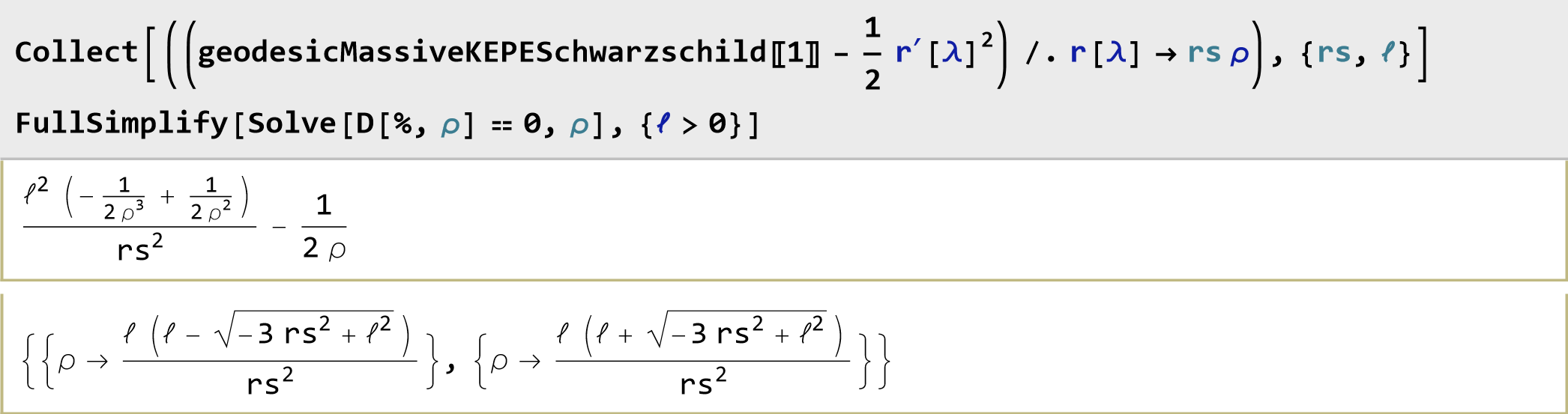}
\boldtitle{Schwarzschild Geodesics -- Null particles} \qquad High frequency gravitational and electromagnetic waves are usually modeled as particles moving along null geodesics. This means their affinely parametrized
Lagrangian is zero.
\includecode{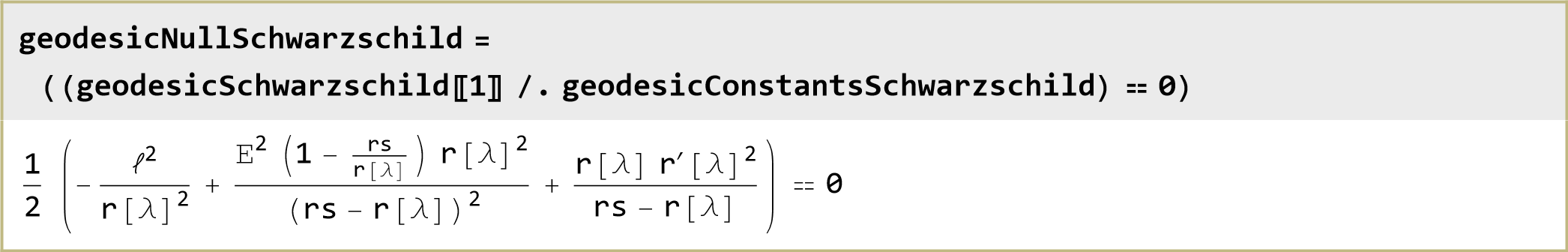}
This allows us to solve dr/d$\lambda $ in terms of \textit{r}.
\includecode{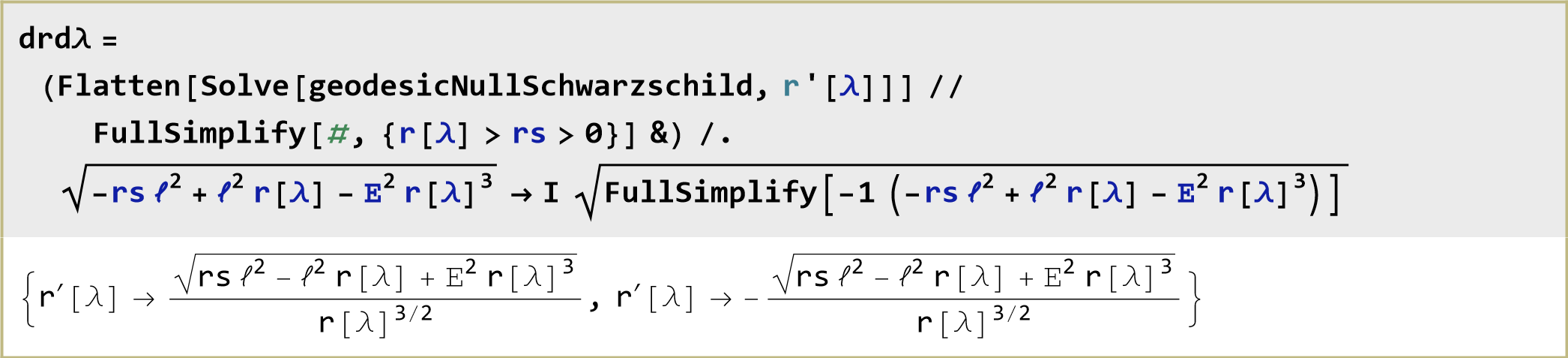}
Additionally, we may re-arrange this null Lagrangian condition into a conservation of kinetic \(\left.r'[\lambda ]^2\right/2\) plus potential energy.
\createlabel{geodesicNullKEPESchwarzschild}
\includecode{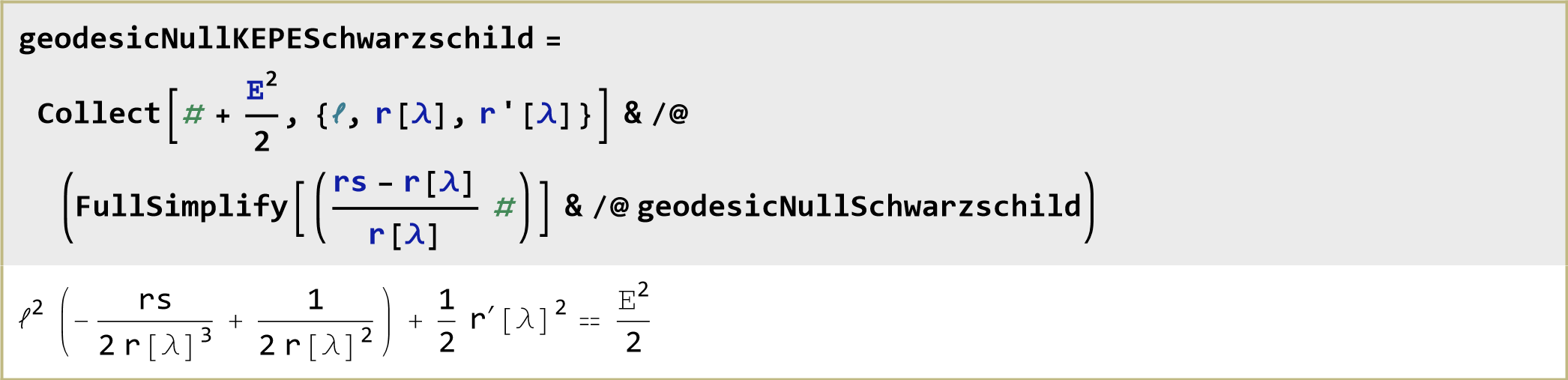}
Let us see that the potential not only has a single turning point, it is necessarily a local maximum.
\includecode{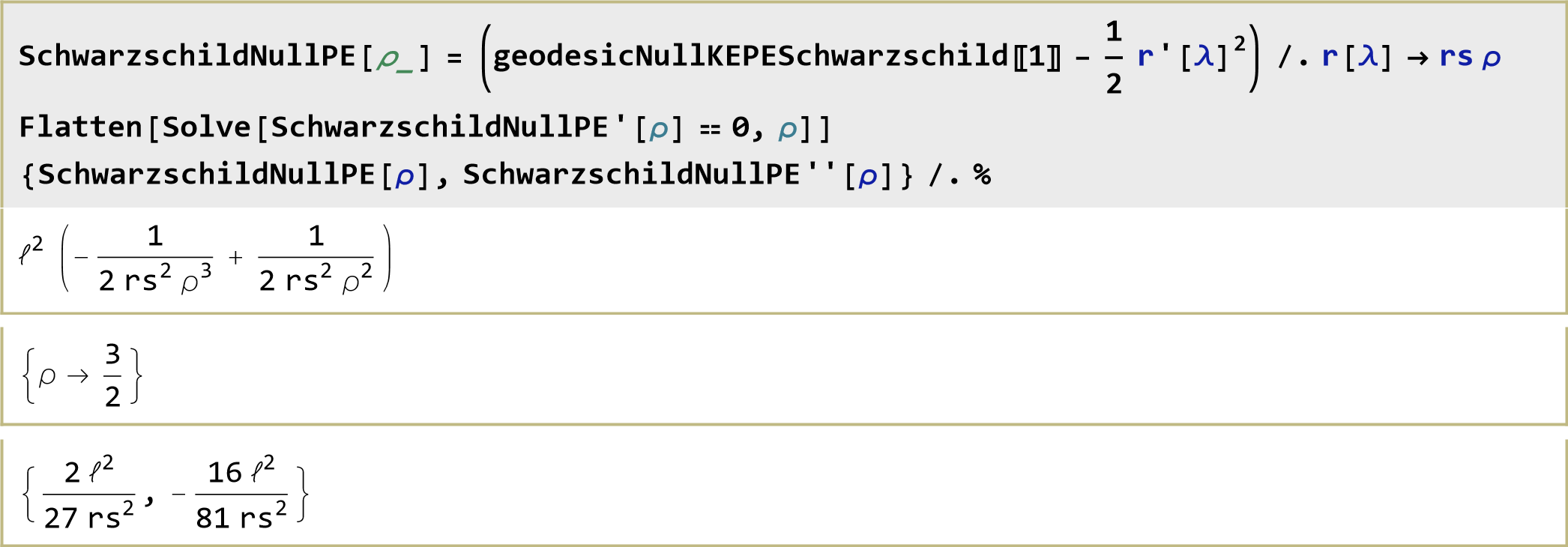}
The potential intercepts the $\rho $-axis only at $\rho $ = 1.
\includecode{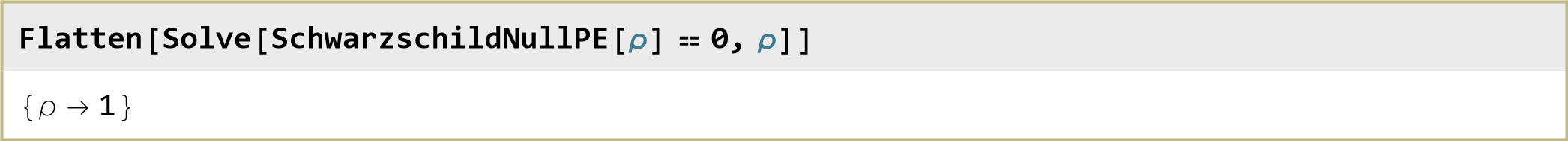}
Up to an overall scale ($\ell $ / \(r_s\)\()^2\), the null geodesic radial potential may be plotted as follows. The horizontal line is (2/27)($\ell$ / \(r_s\)\()^2\).
\createlabel{theplot}
\includecode{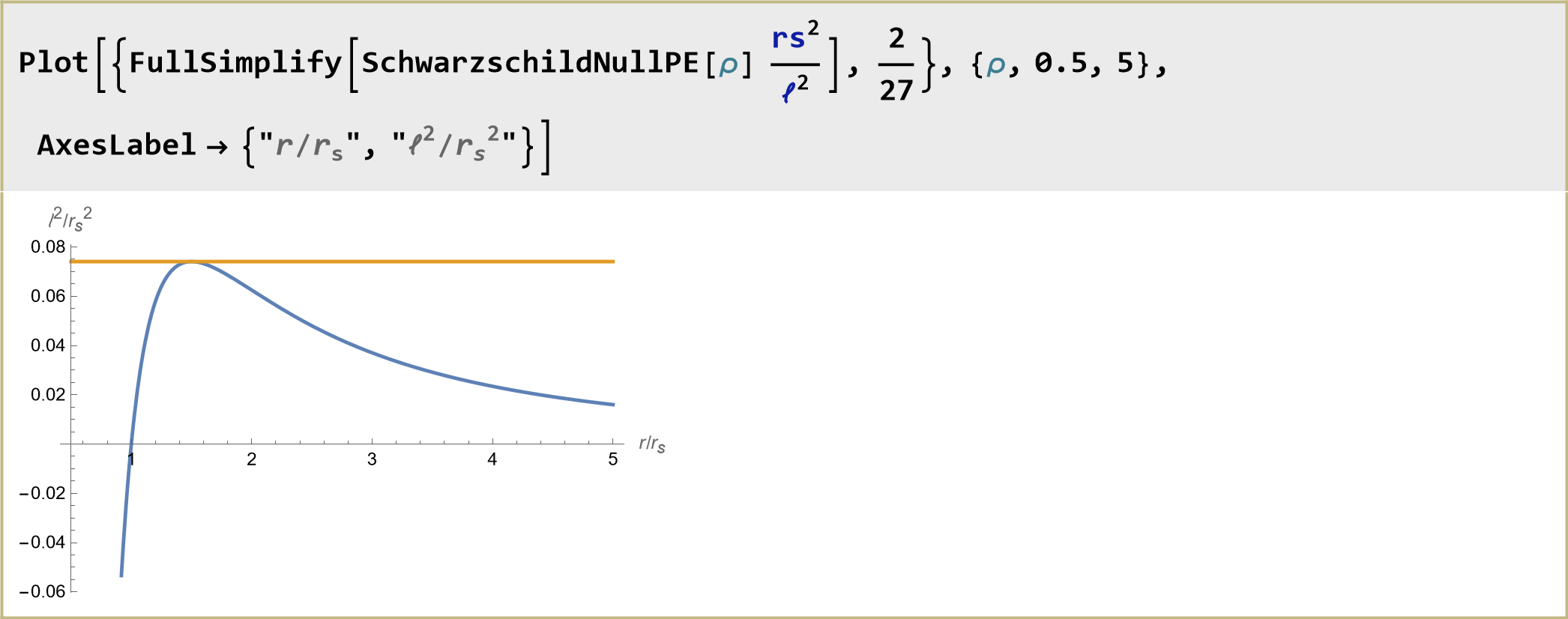}
\boldtitle{Schwarzschild Horizon}
For \textit{r} $>$ \(r_s\), { }$\langle $\(\partial _t\left|\partial _t\right.\)$\rangle $ = \(g_{00}>0\) and $\langle $\(\partial _r\left|\partial_r\right.\)$\rangle $ = \(g_{\text{rr}}<0\); whereas for \textit{r} $<$ \(r_s\), $\langle $\(\partial _t\left|\partial _t\right.\)$\rangle $ = \(g_{00}<0\) and $\langle $\(\partial _r\left|\partial _r\right.\)$\rangle $ = \(g_{\text{rr}}>0\). 
\includecode{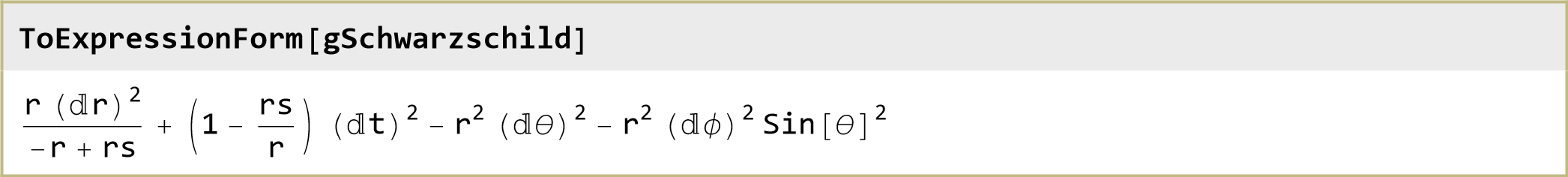}
This tells us the meaning of the coordinates (\textit{t}, \textit{r}) changes when crossing the \textit{r} = \(r_s\) surface. For \textit{r}
$>$ \(r_s\), \textit{t} is the timelike coordinate while \textit{r} is a spatial one; but for \textit{r} $<$ \(r_s\), \textit{r} becomes the
timelike coordinate while \textit{t} is now a spatial one. In particular, going forward in time when { }\textit{r} $<$ \(r_s\), means \textit{r} is \textit{decreasing} from \(r_s\) to \(0^+\).

We may refer back to both the\uselabel{geodesicNullKEPESchwarzschild}{conservation law}and the\uselabel{theplot}{plot}of the {``}potential{''} experienced by null geodesics. As long as \(E^2\)/2 is large enough, it would appear as though the \textit{r}[$\lambda $] trajectory can decrease from large positive values all the way to \(0^+\); or increase from very small values, less than \(r_s\), to very large ones. However, while \textit{r} can decrease past \(r_s\) with time moving forward; to go climb out of the \textit{r} $<$ \(r_s\) portion of the {``}potential well{''} would correspond to traveling back in time. This is the primary obstacle preventing all future-pointing geodesics -- even null ones -- from getting out of the \textit{r} = \(r_s\) surface from \textit{r} $<$ \(r_s\) to \textit{r} $>$ \(r_s\). This \textit{r} = \(r_s\) surface is dubbed the horizon.

\boldtitle{Schwarzschild -- Radial Affinely Parametrized Null Geodesic} \qquad Notice the\uselabel{theplot}{potential}is strictly non-negative for \textit{r$\, $ $\geq $} \(r_s\). Because kinetic plus potential = \(\left.E^2\right/2 \geq  0\)--it is also strictly non-negative -- that means, for a given pair of (E, $\ell $), kinetic energy is maximized by setting angular momentum to zero. In turn, this leads us to the solution to the affinely parametrized radial null geodesics.
\includecode{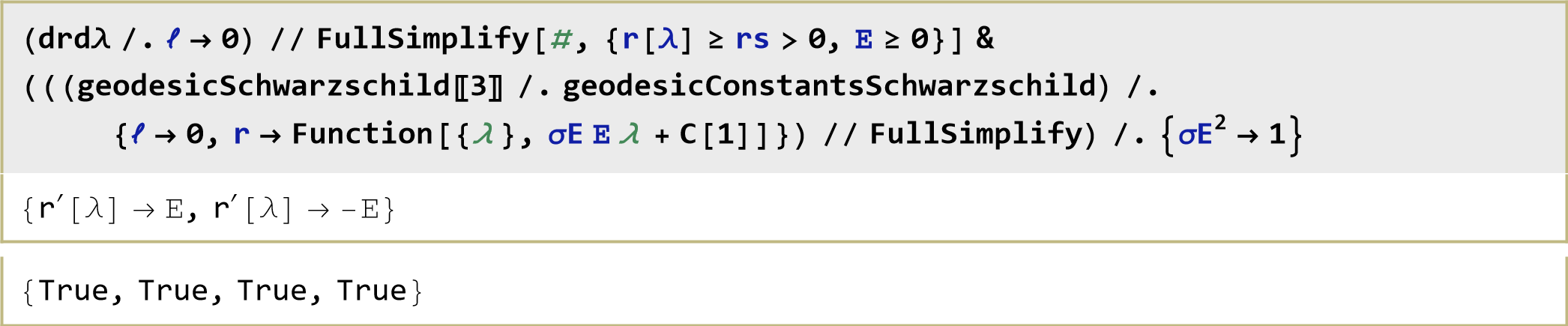}
\boldtitle{Schwarzschild -- Radial Null Geodesic Vector Field} \qquad Let us solve for the radial null geodesic vector field \(q^{\mu }\) by first superposing \(\partial _t\) and \(\partial _r\).
\includecode{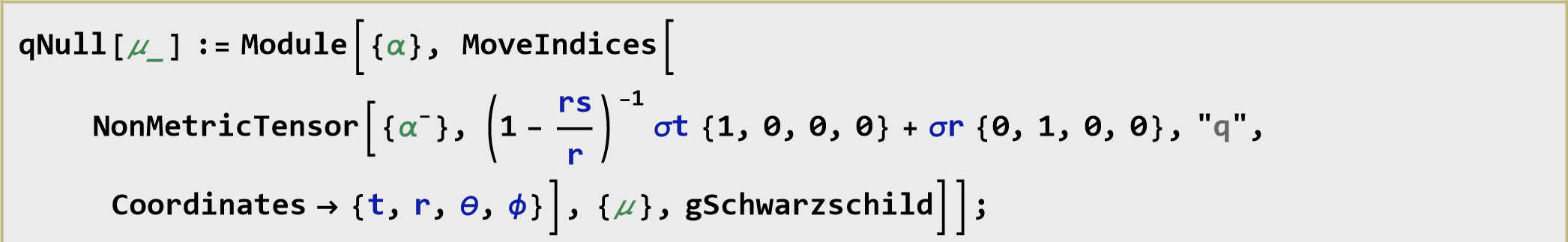}
For \textit{q} to be null, we see that \(\text{$\sigma $t}^2\) = \(\text{$\sigma $r}^2\). 
\includecode{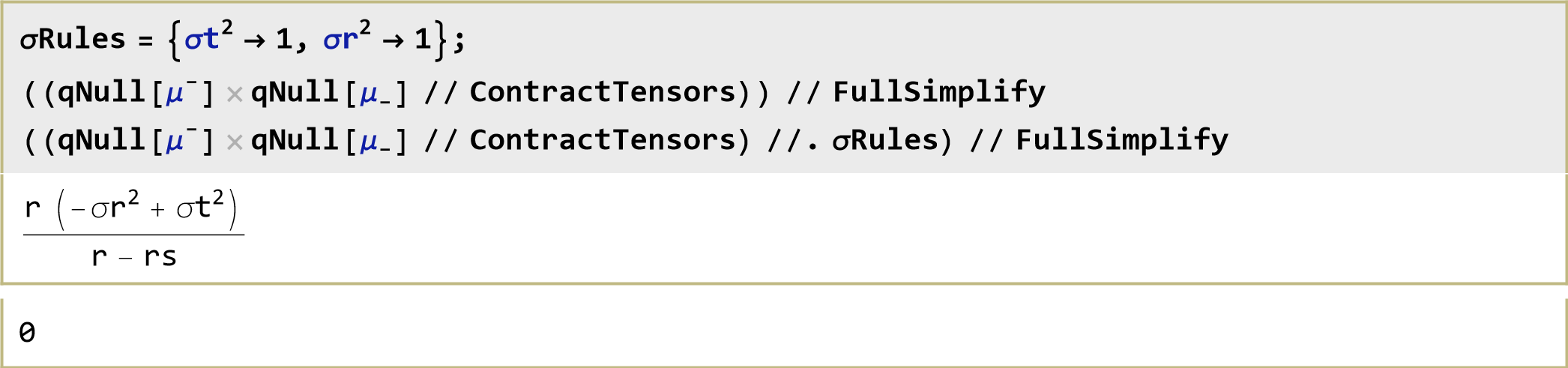}
To justify why the $\sigma ${'}s are constants, let us verify that the result also solves the geodesic equation \(q^{\sigma }\)\(\nabla _{\sigma}\)\(q^{\alpha }\) = 0.
\includecode{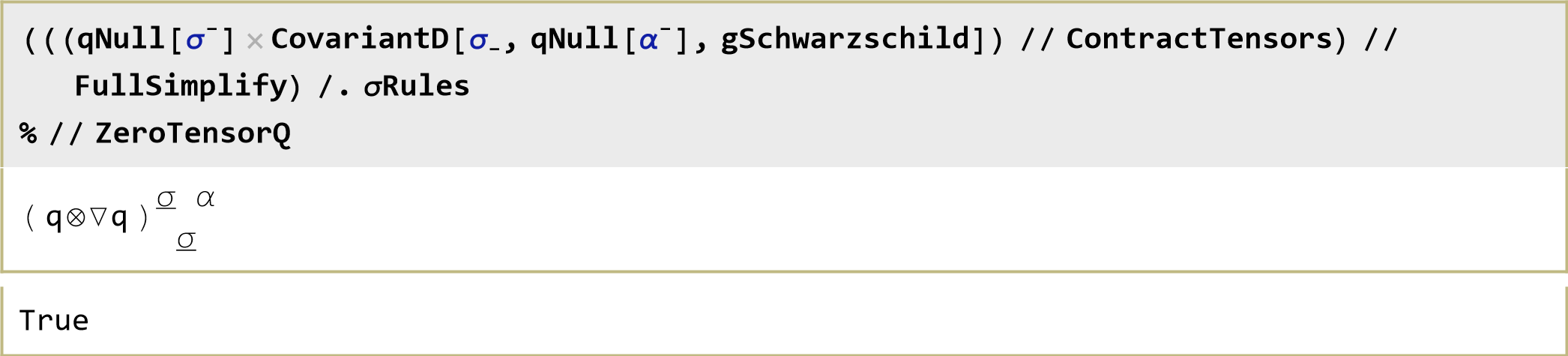}
These radial null geodesics turn out to be the gradient of some scalar \textit{u} by the Poincare lemma, because the exterior derivative of their corresponding 1-form is zero. 
\includecode{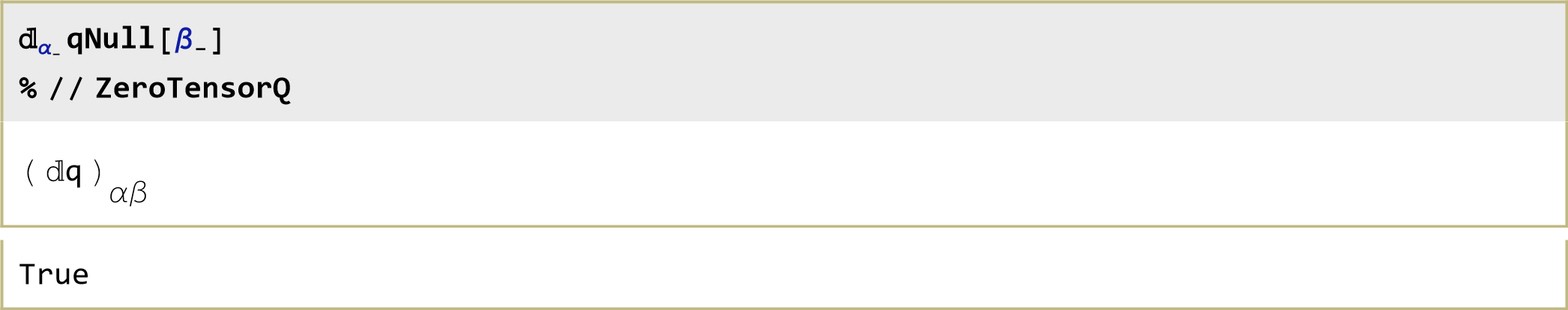}
To solve for \textit{u} such that \((du)_{\beta }\) = \(q_{\beta }\), we invoke PotentialForm.
\includecode{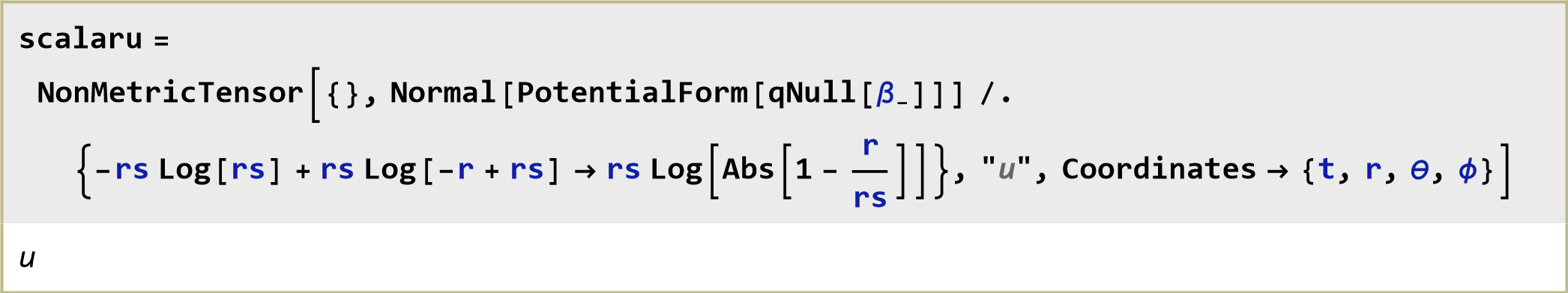}
This leads us to null coordinates
\[
x^{\pm } \equiv  t\text{  }\pm  \left(r + r_s \text{Log}\left|1-\frac{r}{\text{rs}}\right|\right)
\]
for the $(t,r)-$sub metric of the Schwarzschild geometry; specifically,
\[
(1 - r_s/r) \dd x^+ \dd x^- = (1 - r_s/r) (\dd t)^2 - (\dd r)^2/(1 - r_s/r).
\]
ExteriorD[$\tau \_$Tensor] can also be entered as d$\tau $, whereby the extra index would be generated by \TC.
\includecode{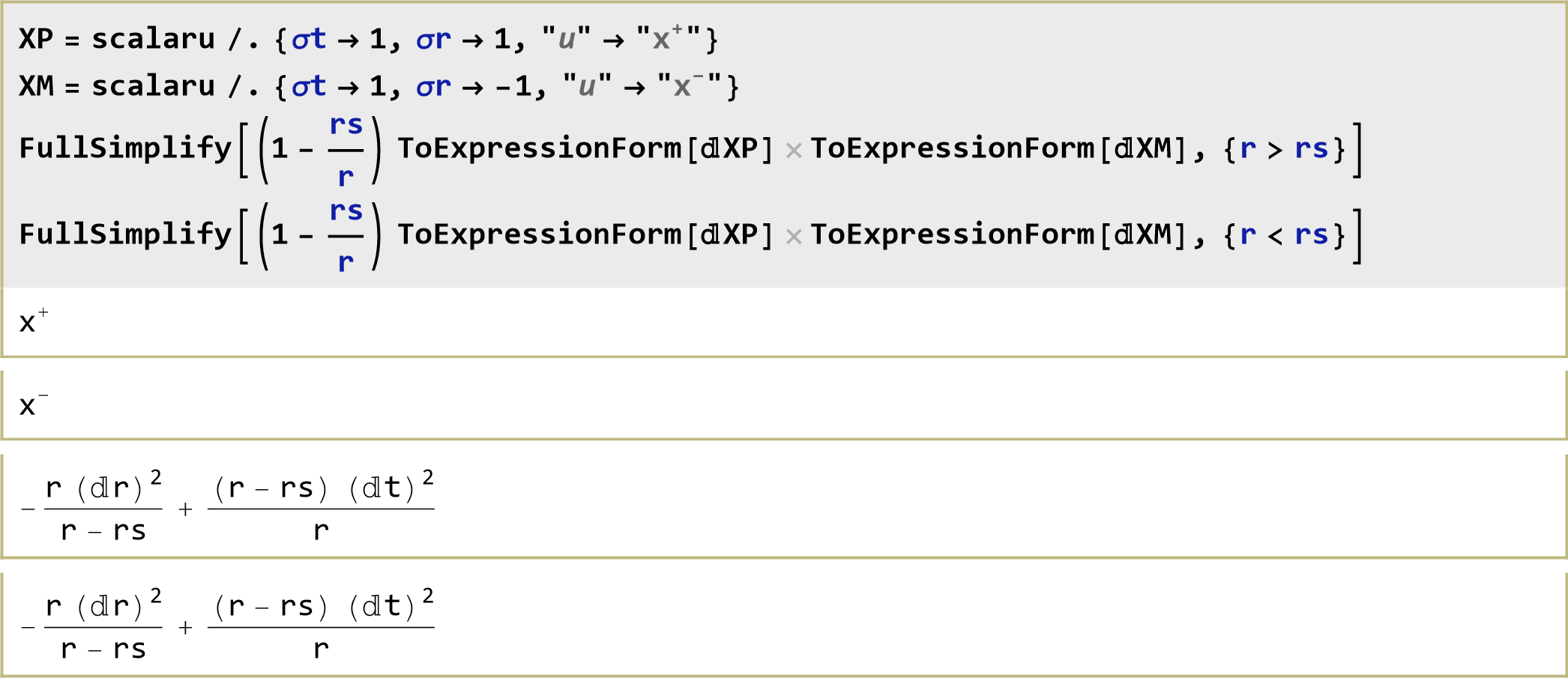}
\boldtitle{Weak Field Light Deflection} \qquad Consider a high frequency photon approaching the Sun from infinity, making its closest approach at \textit{r }= \textit{b}, before flying off to infinity again. Let us now use the above geodesic dynamics to solve for its net deflection angle. The Sun is a weak gravity source in that \(r_s\)/\textit{r }$\ll $ 1 -- the Schwarzschild radius \(r_s\) of the Sun is $\sim $3 km, much smaller than its physical radius of $7 \times 10^5$ km (see below) -- allowing us to expand in powers of \(r_s\), the shortest length scale of the problem at hand.

But, first, we note that d$\phi $/dr is related to dr/d$\lambda $ and \textit{r} itself via the following relation:
\includecode{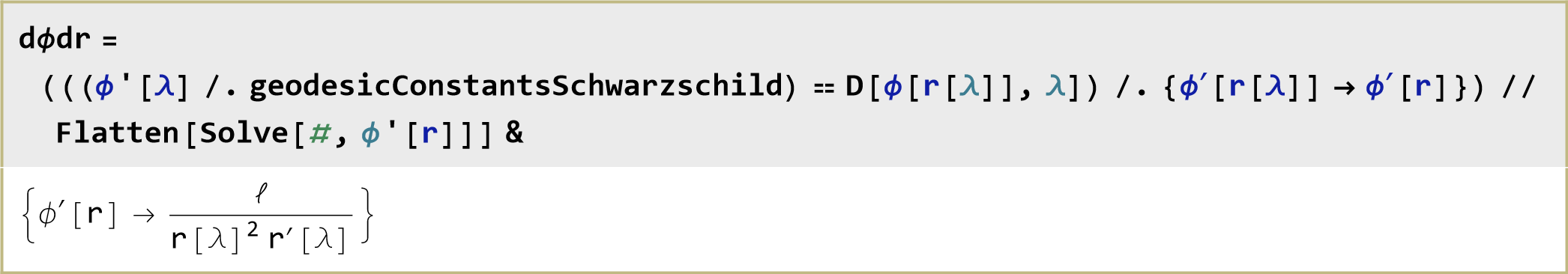}
If we focus on the incoming portion of the trajectory, $r'[\lambda] \leq 0$ because the radius is decreasing from infinity to \textit{b}.
Imposing the null Lagrangian condition to replace $r'[\lambda]$ then yields the following.
\includecode{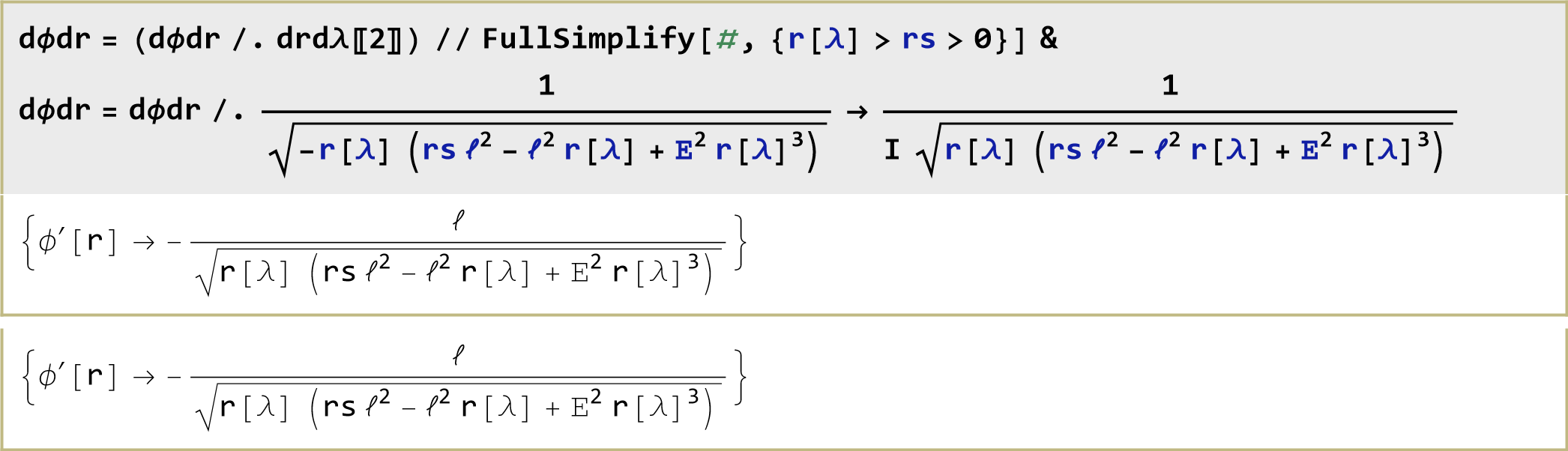}
This means we may integrate this expression for d$\phi $/dr with respect to $r$ to obtain $\Delta \phi $, the total change in azimuthal angle.

Now, since \textit{r }= \textit{b} is the closest approach, dr/d$\lambda $ = 0 there. This allows us to solve $\ell $ for \textit{b}. Following
which, we employ it in d$\phi $/dr and expand in powers of \(r_s\).
\includecode{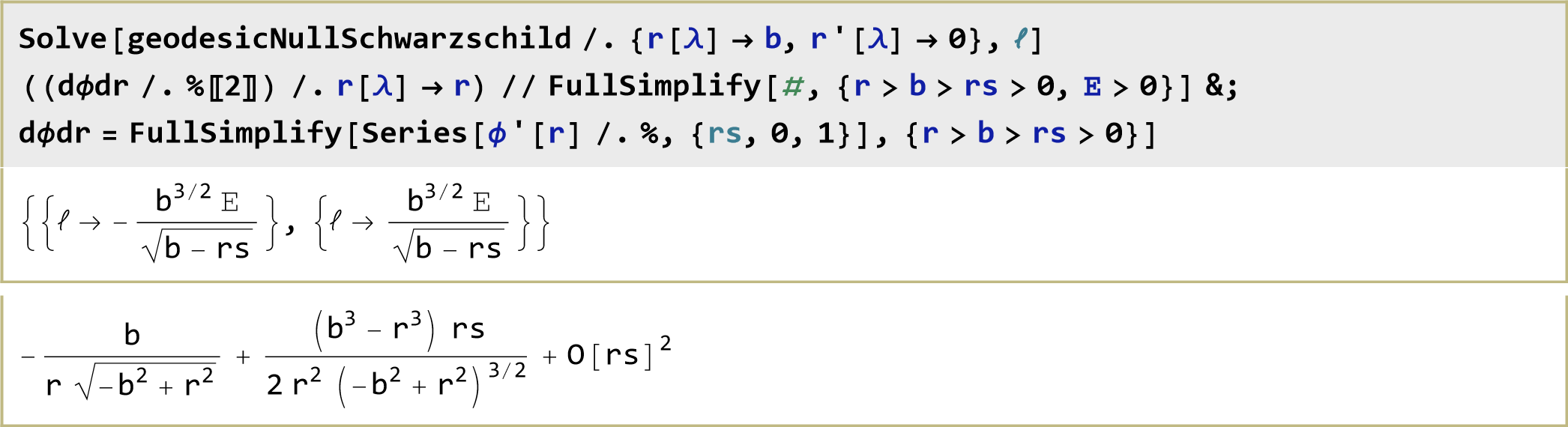}
Integrating over \textit{r} $\in $ $\{\infty $, \(b\)$\}$, describing the photon coming in from infinity and reaching \textit{r} = \textit{b}.
\includecode{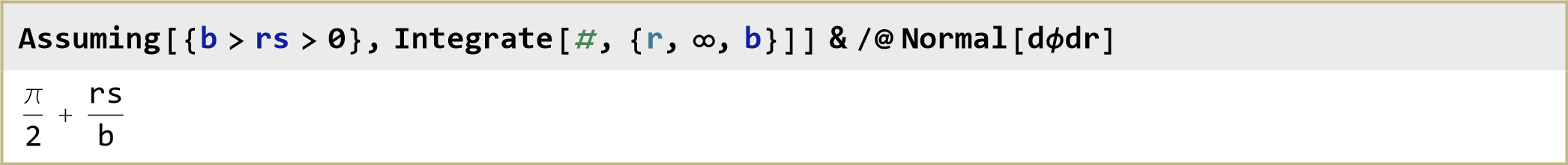}
The total change in azimuthal angle is therefore twice of this -- namely, the sum of the contribution from both the incoming and outgoing legs:
$\Delta \phi $ = $\pi $ + 2 \(\left.r_s\right/b.\) Had there been no gravity at all, the photon would sweep out a straight line and $\Delta \phi
$ = $\pi $. Therefore, the deflection angle is 2\(\left.r_s\right/b.\) Recalling \(r_s\equiv 2G_N M_{\text{Sun}}\) and setting \textit{b} to be
the physical radius of the Sun -- see below -- we obtain a value of 1.75 arc seconds.
\includecode{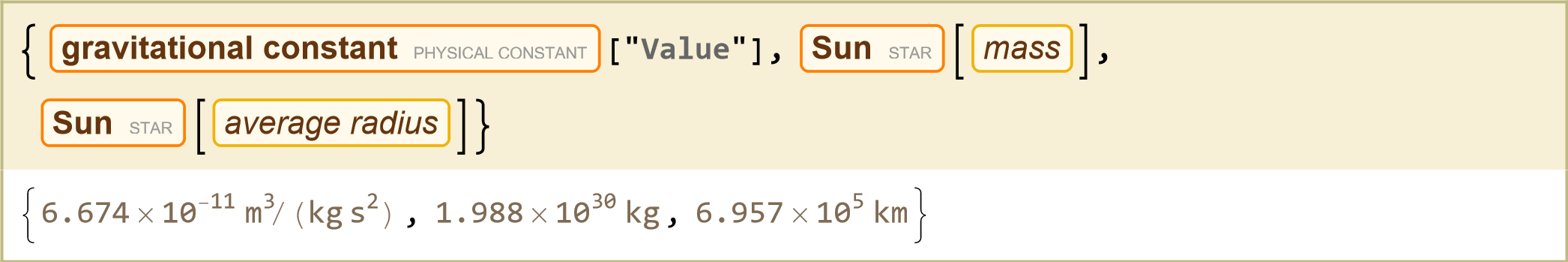}
We use speed of light = 299,792,458 meters/second = 1 to convert 2\(\left.r_s\right/b\) $\to $ 4 \(G_N\) \(M_{\text{Sun}}/R_{\text{radius}}\) to
arc-seconds.
\includecode{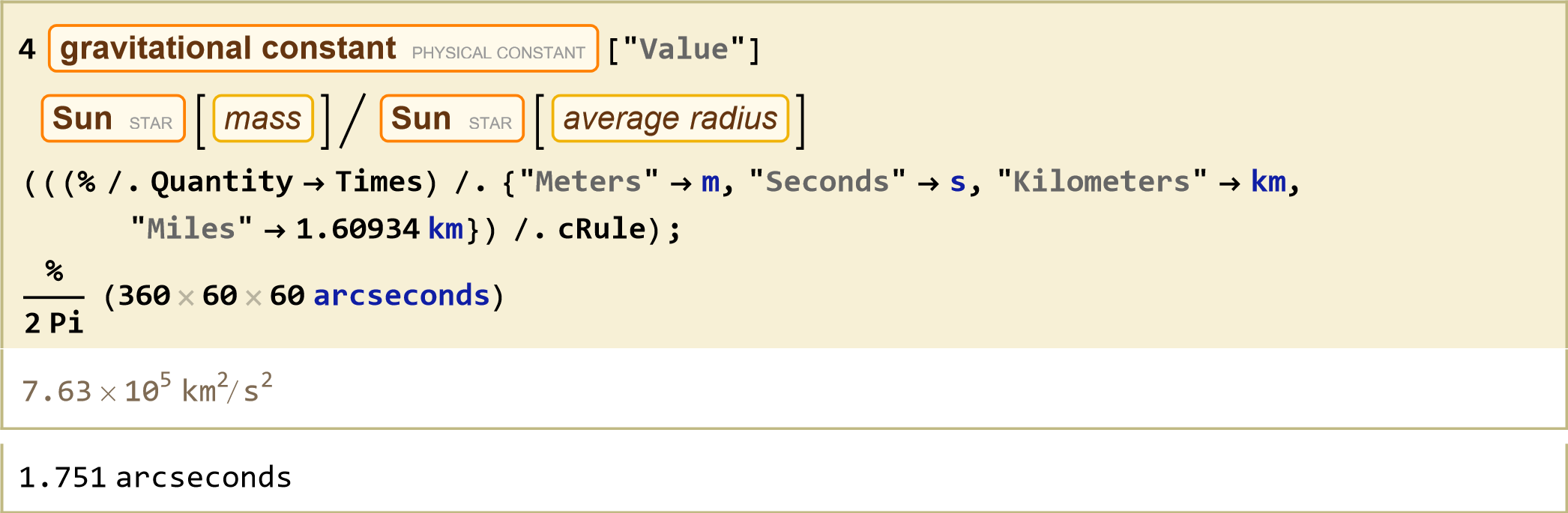}

\subsection{Kerr}

The Kerr metric describes a rotating black hole satisfying the vacuum Einstein{'}s equations \(G_{\alpha \beta }=0\) with zero cosmological constant.
\includecode{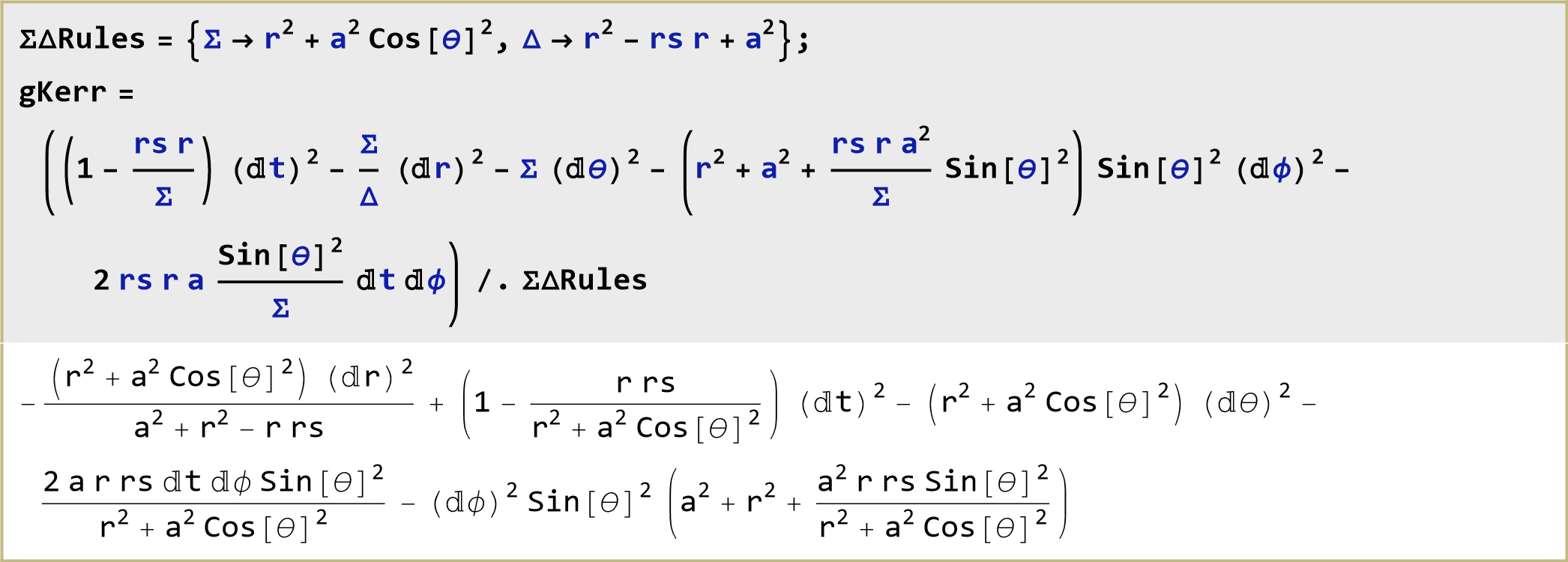}
It turns out that Computer Algebra does not handle trigonometric functions efficiently, so to verify Einstein{'}s equations we shall switch from
$\theta $ to $\chi $ $\equiv $ cos $\theta $.
\includecode{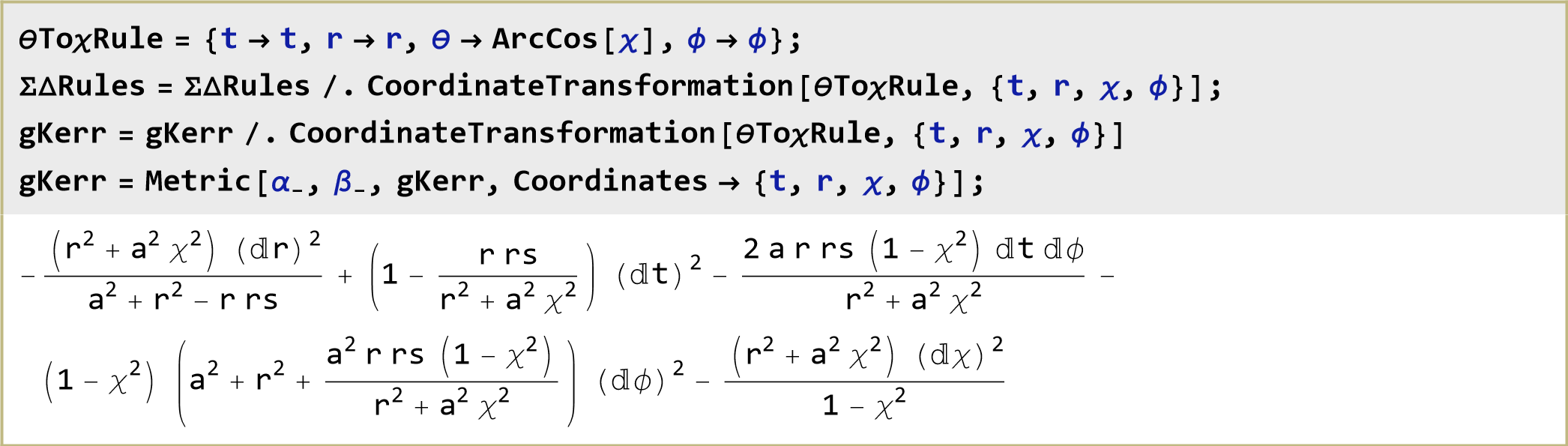}
The Einstein tensor is indeed zero.
\includecode{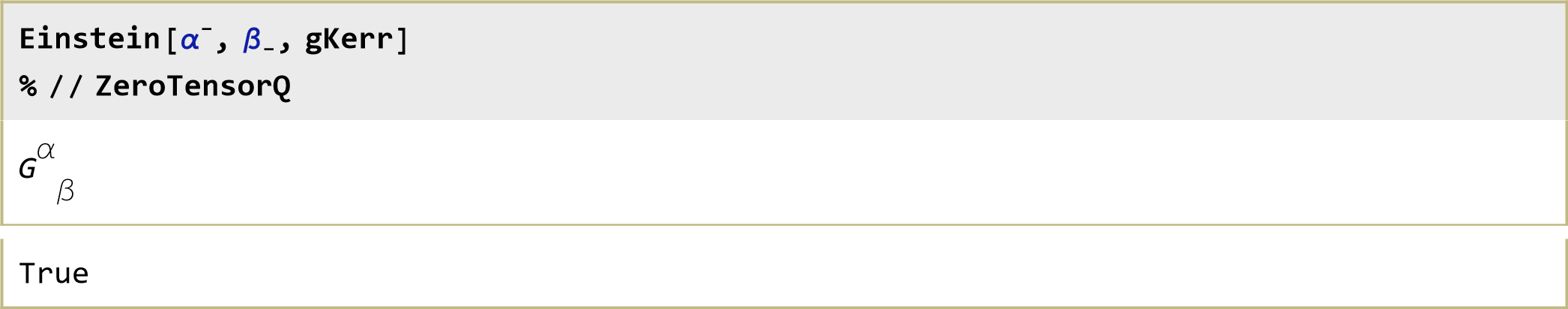}
\boldtitle{Kerr Isometries} \qquad The Kerr metric enjoys time \textit{t}- and rotation $\phi $-translation symmetry. The associated Killing vectors are as follows.
\includecode{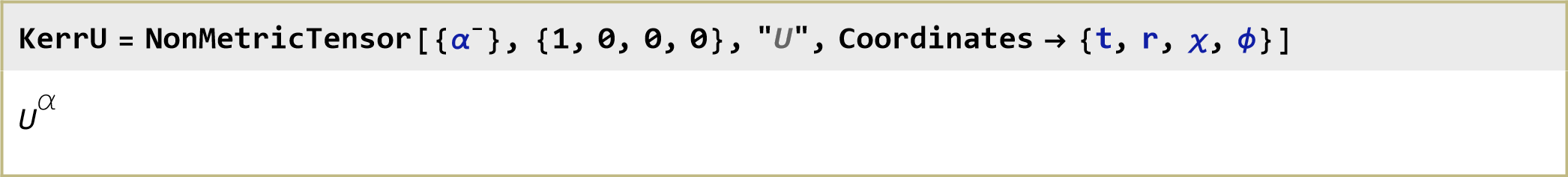}
\includecode{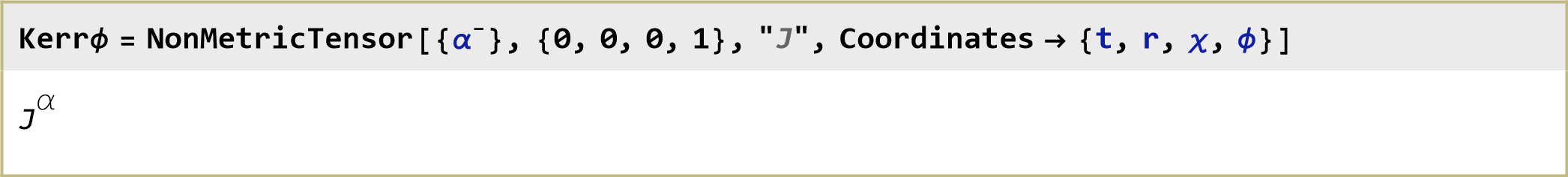}
Let us verify their Killing equations.
\includecode{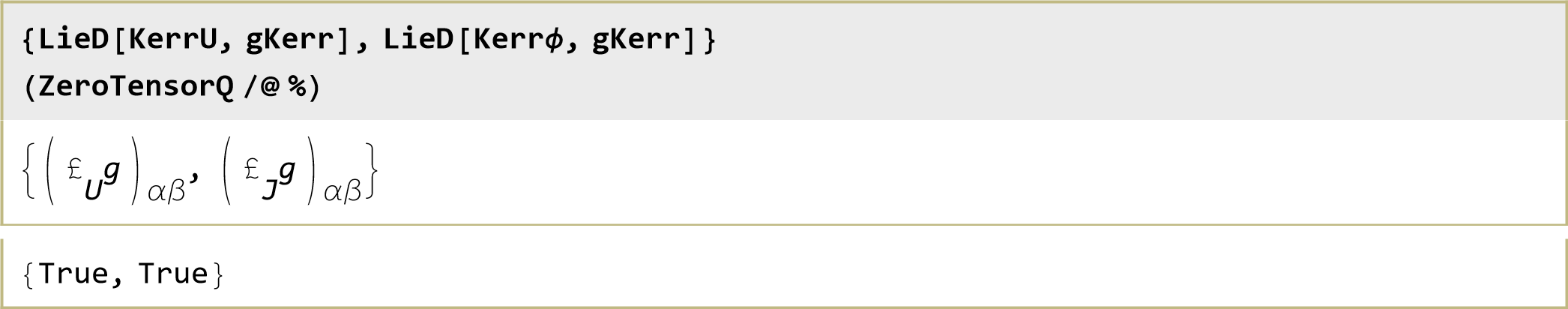}
What is less obvious is the existence of a Killing tensor \(K_{\mu \nu }\) defined by \(\nabla _{\{\alpha }K_{\mu \nu \}}\) = 0 -- if \(q^{\mu
}\) were tangent to some geodesic worldline, obeying \(q^{\sigma }\nabla _{\sigma }q^{\alpha }=0\), we may then construct the conserved quantity
\(K_{\mu \nu }q^{\mu }q^{\nu }\). 

First we define the two vectors $\ell $ and \textit{n}.
\includecode{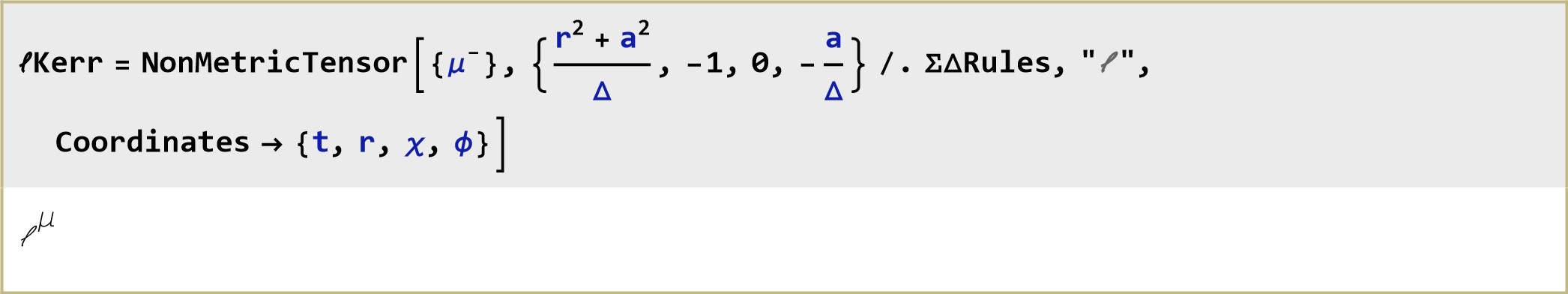}
\includecode{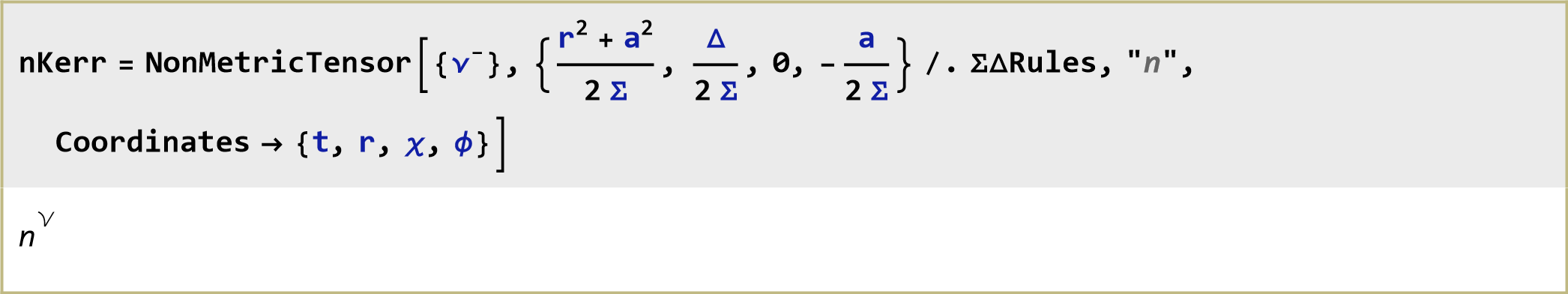}
The Killing tensor itself is then the following object.
\includecode{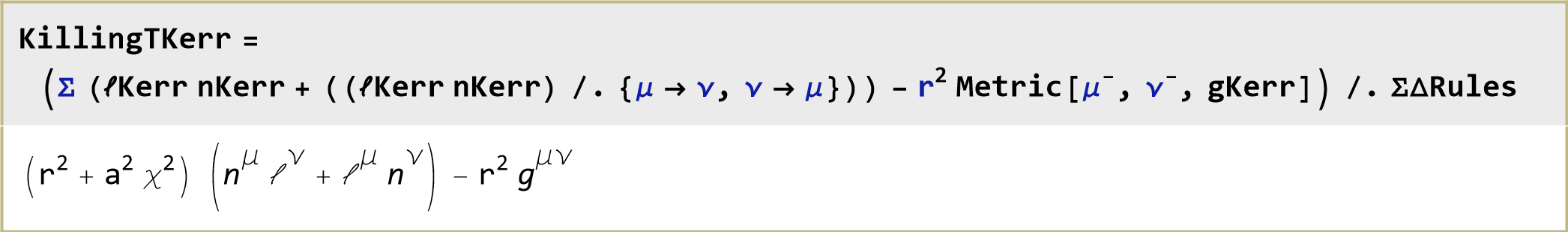}
The key property is that the fully symmetrized first covariant derivative of the Killing tensor is zero.
\includecode{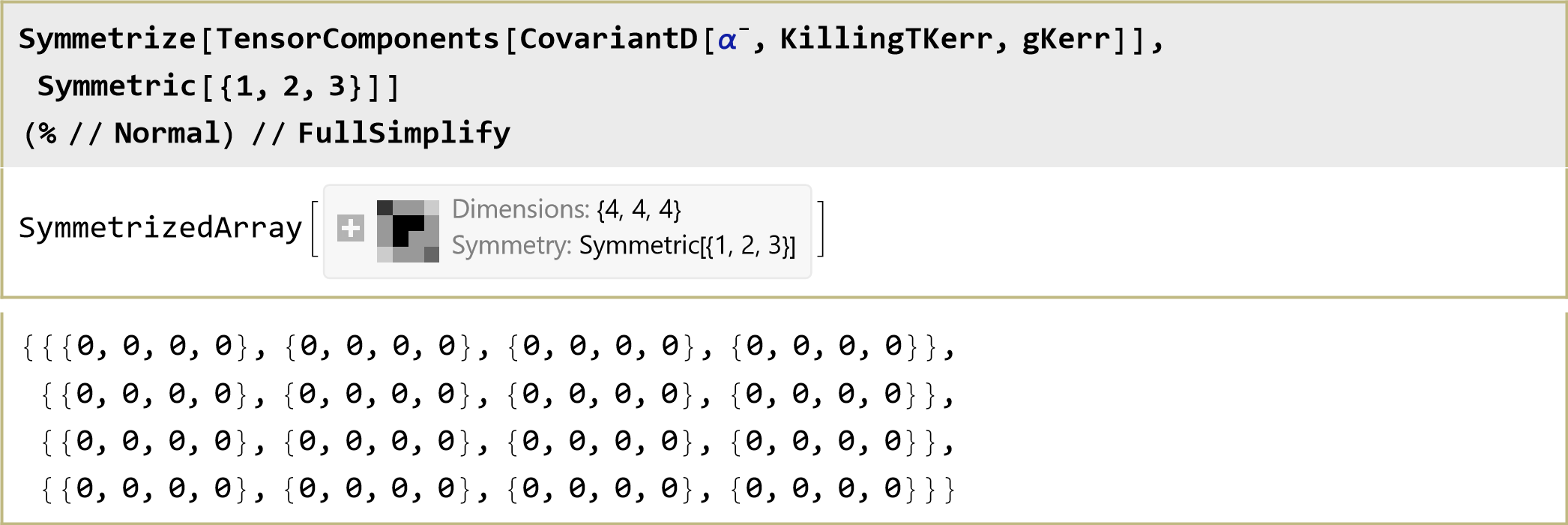}
\boldtitle{Kerr -- Geometric Curvature} \qquad The {``}square{''} of the Kerr Riemann tensor reads as follows. Geometric curvature blows up on a {``}ring{''} defined by \textit{r} = 0 and $\chi$ = cos $\theta $ = 0 (i.e., $\theta $ = $\pi $/2).
\includecode{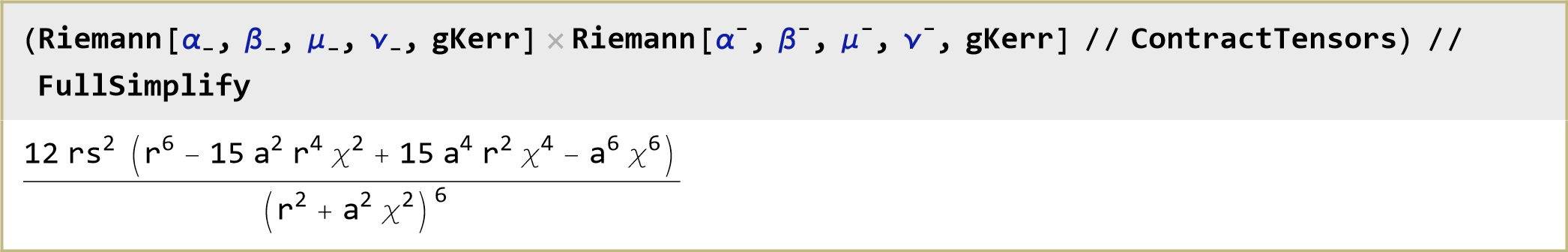}
Since Kerr is a vacuum spacetime, its Weyl and Riemann tensors are the same.
\includecode{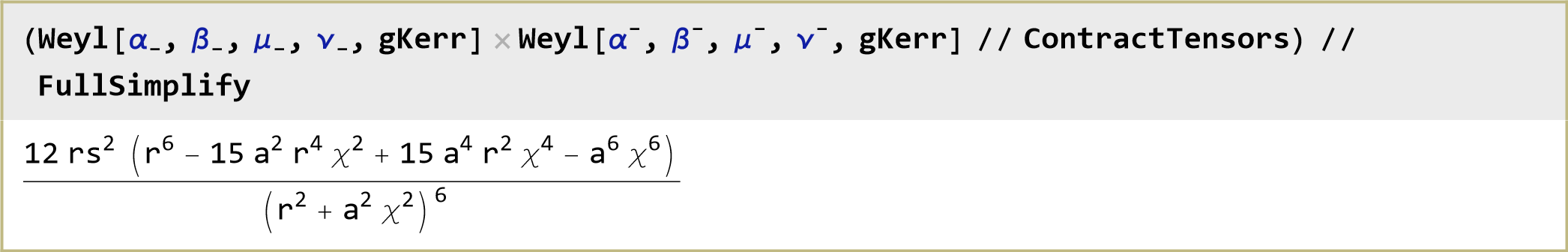}
We may also examine the individual Riemann tensor components in an orthonormal basis. We shall use the following inverse orthonormal frame fields; notice they are no longer simply the straightforward {``}square root{''} of the metric, because the latter is now non-diagonal -- in particular, there is mixing between the (\textit{t}, $\phi $)-components.
\includecode{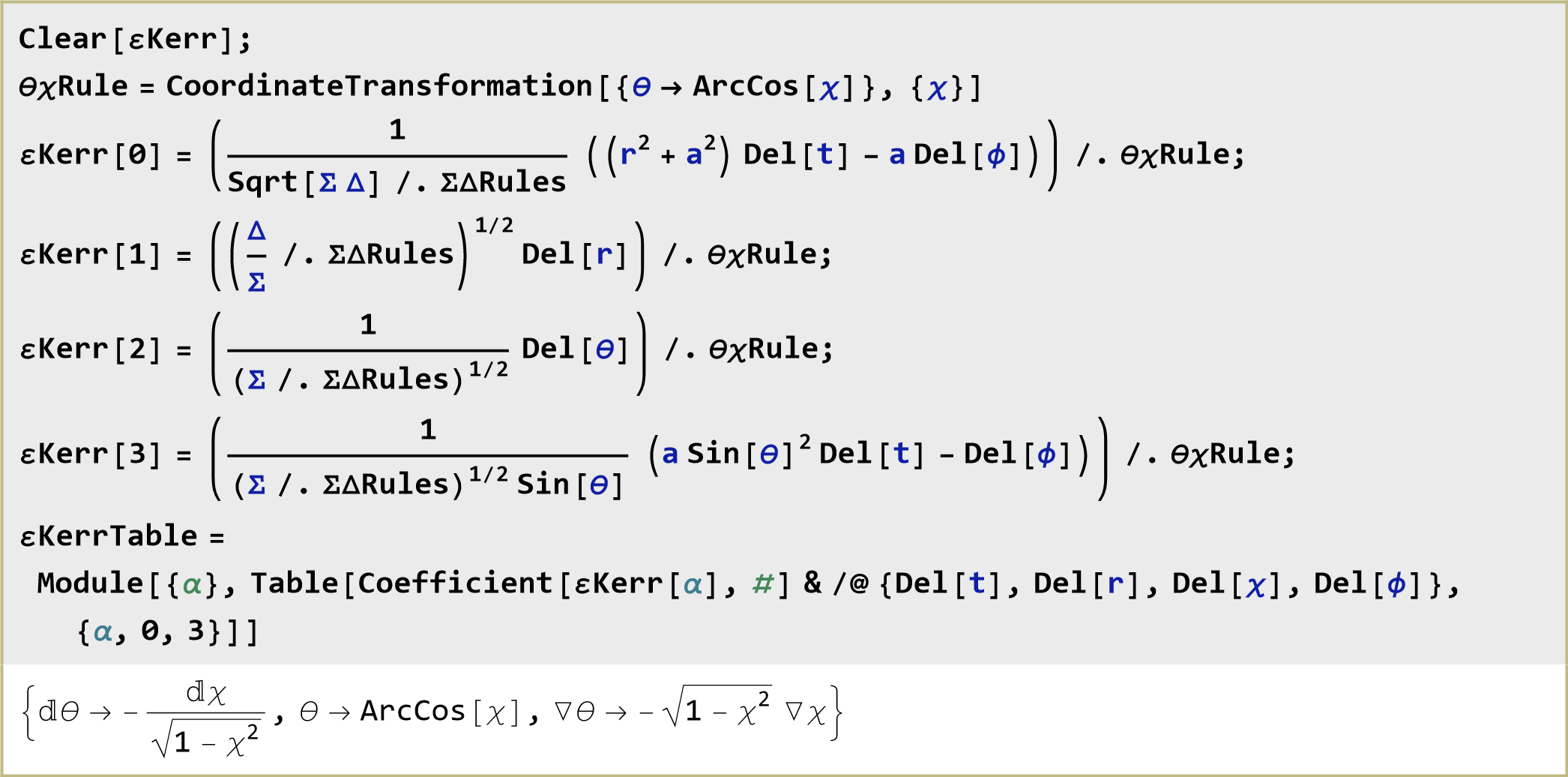}
\sizeincludecode{0.8}{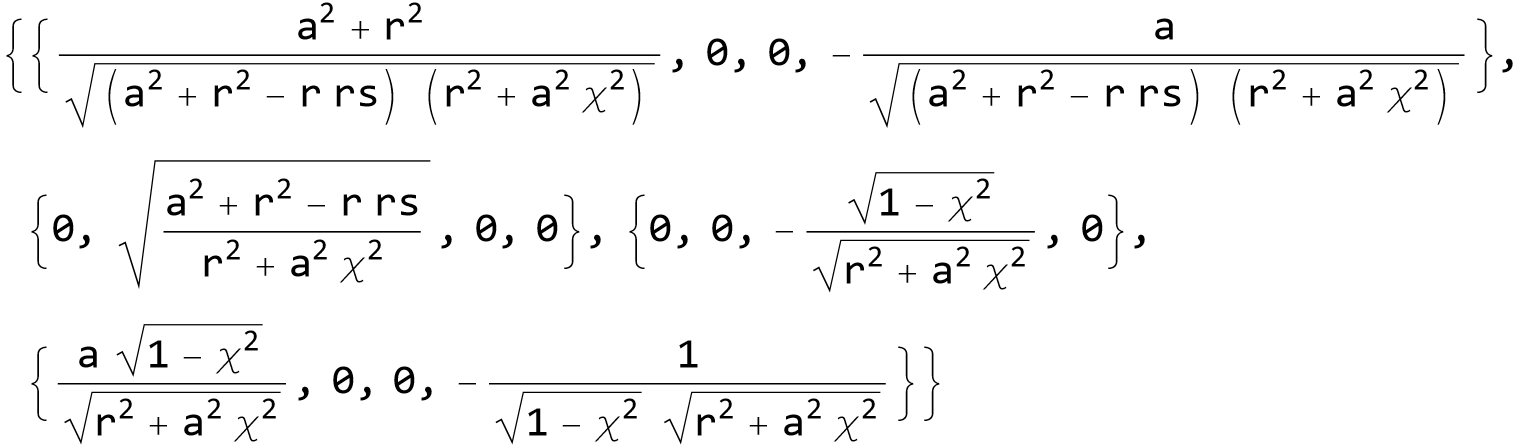}
We append them to gKerr using OrthonormalFrameField. Given a set of orthonormal frame fields (or, its inverse) and a FlatMetric, \TC\ would
automatically compute the inverse orthonormal frame fields (or, the orthonormal frame fields) before storing both sets of frame fields into the metric Tensor object.
\includecode{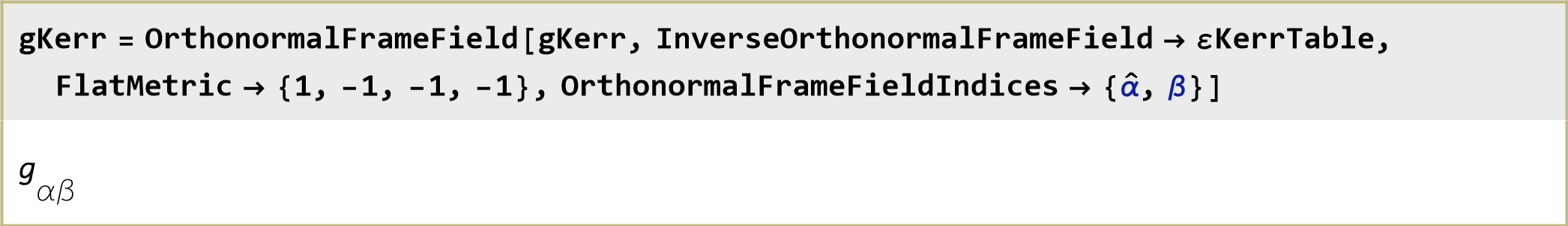}
The inverse orthonormal frame fields should recover the inverse metric in the following manner.
\includecode{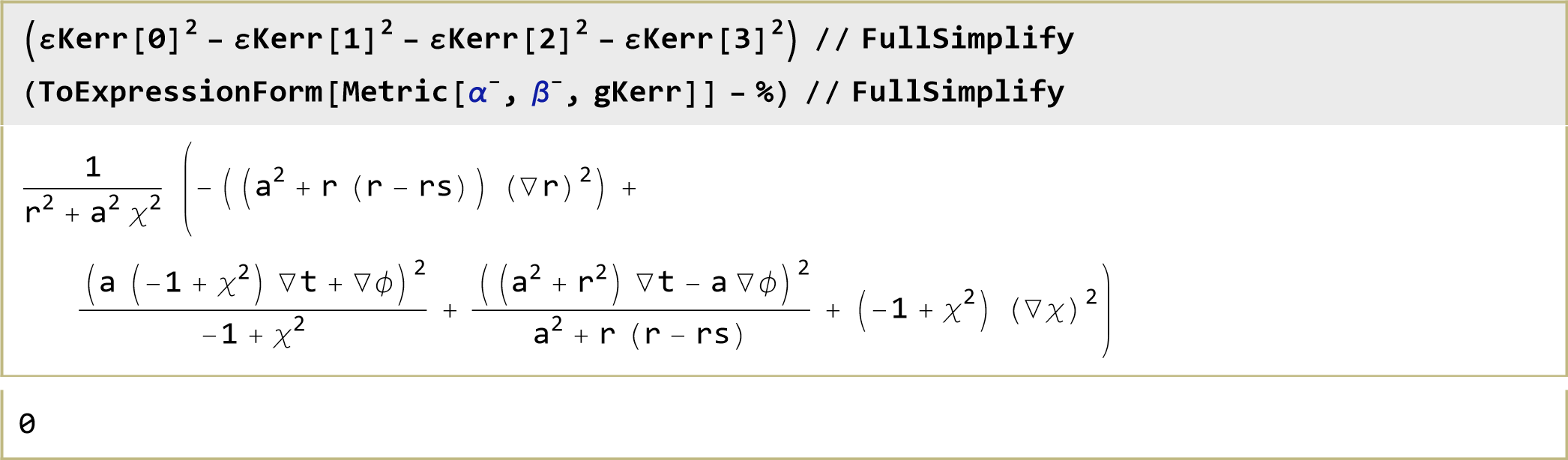}
OrthonormalFrameFieldQ allows us to carry out this verification for both the orthonormal frame fields and their inverses; that they do indeed recover, respectively, the metric and its inverse:
\begin{align*}
	\eta _{\hat{\alpha }\hat{\beta }}\varepsilon ^{\hat{\alpha }}{}_{\mu } \varepsilon ^{\hat{\beta }}{}_{\nu } &= g_{\mu\nu }\quad \\
	\eta ^{\hat{\alpha }\hat{\beta }}\varepsilon_{\hat{\alpha }}{}^{\mu } \varepsilon_{\hat{\beta }}{}^{\nu } &= g^{\mu\nu }
\end{align*}
If both of these two equations return True, OrthonormalFrameFieldQ will return a single True. If one or both of them does not return True, a List of these two sets of equations would be displayed.
\includecode{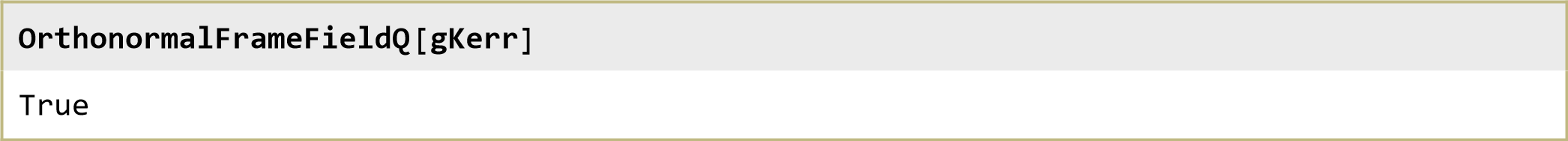}
Let us now examine the Kerr \(R_{\hat{0}\hat{i}\hat{0}\hat{j}}\) components.
\includecode{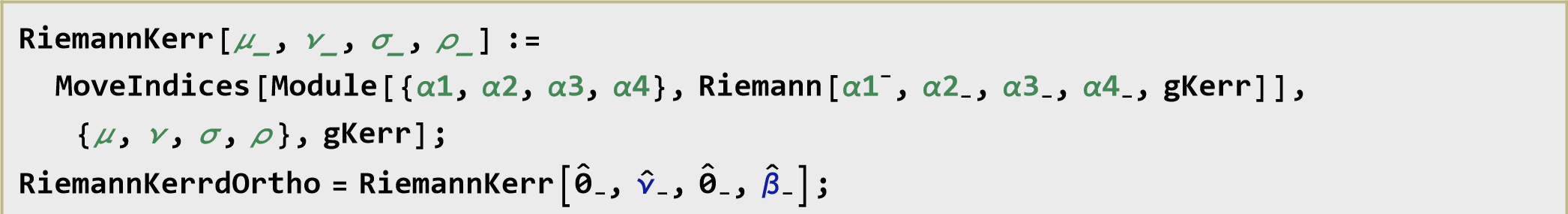}
\begin{figure}[H]
	\includegraphics[width=\textwidth]{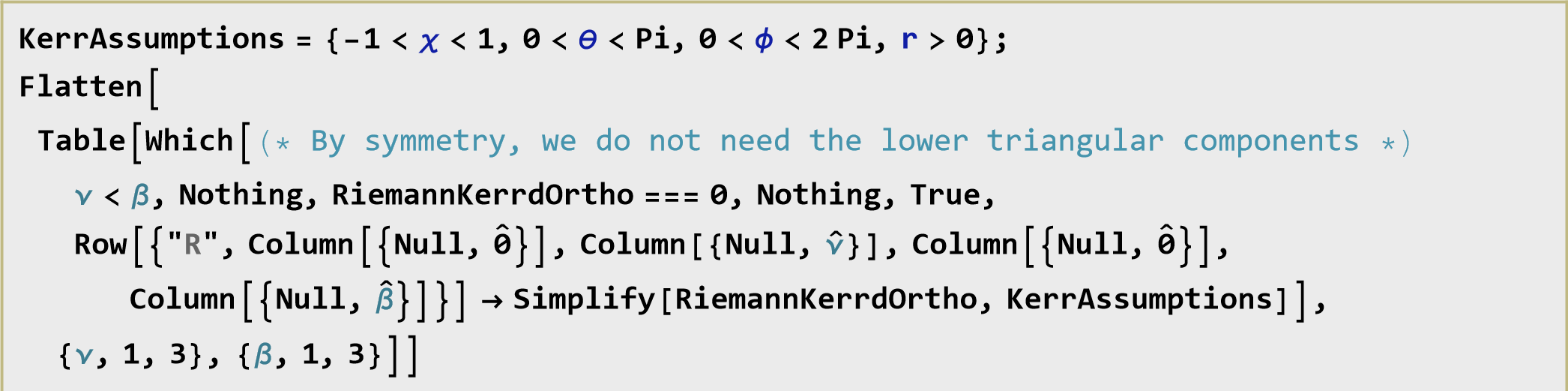}
	\includegraphics[width=\textwidth]{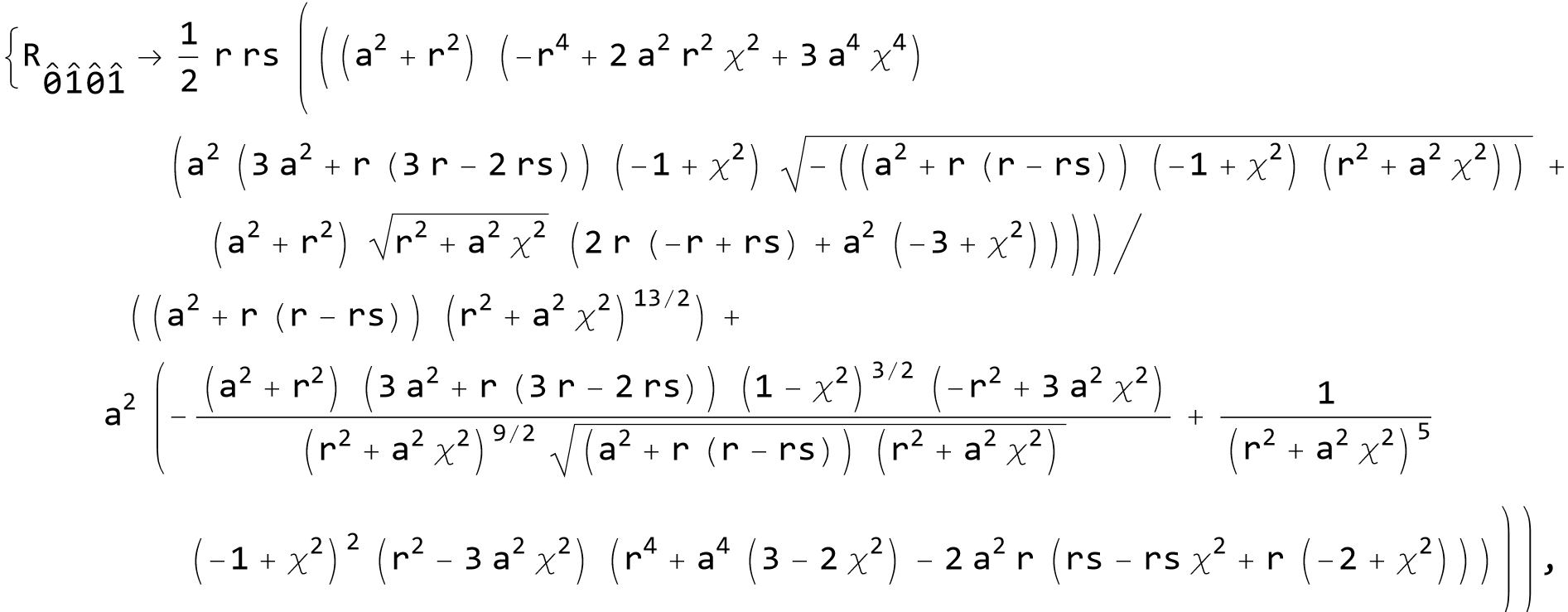}
	\includegraphics[width=\textwidth]{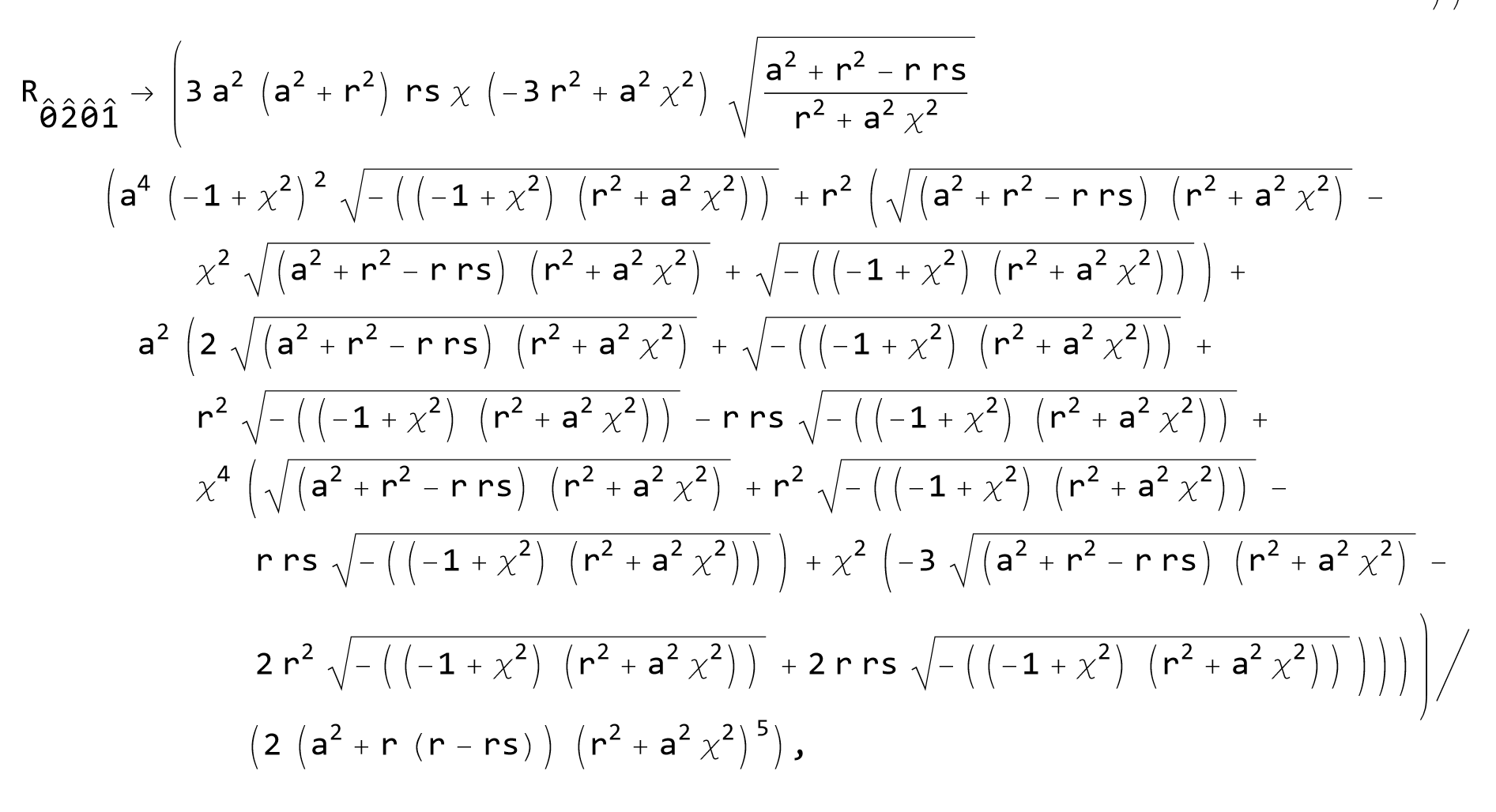}
\end{figure}
\includecode{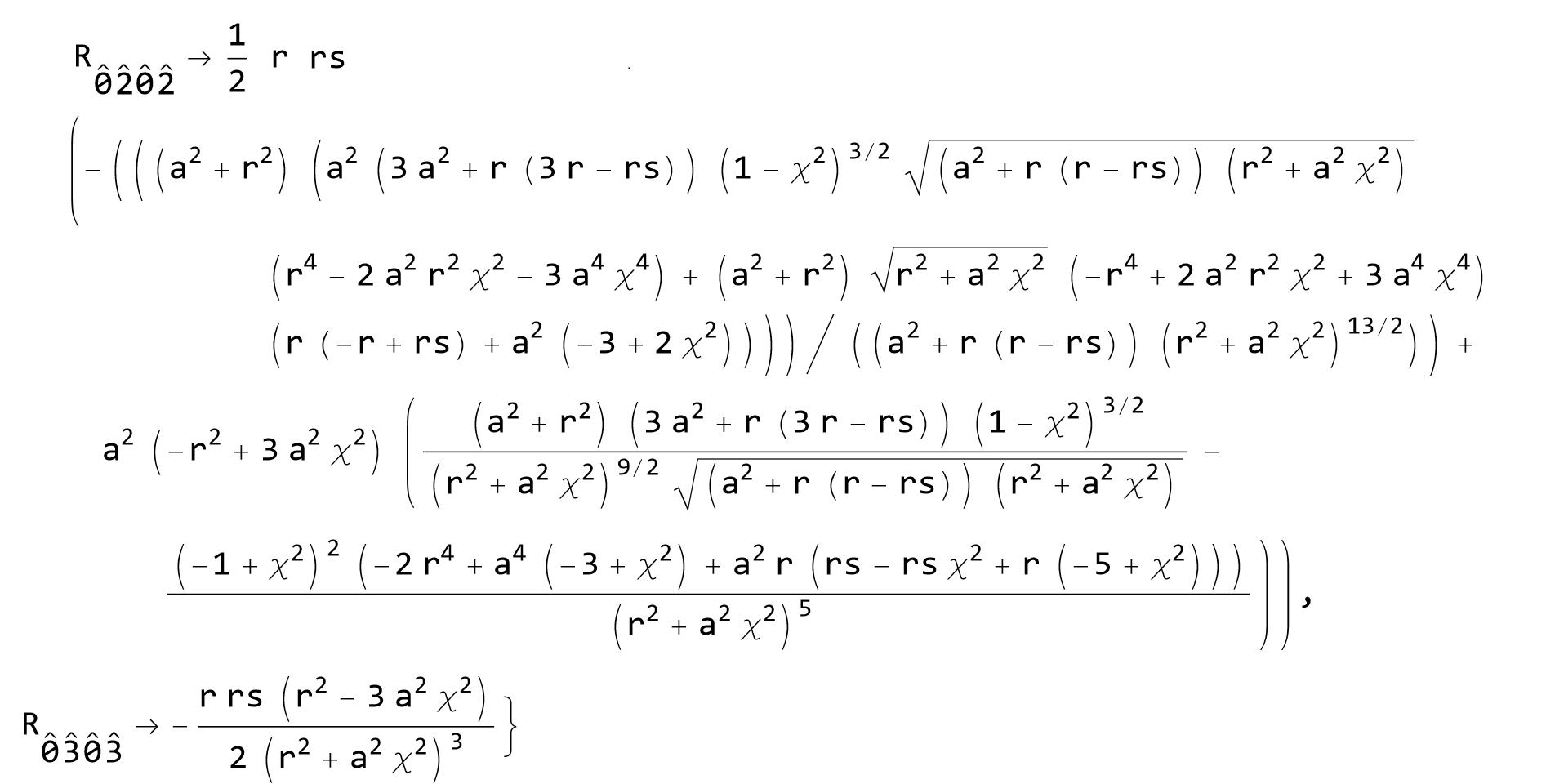}

\section{Summary, Discussion and Future Directions}
\label{Section_Summary}
We have in this article studied a variety of fundamental physics and multi-variable calculus problems -- with a heavy emphasis on symmetry principles -- using \TC, so as to provide a practical introduction to its basic functionality and features. The basic ingredient is the Tensor object, which stores all relevant information of a given tensor -- metric or otherwise -- in the form of Rule's. All other operations such as CovariantD, LieD, CoordinateTransformations, ContractTensors, etc. act upon it directly or indirectly. Additionally, there are functions such as TensorComponents, ToExpressionForm, ZeroTensorQ, etc. that are non-geometric in nature but aid in their manipulation or computation.

We consider this to be version 1 of \TC; and hope to continue its development in the foreseeable future. There are at least two major upgrades we wish to see in a future version.

The first is the ability to accommodate different types of indices and, hence, tensor components. Currently, only spacetime coordinate and orthonormal indices are allowed; but, for instance, spinor indices ought to be allowed for (3+1)D spacetimes. They are also physically important for fermionic systems. Moreover, it is sometimes advantageous to decompose a spacetime index into time versus space; or, a space index into two or more orthogonal sub-spaces. Fundamental physics also demands that we are able to proper account for ``color" indices occurring within non-Abelian gauge theory.

The second is the ability to deal with abstract tensors, as opposed to the concrete ones we have dealt with throughout this article. This includes simplifying or manipulating complicated tensoorial expressions based on the underlying index (anti)symmetries, a highly non-trivial task. Closely related to this is the ability to carry out perturbation theory of tensorial expressions at the abstract level; of importance to post-Newtonian/Minkowskian, cosmological, and black hole physics.

\appendix
\section{Acknowledgements}

WHC, YZC, and VV were funded by NSTC Taiwan grant 113-2112-M-008-013.

\section{History of \TC\ and Statement of Author Contributions}

The zeroth version of \TC\ was written by Yi-Zen Chu during the 2010{'}s. For his Master{'}s thesis project in the early 2020{'}s, Wei-Hao
Chen substantially expanded upon the functionality of \TC; and this current version was the result of his work. Vaidehi Varma worked with
Wei-Hao Chen and Yi-Zen Chu to document the code and write the accompanying manual, which shall be released shortly.



\begin{thebibliography}{99}

\bibitem{MMA}

Wolfram Research, Inc., Mathematica, Version 13.3, Champaign, IL (2025).

\bibitem{TC}
http://www.stargazing.net/yizen/Tensoria.html
\ and \
https://github.com/tensoria/TensoriaCalc

\bibitem{Shtabovenko:2023idz}
V.~Shtabovenko, R.~Mertig and F.~Orellana,
``FeynCalc 10: Do multiloop integrals dream of computer codes?,''
Comput. Phys. Commun. \textbf{306}, 109357 (2025)
doi:10.1016/j.cpc.2024.109357
[arXiv:2312.14089 [hep-ph]].

\bibitem{Shoshany:2021iuc}
B.~Shoshany,
``OGRe: An Object-Oriented General Relativity Package for Mathematica,''
J. Open Source Softw. \textbf{6}, 3416 (2021)
doi:10.21105/joss.03416
[arXiv:2109.04193 [cs.MS]].

\bibitem{xAct}
https://www.xact.es/

\bibitem{Gravitas}
J.~Gorard,
``Computational General Relativity in the Wolfram Language using Gravitas I: Symbolic and Analytic Computation,''
[arXiv:2308.07508 [gr-qc]].

J.~Gorard,
``Computational General Relativity in the Wolfram Language using Gravitas II: ADM Formalism and Numerical Relativity,''
[arXiv:2401.14209 [gr-qc]].

\bibitem{Jackson:1998nia}
J.~D.~Jackson,
``Classical Electrodynamics,''
Wiley, 1998,
ISBN 978-0-471-30932-1

\end{thebibliography}
\end{document}